\documentstyle[12pt]{article}
\newcommand{\ictimes}{\mbox{$\times \hspace{-1em}\supset$}}
\newcounter{lyter}[equation]
\title{General Superfield Quantization Method. \protect \\
II. General Superfield Theory of Fields: \protect \\
Hamiltonian Formalism}
\author{A.A. Reshetnyak\thanks{E-mail: reshetnyak@ssti.ru}}
\date{\it Department of Mathematics, Seversk State Technological Institute,
\protect \\
Seversk, {\rm 636036}, Russia}
\begin{document}
\maketitle
\begin{abstract}
In the framework of started in Ref.[1] construction procedure of the
general superfield quantization method for gauge theories in
Lagrangian formalism
the rules for Hamiltonian formulation of general superfield theory of fields
(GSTF) are introduced and are on the whole considered.

Mathematical means developed in [1] for Lagrangian formulation of GSTF are
extended to use in Hamiltonian one. Hamiltonization
for Lagrangian formulation of GSTF via Legendre
transform of superfunction $S_{L}\bigl({\cal A}(\theta),
{\stackrel{\ \circ}{\cal A}}(\theta),\theta\bigr)$ with
respect to ${\stackrel{\;\circ}{{\cal A}^{\imath}}}(\theta)$ is considered.
As result on the space $T^{\ast}_{odd}{\cal M}_{cl}\times \{\theta\}$
parametrized by classical superfields ${\cal A}^{\imath}(\theta)$,
superantifields
${\cal A}^{\ast}_{\imath}(\theta)$ and odd Grassmann variable $\theta$ the
superfunction
 $S_{H}({\cal A}(\theta),{\cal A}^{\ast}(\theta),\theta)$ is defined. Being
equivalent to different types of Euler-Lagrange equations the distinct
Hamiltonian systems are investigated. Translations along $\theta$ for
superfunctions on $T^{\ast}_{odd}{\cal M}_{cl} \times \{\theta\}$
being associated with these systems are studied.
Various types of antibrackets and differential operators acting on
$C^{k}\bigl(T^{\ast}_{odd}{\cal M}_{cl} \times \{\theta\} \bigr)$ are
considered.

Component (on $\theta$)
formulation for GSTF quantities and operations is produced. Analogy between
ordinary Hamiltonian classical mechanics and GSTF in Hamiltonian formulation
is proposed.
Realization of the GSTF general scheme is demonstrated on  6 models.
\end{abstract}
{\large PACS codes: 03.50.-z, 11.10.Ef, 11.15.-q \protect 
Los Alamos database number: hep-th/0303262}
\renewcommand{\thesection}{\Roman{section}}
\renewcommand{\thesubsection}{\thesection.\arabic{subsection}}
\section{Introduction}
\renewcommand{\theequation}{\arabic{section}.\arabic{equation}}
\renewcommand{\thelyter}{\alph{lyter}}

The problem of construction of the general superfield quantization method
(GSQM) for general gauge theories in Lagrangian formalism, in which it  is
possible to establish a superfield realization for BRST symmetry type
transformations [2] and which includes  the BV quantization method [3] on
component (on $\theta$) level, can not be adequately and noncontradictory
decided without the preliminary creation of GSTF. The latter contains the all
information on ways of the superfield formulation for concrete models of GSTF.
Construction of the Lagrangian formulation for GSTF, as the 1st stage
in creating of GSQM, was realized in paper [1]. In that work both preliminary
historical
excursus on the ways of development of the quantization methods for gauge
theories on the BRST symmetry basis and motivations for origin of the
GSQM construction problem were considered.
In this connection, nevertheless it is appropriate to note the papers [4]
where
one had been made one from the first indications on the Grassmann variable
$\theta$ ($\theta^2=0$) role  for generalization of the differential equations
concept, being interpreted as the odd time, and Refs.[5]
describing the  possibility to use that variable for BV method formulation.
By the following stage
after Lagrangian
formulation for GSTF [1] being by composite part of GSQM in the Lagrangian
formalism (in the usual sense) it appears the creation of the Hamiltonian
formulation for GSTF
that represents the investigation subject of this paper.

In Sec.II  having a preparatory character the possibility
for determination of the Euler-Lagrange equations integrals is considered
together with
introduction of special constraints system for one of integrals in the
Lagrangian formulation for GSTF.
Moreover a sequence of algebraic and
analytic questions of the mathematical means from Ref.[1] are transferred
with some modifications  onto superalgebra of superfunctions
$C^{k}\bigl(T_{odd}(T^{\ast}_{odd}{\cal M}_{cl})\times \{\theta\}\bigr),
k \leq \infty$.
Creating of the Hamiltonian formulation for GSTF on a basis of
realization of the Legendre transform of superfunction
$S_{L}\bigl({\cal A}(\theta),{\stackrel{\ \circ}{{\cal
A}}}(\theta),\theta) \equiv S_{L}(\theta) \in C^{k}\bigl(T_{odd}{\cal
M}_{cl}\times \{\theta\}\bigr)$ with respect to superfields
${\stackrel{\;\circ}{{\cal A}^{ \imath}}}(\theta)$ and construction of
$S_{H}({\cal A}(\theta),{\cal A}^{\ast}(\theta),\theta) \equiv S_{H}(\theta)
\in C^{k}\bigl(T^{\ast}_{odd}{\cal M}_{cl}\times \{\theta\} \bigr)$ together
with obtaining  of dynamical equations in $T^{\ast}_{odd}{ \cal M}_{cl}$
essentially reflecting the properties of $S_{H}(\theta)$ are realized in
Sec.III.

Sec.IV is devoted to detailed study of the properties and relations among
different Lagrangian and Hamiltonian systems of ordinary differential
equations (ODE) with respect to differentiation on $\theta $ describing an
arbitrary model of GSTF with various modifications.

Important role of
the 2nd order odd (with respect to $\varepsilon_{P}$ and $\varepsilon$
gradings) operator $\Delta^{cl}(\theta)$ acting on
$C^{k}\bigl(T^{\ast}_{odd}{\cal M}_{cl}\times \{\theta\}\bigr)$ and
nontrivial master equations in
$T^{\ast}_{odd}{\cal M}_{cl}\times \{\theta\}$ together with
translation transformations on $\theta$ along integral curves of different
Hamiltonian systems (and their some modifications) for arbitrary
superfunctions given on $T^{\ast}_{odd}{\cal M}_{cl}\times \{\theta\}$
are described in Sec.V. In connection with these transformations
the key role of  solvability conditions fulfilment for the Hamiltonian
systems  is demonstrated here as well.
Systematic investigation of the gauge theories in the Hamiltonian
formulation for GSTF including the solvable Hamiltonian systems together with
introduction of detailed concepts on constraints, representations on the
gauge theories, gauge transformations of general and special types are
carried out in Sec.VI.

Sec.VII is devoted to extension of the superalgebra {\boldmath${\cal
A}_{cl}$} [1] of the 1st order differential operators acting on the
superfunctions from $C^{k}\bigl(T_{odd}{\cal M}_{cl} \times \{\theta\}\bigr)$
to superalgebra {\boldmath${\cal B}_{cl}$} of the 1st and 2nd orders
ones defined on $C^{k}\bigl(T^{\ast}_{odd}{\cal M}_{cl}\times
\{\theta\}\bigr)$. The different types of antibrackets being generated by the
2nd order odd operators  are studied here as well. The concept
on transformation of operators from {\boldmath${\cal B}_{cl}$} is introduced.

Component (on $\theta$) formulation for quantities and relations of the
Hamiltonian formulation of GSTF is considered in Sec.VIII.

General statements of Secs.II--VIII are demonstrated in Sec.IX on the example
of 6 models, whose superfield Lagrangian formulation had been proposed
in paper [1]. As it had been noted in that work the almost all from
mentioned models appear
by initial ones to constructing of the interacting  superfield (on
$\theta$) Yang-Mills type models in realizing of the gauge principle
[6].

The final propositions and analogy for Hamiltonian formulation of GSTF with
ordinary classical mechanics in the standard (even) Hamiltonian formalism
conclude the work in Sec.X.

All  assumptions, in the framework of which the paper is made, in fact had
been pointed out in the introduction of work [1]. In the paper unless
otherwise stated it is used the  system of notations suggested in Ref.[1].
\section{Special Properties of the Lagrangian Formulation for
GSTF. Superalgebra $C^{k}\bigl(T_{odd}(T^{\ast}_{odd}{\cal M}_{cl})\times
\{\theta\}\bigr)$}
\subsection{Integrals of Lagrangian System}

E.Noether's theorem [7] analog applied to translation along $\theta$,
being one from the coordinates of the superspace
${\cal M} = \{(z^{a},\theta)\}$ [1], on a constant parameter $\mu$ $\in$
${}^{1}\Lambda_{1}(\theta)$ being considered in question as continuous
transformation of global symmetry for integrand in
\begin{eqnarray}
Z[{\cal A}]= \int d\theta S_L\bigl({\cal A}(\theta), {\stackrel{\ \circ}{
\cal A}}(\theta), \theta\bigr) 
\end{eqnarray}
permits one to find the integral of motion for Euler-Lagrange equations
\begin{eqnarray}
\frac{\delta_l Z[{\cal A}]}{\delta{\cal
A}^{\imath}(\theta)} = \left(\frac{\partial_l \phantom{xxx}}{
\partial{\cal A}^{\imath}(\theta)}
 -(-1)^{\varepsilon_{\imath}}\frac{d}{d\theta}\frac{\partial_l
\phantom{xxx}}{{\partial{\stackrel{\ \circ}{\cal
A}}{}^{\imath}(\theta)}}\right) S_{L}(\theta) \equiv
{\cal L}_{\imath}^{l}(\theta) S_{L}(\theta) = 0,\;\imath=1,\ldots,n\;,
\end{eqnarray}
or one for equivalent system 2n ODE of the 2nd order with respect to
differentiation on $\theta$
\renewcommand{\theequation}{\arabic{section}.\arabic{equation}\alph{lyter}}
\begin{eqnarray}
\setcounter{lyter}{1}
{} & {} &
{\stackrel{\,\circ\circ}{\cal A}}{}^{\jmath}(\theta) \displaystyle\frac{
\partial^{2}_{l} S_{L}(\theta) \phantom{xxxx}}
{{\partial{\stackrel{\ \circ}{\cal
A}}{}^{\imath}(\theta)}{\partial{\stackrel{\ \circ}{\cal A}}{}^{\jmath}
(\theta)}} = 0\;, \\ 
\setcounter{equation}{3}
\setcounter{lyter}{2}
{} & {} & {\Theta}_{\imath}\bigl( {{\cal A}}(\theta), {\stackrel{\
\circ}{{\cal A}}}(\theta), \theta \bigr) \equiv \displaystyle\frac{
\partial_l S_{L}(\theta) }{\partial{\cal A}^{\imath}(\theta)}
 -(-1)^{\varepsilon_{\imath}}\left(\frac{\partial\phantom{x}}{\partial\theta}
  +{\stackrel{\circ}{U}_{+}}(\theta) \right)
\frac{ \partial_l S_{L}(\theta)}{{\partial{\stackrel{\ \circ}{\cal A}}{}^{
\imath}(\theta)}} = 0\;, 
\end{eqnarray}
in the following form
\begin{eqnarray}
\setcounter{lyter}{1}
S_E\bigl({\cal A}(\theta), {\stackrel{\ \circ}{\cal A}}(\theta),\theta\bigr)=
{\stackrel{\ \circ}{\cal A}}{}^{\imath}(\theta)
\frac{ \partial_l S_{L}(\theta)}{{\partial{\stackrel{\ \circ}{\cal A}}{}^{
\imath}(\theta)}} - S_L(\theta)\;, \\
\setcounter{equation}{4}
\setcounter{lyter}{2}
\frac{d S_E(\theta)}{d\theta\phantom{xxxx}}_{\mid {\cal L}_{\imath}^{l}(
\theta)
S_{L}(\theta) = 0} = - \frac{\partial S_L(\theta)}{\partial \theta
\phantom{xxxx}} -
2{\stackrel{\ \circ}{\cal A}}{}^{\imath}(\theta)
\frac{\partial_l S_{L}(\theta)}{\partial{{\cal A}^{\imath}}(\theta)}=0\;.
\end{eqnarray}
The fulfilment of relation (2.4b) is the nontrivial fact. In
particular, in absence of the explicit dependence of $S_{L}(\theta)$ on
$\theta$ (that corresponds to above-mentioned symmetry transformation)
\renewcommand{\theequation}{\arabic{section}.\arabic{equation}}
\begin{eqnarray}
\tilde{P}_1(\theta) S_L\bigl({\cal A}(\theta), {\stackrel{\ \circ}{\cal A}}(
\theta), \theta\bigr)=0 \Longleftrightarrow \frac{\partial S_L(\theta)}{
\partial \theta\phantom{xxxx}}=0\;, 
\end{eqnarray}
Eq.(2.4b) is reduced to
\begin{eqnarray}
{\stackrel{\ \circ}{\cal A}}{}^{\imath}(\theta)
\frac{\partial_l S_{L}(\theta)}{\partial{{\cal A}^{\imath}}(\theta)}_{
\mid {\cal L}_{\imath}^{l}(\theta) S_{L}(\theta) = 0}=0\;. 
\end{eqnarray}
By the sufficient condition for solvability of the last equation is the
fulfilment of the following  system of equations on arbitrary solution
$\bar{\cal A}^{\imath}(\theta)$ for system (2.3)
\renewcommand{\theequation}{\arabic{section}.\arabic{equation}\alph{lyter}}
\begin{eqnarray}
\setcounter{lyter}{1}
{} & {} & {\stackrel{\ \circ}{\cal A}}{}^{A_1}(\theta)_{\mid \bar{\cal A}(
\theta)}=0,\;
a=(A_1,A_2),\;A_1=1,\ldots,n_1,\;n_1=(n_{1+},n_{1-})\;, \\
\setcounter{equation}{7}
\setcounter{lyter}{2}
{} & {} & \frac{\partial_l S_{L}(\theta)}{\partial{{\cal A}^{A_2}}(
\theta)}_{\mid
\bar{\cal A}(\theta)}=0,\;A_2=n_1 +1,\ldots,n\;. 
\end{eqnarray}
Indices $A_{1}$, $A_{2}$, generally speaking, will be able to destroy the
locality and covariance  of the corresponding
superfunctions, being  by elements of representation $T_{J} \equiv T$
space, with respect to index $\imath$  corresponding to the
restriction $T_{\vert \bar{J}}$ of the superfield (on
$\theta$) representation $T$ of supergroup $J = \bar{J} \times P$ onto
subsupergroup $\bar{J}$ [1].
By one from the possible choices for system (2.7a,b) parametrized by
discrete value $n_{1 }$ it appears the setting $n_{1} = 0\, (a = \imath)$
\renewcommand{\theequation}{\arabic{section}.\arabic{equation}}
\begin{eqnarray}
\frac{\partial_l S_{L}(\theta)}{\partial{{\cal A}^{\imath}}(\theta)}_{\mid
\bar{\cal A}(\theta)}=0\;. 
\end{eqnarray}
Thus in order to the superfunction $S_{E}(\theta)$ should appear by
integral for Eqs.(2.3) in fulfilling of Eq.(2.5) it is necessary to reduce
the system (2.3) to one of the 2nd order on $\theta$ 3n ODE
\renewcommand{\theequation}{\arabic{section}.\arabic{equation}\alph{lyter}}
\begin{eqnarray}
\setcounter{lyter}{1}
{} & {} &
{\stackrel{\,\circ\circ}{\cal A}}{}^{\jmath}(\theta) \displaystyle\frac{
\partial^{2}_{l} S_{L}\bigl({\cal A}(\theta),{\stackrel{\ \circ}{\cal A}}(
\theta)\bigr)}{{\partial{\stackrel{\ \circ}{\cal
A}}{}^{\imath}(\theta)}{\partial{\stackrel{\ \circ}{\cal A}}{}^{\jmath}
(\theta)}\phantom{xxx}} = 0\;, \\ 
\setcounter{equation}{9}
\setcounter{lyter}{2}
{} & {} & {\Theta}_{\imath}\bigl( {{\cal A}}(\theta), {\stackrel{\
\circ}{{\cal A}}}(\theta) \bigr) = 0\;, \\
\setcounter{equation}{9}
\setcounter{lyter}{3}
{} & {} &{\stackrel{\ \circ}{\cal A}}{}^{A_1}(\theta) =
{\Theta}_{\imath}(\theta)
\lambda_1{}^{A_1 \imath}\bigl({{\cal A}}(\theta), {\stackrel{\ \circ}{{
\cal A}}}(\theta) \bigr),\;\varepsilon(\lambda_1{}^{A_1 \imath}) =
\varepsilon_{A_1} + \varepsilon_{\imath} + 1\;, \\
\setcounter{equation}{9}
\setcounter{lyter}{4}
{} & {} &\frac{\partial_l S_{L}\bigl({\cal A}(\theta), {\stackrel{\ \circ}{
\cal A}}(\theta) \bigr)}{\partial {\cal A}^{A_2}(\theta)\phantom{xxxxxxx}} =
{\Theta}_{\jmath}(\theta)
\lambda_{2{}{A_2}}{}^{\jmath}\bigl({\cal A}(\theta), {\stackrel{\ \circ}{
\cal A}}(\theta) \bigr),\;\varepsilon(\lambda_{2{}A_2}{}^{\jmath}) =
\varepsilon_{A_2} + \varepsilon_{\jmath}\;,  
\end{eqnarray}
called further the extended Lagrangian system (ELS). Subsystem (2.9c,d)
appears
by additional constraints to differential constraints in Lagrangian formalism
(DCLF) (2.9b) [1]. For $n_{1}= 0$, $\lambda_{\imath}{}^{\jmath}\bigl({\cal A}
(\theta), {\stackrel{\ \circ}{\cal A}}(\theta)\bigr) = 0$ and absence of
explicit dependence upon $\theta $ for $S_{L}(\theta)$ the constraints
(2.8) coincide with DCLF corresponding to the GSTF model  which is the
natural system from corollary 2.2 of theorem 2 of the Ref.[1]. The presence
of arbitrary superfunctions $\lambda_{1}^{A_1{}\imath}(\theta)$, $\lambda_{
2{}A_2}{}^{\jmath}(\theta)$ $\in$ $C^{k}\bigl(T_{odd}{\cal M}_{cl}\bigr)$ in
Eqs.(2.9c,d) reflects the fact of the Eqs.(2.7) fulfilment on the solutions
for LS (2.3).

In general case the equation (2.4b) is always fulfilled without
additional constraints (2.7) for following choice of explicit
dependence of  $S_{L}(\theta)$ upon $\theta$
\renewcommand{\theequation}{\arabic{section}.\arabic{equation}}
\begin{eqnarray}
\tilde{P}_1(\theta) S_L\bigl({\cal A}(\theta), {\stackrel{\ \circ}{\cal A}}(
\theta), \theta\bigr) = - \theta \left(2{\stackrel{\circ}{{\cal A}^{\imath
}}}(\theta)\frac{\partial_l S_{L}
\bigl({\cal A}(\theta), {\stackrel{\ \circ}{\cal A}}(\theta), \theta\bigr)}{
\partial{\cal A}^{\imath}(\theta)\phantom{xxxxxxxxx}}\right),
\end{eqnarray}
where as in (2.5) the quantity $\tilde{P}_{1}(\theta)$ is one from the system
of projectors $\{\tilde{P}_{a}(\theta)$, $U(\theta)\}, a=0,1$ acting on
$C^{k}\bigl(T_{odd}{\cal M}_{cl}\times \{\theta\}\bigr)$ [1]. The consistency
of Eq.(2.4b) and its solution (2.10) follows from the fact that for
the system of the 1st order with respect to $\theta$ n ODE of the form
\begin{eqnarray}
h^{\imath}\bigl({\cal A}(\theta), {\stackrel{\ \circ}{\cal A}}(\theta),\theta
\bigr) = 0,\ \  h^{\imath}(\theta) \in
C^{k}\bigl(T_{odd}{\cal M}_{cl} \times \{\theta\}\bigr) 
\end{eqnarray}
it is necessary to fulfill the solvability conditions [1] which have
the form  of the 1st or 2nd orders on $\theta$  system of 2n ODE
\begin{eqnarray}
h^{\imath}\bigl({\cal A}(\theta), {\stackrel{\ \circ}{\cal A}}(\theta),\theta
\bigr) = 0,\ \
\frac{d}{d\theta}h^{\imath}\bigl({
\cal A}(\theta), {\stackrel{\ \circ}{\cal A}}(\theta),\theta \bigr) = 0\;.
\end{eqnarray}
Equivalently, in place of the 1st subsystem in Eqs.(2.12) the following
system can be written (no longer in the superfield form with respect to
superfield representation $T$ [1])
\begin{eqnarray}
h^{\imath}\bigl(P_0{\cal A}(\theta), {\stackrel{\ \circ}{\cal A}}(\theta),0
\bigr) = 0\;. 
\end{eqnarray}
It means in fact, that in Eq.(2.4b) in front of $\frac{\partial_{l}S_{L}(
\theta)}{\partial {\cal A}^{\imath}(\theta)}$ the projector $P_{0}(\theta)$
(from system $P_{a}(\theta)$, $a = 0,1$ being by projectors on
$C^{k}\bigl(T_{odd}{\cal M}_{cl}\times \{\theta\}\bigr)$ as well [1]) can be
written.

The comparison of the constraints (2.9c,d), relations (2.4a,b) and Eqs.(2.5),
(2.6) permits one to represent the Eqs.(2.6) in the whole $T_{odd}{\cal
M}_{cl}$ in the form
\begin{eqnarray}
{\stackrel{\ \circ}{\cal A}}{}^{\imath}(\theta)
\frac{\partial_l S_{L}\bigl(
{\cal A}(\theta),{\stackrel{\ \circ}{\cal A}}(\theta)\bigr)}{\partial{{
\cal A}^{\imath}}(\theta)\phantom{xxxxxxx}} =
\Theta_{\jmath}\bigl({\cal A}(\theta),{\stackrel{\ \circ}{\cal A}}(\theta)
\bigr)
K^{\jmath}\bigl({\cal A}(\theta),{\stackrel{\ \circ}{\cal A}}(\theta)\bigr)
= \left({\cal L}_{\jmath}^{l}(\theta) S_{L}(\theta)\right)K^{\jmath
}(\theta)\;. 
\end{eqnarray}
Eq.(2.14) imposes definite restrictions on
$\lambda_{1}^{A_1 \imath}(\theta)$, $\lambda_{2{}A_2}{}^{\jmath}(\theta)$
in (2.9c,d)
\begin{eqnarray}
K^{\jmath}\bigl({\cal A}(\theta),{\stackrel{\ \circ}{\cal A}}(\theta)\bigr) =
\lambda_1{}^{A_1 \jmath}\bigl({\cal A}(\theta), {\stackrel{\ \circ}{
\cal A}}(\theta)\bigr)\frac{\partial_l S_{L}(\theta)}{\partial {\cal A}^{
A_1}(\theta)} +
\lambda_{2{}{A_2}}{}^{\jmath}\bigl({\cal A}(\theta), {\stackrel{\ \circ}{
\cal A}}(\theta) \bigr){\stackrel{\ \circ}{\cal A}}{}^{A_2}(\theta)\;.
\end{eqnarray}
Consider  two possible limiting cases of the Eq.(2.14) fulfilment.
Namely, having put the constraints (2.9c) for $n_{1} = n$  and $\lambda_{
1}^{A_1\imath}(\theta) = 0$ satisfying  to equations
\renewcommand{\theequation}{\arabic{section}.\arabic{equation}\alph{lyter}}
\begin{eqnarray}
\setcounter{lyter}{1}
{\stackrel{\ \circ}{\cal A}}{}^{\imath}(\theta) = 0\;, 
\end{eqnarray}
obtain their solution [1] in the whole $T_{odd}{\cal M}_{cl}$
\begin{eqnarray}
\setcounter{equation}{16}
\setcounter{lyter}{2}
{\cal A}^{\imath}(\theta) = P_0(\theta){\cal A}^{\imath}(\theta)
= {\cal A}^{\imath}(0)
\;. 
\end{eqnarray}
It means the superfunction $S_{L}(\theta )$ does not depend
on ${\stackrel{\ \circ}{\cal A}}{}^{\imath}(\theta)$ and has the form
\renewcommand{\theequation}{\arabic{section}.\arabic{equation}}
\begin{eqnarray}
S_{L}\bigl({\cal A}(\theta),{\stackrel{\ \circ}{\cal A}}(\theta)\bigr) \equiv
S_L({\cal A}(\theta)) = S_L({\cal A}(0))\;. 
\end{eqnarray}
Together with (2.17) and identical fulfilment of Eq.(2.4b) the following
equality is valid
\begin{eqnarray}
S_{E}({\cal A}(\theta)) = - S_{L}({\cal A}(\theta))\;. 
\end{eqnarray}

Another variant is derived in fulfilling of the
condition (2.5) and of  constraints (2.9c,d) for $n_{1}= 0$ for any
configuration of ${\cal A}^{\imath}(\theta) \in {\cal M}_{cl}$.
This case is realized by means of identities
\begin{eqnarray}
 S_{L},_{\imath}(\theta) = 0\;. 
\end{eqnarray}
An arbitrary differential consequence of the last relation being obtained in
calculating of the $l$ order derivative ($l = 0,1,\ldots$) of Eq.(2.19)  with
respect to ${\cal A}^{\imath}(\theta)$ means the independence of the
superfunction
$S_{L}(\theta)$ on superfields ${\cal A}^{\imath}(\theta)$ in the whole
$T_{odd}{\cal M}_{cl}$
\begin{eqnarray}
S_{L}\bigl({\cal A}(\theta),{\stackrel{\ \circ}{\cal A}}(\theta)\bigr) =
S_{L}\bigl({\stackrel{\ \circ}{\cal A}}(\theta)\bigr)\;. 
\end{eqnarray}
It follows from (2.20) the superfunctions $\Bigl(\frac{\partial_{l} S_{L}
({\stackrel{\ \circ}{\cal A}}(\theta))}{\partial
{\stackrel{\ \circ}{\cal A}}{}^{\imath}(\theta)\phantom{xxxx}}\Bigr)$
appear not only by
integrals of motion for system (2.3) but are constant for any value of
${\cal A}^{\imath}(\theta)$ by virtue of
${\stackrel{\,\circ\circ}{\cal A}}{}^{\imath}(\theta) \equiv 0$. The
last fact can be interpreted as well with help of Noether's theorem.
Namely, the invariance of $S_{L}(\theta)$ with respect to symmetry
transformation, being by translation of the superfield ${\cal A}^{\imath}(
\theta) \to {{\cal A}'}^{\imath}(\theta) = {\cal A}^{\imath}(\theta ) +
C^{\imath}(\theta)$ on an arbitrary constant superfield $C^{\imath}(\theta)$
($\varepsilon(C^{\imath}) = \varepsilon_{\imath}$),
corresponds to existence of integrals for Eqs.(2.3), i.e. to conservation
of the "momenta" $\Bigl(\frac{\partial_{l} S_{L}
({\stackrel{\ \circ}{\cal A}}(\theta))}{\partial
{\stackrel{\ \circ}{\cal A}}{}^{\imath}(\theta)\phantom{xxxx}}\Bigr)$.
The condition (2.5) have been made use for the last conclusion as well.

\noindent
\underline{\bf Remark:} The definition of the symmetry transformations,
Noether's Theorem formulation and its proof in the context of GSQM are the
separate problems and are eliminated outside the paper's scope.

The independence of the
"momenta" upon $\theta$ implies together with (2.19) the identical fulfilment
of Eqs.(2.3) meaning formally that
almost any superfields ${\cal A}^{\imath}(\theta)$ are the solutions for LS
(2.3) and under condition
\begin{eqnarray}
{\rm rank}{}S_{L}''(\theta) = n,\;
S_{L}''(\theta) \equiv \left\|(S_{L}'')_{\jmath\imath}\bigl({
\cal A}(\theta), {\stackrel{\ \circ}{\cal A}}(\theta), \theta \bigr)\right\|
\equiv
\left\|{\frac{\partial_l \phantom{xxx}}{\partial{
\stackrel{\ \circ}{\cal A}}{}^{\jmath}(\theta)}
\frac{\partial_l S_L(\theta)\phantom{x}}{\partial
{\stackrel{\ \circ}{\cal A}}{}^{\imath}(\theta)}}\right\|,
\end{eqnarray}
exactly the any ${\cal A}^{\imath}(\theta)$ are the same.

From these two cases of  Eq.(2.6) fulfilment  for any superfield
${\cal A}^{\imath}(\theta)$ it follows that by the sufficient condition for
its resolution is the existence of the solution for system (2.7) for
arbitrary configuration ${\cal A}^{\imath}(\theta)$ and $n_{1}$. The latter
implies the identical vanishing of $\lambda_{1}^{A_1 \imath}(
\theta)$, $\lambda_{2{}A_2}{}^{\imath}(\theta)$ in (2.9c,d) together with
constraints (2.7) themselves, i.e. their fulfilment
for any ${\cal A}^{\imath}(\theta) \in {\cal M}_{cl}$.

Thus, the validity of Eq.(2.6) for any ${\cal A}^{\imath}(\theta)$ imposes
the strong restrictions on the structure of the superfunction $S_{L}(\theta)$.
Indicate in this case the necessary condition on the structure of
$S_{E}\bigl({\cal A}(\theta), {\stackrel{\ \circ}{\cal A}}(
\theta) \bigr)$ (2.4a) and therefore for $S_{L}(\theta)$
\begin{eqnarray}
{\rm rank}\left\|\displaystyle\frac{\partial_r \phantom{xxx}}{\partial
\tilde{\Gamma}^p(\theta)}
\frac{\partial_l S_E(\theta)}{\partial \tilde{\Gamma}^q(\theta)}
\right\|_{
\mid\frac{\partial_l S_E(\theta)}{\partial\tilde{\Gamma}^p(\theta)\phantom{x}
}=0} \leq n,\;p,q=1,\ldots,2n,\;
\tilde{\Gamma}^p(\theta) = \left({\cal A}^{\imath}(\theta),
\displaystyle\frac{\partial_l S_L(\theta)}{\partial{\stackrel{\ \circ}{
\cal A}}{}^{\imath}(\theta)}\right).
\end{eqnarray}
Except for Noether's integral for LS the another (non Noetherian) one exists
of the form
\begin{eqnarray}
V_E\bigl({\cal A}(\theta), {\stackrel{\ \circ}{\cal A}}(\theta),
\theta\bigr)
= {\stackrel{\circ}{{\cal A}^{\imath}}}(\theta)
\frac{\partial_l S_L(\theta)}{\partial{\stackrel{\ \circ}{
\cal A}}{}^{\imath}(\theta)} + S_L(\theta),\;
{\frac{d V_E(\theta)}{d\theta\phantom{xxxx}}}_{
\mid {\cal L}_{\imath}^{l}(
\theta) S_{L}(\theta) =0}= \frac{\partial S_L(\theta)}{\partial \theta
\phantom{xxxx}}=0\;, 
\end{eqnarray}
that is equivalent to realization of the only condition (2.5) without
relationships (2.7), (2.22).

Consider the problem of reduction of the 2nd order
on $\theta$ system of 2n ODE  (2.3) to the normal form (NF), i.e. to the
system of solvable ODE with respect to higher derivatives: superfields
${\stackrel{\,\circ\circ}{\cal A}}{}^{\imath}(\theta)$ (formally).
It is possible in $T_{odd}{\cal M}_{cl} \times \{\theta\}$ if and only if
the condition (2.21)  on the rank of supermatrix $K(\theta
)$ is fulfilled almost everywhere in $T_{odd}{\cal M}_{cl}$.
The comparison of the conditions (2.21) with (2.22) leads to very strong
restriction on the structure of $S_{L}(\theta)$, namely to its
independence upon superfields ${\cal A}^{\imath}(\theta)$.
Therefore for realizing of the Eq.(2.6) on the given
stage of study of the classical theory properties  confine
ourselves by conditions (2.5)--(2.7) instead of (2.22).

Having suggested the realization of (2.21) the formulated problem
can be solved by means of Legendre transform of $S_{L}(\theta)$ with respect
to superfields ${\stackrel{\ \circ}{\cal A}}{}^{
\imath}(\theta)$. To this
end, introduce the additional to ${\cal A}^{\imath}(\theta)$,
${\stackrel{\ \circ}{\cal A}}{}^{\imath}(\theta)$
superfields of $\bar{J}$ (Lorentz) type [1] ${\cal A}^{\ast}_{\imath
}(\theta)$
as the superfunctions over $\Lambda_{D|Nc + 1}(z^{a},\theta;{\bf K})$.
The last set is the Berezin superalgebra over number field ${\bf K}$
(${\bf R}$ or ${\bf C}$) being by Grassmann algebra with ($D+Nc$) even with
respect to $\varepsilon_P$ grading (from latter only $D$ are even with
respect to $\varepsilon_{\bar{J}}$ grading
and other $Nc$ are odd with respect to $\varepsilon_{\bar{J}}$ one)
generating elements $z^{a}$ and with one odd with respect to $
\varepsilon_{P}$ grading one $\theta$ [1]. ${\cal A}^{\ast}_{\imath}(
\theta)$ are transformed with respect to the supergroup $J$
superfield representation $T^{\ast}$
which is connected by a special form with representation\footnote{as one
had already been noted in Ref.[1] the separate work would be devoted
to detailed study of (ir)reducible supergroup $J$ superfield representations}
$T$ (conjugate to $T$ with respect to a some bilinear form).
Grading properties of those superfields, called further as
superantifields, and their derivatives on $\theta$:
${\stackrel{\circ}{{\cal A}^{\ast}_{\imath}}}(\theta)$ $\equiv$ $\frac{
d {\cal A}^{\ast}_{\imath}(\theta)}{d \theta\phantom{xxx}}$
are written as follows
\begin{eqnarray}
(\varepsilon_P, \varepsilon_{\bar{J}}, \varepsilon) {\cal A}^{\ast}_{\imath}(
\theta) = (\varepsilon_P({\stackrel{\circ}{{\cal A}^{\ast}_{\imath}}}(\theta))
+ 1, \varepsilon_{\bar{J}}({\stackrel{\circ}{{\cal A}^{\ast}_{\imath}}}(
\theta)), \varepsilon({\stackrel{\circ}{{\cal A}^{\ast}_{\imath}}}(\theta))
+ 1) = (1, \varepsilon_{\imath}, \varepsilon_{\imath} + 1)\;. 
\end{eqnarray}
\sloppy
\begin{sloppypar}
In ac\-cor\-dan\-ce with ter\-mi\-no\-lo\-gy of the work [1] the
su\-per\-an\-ti\-fi\-elds
${\cal A}^{\ast}_{\imath}(\theta)$, ${\stackrel{\circ}{{\cal A}^{\ast}_{
\imath}}}(\theta)$ be\-long to
 $\tilde{\Lambda}_{D|Nc + 1}(z^{a}, \theta;{\bf K})$,
${\bf K} = ({\bf R}$ or ${\bf C})$
be\-ing by su\-per\-al\-geb\-ra of the su\-per\-fun\-cti\-ons, to be
trans\-for\-med with res\-pect to
su\-per\-gro\-up $J$ su\-per\-fi\-eld re\-pre\-sen\-ta\-ti\-ons  and
de\-fi\-ned on $\Lambda_{D|Nc +1}(z^{a},\theta;{\bf K})$.
\end{sloppypar}
\subsection{Elements of Algebra and Analysis on $T_{odd}(T^{\ast}_{odd}
{\cal M}_{cl}) \times  \{\theta \}$}

The transformation laws for superantifields
${\cal A}^{\ast}_{\imath}(\theta)$,
${\stackrel{\circ}{{\cal A}^{\ast}_{\imath}}}(\theta)$
with respect to action of the supergroup $J$ = $(\overline{M} \ictimes
\bar{J}_{\tilde{A}}) \times P$ [1]
(ir)reducible superfield representation $T^{\ast}$  operators
have the form
\renewcommand{\theequation}{\arabic{section}.\arabic{equation}\alph{lyter}}
\begin{eqnarray}
\setcounter{lyter}{1}
{} & {} &{\cal A}^{\ast}_{\imath}(\theta) \mapsto {{\cal A}^{\ast}
_{\imath}}'(\theta')
= \left(T^{\ast}(e,\tilde{g}){\cal A}^{\ast}\right)_{\imath}(\theta),\;
e \in \overline{M},\;\tilde{g} \in \bar{J}_{\tilde{A}}\;,\\ 
\setcounter{equation}{25}
\setcounter{lyter}{2}
{} & {} &{\stackrel{\ \circ}{\cal A}}{}^{\ast}_{\imath}(\theta) \mapsto
{\stackrel{\ \circ}{\cal A}}{}^{\ast}_{\imath}{}'(\theta')=
\Bigl(T^{\ast}(e,\tilde{g}){\stackrel{\ \circ}{\cal A}}{}^{\ast}\Bigr){}_{
\imath}(\theta)
\;, \\ 
\setcounter{lyter}{1}
{} & {} &{\cal A}^{\ast}_{\imath}(\theta) \mapsto {{\cal A}^{\ast}_{
\imath}}'(\theta)
= \left(T^{\ast}(e,\tilde{g}){\cal A}^{\ast}\right)_{\imath}(T(h^{-1}(\mu))
\theta) = \left(T^{\ast}(e,\tilde{g}){\cal A}^{\ast}\right)_{\imath}(\theta
- \mu) \;,\\ 
\setcounter{equation}{26}
\setcounter{lyter}{2}
{} & {} &{\stackrel{\ \circ}{\cal A}}{}^{\ast}_{\imath}(\theta) \mapsto
{\stackrel{\ \circ}{\cal A}}{}^{\ast}_{\imath}{}'(\theta)=
\Bigl(T^{\ast}(e,\tilde{g}){\stackrel{\ \circ}{\cal A}}{}^{\ast}\Bigr){}_{
\imath}(\theta),\;h(\mu) \in P,\;\mu \in {}^1\Lambda_1(\theta) 
\end{eqnarray}
and realize the finite-dimensional (2.25) and infinite-dimensional (2.26)
superfield representations  respectively.

Under permutation of two and more elements from $\Lambda_{D|Nc + 1}
(z^{a},\theta; \mbox{\boldmath$K$})$ the sign ("$+$" or "$-$") arises being
dictated by their $\varepsilon$(!) grading.

\begin{sloppypar}
Starting from supermanifolds ${\cal M}_{cl}$ and $T_{odd}{\cal M}_{cl}$
let us formally construct the following supermanifolds $T^{\ast}_{odd}{\cal
M}_{cl}$, $T^{\ast}_{odd}{\cal M}_{cl} \times \{\theta\}$, $T_{odd}\bigl(
T^{\ast}_{odd}{\cal M}_{cl}\bigr)$ and $T_{odd}\bigl(T^{\ast}_{odd}{\cal M
}_{cl}\bigr) \times \{\theta\}$ parametrized by local coordinates
$\Gamma^{p}(\theta)$ $\equiv$
$({\cal A}^{\imath}(\theta )$, ${\cal A}_{\imath}^{\ast}(\theta))$,
$(\Gamma^{p}(
\theta), \theta)$, $\bigl(\Gamma^{p}(\theta)$, ${\stackrel{\circ\ }{\Gamma^{
p}}}(\theta)$ $\equiv$ $\bigl({\stackrel{\;\circ}{{\cal A}^{\imath}}}(
\theta)$, ${\stackrel{\circ}{{\cal A}^{\ast}_{\imath}}}(\theta)\bigr)\bigr)$
and
$\bigl(\Gamma^{p}(\theta)$, ${\stackrel{\circ\ }{\Gamma^{p}}}(\theta)$,
$\theta
\bigr)$, $p=1,\ldots,2n$ respectively with ${\cal A}^{\imath}(\theta)$ $\in$
${\cal M}_{cl}$.
\end{sloppypar}

\begin{sloppypar}
In complete analogy with Ref.[1] define the superalgebras of superfunctions
${\bf K}[[T_{odd}$ $(T^{\ast}_{odd}{\cal M}_{cl})$ $\times$ $\{\theta \}]]$ $
\supset$
${\bf K}[[T^{\ast}_{odd}{\cal M}_{cl}$ $\times$ $\{\theta\}]]$ determined on
$T_{odd}(T^{\ast}_{odd}{\cal M}_{cl})$ $\times$ $\{\theta\}$,
$T^{\ast}_{odd}{\cal M}_{cl}$ $\times$ $\{\theta\}$
respectively and being by formal power series with respect to
generating elements $\Gamma^{p}(\theta)$, ${\stackrel{\circ\ }{\Gamma^{p}}}(
\theta)$, $\theta$ and
$\Gamma^{p}(\theta )$, $\theta$ correspondingly (${\bf K}$ =
${\bf R}$ or ${\bf C}$). In
particular, these sets contain superalgebras of superfunctions
${\bf K}[T_{odd}$ $(T^{\ast}_{odd}{\cal M}_{cl})$ $\times$ $\{\theta\}]$
$\supset$ ${\bf K}[T^{\ast}_{odd}{\cal M}_{cl}$ $\times$ $\{\theta\}]$
respectively appearing by finite polynomials with respect to $\Gamma^{p}(
\theta)$, ${\stackrel{\circ\ }{\Gamma^{p}}}(\theta)$. For any superfunction
${\cal F}(\theta)$ $\equiv$
${\cal F}\bigl(\Gamma(\theta), {\stackrel{\circ}{\Gamma}}(\theta),
\theta\bigr)$
$\in$ ${\bf K}[[T_{odd}$ $(T^{\ast}_{odd}{\cal M}_{cl})$ $\times$ $\{\theta
\}]]$ the transformation laws are valid, in acting of representation
$\tilde{T}$ constructed from $T$, $T^{\ast}$  on the
indicated superalgebra, following from (2.25), (2.26) respectively
\end{sloppypar}
\renewcommand{\theequation}{\arabic{section}.\arabic{equation}}
\begin{eqnarray}
{\cal F}\left(\Gamma'(\theta'), \frac{d \Gamma'(\theta')}{d\theta'\phantom{x
x}}, \theta'\right)
= {\cal F}\left(\Bigl(\bigl(T\oplus T^{\ast}\bigr)(e,\tilde{g})\Gamma
\Bigr)(\theta), \Bigl(\bigl(T\oplus T^{\ast}\bigr)(e,\tilde{g})
{\stackrel{\circ}{\Gamma}}\Bigr)(\theta),\theta\right), \\ 
{\cal F}\left(\Gamma'(\theta), {\stackrel{\circ}{\Gamma}}{}'(\theta),
\theta\right)
= {\cal F}\left(\Bigl(\bigl(T \oplus T^{\ast}\bigr)(e,\tilde{g})\Gamma
\Bigr)(\theta - \mu), \Bigl(\bigl(T\oplus T^{\ast}\bigr)(e,\tilde{g}){
\stackrel{\circ}{\Gamma}}\Bigr)(\theta),\theta - \mu\right). 
\end{eqnarray}
By definition ${\cal F}(\theta)$ is expanded in the formal power series (in
finite sum for polynomials  corresponding to local superfunctions) in
powers of $\Gamma^{p}(\theta)$, ${\stackrel{\circ\ }{\Gamma^{p}}}(\theta)$
\begin{eqnarray}
{} & {\cal F}\bigl({\cal A}(\theta), {\stackrel{\ \circ}{\cal A}}(\theta),
{\cal A}^{\ast}(\theta),{\stackrel{\circ}{{\cal A}^{\ast}}}(\theta),
\theta\bigr)
= \displaystyle\sum\limits_{l=0}\frac{1}{l!}
\hspace{-0.9em}\vec{\hspace{1em}\stackrel{\hspace{-0.8em}\circ}{{\cal A}^{
\ast}_{(\jmath)_l}}}(\theta)
{\cal F}^{(\jmath)_l}\bigl({\cal A}(\theta), {\stackrel{\ \circ}{\cal A}}(
\theta),{\cal A}^{\ast}(\theta),\theta\bigr) =  & {}  \nonumber \\
{} &  \displaystyle\sum\limits_{m,l=0}\frac{1}{m!l!}\hspace{-0.9em}
\vec{\hspace{1em}\stackrel{\hspace{-0.8em}\circ}{{\cal A}^{
\ast}_{(\jmath)_l}}}(\theta)\vec{{\cal A}}^{\ast}_{(\imath)_m}(\theta)
{\cal F}^{(\jmath)_l (\imath)_m}
\bigl({\cal A}(\theta), {\stackrel{\ \circ}{\cal A}}(\theta),\theta\bigr)\;,
& {} 
\end{eqnarray}
where the notations are introduced
\begin{eqnarray}
{} &
\vec{\hspace{1em}{\stackrel{\hspace{-0.8em}\circ}{{\cal A}^{
\ast}_{(\jmath)_l}}}}(\theta)\equiv
\prod\limits_{p=1}^{l}
{\stackrel{\circ\ }{{\cal A}^{\ast}_{\jmath_p}}}(\theta),\;
{\cal F}^{(\jmath)_l}\bigl( \Gamma(\theta), {\stackrel{\ \circ}{\cal A}}
(\theta),\theta \bigr) \equiv {\cal F}^{ \jmath_1 \ldots \jmath_l}\bigl(
\Gamma(\theta), {\stackrel{\ \circ}{\cal A}}(\theta), \theta \bigr)\;,
\nonumber \\
{} & \vec{{\cal A}}^{\ast}_{(\imath)_m}(\theta) \equiv \prod\limits_{p=1}^{m}
{\cal A}^{\ast}_{\imath_p}(\theta),\;
{\cal F}^{(\jmath)_{l} (\imath)_{m}}\bigl({\cal A}(\theta), {\stackrel{\
\circ}{\cal A}}(\theta),\theta\bigr)
 \equiv {\cal F}^{\;\jmath_1 \ldots
\jmath_l \imath_1 \ldots \imath_m }\bigl({\cal A}(\theta), {\stackrel{\
\circ}{\cal A}}(\theta),\theta\bigr)\;. & {} 
\end{eqnarray}
Coefficients of decomposition  in (2.29) appear themselves by superfunctions.
Moreover they are expanded in power series with respect to
${\cal A}^{\imath}(\theta)$,
${\stackrel{\circ}{{\cal A}^{\imath}}}(\theta)$  as well and
${\cal F}^{(\jmath)_l (\imath)_m}\bigl({\cal A}(\theta), {\stackrel{\
\circ}{\cal A}}(\theta),\theta\bigr)$
$\in$ ${\bf K}[[T_{odd}{\cal M}_{cl}$ $\times$ $\{\theta\}]]$ [1].
In addition the coefficients
possess the following properties of  generalized symmetry, for instance, for
${\cal F}^{(\jmath)_l (\imath)_m}\bigl({\cal A}(\theta)$, ${\stackrel{\
\circ}{\cal A}}(\theta),\theta\bigr)$
\begin{eqnarray}
{} & {\cal F}^{(\jmath)_l (\imath)_m}\bigl({\cal A}(\theta), {\stackrel{\
\circ}{\cal A}}(\theta),\theta\bigr)
= (-1)^{(\varepsilon_{\imath_s} + 1)(
\varepsilon_{\imath_{s-1}} + 1)}
{\cal F}^{(\jmath)_l \imath_1 \ldots \imath_s \imath_{s-1} \ldots \imath_m}
\bigl({\cal A}(\theta), {\stackrel{\ \circ}{\cal A}}(\theta),\theta\bigr) = &
{}  \nonumber \\
{} & (-1)^{\varepsilon_{\jmath_r} \varepsilon_{\jmath_{r-1}}}
{\cal F}^{\jmath_1 \ldots \jmath_r \jmath_{r-1} \ldots \jmath_l (\imath)_m }
\bigl({\cal A}(\theta), {\stackrel{\ \circ}{\cal A}}(\theta),\theta\bigr),\;
s = \overline{2,m},\;r = \overline{2,l}\;.  & {} 
\end{eqnarray}
Introducing the operations of differentiation with respect to superfields
$\Gamma^{p}(\theta )$, ${\stackrel{\circ\ }{{\Gamma}^{p}}}(\theta)$ on
${\bf K}[[T_{odd}(T^{\ast}_{odd}{\cal M}_{cl}) \times \{\theta \}]]$
one can convert the last  superalgebra in one of the
$k$-times differentiable superfunctions
$C^{k}\bigl(T_{odd}(T^{\ast}_{odd}
{\cal M}_{cl}) \times \{\theta\}\bigr)$ $\equiv$ $D_{cl}^{k}, k\leq \infty$.
In equipping of
${\bf K}[[T_{odd}(T^{\ast}_{odd}{\cal M}_{cl}) \times \{\theta \}]]$
by a some norm one can convert this set into functional space
regarding the series in (2.29) by convergent with respect to above norm and
operations of differentiation on
$\Gamma^{p}(\theta )$, ${\stackrel{\circ\ }{{\Gamma}^{p}}}(\theta)$
by commutative with sign of sum.

Regarding that $D_{cl}^{k}$ and $C^{k\ast}\times \{\theta\}\equiv
C^{k}\bigl(T^{\ast}_{odd}{\cal M}_{cl} \times \{\theta\}\bigr)$
$\subset$
$D_{cl}^{k}$ are supplied by above-mentioned structure of norm
and by corresponding convergence of series (2.29) for arbitrary
${\cal F}(\theta)$ $\in$ $D_{cl}^{k}$ we assume to be valid the following
expansion in the functional Taylor's series in powers of
$\delta\Gamma^{p}(\theta)$ = $\bigl(\Gamma^{p}(\theta) - \Gamma_{
0}^{p}(\theta)\bigr)$, $\delta{\stackrel{\circ\ }{{\Gamma}^{p}}}(\theta)$ =
$\bigl({\stackrel{\circ\ }{{\Gamma}^{p}}}(\theta) - {\stackrel{\circ\ \,}{{
\Gamma}_{0}^{p}}}(\theta)\bigr)$ in a some neighbourhood (possibly in
the whole $T_{odd}(T^{\ast}_{odd}{\cal M}_{cl}) \times \{\theta\}$) of
$\Gamma_{0}^{p}(\theta)$, ${\stackrel{\circ\ \,}{{\Gamma}_{0}^{p}}}(\theta)$
(write this fact only for
${\cal A}^{\ast}_{0{}\imath}(
\theta)$, ${\stackrel{\ \circ}{\cal A}}{}^{\ast}_{0{}\imath}(\theta)$)
\begin{eqnarray}
{} & \hspace{-2em}{\cal F}\left( \Gamma(\theta),
{\stackrel{\circ}{\Gamma}}(\theta), \theta \right) =
 \displaystyle\sum\limits_{m,l=0}\displaystyle\frac{1}{m!l!}
{\delta\hspace{-0.9em}\vec{\hspace{1em}\stackrel{\hspace{-0.8em}\circ}{{\cal
A}^{\ast}_{(\jmath)_l}}}}(\theta) \delta\vec{{\cal A}}^{\ast}_{(\imath
)_m}(\theta) {\cal F}^{(\jmath)_l (\imath)_m} \left(
{\cal A}(\theta), {\stackrel{\ \circ}{{\cal A}}}(\theta), {\cal A}^{\ast}_{
0}(\theta), {
\stackrel{\circ}{{\cal A}^{\ast}_{0}}}(\theta), \theta \right) \equiv
& {}  \nonumber \\
{} & \hspace{-2em}\displaystyle\sum\limits_{m,l=0}
\displaystyle\frac{1}{m!l!}
{\delta\hspace{-0.9em}
{\vec{\hspace{1em}\stackrel{\hspace{-0.8em}\circ}{{\cal A}^{\ast}_{(\jmath)_l}}}}}(\theta)
\delta\vec{\cal A}^{\ast}_{(\imath)_m}( \theta)
\left( \prod\limits_{p=0}^{m-1}
\displaystyle\frac{{\partial}_l \phantom{xxxxxx}}{\partial {\cal
A}^{\ast}_{\imath_{m-p}}(\theta)} \prod\limits_{q=0}^{l-1}
\displaystyle\frac{{\partial}_l
\phantom{xxxxx}}{{\partial{\stackrel{\ \circ}{\cal
A}}{}^{\ast}_{\jmath_{l-q}}(\theta)}}{\cal F}\left( \Gamma(\theta),
{\stackrel{\circ}{\Gamma}}(\theta), \theta \right) \right)_{ \mid {\cal
A}^{\ast}_{0}(\theta), {\stackrel{\circ}{{\cal A}^{\ast}_{0}}}(\theta)}.
& {}
\end{eqnarray}
For coefficient's superfunctions
${\cal F}^{(\jmath)_l (\imath)_m}\bigl({\cal A}(\theta), {\stackrel{\
\circ}{{\cal A}}}(
\theta), {\cal A}^{\ast}_{0}(\theta), {\stackrel{\circ}{{\cal A}^{\ast}_{
0}}}(\theta), \theta\bigr)$
the properties (2.31) and expansion in Taylor's series in powers of
$\delta{\cal A}^{\imath}(\theta)$, $\delta{\stackrel{\;
\circ}{{\cal A}^{\imath}}}(\theta)$ hold [1]. Notations of the form (2.30)
have been made use in
(2.32) and the left partial superfield derivatives with respect to
${\cal A}^{\ast}_{\imath}(\theta)$, ${\stackrel{\circ}{{
\cal A}^{\ast}_{\imath}}}(\theta)$ for fixed $\theta $ were introduced
nontrivially acting on the ${\cal F}(\theta) \in D^k_{cl}$ only at
coinciding $\theta $. Their nonzero action
on  ${\cal A}^{\ast }_{\jmath}(\theta)$ and ${\stackrel{\circ}{{\cal A}^{
\ast}_{\jmath}}}(\theta)$ reads as follows
\begin{eqnarray}
\frac{{\partial}_l {\cal
A}^{\ast}_{\jmath}(\theta)}{\partial {\cal A}^{\ast}_{\imath}(\theta)} =
{\delta}_{\jmath}{}^{ \imath},\  \frac{{\partial}_l
{\stackrel{\ \circ}{\cal A}}{}^{\ast}_{\jmath}(\theta)}{{\partial{\stackrel{\
\circ}{\cal A}}{}^{\ast}_{\imath}(\theta)}} =
\delta_{\jmath}{}^{\imath}\;.
\end{eqnarray}
At last, according to Ref.[1] one can use the combination of decompositions
(2.29) and (2.32) regarding, for instance, ${\cal F}(\theta)$ is expanded with
respect to ${\cal A}^{\ast }_{\imath}(\theta)$, ${\stackrel{\circ\ }{
\Gamma^{p}}}(\theta)$ as a polynomial and in Taylor's series in powers of
$\delta{\cal A}^{\imath}(\theta)$ in a
neighbourhood of ${\cal A}^{\imath}_{0}(\theta)$. It should be noted
that for local superfunctions the  series (2.32) (or mentioned
combination of decompositions (2.29), (2.32)) pass into finite sum.
The action of projector's systems $P_{a}(\theta)$, $a = 0,1$ and
$\{\tilde{P}_{b}(\theta), U(\theta), V(\theta)\}$, $b = 0,1$ are
naturally extended onto $D^k_{cl}$ decomposing the last set in direct sum
\begin{eqnarray}
{} & D^k_{cl} = C^{k}\bigl(P_0(T_{odd}(T^{\ast}_{odd}{\cal M}_{cl}))\bigr)
\oplus C^{k}\bigl(P_1(T_{odd}(T^{\ast}_{odd}{\cal M}_{cl}))\bigr) \oplus & {}
\nonumber \\
{} & \oplus C^{k}\bigl(P_0(T_{odd}(T^{\ast}_{odd}{\cal M}_{cl}))\times \{
\theta\}\bigr) \equiv {}^{0,0}D^k_{cl} \oplus {}^{1,0}D^k_{cl} \oplus {}^{0,1
}D^k_{cl}\;. 
\end{eqnarray}
By invariant subsuperspaces in $D^k_{cl}$ with respect to action of
projectors $\{\tilde{P}_{a}(\theta), U(\theta), V(\theta)\}$ are the following
ones
\begin{eqnarray}
\tilde{P}_0(\theta)D^k_{cl} = {}^{0,0}D^k_{cl},\;\tilde{P}_1(\theta)D^k_{cl}
= {}^{0,1}D^k_{cl},\;(U+V)(\theta)D^k_{cl} = {}^{1,0}D^k_{cl}\;,
\end{eqnarray}
from which the only ${}^{0,0}D^k_{cl}$ is the nontrivial subsuperalgebra,
whereas ${}^{1,0}D^k_{cl}$ and ${}^{0,1}D^k_{cl}$ are  nilpotent
ideals in $D^k_{cl}$. To decompose ${}^{1,0}D^k_{cl}$ into a direct
sum of subsuperalgebras it is necessary to make use in explicit form the
superfield-superantifield structure of the coordinates
$\Gamma^{p}(\theta) = ({\cal A}^{\imath}(\theta ),
{\cal A}_{i}^{\ast}(\theta))$ while in decomposition (2.35) that
polarization was not taken into account. Preliminarily indicate  the
corresponding definite gradings for superfields $\Gamma^{p}(\theta)$,
${\stackrel{\circ\ }{\Gamma^{p}}}(\theta)$ in ignoring of the
superfield-superantifield structure
\begin{eqnarray}
(\varepsilon_P, \varepsilon_{\bar{J}}, \varepsilon) \Gamma^{p}(\theta) =
(\varepsilon_P({\stackrel{\circ\ }{\Gamma^{p}}}(\theta)) + 1,
\varepsilon_{\bar{J}}({\stackrel{\circ\ }{\Gamma^{p}}}(\theta)),
\varepsilon({\stackrel{\circ\ }{\Gamma^{p}}}(\theta)) + 1)
 =
\bigl(\varepsilon_P(\Gamma^p), \varepsilon_{\bar{J}}(\Gamma^p),
\varepsilon_p) \;.
\end{eqnarray}
Projector ${\cal W}(\theta)$ with respect to coordinates $\Gamma^{p}(\theta)$
\begin{eqnarray}
{\cal W}(\theta)= (U + V)(\theta) 
\end{eqnarray}
is given to be covariant, and  ${}^{1,0}D^k_{cl}$ with respect to action
of projectors $\{\tilde{P}_{a}(\theta), {\cal W}(\theta)\}$ are not reduced
further.
Choosing the coordinates on $T^{\ast}_{odd}{\cal M}_{cl}$ in the form
$({\cal A}^{\imath}(\theta), {\cal A}^{\ast}_{\imath}(\theta))$ we get
\begin{eqnarray}
{}^{1,0}D^k_{cl} = {}^{1,0,0}D^k_{cl} \oplus {}^{0,1,0}D^k_{cl} \equiv
 U(\theta)D^k_{cl} \oplus  V(\theta)D^k_{cl}\;.
\end{eqnarray}
System of even, with respect to $\varepsilon_P, \varepsilon_{\bar{J}},
\varepsilon$ gradings,
projectors $\{\tilde{P}_{a}(\theta)$, $U(\theta)$, $V(\theta)\}$ or
$\{\tilde{P}_{a}(\theta)$, ${\cal W}(\theta)\}$ are described by relations
being analogous to one $\{\tilde{P}_{a}(\theta), U(\theta)\}$ in Ref.[1]
with multiplication table and completeness relation respectively
\begin{eqnarray}
\begin{array}{l|cccc}
{} & \tilde{P}_a & U &  V & {\cal W}  \\ \hline
\tilde{P}_b & \delta_{ab}\tilde{P}_b & 0 & 0 & 0  \\
U & 0 &  U & 0 & U  \\
 V & 0 & 0 &  V &  V  \\
{\cal W} & 0 & U &  V & {\cal W}
\end{array},\ \sum\limits_a \tilde{P}_a(\theta) + {\cal W}(\theta) = 1\;.
\end{eqnarray}
Systems $P_{a}(\theta)$ and $\{\tilde{P}_{a}(\theta), {\cal W}(\theta)\}$ on
$T^{\ast}_{odd}{\cal M}_{cl}$  are connected by means of relations
\begin{eqnarray}
P_0(\theta) = \tilde{P}_0(\theta),\;\tilde{P}_1(\theta) + {\cal W}(\theta) =
P_1(\theta)\;. 
\end{eqnarray}
Projectors $\tilde{P}_1(\theta)$, $U(\theta)$, $V(\theta)$ are  derivations
on $D^k_{cl}$, whereas
for $\tilde{P}_{0}(\theta)$, $P_0(\theta)$ the following rule in acting on
the product of any ${\cal F}(\theta), {\cal J}(\theta)
\in D^k_{cl}$ is valid
\begin{eqnarray}
D_0(\theta)\bigl({\cal
F}(\theta)\cdot{\cal J}(\theta)\bigr) = \bigl(D_0(\theta){\cal
F}(\theta)\bigr) \bigl(D_0(\theta){\cal J}(\theta)\bigr),\ D_0 \in \{P_0,
\tilde{P}_0\} \;. 
\end{eqnarray}
Any ${\cal F}(\theta) \in D^k_{cl}$ is
decomposed onto the component functions (not being by elements from a
superspace of superfield representation $\tilde{T}$)
\begin{eqnarray}
{} & {} &\hspace{-2em}{\cal F}\bigl(\Gamma(\theta),
{\stackrel{\circ}{\Gamma}}(\theta), \theta \bigr) = \tilde{P}_0
(\theta){\cal F}(\theta) +  U(\theta){\cal F}(\theta) + V(\theta){\cal F}(
\theta) +
\tilde{P}_1(\theta){\cal F}(\theta) \equiv
{\cal F}\bigl( P_0\Gamma(\theta), {\stackrel{\circ}{\Gamma}}(
\theta),0\bigr)+  \nonumber \\
{} & {} & \hspace{-2em} P_1(\theta) {\cal F}\bigl({\cal A}(\theta),
P_0{\cal A}^{\ast}(\theta),
{\stackrel{\circ}{\Gamma}}(\theta), 0 \bigr) +
P_1(\theta) {\cal F}\bigl(P_0{\cal A}(\theta), {\cal A}^{\ast}(
\theta), {\stackrel{\circ}{\Gamma}}(\theta), 0 \bigr) +
{\cal F}\bigl( P_0\Gamma(\theta), {\stackrel{\circ}{\Gamma}}(\theta),
\theta \bigr) 
\end{eqnarray}
from subsuperalgebras ${}^{0,0}D^k_{cl}$, ${}^{1,0,0}D^k_{cl}$, ${}^{0,1,
0}D^k_{cl}$, ${}^{0,1}D^k_{cl}$ respectively.

The matrix realization of elements from $D^k_{cl}$ (as the vector superspace
elements) in the form of column-vector from 4 components
\begin{eqnarray}
{\cal F}(\theta) \mapsto
(\tilde{P}_0 {\cal F}, U{\cal F}, V{\cal F}, \tilde{P}_1{\cal F})^{T}
\equiv
({\cal F}^{0,0}, {\cal F}^{1,0,0}, {\cal F}^{0,1,0}, {\cal F}^{0,1})^T
\;,
\end{eqnarray}
permits to represent the projectors in the form of $4\times 4$ matrices
\begin{eqnarray}
\tilde{P}_a(\theta) = {\rm diag}(\delta_{a0},0,0,\delta_{a1}),\;
U(\theta) = {\rm diag}(0,1,0,0),\;V(\theta) = {\rm diag}(0,0,1,0)\;
.
\end{eqnarray}
The analytic representation for ${\cal F}(\theta)$ by means of relations
(2.32), (2.42) leads to validity  of the projectors realization  as the 1st
order differential operators under their action on  $D^k_{cl}$
\begin{eqnarray}
\tilde{P}_0(\theta) = 1 - \theta\frac{\partial}{\partial \theta},\;
U(\theta) = P_1(\theta){\cal A}^{\imath}(\theta)\frac{\partial_l \phantom{xx
x}}{\partial {\cal A}^{\imath}(\theta)},\;
V(\theta) = P_1(\theta){\cal A}^{\ast}_{\imath}(\theta)\frac{\partial_l
\phantom{xxx}}{\partial {\cal A}^{\ast}_{\imath}(\theta)}\;, \nonumber \\
\tilde{P}_1(\theta) = \theta\frac{\partial}{\partial \theta},\;
 {\cal W}(\theta) = P_1(\theta)\Gamma^p(\theta)
\displaystyle\frac{\partial_l\phantom{xxx}}{\partial\Gamma^p(
\theta)}\footnotemark ,\;
\displaystyle\frac{\partial_l\phantom{xxx}}{\partial\Gamma^p(\theta)} =
\left(\displaystyle\frac{\partial_l\phantom{xxx}}{\partial{\cal A}^{\imath}(
\theta)}, \frac{\partial_l\phantom{xxx}}{\partial {\cal A}^{\ast}_{\imath}(
\theta)}\right). 
\end{eqnarray}
\footnotetext{$P_1(\theta)\Gamma^p(\theta) = (P_1(\theta){\cal A}^{
\imath}(\theta), P_1(\theta){\cal A}^{\ast}_{\imath}(\theta))$ is
understood as the
indivisible object for partial differentiation with respect to $\theta$:
$\frac{\partial}{\partial\theta}P_1(\theta)\Gamma^p(\theta) = 0$, but
$\frac{\partial}{\partial\theta}(\theta {\stackrel{\circ\ }{\Gamma^p}}(
\theta)) = {\stackrel{\circ\ }{\Gamma^p}}(\theta)$.}
\vspace{-2ex}

\noindent
The left partial superfield derivative with respect to superfield
$\Gamma^{p}(\theta)$ for fixed $\theta$ is introduced in (2.45).
The connection
between derivatives $\frac{d}{d\theta}$ and $\frac{\partial}{\partial\theta}$
under their action on the elements from $D^k_{cl}$ is established by the
formula (being by continuation of the corresponding one given on
$C^k\bigl(T_{odd}{\cal M}_{cl} \times \{\theta \}\bigr)$ [1])
\begin{eqnarray}
\displaystyle\frac{d}{d\theta} =
\frac{\partial}{\partial\theta} + {\stackrel{\circ\ }{{\Gamma}^{p}}}(
\theta) P_0(\theta)\frac{\partial_l
\phantom{xxx}}{\partial{\Gamma}^{p}(\theta)} \equiv
\frac{\partial}{\partial\theta} +
P_0(\theta){\stackrel{\circ}{\cal W}}(\theta),\ {\stackrel{\circ}{\cal W}}(
\theta) =
\left[\displaystyle\frac{d}{d\theta}, {\cal W}(\theta)\right]_s. 
\end{eqnarray}
Class $C_{FH,cl}$ of regular (analytic) over ${\bf K}$ superfunctionals
on $T_{odd}(T^{\ast}_{odd}{\cal M}_{cl}) \times \{\theta\}$ as continuation
of the class $C_{F}$ of superfunctionals on $T_{odd}{\cal M}_{cl} \times
\{\theta\}$ [1] is defined by the formula
\begin{eqnarray}
\hspace{-1em}F_{H,cl}[{\Gamma}] \equiv F_{H,cl}[{\cal A}, {\cal A}^{\ast}] =
\hspace{-0.1em}\int \hspace{-0.2em}d\theta {\cal F}\bigl(\Gamma
(\theta), {\stackrel{\circ}{\Gamma}}(\theta), \theta\bigr) \equiv
\frac{d}{d\theta}{\cal F}(\theta),\; F_{H,cl}[{\Gamma}] \in C_{FH,cl}, {\cal
F}(\theta) \in D^{k}_{cl}. 
\end{eqnarray}
It is evident that $F_{H,cl}$ is given with accuracy up to
$P_{0}(\theta){\cal F}(\theta)$ component part of ${\cal F}(\theta)$.
Since the operator $\frac{d}{d\theta}$ does not lead out
${\cal F}(\theta)$ from $D^k_{cl}$, then
$F_{H,cl}[\Gamma]$ belongs to $D^k_{cl}$. Besides  $F_{H,cl}[\Gamma]$ is a
scalar with respect to action of $\tilde{T}_{\mid P}$ operators,
being by restriction of representation $\tilde{T}$ operators onto
$P$, if ${\cal F}(\theta)$ is transformed by the rules (2.27) or
(2.28).

An analog of the basic lemma of variational calculus [1] is valid in this
case,
from which it follows the connection for variational superfield derivatives
of $F_{H}[\Gamma]$ with respect to $\Gamma^p(\theta)$ with partial
superfield ones of its density
${\cal F}(\theta) \in D^k_{cl}$ with respect to $\Gamma^{p}(\theta)$,
${\stackrel{\circ\ }{\Gamma^{p}}}(\theta)$.
For instance, obtain
for the  left and right derivatives with respect to superantifields the
formulae respectively
\renewcommand{\theequation}{\arabic{section}.\arabic{equation}\alph{lyter}}
\begin{eqnarray}
\setcounter{lyter}{1}
{} & \displaystyle\frac{\delta_l F_{H}[{\Gamma}]}{\delta
{\cal A}^{\ast}_{\imath}(\theta)}=  \displaystyle\frac{ \partial_l {\cal
F}(\theta)}{\partial{\cal A}^{\ast}_{\imath}(\theta)}
 +(-1)^{\varepsilon_{\imath}}\frac{d}{d\theta} \frac{\partial_l {\cal
F}(\theta)}{{\partial{\stackrel{\ \circ}{\cal
A}}{}^{\ast}_{\imath}(\theta)}} \equiv {\cal L}^{\ast{}\imath}_l(\theta)
{\cal F}(\theta)\;, & {}\\ 
\setcounter{equation}{48}
\setcounter{lyter}{2}
{} & \displaystyle\frac{\delta_r F_{H}[{\Gamma}]}{\delta{\cal
A}^{\ast}_{\imath}(\theta)}= \left[\frac{\partial_r \phantom{xxxx}}{\partial{
\cal A}^{\ast}_{\imath}(\theta)} + (-1)^{\varepsilon_{\imath}}\frac{d_r}{d
\theta}\frac{\partial_r \phantom{xxxxxxxx}} {\partial
\left(\displaystyle\frac{d_r{{\cal
A}}^{\ast}_{\imath}(\theta)}{d\theta\phantom{xxxx}}\right)}\right]{\cal F}(
\theta) \equiv
 {\cal L}^{\ast{}\imath}_{r}(\theta){\cal F}(\theta)\;. & {} 
\end{eqnarray}
The connection of above left and right derivatives is established by the
relationship
\renewcommand{\theequation}{\arabic{section}.\arabic{equation}}
\begin{eqnarray}
{} &
\displaystyle\frac{\delta_l F_{H}[{\Gamma}]}{\delta{\cal A}^{\ast}_{
\imath}(\theta)}=
(-1)^{(\varepsilon_{\imath} + 1)\varepsilon(F_{H})} \frac{\delta_r F_{
H}[{\Gamma}]}{\delta{\cal A}^{\ast}_{\imath}(\theta)}\;. & {} 
\end{eqnarray}
Superfield derivatives have the following table of Grassmann parities
according to (2.24), (2.36)
\begin{eqnarray}
\begin{array}{lccccccc}
{} &
\displaystyle\frac{\delta_l \phantom{xxx}}{\delta{\cal
A}^{\ast}_{\imath}(\theta)} & \displaystyle\frac{\partial_l
\phantom{xxx}}{\partial{\cal A}^{\ast}_{\imath}(\theta)} &
 \displaystyle\frac{\partial_l \phantom{xxx}
 }{{\partial{\stackrel{\ \circ}{\cal A}}{}^{\ast}_{\imath}(\theta)}} &
\displaystyle\frac{\delta_l \phantom{xxx}}{\delta{\Gamma}^{p}(\theta)} &
\displaystyle\frac{\partial_l
\phantom{xxx}}{\partial{\Gamma}^{p}(\theta)} &
 \displaystyle\frac{\partial_l \phantom{xxx}
 }{{\partial{\stackrel{\circ}{\Gamma}}{}^{p}(\theta)}} &{}\\
\varepsilon_{P} & 0 & 1 & 0 & \varepsilon_{P}(\Gamma^p) + 1  & \varepsilon_{
P}(\Gamma^p) & \varepsilon_{P}(\Gamma^p) + 1  &{}\\
\varepsilon_{\bar{J}} &
\varepsilon_{\imath} & \varepsilon_{\imath} & \varepsilon_{\imath} &
\varepsilon_{\bar{J}}(\Gamma^p) & \varepsilon_{\bar{J}}(\Gamma^p) &
\varepsilon_{\bar{J}}(\Gamma^p) &{}\\
\varepsilon  & \varepsilon_{\imath} & \varepsilon_{\imath} + 1 &
\varepsilon_{\imath}  & \varepsilon_p + 1 & \varepsilon_p & \varepsilon_p + 1
&.
\end{array}  
\end{eqnarray}
The $n$th superfield variational derivative with respect to superantifields
${\cal A}^{\ast }_{\imath_1}(\theta_{1})$,\ldots, ${\cal A}^{\ast}_{
\imath_k}(\theta_{k})$ of superfunctional $F_{H}[\Gamma]$ is expressed in
terms of
partial superfield ones with respect to ${\cal A}^{\ast }_{\imath_1}(
\theta)$, \ldots, ${\cal A}^{\ast}_{\imath_k}(\theta)$, ${\stackrel{\ \circ}{
\cal A}}{}^{\ast}_{\imath_1}(\theta)$,\ldots, ${\stackrel{\ \circ}{\cal A}}{
}^{\ast}_{\imath_k}(\theta)$ of its density ${\cal F}(\theta)$ by the formula
\renewcommand{\theequation}{\arabic{section}.\arabic{equation}\alph{lyter}}
\begin{eqnarray}
\setcounter{lyter}{1}
{} & \left(\displaystyle\prod\limits_{l=0}^{k-1}
\frac{\delta_l \phantom{xxxxxxx}}{\delta{\cal
A}^{\ast}_{\imath_{k-l}}(\theta_{k-l})}\right)F_H[\Gamma] \equiv
\displaystyle\frac{\delta_l^k F_H[\Gamma]
\phantom{xxxxxxxxx}}{\delta{\cal A}^{\ast}_{\imath_1}(\theta_1) \ldots
\delta{\cal A}^{\ast}_{\imath_k}(\theta_k)} \equiv F_{H}^{(\imath)_k}[\Gamma,
\vec{\theta}_k]= & {}\nonumber \\
{} & \Bigl( \displaystyle\prod\limits_{l=0}^{k-2}{\cal
L}^{\ast{}\imath_{k-l}}_{l}(\theta_k)\delta(\theta_k - \theta_{k-l-1})\Bigr)
{\cal L}^{\ast{}\imath_1}_{l}(\theta_k){\cal F}\bigl(\Gamma(\theta_k),
{\stackrel{\circ}{\Gamma}}(\theta_k), \theta_k \bigr)\;, & {} \\
\setcounter{equation}{51}
\setcounter{lyter}{2}
{} & \delta({\theta}' -\theta)= {\theta}'
-\theta\;,\;\displaystyle\int d{\theta}'\delta({\theta}' -\theta)y(\theta')=
y(\theta)\;,\;\vec{\theta}_k \equiv \theta_1, \ldots,\theta_k\;. &{}
\end{eqnarray}
By means of the last relation the superfield variational derivative with
respect to
${\cal A}^{\ast}_{\imath}(\theta)$ of arbitrary superfunction ${\cal F}
\bigl(\Gamma(\theta'),{\stackrel{\circ}{\Gamma}}(\theta'),\theta'\bigr)$ is
defined for necessarily(!) not coinciding values
of $\theta$ and $\theta'$. For its calculation from arbitrary ${\cal F}
\bigl(\Gamma(\theta'),{\stackrel{\circ}{\Gamma}}(\theta'),\theta';{\vec{
\theta}}_k\bigr)$ $\in$ $D^k_{cl}\times \{\vec{\theta}_{k}\}$ it is sufficient
to know the values of the following derivatives arising from (2.33), (2.48),
(2.51a)
\renewcommand{\theequation}{\arabic{section}.\arabic{equation}}
\begin{eqnarray}
\frac{\delta_l {\cal
A}^{\ast}_{\jmath}(\theta')}{\delta{\cal A}^{\ast}_{\imath}(\theta)\phantom{
x}} =
(-1)^{\varepsilon_{\jmath}+1} \delta_{\jmath}{}^{\imath}\delta(\theta -
\theta'),\  \frac{\delta_l \left(\displaystyle\frac{d{\cal
A}^{\ast}_{\jmath}(\theta')}{d\theta'\phantom{xxx}}\right)}{\delta{\cal A}^{
\ast}_{\imath}(\theta)\phantom{xxxx}}
= \delta_{\jmath}{}^{\imath}\ .
\end{eqnarray}
\underline{\bf Remark:} Further  all derivatives
with respect to ${\cal A}^{\ast}_{\imath}(\theta)$, ${\stackrel{\circ}{{
\cal A}^{\ast}_{\imath}}}(\theta)$ are considered (under omission)
by the left and the right ones with respect to these superantifields
are labeled by the sign "r".
\section{Basic Statements of the  Hamiltonian Formulation \protect \\
for GSTF}
\setcounter{equation}{0}

Define on $T_{odd}(T^{\ast}_{odd}{\cal M}) \times \{\theta\}$
the superfunction $S_{H,L} \in D^k_{cl},\,k \leq \infty$ by the
relation
\renewcommand{\theequation}{\arabic{section}.\arabic{equation}\alph{lyter}}
\begin{eqnarray}
\setcounter{lyter}{1}
{} & S_{H,L}(\theta) =
S_{H,L}\bigl({\cal A}(\theta), {\stackrel{\ \circ}{
\cal A}}(\theta), {\cal A}^{\ast}(\theta), \theta \bigr) =
{\stackrel{\circ}{{\cal A}^{\imath}}}(\theta){\cal A}^{\ast}_{\imath}(
\theta) -
S_L\bigl({\cal A}(\theta), {\stackrel{\ \circ}{\cal A}}(\theta),
\theta\bigr)\;,& {} \\
\setcounter{equation}{1}
\setcounter{lyter}{2}
{} & (\varepsilon_P, \varepsilon_{\bar{J}}, \varepsilon) S_{H,L}(\theta) =
(0,0,0)\;. 
\end{eqnarray}
Having considered the problem on extremum for $S_{H,L}(\theta)$ with
respect to superfields ${\stackrel{\ \circ}{\cal A}}{}^{\imath}(\theta)$ we
arrive to standard Legendre transform of $S_{L}(
\theta)$ with respect to ${\stackrel{\ \circ}{\cal A}}{}^{\imath}(\theta)$
being defined by the relationships
\renewcommand{\theequation}{\arabic{section}.\arabic{equation}}
\begin{eqnarray}
{\cal A}^{\ast}_{\imath}(\theta) =
\frac{\partial_l S_L\bigl({\cal A}(\theta),
{\stackrel{\ \circ}{\cal A}}(\theta), \theta\bigr)}{\partial
{\stackrel{\ \circ}{\cal A}}{}^{\imath}(\theta)\phantom{xxxxxxxxx}},\;
S_H({\cal A}(\theta), {\cal A}^{\ast}(\theta), \theta) =
\begin{array}{c}
{} \\
{\rm extr} \\
{\scriptstyle{\stackrel{\ \circ}{\cal A}}{}^{\imath}(\theta)}
\end{array}
S_{H,L}\bigl({\cal A}(\theta), {\stackrel{\ \circ}{
\cal A}}(\theta), {\cal A}^{\ast}(\theta), \theta \bigr)\; , 
\end{eqnarray}
expressing by virtue of (2.21) ${\stackrel{\ \circ}{\cal A}}{}^{\imath}(
\theta)$ in terms of generalized "momenta" -- superantifields
${\cal A}^{\ast}_{\imath}(\theta)$:
${\stackrel{\ \circ}{\cal A}}{}^{\imath}(\theta)$ = ${\stackrel{\ \circ}{
\cal A}}{}^{\imath}\bigl({\cal A}(\theta), {\cal A}^{\ast}(\theta),
\theta \bigr)$. It follows from expressions (3.1), (3.2) the consequences
for Legendre transform
\begin{eqnarray}
{} & {} & S_H({\cal A}(\theta), {\cal A}^{\ast}(\theta), \theta) =
{\stackrel{\ \circ}{\cal A}}{}^{\imath}\bigl({\cal A}(\theta), {\cal A}^{
\ast}(\theta),
\theta\bigr){\cal A}^{\ast}_{\imath}(\theta) -
S_{L}\bigl({\cal A}(\theta),
{\stackrel{\ \circ}{\cal A}}({\cal A}(\theta),
{\cal A}^{\ast}(\theta), \theta), \theta \bigr)
\;, \\ 
{} & {} &\displaystyle\frac{d_r {\cal A}^{\imath}(\theta)}{d\theta
\phantom{xxxx}} =
\frac{\partial S_H(\theta)}{\partial{\cal A}^{\ast}_{\imath}(\theta)},\;
\displaystyle\frac{\partial S_H(\theta)}{\partial{\cal A}^{\imath}(
\theta)} = - \frac{\partial S_L(\theta)}{\partial{\cal A}^{\imath}(\theta)},\;
\frac{\partial S_H(\theta)}{\partial \theta\phantom{xxxx}} = - \frac{\partial
S_L(\theta)}{\partial \theta\phantom{xxxx}}\;. 
\end{eqnarray}
\underline{\bf Definition:} Call the system of the 1st order on $\theta$
3n  ODE, 2n from which are in NF
\renewcommand{\theequation}{\arabic{section}.\arabic{equation}\alph{lyter}}
\begin{eqnarray}
\setcounter{lyter}{1}
{} & {} & \displaystyle\frac{d_r {\cal A}^{\ast}_{\imath}(\theta)}{d\theta
\phantom{xxxx}} = (-1)^{\varepsilon_{\imath} + 1}
\frac{\partial S_H({\cal A}(\theta), {\cal A}^{\ast}(\theta), \theta)}{
\partial{\cal A}^{\imath}(\theta)\phantom{xxxxxxxxxx}}\;, \\
\setcounter{equation}{5}
\setcounter{lyter}{2}
{} & {} & \displaystyle\frac{d_r {\cal A}^{\imath}(\theta)}{d\theta
\phantom{xxxx}} =
\frac{\partial S_H({\cal A}(\theta), {\cal A}^{\ast}(\theta), \theta)}{
\partial{\cal A}^{\ast}_{\imath}(\theta)\phantom{xxxxxxxxxx}}\;, \\
\setcounter{equation}{5}
\setcounter{lyter}{3}
{} & {} & \Theta_{\imath}\bigl({\cal A}(\theta), {\stackrel{\
\circ}{\cal A}}(\theta),
\theta\bigr)_{\mid {\stackrel{\ \circ}{\cal A}}(\theta) =
{\stackrel{\;\circ}{\cal A}}({\cal A}(\theta),
{\cal A}^{\ast}(\theta), \theta)} \equiv
\Theta^H_{\imath}\bigl({\cal A}(\theta), {\cal A}^{\ast}(\theta), \theta\bigr)
 = 0 
\end{eqnarray}
the generalized Hamiltonian system (GHS) being defined by superfunction
$S_{H}(\theta)$ (3.3) and the subsystem of
2n ODE (3.5a,b) is called the  Hamiltonian system (HS).

DCLF (2.3b) expressed in terms of $T^{\ast}_{odd}{\cal M}_{cl}$ coordinates
in Eqs.(3.5c) are according to definition
the system of n algebraic (i.e. of the 0 order with respect to operators
of differentiation on $\theta$) equations on 2n unknowns $\Gamma^{p}(
\theta)$ for any values of $\theta$. Call the subsystem
(3.5c) the generalized constraints in Hamiltonian formalism (GCHF). In the
same way just as  DCLF for Euler-Lagrange equations restrict the
formulation
of Cauchy problem in $T_{odd}{\cal M}_{cl} \times \{\theta\}$ [1] the
GCHF have the analogous meaning with respect to HS (3.5a,b).

Under fulfilment of the condition
\renewcommand{\theequation}{\arabic{section}.\arabic{equation}}
\begin{eqnarray}
{\rm deg}_{{\cal A}^{\ast}(\theta)}\Theta^H_{\imath}(\Gamma(\theta), \theta)
= 0 
\end{eqnarray}
let us call $\Theta^{H}_{\imath}({\cal A}(\theta), \theta)$ the holonomic
constraints in Hamiltonian formalism (HCHF).

The integer-valued functions ${\rm deg}_{{\cal A}^{\ast}(\theta)}$
written in (3.6) appear according to Ref.[1]  by functions of degree
on $D^k_{cl}$ with respect to generating elements
${\cal A}^{\ast}_{\imath}(\theta)$. In addition we assume the following
functions given on $D^k_{cl}$
\begin{eqnarray}
{} & {\min{\rm deg}}_{C(\theta)}, {\rm deg}_{C(\theta)}: D^k_{cl} \to \mbox{
\boldmath$N_0$}\;, & {} \nonumber \\
{} & C(\theta) \in \{
{\cal A}(\theta), {\stackrel{\ \circ}{\cal A}}(\theta),
{\cal A}^{\ast}(\theta), {\stackrel{\ \circ}{\cal A}}{}^{\ast}(\theta),
\Gamma(\theta), {\stackrel{\circ}{\Gamma}}(\theta),
{\cal A}(\theta){\stackrel{\ \circ}{\cal A}}(\theta),
{\cal A}^{\ast}(\theta){\stackrel{\ \circ}{\cal A}}{}^{\ast}(\theta),
\Gamma(\theta){\stackrel{\circ}{\Gamma}}(\theta),\ldots \}  & {} 
\end{eqnarray}
and called the least degree and degree of corresponding superfunctions
with respect to $C(\theta)$.

To be solvable GHS (3.5) must satisfy to solvability conditions of the type
(2.12) written additionally to GHS. Therefore the solvable GHS
is equivalent to the system of the 1st order on $\theta$  5n ODE
consisting, in addition to (3.5), of the subsystem
\begin{eqnarray}
\frac{d_r}{d\theta}\left(\frac{\partial S_H(\theta)}{
\partial {\cal A}^{\ast}_{\imath}(\theta)}\right)= 0,\
\frac{d_r}{d\theta}\left(\frac{\partial S_H(\theta)}{
\partial {\cal A}^{\imath}(\theta)}(-1)^{\varepsilon_{\imath} + 1}\right)=
0\;,   
\end{eqnarray}
being valid by virtue of identity ${\stackrel{\,\circ\circ}{\cal A}}{}^{
\imath}(\theta) = {\stackrel{\,\circ\circ}{\cal A}}{}^{\ast}_{\jmath}(\theta)
= 0$.

At last note that $S_{H}(\theta)$ (3.3) is a scalar with
respect to action of the superfield representation $\tilde{T}_{\mid
\bar{J}}$ operators  and is transformed with respect to
$\tilde{T}_{\mid P}$ ones by the rule (2.26), (2.28)
\begin{eqnarray}
S_H\left({\cal A}'(\theta), {{\cal A}^{\ast}}'(\theta),
\theta\right)
& = & S_H\left(T\bigl(h^{-1}(\mu)\bigr){\cal A}(\theta),
T\bigl(h^{-1}(\mu)\bigr){\cal A}^{\ast}(\theta),
T\bigl(h^{-1}(\mu)\bigr)\theta\right) = \nonumber \\
{} & {} & \phantom{=}
S_H\left({\cal A}(\theta - \mu), {{\cal A}^{\ast}}(\theta - \mu),\theta -
\mu\right) \equiv S_H(\theta')\;, 
\end{eqnarray}
so that
\begin{eqnarray}
\delta S_H(\theta) = S_H(\theta') - S_H(\theta)= - \mu \frac{d S_H(\theta)}{d
\theta\phantom{xxx}}= - \mu \left[\frac{\partial}{\partial \theta} + P_0(
\theta){\stackrel{\;\circ}{\cal W}}(\theta)\right] S_H(\theta). 
\end{eqnarray}
\underline{\bf Statement 3.1} (equivalence of GHS (3.5) and LS (2.3))

\noindent
GHS (3.5) is equivalent to system of the 2nd order on $\theta$ 2n ODE (2.3)
under following admissible formulation of the initial conditions [1] (Cauchy
problem) for GHS in $T^{\ast}_{odd}{\cal M}_{cl} \times \{\theta\}$ for
$\theta=0$ determining the integral curve $\overline{\Gamma}{}^p(\theta)$ for
GHS
\renewcommand{\theequation}{\arabic{section}.\arabic{equation}\alph{lyter}}
\begin{eqnarray}
\setcounter{lyter}{1}
{} & {} & P_0(\theta){\cal A}^{\ast}_{\imath}(\theta) =
{\cal A}^{\ast}_{\imath}(\theta)_{\mid\theta=0}
= \left(P_0(\theta)\frac{\partial_l S_L(
\theta)}{\partial{\stackrel{\ \circ}{\cal A}}{}^{\imath}(\theta)}\right)_{\mid
\{{\stackrel{\ \circ}{\cal A}}{}^{\imath}(0) = \bar{\stackrel{\;\circ}{
\cal A}\,}{}^{\imath}(0),
{\cal A}^{\imath}(0) = {\bar{\cal A}}^{\imath}(0)\}}\equiv \bar{{\cal A}^{
\ast}_{\imath}}(0)\;,\\ 
\setcounter{equation}{11}
\setcounter{lyter}{2}
{} & {} & P_0(\theta){\cal A}^{\imath}(\theta) =
{\cal A}^{\imath}(\theta)_{\mid\theta=0} = {\bar{\cal A}}^{\imath}(0)\;,\\
\setcounter{equation}{11}
\setcounter{lyter}{3}
{} & {} & \Theta^H_{\imath}(\overline{\Gamma}(\theta),\theta) = 0\;. 
\end{eqnarray}
In this case the Cauchy problem for LS (2.3) in
$T_{odd}{\cal M}_{cl}\times\{\theta\}$ means the integral curve for Eqs.(2.3)
$\bar{\cal A}^{\imath}(\theta)$ in $T_{odd}{\cal M}_{cl}\times\{\theta\}$
for $\theta = 0$
goes through point $\Bigl(\bar{\cal A}^{\imath}(0),
\bar{\stackrel{\;\circ}{{\cal A}^{\imath}}\,}(0)\Bigr)$ [1]
\renewcommand{\theequation}{\arabic{section}.\arabic{equation}}
\begin{eqnarray}
{\cal A}^{\imath}(\theta)_{\mid\theta=0}  = {\bar{\cal A}}^{\imath}(0),\;
{\stackrel{\ \circ}{\cal A}}{}^{\imath}(\theta)_{\mid\theta=0} =
\bar{\stackrel{\ \circ}{\cal A}\;}{}^{\imath}(0)  
\end{eqnarray}
and must satisfy to DCLF:
$\Theta_{\imath}\Bigl({\bar{\cal A}}^{\imath}(\theta), {\stackrel{\;
\circ}{\bar{\cal A}^{\imath}}}(\theta), \theta\Bigr) = 0$.

Proof of the Statement 3.1  will be developed later in fulfilling of the
condition (2.21).

\noindent
\underline{\bf Remarks:}

\noindent
{\bf 1)} The question on formulation of the Cauchy problem is more
complicated for
field theory since both GHS (3.5) and LS (2.3) with respect to indices of
supergroup $\bar{J}$ representation have the nontrivial (hidden in this work
in the condensed notations) functional structure;

\noindent
{\bf 2)} system (3.5) can be obtained from problem on conditional extremum
for superfunctional
\begin{eqnarray}
Z_{H}[\Gamma] \equiv Z_{H}[{\cal A}, {\cal A}^{\ast}] = \displaystyle\int
d\theta \Bigl({\stackrel{\ \circ}{\cal A}}{}^{\imath}(\theta) {\cal A}^{
\ast}_{\imath}(\theta) - S_{H}({\cal A}(\theta), {\cal A}^{\ast}(
\theta), \theta)\Bigr),\;Z_H[\Gamma] \in C_{FH,cl}  
\end{eqnarray}
in fulfilling of GCHF (3.5c), or equivalently from problem on
unconditional extremum for the extended superfunctional
\renewcommand{\theequation}{\arabic{section}.\arabic{equation}\alph{lyter}}
\begin{eqnarray}
\setcounter{lyter}{1}
Z^{(1)}_{H}[\Gamma,D] = \displaystyle\int
d\theta \Bigl({\stackrel{\ \circ}{\cal A}}{}^{\imath}(\theta){\cal A}^{
\ast}_{\imath}(\theta) - S^{(1)}_{H}(\Gamma(\theta), D(\theta), \theta)\Bigr)
\;,\\ 
\setcounter{equation}{14}
\setcounter{lyter}{2}
S^{(1)}_{H}(\Gamma(\theta), D(\theta), \theta) = S_{H}(\Gamma(\theta),\theta)
- D^{\imath}(\theta)\Theta^H_{\imath}(\Gamma(\theta),\theta)\;,
\end{eqnarray}
in setting of the additional superfields $D^{\imath}(\theta)$ to be equal
to 0 after  variation. Superfields $D^{\imath}(\theta)$ play the standard
role of Lagrange multipliers for GCHF (3.5c)
extending $T_{odd}(T^{\ast}_{odd}{\cal M}_{cl})$, being transformed with
respect to representation $T$ and having the same
$\varepsilon_{P}$, $\varepsilon_{\bar{J}}$, $\varepsilon$ gradings as for
${\cal A}^{\imath}(\theta)$ ($
(\varepsilon_{P}, \varepsilon_{\bar{J}}, \varepsilon)D^{\imath}(\theta) =
(0, \varepsilon_{\imath}, \varepsilon_{\imath})$).

Superfunction $S_{H}(\theta)$ is the expression of the integral (2.4a)
for Eqs.(2.3) in terms of
coordinates on $T^{\ast}_{odd}{\cal M}_{cl} \times \{\theta \}$
\renewcommand{\theequation}{\arabic{section}.\arabic{equation}}
\begin{eqnarray}
S_{H}(\Gamma(\theta), \theta) = S_{E}\bigl({\cal A}(\theta),
{\stackrel{\ \circ}{\cal A}}(\theta), \theta\bigr)_{\mid {\stackrel{\ \circ}{
\cal A}}(\theta) = {\stackrel{\ \circ}{\cal A}}({\cal A}(\theta), {\cal A}^{
\ast}(\theta), \theta)}\;.  
\end{eqnarray}
The translation of the superfield $\Gamma^{p}(\theta)$ along $\theta$ on a
constant parameter $\mu \in {}^{1}\Lambda_{1}(\theta)$
\begin{eqnarray}
\delta_{\mu}\Gamma^{p}(\theta) = \Gamma^p(\theta + \mu) - \Gamma^p(\theta)=
\mu{\stackrel{\circ\ }{{\Gamma}^p}}(\theta) = \frac{d_r\Gamma^{p}(\theta)}{d
\theta\phantom{xxx}}\mu 
\end{eqnarray}
for $S_{H}(\theta)$ has the form (3.10) (further denote
$(-\delta S_{H}(\theta))$ in (3.10) as $\delta_{\mu} S_{H}(\theta)$).
Expression (3.10) on the solution $\overline{\Gamma}{}^{p}(\theta)$ of GHS
(3.5) is given by the formulae taking representation
(2.46) for ${\stackrel{\circ}{\cal W}}(\theta)$ into account
\begin{eqnarray}
\delta_\mu S_H(\theta)_{\mid \overline{\Gamma}(\theta)} & = & \mu
\left[\frac{\partial
S_H(\overline{\Gamma}(\theta), \theta)}{\partial \theta\phantom{xxxxxxxx}} -
P_0(\theta)
\left(\frac{\partial_r S_H(\overline{\Gamma}(\theta),\theta)}{\partial
\Gamma^q(
\theta)\phantom{xxxxx}}\omega^{qp}(\theta)\frac{\partial_l S_H(
\overline{\Gamma}(\theta),\theta)}{
\partial \Gamma^p(\theta)\phantom{xxxxx}}\right)\right] \equiv \nonumber \\
{} & {} & \mu  \left[\frac{\partial
S_H(\overline{\Gamma}(\theta), \theta)}{\partial \theta\phantom{xxxxxxxx}} -
P_0(\theta)
\left(S_H(\overline{\Gamma}(\theta),\theta), S_H(\overline{\Gamma}(
\theta),\theta)
\right)_{\theta}\right], 
\end{eqnarray}
being true on the solutions for HS (3.5a,b) as well.

Vanishing of the expression (3.17) means by virtue of (2.4b), (3.4)
the fulfilment on the solutions for GHS (3.5) of the following equation
\begin{eqnarray}
\frac{\partial_r
S_H(\overline{\Gamma}(\theta), \theta)}{\partial \theta\phantom{xxxxxxxx}} +
P_0(\theta)
\left(S_H(\overline{\Gamma}(\theta),\theta), S_H(\overline{\Gamma}(\theta),
\theta)\right)_{\theta} = 0\;.
\end{eqnarray}
Eq.(3.18) under validity of the condition (2.5) with regard for
(3.4) is reduced to the system
\renewcommand{\theequation}{\arabic{section}.\arabic{equation}\alph{lyter}}
\begin{eqnarray}
\setcounter{lyter}{1}
{} & \tilde{P}_1(\theta)S_H(\Gamma(\theta),\theta) = 0\;\Longleftrightarrow\;
S_H(\Gamma(\theta),\theta) = S_H(\Gamma(\theta), 0) \equiv
S_H(\Gamma(\theta))\;,{} & \\ 
\setcounter{equation}{19}
\setcounter{lyter}{2}
{} & \left(S_H(\overline{\Gamma}(\theta)), S_H(\overline{\Gamma}(\theta))
\right)_{\theta} = 0\;. 
\end{eqnarray}
By sufficient condition for Eq.(3.19b) resolution is in force of (2.7)
the fulfilment on the solutions for GHS of the following system
\renewcommand{\theequation}{\arabic{section}.\arabic{equation}}
\begin{eqnarray}
\frac{\partial S_H(\Gamma(\theta))}{\partial {\cal A}^{\ast}_{
A_1}(\theta)\phantom{x}}_{\mid \overline{\Gamma}(\theta)} = 0,\
\frac{\partial_l S_H(\Gamma(\theta))}{\partial {\cal A}^{
A_2}(\theta)\phantom{xx}}_{\mid \overline{\Gamma}(\theta)} = 0,\
A_1 = 1,\ldots,n_1,\; A_2 =
n_1+1,\ldots,n\;. 
\end{eqnarray}
With join of the superfields $( - {\cal A}^{\ast}_{A_1}(\theta), {\cal
A}^{A_2}(\theta))$ into uniform superfield $\varphi^{a}(\theta)$ (being
transformed with respect to a some
subrepresentation $T_{\varphi}$ from $T\oplus T^{\ast}$) now without
definite $\varepsilon_{P}$ parity the system (3.20) will be written in
the form
\renewcommand{\theequation}{\arabic{section}.\arabic{equation}}
\begin{eqnarray}
\frac{\partial_l S_H(\Gamma(\theta))}{\partial
\varphi^a(\theta)\phantom{xxx}}_{\mid \overline{\Gamma}(\theta)}
 = 0,\;
\Gamma^p(\theta) = (\varphi^a(\theta), \varphi^{\ast}_a(\theta)) = ((-{{\cal
A}^{\ast}_{A_1}}(\theta), {{\cal A}^{A_2}}(\theta)), ({{\cal A}^{A_1}}(
\theta),
{{\cal A}^{\ast}_{A_2}}(\theta))). 
\end{eqnarray}
System (3.21)   according to (2.8) and (3.4) takes the form for $n_{1} = 0$
\begin{eqnarray}
\frac{\partial_l S_H(\Gamma(\theta))}{\partial {\cal A}^{\imath}(\theta)
\phantom{xxx}}_{\mid \overline{\Gamma}(\theta)} = 0\;. 
\end{eqnarray}
The designation for odd Poisson bracket (antibracket) $(\ ,\ )_{\theta}$ as a
bilinear differential mapping on $D^k_{cl}$
is introduced in the relations (3.17) -- (3.19) having the form
in coordinates ${\cal A}^{\imath}(\theta), {\cal A}^{\ast}_{\imath}(\theta)$
and $\Gamma^{p}(\theta)$ for arbitrary ${\cal F}(\theta),
{\cal J}(\theta) \in D^k_{cl}$
\begin{eqnarray}
({\cal F}(\theta), {\cal J}(\theta))_{\theta} & = &
\frac{\partial {\cal F}(\theta)}{\partial {\cal A}^{\imath}(\theta)}
\frac{\partial {\cal J}(\theta)}{\partial {\cal A}^{\ast}_{\imath}(\theta)} -
\frac{\partial_r {\cal F}(\theta)}{\partial {\cal A}^{\ast}_{\imath}(\theta)
\phantom{x}}
\frac{\partial_l {\cal J}(\theta)}{\partial {\cal A}^{\imath}(\theta)
\phantom{x}} =
\frac{\partial_r {\cal F}(\theta)}{\partial \Gamma^p(\theta)}
\omega^{pq}(\theta)
\frac{\partial_l {\cal J}(\theta)}{\partial {\Gamma}^q(\theta)} =
\nonumber \\
{} & = & \frac{\partial {\cal F}(\theta)}{\partial {\cal A}^{\imath}(\theta)}
\frac{\partial {\cal J}(\theta)}{\partial {\cal A}^{\ast}_{\imath}(\theta)} -
(-1)^{(\varepsilon({\cal F}) + 1)(\varepsilon({\cal J}) + 1)}
\frac{\partial {\cal J}(\theta)}{\partial {\cal A}^{\imath}(\theta)}
\frac{\partial {\cal F}(\theta)}{\partial {\cal A}^{\ast}_{\imath}(\theta)}\;
, 
\end{eqnarray}
where
\begin{eqnarray}
{} & {} & \hspace{-4em} \omega^{pq}(\theta)= (\Gamma^p(\theta),
\Gamma^q(\theta))_{\theta} =
P_0(\theta)\omega^{pq}(\theta),\;
\omega^{pq}(\theta) = -(-1)^{(\varepsilon_p + 1)(\varepsilon_q +
1)}\omega^{qp}(\theta),\;
\left\|\omega^{pq}(\theta)\right\|=   \nonumber \\
{} &{} &\hspace{-4em}
\left\|
\begin{array}{cc}
0_n  & 1_n\\
-1_n & 0_n
\end{array}
\right\| ,\;
 (\varepsilon_P, \varepsilon_{\bar{J}}, \varepsilon)
\omega^{pq}(\theta) = (\varepsilon_P(\Gamma^p) + \varepsilon_P(\Gamma^q) + 1,
\varepsilon_{\bar{J}}(\Gamma^p) + \varepsilon_{\bar{J}}(\Gamma^q),
\varepsilon_p + \varepsilon_q +1)\;. 
\end{eqnarray}
In the 3rd equality in (3.23) it is assumed that ${\cal F}(\theta)$,
${\cal J}(\theta)$ are to be homogeneous with respect to $\varepsilon$
grading. Properties of
$\varepsilon_{P}, \varepsilon_{\bar{J}}, \varepsilon$ parities for
antibracket read as follows
\begin{eqnarray}
(\varepsilon_P, \varepsilon_{\bar{J}}, \varepsilon)
({\cal F}(\theta), {\cal J}(\theta))_{\theta} = \bigl(\varepsilon_P({\cal F})
+ \varepsilon_P({\cal J}) + 1, \varepsilon_{\bar{J}}({\cal F}) + \varepsilon_{
\bar{J}}({\cal J}), \varepsilon({\cal F}) + \varepsilon({\cal J}) + 1\bigr)
\;. 
\end{eqnarray}
Under equivalent parametrization for $T^{\ast}_{odd}{\cal M}_{cl}$ by
coordinates $\Gamma^{p}(\theta) = (\varphi^{a}(\theta), \varphi^{\ast}_{a}(
\theta))$ (3.21) the antibracket (3.23) takes the form
\renewcommand{\theequation}{\arabic{section}.\arabic{equation}\alph{lyter}}
\begin{eqnarray}
\setcounter{lyter}{1}
({\cal F}(\theta), {\cal J}(\theta))_{\theta}^{(\varphi, \varphi^{\ast})} =
\left(
\frac{\partial {\cal F}(\theta)}{\partial \varphi^a(\theta)}
\frac{\partial {\cal J}(\theta)}{\partial \varphi^{\ast}_a(\theta)} -
\frac{\partial_r {\cal F}(\theta)}{\partial \varphi^{\ast}_a(\theta)
\phantom{x}}
\frac{\partial_l {\cal J}(\theta)}{\partial \varphi^a(\theta)\phantom{x}}
\right)\;,
\\
\setcounter{equation}{26}
\setcounter{lyter}{2}
\varepsilon_P(\varphi^a, \varphi^{\ast}_a) =
\left\{
\begin{array}{l}
(1,0),\;a = 1,\ldots,n_1\;, \\
(0,1),\;a = n_1+1,\ldots,n\;.
\end{array} \right.  
\end{eqnarray}
Index $(\varphi,\varphi^{\ast})$ on the antibracket
(3.26a) denotes the local coordinates with respect to which one is calculated.
By omission we suppose that $({\cal F}(\theta),{\cal J}(\theta))_{\theta}$
is calculated in the initial
coordinates $({\cal A}^{\imath}(\theta), {\cal A}^{\ast}_{\imath}(\theta))$.
Antibracket  satisfies to standard properties of an generalized
antisymmetry, Leibnitz rule and Jacobi identity for arbitrary
${\cal F}(\theta)$, ${\cal J}(\theta)$, ${\cal K}(\theta) \in D^k_{cl}$
with definite $\varepsilon$ parity
\begin{eqnarray}
\setcounter{lyter}{1}
{} & {} & \hspace{-3em}({\cal F}(\theta), {\cal J}(\theta))_{\theta} = -
(-1)^{(\varepsilon({\cal F}) + 1)(\varepsilon({\cal J}) + 1)}({\cal J}(
\theta), {
\cal F}(\theta))_{\theta}\;, \\
\setcounter{equation}{27}
\setcounter{lyter}{2}
{} & {} & \hspace{-3em}({\cal F}(\theta) \cdot {\cal J}(\theta), {\cal K}(
\theta))_{\theta}
= {\cal F}(\theta)({\cal J}(\theta), {\cal K}(\theta))_{\theta}
 + (-1)^{\varepsilon({\cal J})(\varepsilon({\cal K}) + 1)}({\cal F}(\theta),
 {\cal K}(\theta))_{\theta}{\cal J}(\theta)\;, \\ 
\setcounter{equation}{27}
\setcounter{lyter}{3}
{} & {} & \hspace{-3em}(({\cal F}(\theta), {\cal J}(\theta))_{\theta}, {\cal
K}(\theta))_{
\theta}(-1)^{(\varepsilon({\cal F}) + 1)(\varepsilon({\cal K}) + 1)} +
\mbox{cycl. perm.}({\cal F}(\theta), {\cal J}(\theta), {\cal K}(\theta)) = 0
\;.
\end{eqnarray}
Summarizing the developed investigation of the properties for
$S_H(\Gamma(\theta), \theta)$  let us formulate the results in the form of

\noindent
\underline{\bf Statement 3.2}

\noindent
In order to the superfunction $S_H(\Gamma(\theta))$ not explicitly
depending upon $\theta$ would be by integral for GHS (3.5) it is sufficient
to fulfill of the following system of equations (being algebraic with respect
to $\theta$ on solutions for GHS)
\renewcommand{\theequation}{\arabic{section}.\arabic{equation}}
\begin{eqnarray}
{(\varphi^{\ast}_a(\theta), S_H(\Gamma(\theta)))_{\theta}}_{\mid \overline{
\Gamma}(\theta)} = 0\footnotemark\,. 
\end{eqnarray}
\footnotetext{the proof of the fact that condition (3.28) is necessary for
realization of Eq.(3.19b) appears by  more technically complicated problem}
\underline{\bf Remarks:}

\noindent
{\bf 1)} System (3.28) or (3.21) considered for any values of superfields
$\Gamma^p(\theta)$ may be interpreted as a consequence from Noether's theorem
of the
fact that translation transformation  on a constant superfield $d^a(\theta)$:
$\varphi^a(\theta) \to \tilde{\varphi}^a(\theta) = \varphi^a(\theta) + d^a(
\theta)$ $(\varepsilon(d^a) = \varepsilon(\varphi^a))$ is a symmetry
transformation for $S_{H}(\Gamma(\theta))$ and therefore in this case
the relation is true
\begin{eqnarray}
S_H(\Gamma(\theta)) = S_H(\varphi^{\ast}(\theta))\;.
\end{eqnarray}
Then the set of superantifields $\varphi^{\ast}_a(\theta)$ is  conserved on
solutions for GHS
\begin{eqnarray}
{\stackrel{\,\circ}{\varphi}}{}^{\ast}_a(\theta)_{\mid
\overline{\Gamma}(\theta)} = 0
\;; 
\end{eqnarray}
{\bf 2)} Statement 3.2 in fact permits one to claim that $S_H(\Gamma(\theta))$
appears by the integral for only Hamiltonian part (3.5a,b) of the GHS
under conditions (3.28).

Thus $S_H(\Gamma(\theta))$  in fulfilling of (3.19a) is the integral
of the 1st order on $\theta$ system 4n ODE
\renewcommand{\theequation}{\arabic{section}.\arabic{equation}\alph{lyter}}
\begin{eqnarray}
\setcounter{lyter}{1}
{} & {} & \frac{d_r \Gamma^p(\theta)}{d\theta \phantom{xxx}} =
\omega^{pq}(\theta)
\frac{\partial_l S_H(\Gamma(\theta))}{\partial \Gamma^q(\theta)\phantom{xx
x}},\;p,q=1,\ldots,2n\;,
\\
\setcounter{equation}{31}
\setcounter{lyter}{2}
{} & {} & \Theta^H_{\imath}(\Gamma(\theta)) = 0\;,\\ 
\setcounter{equation}{31}
\setcounter{lyter}{3}
{} & {} & \chi^H_a(\Gamma(\theta)) \equiv (\varphi^{\ast}_a(\theta),
S_H(\Gamma(\theta)))_{\theta} = \left(\frac{d_r \Gamma^p(\theta)}{d\theta
\phantom{xxx}} -
(\Gamma^p(\theta), S_H(\theta))_{\theta}\right)\lambda_{pa}(\Gamma(
\theta))\;,
\end{eqnarray}
with arbitrary superfunctions $\lambda_{pa}(\Gamma(\theta)) \in C^{k\ast}
\equiv C^k(T^{\ast}_{odd}{\cal M}_{cl})$
satisfying to the property
\begin{eqnarray}
\varepsilon(\lambda_{pa}(\Gamma(\theta))) = \varepsilon_p + 1 +
\varepsilon(\varphi^a),\;a=(A_1, A_2) = \overline{1,n}\;.
\end{eqnarray}
\underline{\bf Definition:} Systems (3.31) and (3.31a,c) are called the
extended generalized Hamiltonian system (EGHS) and the extended HS
(EHS) respectively.
Subsystem (3.31c) we will call the special constraints in Hamiltonian
formalism (SCHF) and their analog (2.7) as the special constraints in
Lagrangian formalism (SCLF). In fulfilling of the conditions
\renewcommand{\theequation}{\arabic{section}.\arabic{equation}}
\begin{eqnarray}
{\rm deg}_{{\cal A}^{\ast}(\theta)}\chi^H_a(\Gamma(\theta)) = 0,\;
{\rm deg}_{{\stackrel{\ \circ}{\cal A}}(\theta)}\left(\chi^H_a(
\Gamma(\theta))_{\vert {\cal A}^{\ast}_{\imath}(\theta) =
\frac{\partial_l S_L(\theta)}{\partial
{\stackrel{\,\circ}{\cal A}}{}^{\imath}(\theta)\phantom{x}}}\right) \equiv
{\rm deg}_{{\stackrel{\ \circ}{\cal A}}(\theta)}\chi_a\bigl(
{\cal A}(\theta),
{\stackrel{\ \circ}{\cal A}}(\theta)\bigr) = 0\;,   
\end{eqnarray}
call SCHF, SCLF the holonomic SCHF (HSCHF), holonomic SCLF (HSCLF)
respectively.

\nopagebreak
From relations (2.3b) for $\Theta_{\imath}\bigl({\cal A}(\theta), {\stackrel{
\ \circ}{\cal A}}(\theta),\theta\bigr)$ with regard of
(3.4) it follows the explicit representation for GCHF (3.5c) in terms
of $\Gamma(\theta)$ and $S_H(\Gamma(\theta))$
\begin{eqnarray}
\Theta^H_{\imath}(\Gamma(\theta), \theta) & = &  ({\cal A}^{\ast}_{\imath}(
\theta),
S_H(\theta))_{\theta} - (-1)^{\varepsilon_{\imath}}\left({{S''_H}}^{-1}
\right)_{\imath k}(\theta)\Biggl[\frac{\partial}{\partial\theta}
({\cal A}^k(\theta),S_H(\theta))_{\theta}(-1)^{\varepsilon_{k}} +
\nonumber \\
{} & {} &  ({\cal A}^k(\theta),({\cal A}^{\ast}_{\jmath}(\theta),S_H(
\theta))_{\theta})_{\theta}\left[({\cal A}^{\jmath}(\theta),S_H(\theta))_{
\theta} - \displaystyle\frac{d_r {\cal A}^{\jmath}(\theta)}{d\theta\phantom{
xxx}}\right](-1)^{\varepsilon_{\jmath}} -  \nonumber \\
{} & {} &  (-1)^{\varepsilon_{\jmath}}({\cal A}^k(\theta),({\cal A}^{\ast}_{
\jmath}(\theta),S_H(
\theta))_{\theta})_{\theta}({\cal A}^{\jmath}(\theta),S_H(\theta))_{\theta}
\Biggl]\;.
\end{eqnarray}
In obtaining of (3.34)  the formulae following from calculation rules for
composite derivatives and from Legendre transform (3.2) have been made use
in addition to above-mentioned ones
\renewcommand{\theequation}{\arabic{section}.\arabic{equation}\alph{lyter}}
\begin{eqnarray}
\setcounter{lyter}{1}
{} & {} & \hspace{-4em}\frac{\partial_l {\cal F}\bigl(
{\cal A}(\theta), {\cal A}^{\ast}\bigl({\cal A}(\theta),{\stackrel{\ \circ}{
\cal A}}(\theta), \theta\bigr), \theta\bigr)}{\partial
{\stackrel{\ \circ}{\cal A}}{}^{\imath}(\theta)\phantom{xxxxxxxxxxxxxxxxxx}}
= \frac{\partial_l {\cal
A}^{\ast}_{k}(\theta)}{\partial{\stackrel{\ \circ}{\cal A}}{}^{\imath}(
\theta)}\,
\frac{\partial {\cal F}\bigl(
\Gamma(\theta), \theta\bigr)}{\partial {\cal A}^{
\ast}_k(\theta)\phantom{xxxx}},\;\forall {\cal F}(\theta) \in C^{k\ast}
\times\{\theta\}\;,
\\
\setcounter{equation}{35}
\setcounter{lyter}{2}
{} & {} &\hspace{-4em}\displaystyle\frac{\partial_l {\cal
A}^{\ast}_{k}(\theta)}{\partial{\stackrel{\ \circ}{\cal A}}{}^{\imath}(
\theta)} = ({S''_L})_{\imath k}(\theta),\;({S''_H})^{k\jmath}(\Gamma(\theta),
\theta) = \frac{\partial\phantom{xxxx}}{\partial{\cal A}^{\ast}_{k}(\theta)}
\frac{\partial S_H(\Gamma(\theta),\theta)}{\partial{\cal A}^{\ast}_{
\jmath}(\theta)\phantom{xxxx}}\;,
\\
\setcounter{equation}{35}
\setcounter{lyter}{3}
{} & {} & \hspace{-4em}
{(S_L'')}_{\imath k}\bigl({\cal A}(\theta),
{\stackrel{\ \circ}{\cal A}}(\theta),\theta\bigr)
(S_H'')^{k\jmath}(\theta)(-1)^{\varepsilon_{
\jmath} + 1} = \delta_{\imath}{}^{\jmath},\;
({S_H''}^{-1})_{\jmath k}(\theta){(S_H'')}^{k \imath}(\theta) =
\delta_{\jmath}{}^{\imath}\;, \\
\setcounter{equation}{35}
\setcounter{lyter}{4}
{} & {} &\hspace{-4em} ({S_H''}^{-1})_{\jmath k}(\Gamma(\theta),
\theta) = (-1)^{\varepsilon_{\jmath} + 1}{(S_L'')}_{\jmath k}\bigl({\cal A}(
\theta), {\stackrel{\ \circ}{\cal A}}(\theta), \theta\bigr)_{\mid {\stackrel{
\ \circ}{\cal A}}(\theta) = {\stackrel{\ \circ}{\cal A}}(\Gamma(\theta),
\theta)}\;,
\\
\setcounter{equation}{35}
\setcounter{lyter}{5}
{} & {} &\hspace{-4em}{(S_L'')}_{\imath k}(\theta) = (-1)^{(\varepsilon_{
\imath}+1)(\varepsilon_{k}
+1)}{(S_L'')}_{k \imath}(\theta),\;{(S_H'')}^{k \imath}(\theta) =
(-1)^{(\varepsilon_{\imath}+1)(\varepsilon_{k}+ 1)}{(S_H'')}^{\imath k}(
\theta)\;, \nonumber \\
{} & {} & \hspace{-4em}
({S_H''}^{-1})_{\imath k}(\theta) = - (-1)^{\varepsilon_{\imath}
\varepsilon_{k}}({S_H''}^{-1})_{k \imath}(\theta)\;, \\ 
\setcounter{equation}{35}
\setcounter{lyter}{6}
{} & {} &\hspace{-4em} \displaystyle\frac{\delta_l {\cal F}\bigl(
{\cal A}(\theta), {\stackrel{\ \circ}{\cal A}}\bigl({\cal A}(\theta),
{\cal A}^{\ast}(\theta), \theta\bigr), \theta\bigr)}{\delta
{\cal A}^{\jmath}(
\theta_1)\phantom{xxxxxxxxxxxxxxxxx}} =
\frac{\partial_l  \left(\delta(\theta_1 - \theta){\cal F}\bigl(
{\cal A}(\theta_1), {\stackrel{\ \circ}{\cal A}}\bigl({\cal A}(\theta_1),
{\cal A}^{\ast}(\theta_1)
, \theta_1\bigr), \theta_1\bigr)\right)}{\partial{\cal A}^{
\jmath}
(\theta_1)\phantom{xxxxxxxxxxxxxxxxxxxxxxxxxxxxxx}} + \nonumber\\
{} & {} &\hspace{-4em}
\displaystyle\frac{d\phantom{x}}{d\theta_1}\left(
({S_H''}^{-1})_{\jmath k}(\theta_1)
\frac{\partial  \left(\delta(\theta_1 - \theta){\cal F}\bigl(
\Gamma(\theta_1), \theta_1\bigr)\right)}{\partial {\cal A}^{\ast}_{k}(
\theta_1)\phantom{xxxxxxxxxxxxx}}\right) =
\frac{\partial_l  \left(\delta(\theta_1 - \theta){\cal F}\bigl(
\Gamma(\theta_1), \theta_1\bigr)\right)}{\partial{\cal A}^{\jmath}(
\theta_1)\phantom{xxxxxxxxxxxxx}}
- \nonumber \\
{} & {} &\hspace{-4em}
\left(\displaystyle\frac{\partial_l \phantom{xxxx}}{\partial {\cal A}_{
\jmath}(\theta_1)}
\frac{\partial S_H(\theta_1)}{\partial {\cal A}^{\ast}_{l}(\theta_1)}\right)
({S_H''}^{-1})_{l k}(\theta_1)
\frac{\partial \left(\delta(\theta_1 - \theta){\cal F}\bigl(\Gamma(
\theta_1),
\theta_1\bigr)\right)}{\partial {\cal A}^{\ast}_{k}(\theta_1)\phantom{xxx
xxxxxxxxxx}} +
\nonumber \\
{} & {} & \hspace{-4em}
\displaystyle\frac{d\phantom{x}}{d\theta_1}\left(
({S_H''}^{-1})_{\jmath k}(\theta_1)
\frac{\partial  \left(\delta(\theta_1 - \theta)
{\cal F}\bigl(
\Gamma(\theta_1), \theta_1\bigr)\right)}{\partial {\cal A}^{\ast}_{k}(
\theta_1)\phantom{xxxxxxxxxxxx}}\right) \equiv \frac{\delta_l
{\cal F}\bigl(
\Gamma(\theta), \theta\bigr)}{\delta\hat{\cal A}^{\jmath}(\theta_1)
\phantom{xxxx}}\;.   
\end{eqnarray}
One can show the SCHF (3.31c) under definite choice for $\lambda_{pa}(
\Gamma(\theta))$ are equivalent to the SCLF (2.9c,d) out of
solutions for LS (2.3) as well. Analyzing a set of solutions for SCHF
(SCLF) simultaneously with other Eqs.(3.31) ((2.9)) one can pass by means of
addition to
the subsystem (3.31c), ((2.9c,d)), in general, a nonlinear combination of
Eqs.(3.31a) ((2.9b))  to equivalent EGHS (3.31) (ELS(2.9))
in which the superfunctions $\lambda_{pa}(\Gamma(\theta))$ ($
\lambda_1^{{}A_1 \imath}(\theta), \lambda_{2{}A_2}{}^{\imath}(\theta)$)
vanish.

Described transformation of EGHS (3.31) (ELS(2.9)) changing the SCHF
(3.31c) (SCLF (2.9c,d)) themselves, being considered without other
Eqs.(3.31a,b) ((2.9a,b)), does not change a set of solutions for EGHS (ELS)
on the whole.
Under various choice of the SCLF, SCHF being realized
by changing of parameter $n_1$ and
under special structure of the superfunctions $S_L(\theta)$, $S_H(\theta)$ it
is possible to achieve that the Eqs.(2.6), (3.19b) written in the whole
spaces $T_{odd}{\cal M}_{cl}$, $T^{\ast}_{odd}{\cal M}_{cl}$ would be by
double zeros of solutions for LS (2.3), GHS (3.5)  respectively
\renewcommand{\theequation}{\arabic{section}.\arabic{equation}\alph{lyter}}
\begin{eqnarray}
\setcounter{lyter}{1}
{} & \hspace{-2.5em}
{\stackrel{\ \circ}{\cal A}}{}^{\imath}(\theta)
\displaystyle\frac{\partial_l S_{L}(\theta)}{\partial{{\cal A}^{\imath}}(
\theta)}=  \hspace{-0.2em}\left(  \hspace{-0.2em}
\frac{\partial_l S_{L}(\theta)}{\partial{{\cal A}^{A_1}}(\theta)}\chi^{A_1}
(\theta)
+ {\stackrel{\ \circ}{\cal A}}{}^{A_2}(\theta)\chi_{A_2}(\theta)
\hspace{-0.2em}\right)\hspace{-0.2em}
= ({\cal L}^l_{\jmath}(\theta)S_L(\theta))
({\cal L}^l_{k}(\theta)S_L(\theta))d^{\jmath k}(\theta), & {} \\ 
\setcounter{equation}{36}
\setcounter{lyter}{2}
{} & \left(S_H({\Gamma}(\theta)), S_H({\Gamma}(\theta))
\right)^{(\varphi.\varphi^{\ast})}_{\theta} = -2
\displaystyle\frac{\partial_r S_H(\theta)}{\partial\varphi^{\ast}_a(\theta)}
\chi^H_a(\Gamma(\theta))=
\left[\displaystyle\frac{d_r \Gamma^p(\theta)}{d\theta
\phantom{xxx}} - (\Gamma^p(\theta), S_H(\theta))_{\theta}\right]\times & {}
\nonumber \\
{} & \left[\displaystyle\frac{d_r \Gamma^q(\theta)}{d\theta
\phantom{xxx}} - (\Gamma^q(\theta), S_H(\theta))_{\theta}\right]d_{pq}(
\Gamma(\theta))\;, & {} 
\end{eqnarray}
with arbitrary superfunctions $d_{pq}(\Gamma(\theta))$ $\in$ $C^{k\ast}$
and $d^{\jmath k}(\theta)$ $\in$
$C^k\bigl(T_{odd}{\cal M}_{cl}\bigr)$, ($p,q = \overline{1,2n};$ $j,k=
\overline{1,n}$) possessing by the properties
\renewcommand{\theequation}{\arabic{section}.\arabic{equation}}
\begin{eqnarray}
\varepsilon(d_{pq}) = \varepsilon_p + \varepsilon_q + 1,
\;\varepsilon(d^{\jmath k}) = \varepsilon_{\jmath} + \varepsilon_k + 1,\;
d_{pq} = (-1)^{(\varepsilon_p + 1)(\varepsilon_q + 1)}d_{qp},
\;d^{\jmath k} = (-1)^{\varepsilon_{\jmath}\varepsilon_k}d^{
k \jmath}.
\end{eqnarray}
The fact, that the left-hand side of expressions (3.36) are
double zeros for solutions of the systems (2.3), (3.5)
respectively, means the fulfilment on their solutions of the equalities
\begin{eqnarray}
{} & \displaystyle\frac{\partial_l\phantom{xxx}}{\partial{\stackrel{\
\circ}{\cal A}}{}^{\jmath}(\theta)}\left(
{\stackrel{\;\circ}{{\cal A}^{\imath}}}(\theta)
\frac{\partial_l S_{L}(\theta)}{\partial{{\cal A}^{\imath}}(\theta)}\right)_{
\mid {\cal L}^l_{\jmath}(\theta)S_L(\theta)=0} =
\frac{\partial_l\phantom{xxx}}{\partial{\cal A}^{\jmath}(\theta)}\left(
{\stackrel{\;\circ}{{\cal A}^{\imath}}}(\theta)
\frac{\partial_l S_{L}(\theta)}{\partial{{\cal A}^{\imath}}(\theta)}\right)_{
\mid {\cal L}^l_{\jmath}(\theta)S_L(\theta)=0} = 0\;, & {} \\
{} & \displaystyle\frac{\partial_l\phantom{xxx}}{\partial \Gamma^p(\theta)}
{\left(S_H(
\Gamma(\theta)), S_H(\Gamma(\theta))\right)_{\theta}}_{\mid \overline{\Gamma}(
\theta)} = 0\;.
\end{eqnarray}
Note that under definite choice for superfunctions $d_{pq}(\theta)$,
$d^{\jmath k}(\theta)$ the equations (3.36) themselves will become
by equivalent to each other (in sense of change of variables under Legendre
transform) out of solutions for GHS and LS as well.

Being easily obtained from definition of Legendre transform (3.2)
the following relation taking Eqs.(3.4), LS (2.3) in the form (2.2) and
formulae (3.5c) into account
\begin{eqnarray}
\Theta^H_{\imath}(\Gamma(\theta),\theta) = - \left(\frac{d_r {\cal A}^{\ast}_{
\imath}(\theta)}{d\theta\phantom{xxxx}} +
\frac{\partial_l S_H(\theta)}{\partial{\cal A}^{\imath}(\theta)}\right)
\end{eqnarray}
establishes the coincidence of GCHF (3.5c), therefore of DCLF (2.3b) as
well, with
Eqs.(3.5a) from HS. On the solutions for the other half of equations
from HS (3.5b), in fulfilling of the condition (3.19a), the GCHF (3.34)
are defined only in $T^{\ast}_{odd}{\cal M}_{cl}$ and have the more simple
form
\begin{eqnarray}
\Theta^H_{\imath}(\Gamma(\theta)) & = &
({\cal A}^{\ast}_{\imath}(\theta),S_H(\theta))_{\theta} +
(-1)^{\varepsilon_{\imath} + \varepsilon_{\jmath}}({S_H''}^{-1})_{\imath
k}(\theta)({\cal A}^{k}(\theta),({\cal A}^{\ast}_{\jmath}(\theta),
S_H(\theta))_{\theta})_{\theta}\times \nonumber\\
{} & {} & ({\cal A}^{\jmath}(\theta),S_H(\theta))_{\theta}\;. 
\end{eqnarray}
Relationship (3.40) leads to validity of

\noindent
\underline{\bf Statement 3.3} (on equivalence of GHS and HS)

\noindent
GHS (3.5) is equivalent to its proper subsystem -- HS (3.5a,b).

Representation (3.34) is very useful for proof of implication from GHS (3.5)
to LS (2.3) in the Statement 3.1. To this end, it is sufficient to
differentiate with respect to $\theta$ the subsystem (3.5b), next to multiply
the obtained expression on the supermatrix elements $\bigl(S_L''\bigr)_{\imath
\jmath}(\theta)$ taking the formulae (3.35b,c,d) into account, having
got from the left in derived
differential consequence for (3.5b) the Eq.(2.3a) in terms of
coordinates on $T_{odd}{\cal M}_{cl}$ and $S_L(\theta)$.

But from the right with use of the property  for Legendre
transform to be involutory, relations (3.34) and Eqs.(3.5a,b) themselves
the system of DCLF $\Theta_{\imath}\bigl({\cal A}(\theta),{
\stackrel{\ \circ}{\cal A}}(\theta),\theta\bigr)$ will be obtained, which
vanishes by virtue of validity of (3.5c) and (3.40)
\renewcommand{\theequation}{\arabic{section}.\arabic{equation}\alph{lyter}}
\begin{eqnarray}
\setcounter{lyter}{1}
{} & {} & \hspace{-4em}{\stackrel{\,\circ\circ}{\cal A}}{}^{\imath}(\theta)
(S_L'')_{\imath
\jmath}\bigl({\cal A}(\theta),{\stackrel{\ \circ}{\cal A}}(\theta),\theta
\bigr) =
\displaystyle\frac{d}{d\theta}\left(\frac{\partial S_H(\Gamma(\theta),
\theta)}{\partial
{\cal A}^{\ast}_{\imath}(\theta)\phantom{xxxx}}\right)({S_H''}^{-1})_{\imath
\jmath}(\Gamma(\theta),\theta)_{\mid {\cal A}^{\ast}(\theta) = {\cal A}^{\ast}
\bigl({\cal A}(\theta), {\stackrel{\ \circ}{\cal A}}(\theta),\theta\bigr)},
\\
\setcounter{equation}{42}
\setcounter{lyter}{2}
{} &{} &\hspace{-4em}\displaystyle\frac{d}{d\theta}\left(\frac{\partial
S_H(\Gamma(\theta),
\theta)}{\partial
{\cal A}^{\ast}_{\imath}(\theta)\phantom{xxxx}}\right)({S_H''}^{-1})_{\imath
\jmath}(\Gamma(\theta),\theta)_{\mid {\cal A}^{\ast}(\theta) =
{\cal A}^{\ast}
\bigl({\cal A}(\theta), {\stackrel{\ \circ}{\cal A}}(\theta),\theta\bigr)}
=  \nonumber \\
{} & {} &\hspace{-4em} \Theta^H_{\jmath}\bigl({\cal A}(\theta),{\cal A}^{
\ast}\bigl({
\cal A}(\theta), {\stackrel{\ \circ}{\cal A}}(\theta),\theta\bigr), \theta
\bigr)(-1)^{\varepsilon_{\jmath}} \equiv
\Theta_{\jmath}\bigl({\cal A}(\theta),{\stackrel{\ \circ}{
\cal A}}(\theta), \theta\bigr)(-1)^{\varepsilon_{\jmath}} = 0\;.
\end{eqnarray}
Thus one can take for granted the Statement 3.1 entirely (inverse
implication was developed by directly construction of GHS (3.5) from LS
(2.3)). Combining the results of Statements 3.1, 3.3 we arrive to
conclusion on equivalence of HS (3.5a,b) and LS (2.3). It means in setting of
Cauchy problem for HS  the identical fulfilment of the Eqs.(3.11c).
In what follows we will be able to ignore the Eqs.(3.11) in this question.

Relations (3.42) mean the solvability conditions for subsystem (3.5b),
written in the 1st subsystem in (3.8), are identically realized  on
solutions for HS (3.5a)
in force of the formula (3.42). GCHF (3.5c) are  the differential consequence
of (3.5b) coinciding with (3.5a). But solvability of subsystem (3.5a) written
in the 2nd subsystem in (3.8) and being equivalent to solvability of
(2.3b), that with allowance made for (2.12) is given by the equations
\renewcommand{\theequation}{\arabic{section}.\arabic{equation}}
\begin{eqnarray}
\frac{d}{d\theta}\left(\frac{\partial S_L(\theta)}{\partial
{\cal A}^{\imath}(\theta)}\right) = 0\;, 
\end{eqnarray}
being additional to (2.3), is more complicated problem. From
equivalence of HS (3.5a,b) and LS (2.3), Statements 3.1, 3.3  it follows
the easily provable validity of the

\noindent
\underline{\bf Statement 3.4} (on equivalence of EGHS and EHS, of EGHS and
ELS)

\noindent
a) EGHS (3.31) is equivalent to its proper subsystem EHS (3.31a,c);

\noindent
b) EGHS (3.31) is equivalent to ELS in setting of Cauchy problem for EGHS
in $T^{\ast}_{odd}{\cal M}_{cl} \times \{\theta \}$ for $\theta=0$ in
the form (3.11) and integral curve $\overline{\Gamma}{}^p(\theta)$ must
satisfy to the equations
\begin{eqnarray}
\chi^H_{a}(\overline{\Gamma}(\theta)) = 0,\; a=(A_1,A_2)=1,\ldots,n\;.
\end{eqnarray}
Cauchy problem for system (2.9) in the Lagrangian formalism in this case is
formulated in $T_{odd}{\cal M}_{cl} \times \{\theta \}$ and corresponding
integral curve $\bar{\cal A}^{\imath}(\theta)$ satisfies to
DCLF and SCLF
\begin{eqnarray}
{} & {} &\Theta_{\imath}\left(\bar{\cal A}(\theta), {\stackrel{\ \circ}{\bar{
\cal A}}}(\theta)\right) = 0,\
\chi_{a}\left(\bar{\cal A}(\theta), {\stackrel{\ \circ}{\bar{\cal A}}}(
\theta)\right) = 0\;. 
\end{eqnarray}

From the last statement  it follows the subsystem (3.43) is
formally  the solvability conditions for EGHS (3.31) which on the whole in
terms of antibracket have the form
\renewcommand{\theequation}{\arabic{section}.\arabic{equation}\alph{lyter}}
\begin{eqnarray}
\setcounter{lyter}{1}
{} & {} & \displaystyle\frac{d_r}{d\theta}\left(\omega^{pq}(\theta)\frac{
\partial_l S_H(\Gamma(\theta))}{\partial\Gamma^q(\theta)\phantom{xxx}}\right) =
((\Gamma^p(
\theta), S_H(\Gamma(\theta)))_{\theta}, S_H(\Gamma(\theta)))_{\theta} = 0
\;, \\
\setcounter{equation}{46}
\setcounter{lyter}{2}
{} & {} & \displaystyle\frac{d_r}{d\theta}\left[\chi^H_a(\Gamma(\theta)) -
\left(\frac{d_r \Gamma^p(\theta)}{d\theta\phantom{xxxx}} - (\Gamma^p(\theta),
S_H(\Gamma(\theta)))_{\theta}\right)\lambda_{pa}(\Gamma(\theta))\right] =
0\;. 
\end{eqnarray}
In its turn, the subsystem (3.46b) is equivalent to the following one
on solutions for (3.31)
\renewcommand{\theequation}{\arabic{section}.\arabic{equation}}
\begin{eqnarray}
{} & \bigl((\varphi^{\ast}_a(\theta), S_H(\Gamma(\theta)))_{\theta},
S_H(\Gamma(\theta))\bigr)_{\theta} - ((\Gamma^p(
\theta), S_H(\Gamma(\theta)))_{\theta}, S_H(\Gamma(\theta)))_{\theta}
\lambda_{pa}(\Gamma(\theta))(-1)^{\varepsilon_p + \varepsilon(\varphi^a)}
& {}\nonumber \\
{} & - \displaystyle\left[\frac{d_r \Gamma^p(\theta)}{d\theta\phantom{xxxx}}
- (\Gamma^p(
\theta), S_H(\Gamma(\theta)))_{\theta}\right]\left(\lambda_{pa}(\Gamma(
\theta)), S_H(\Gamma(\theta))\right)_{\theta} = 0\;. & {} 
\end{eqnarray}
From explicit form of Eqs.(3.46a), (3.47)  it is obvious that on the
solutions for EGHS and on any solution for (3.46a) it
follows the identical fulfilment of Eqs.(3.46b). In obtaining of the
relations (3.46), (3.47) the identity ${\stackrel{\circ\circ}{\Gamma}}{}^p(
\theta) \equiv 0$ with following formula have been made use
\begin{eqnarray}
\frac{d_r {\cal F}(\Gamma(\theta),\theta)}{d\theta\phantom{xxxxxxxx}}_{\mid
\overline{\Gamma}(\theta)}=\left(\frac{\partial_r {\cal F}(\Gamma(\theta),
\theta)}{\partial\theta
\phantom{xxxxxxxx}} + ({\cal F}(\Gamma(\theta),\theta), S_H(\Gamma(
\theta)))_{\theta}\right){\hspace{-0.7em}\phantom{\Bigr)}}_{\mid
\overline{\Gamma}(\theta)}     
\end{eqnarray}
being valid $\forall {\cal F}(\theta) \in C^{k\ast}\times \{\theta\}$
(and even from $D^k_{cl}$).
In (3.48) through $\overline{\Gamma}(\theta)$  an integral curve of
HS (3.31a) is
denoted as in (3.17)  taking statement 3.3 into consideration. The fact,
that projector $P_0(\theta)$ in (3.48) and, therefore in (3.46), (3.47) was
omitted in comparison with (3.18), for now  means the tending to obtain,
namely, the such expression that is provided by solvability of the relations
(3.46a).

\noindent
\underline{\bf Remarks:}

\noindent
{\bf 1)} Statements 3.3, 3.4 have been obtained with use of not only
Legendre transform but with regard for validity of LS (2.3). One can
give rise to doubt the equivalence of GHS and HS, for instance, for
following structure of superfunction $S_L(\theta)$ satisfying to (2.5)
\begin{eqnarray}
S_L\bigl({\cal A}(\theta), {\stackrel{\ \circ}{\cal A}}(\theta)\bigr) =
T\bigl({\stackrel{\ \circ}{\cal A}}(\theta)\bigr) -
S({\cal A}(\theta))\;. 
\end{eqnarray}
In that case the relationships are valid
\begin{eqnarray}
{} \hspace{-1em}\frac{d}{d\theta}\left(
\frac{\partial_l S_L(\theta)}{
\partial{\stackrel{\ \circ}{\cal A}}{}^{\imath}(\theta)\phantom{x}}\right)
\equiv
\displaystyle\frac{d}{d\theta}\left(\frac{\partial_l
T\bigl({\stackrel{\ \circ}{\cal A}}(\theta)\bigr)}{
\partial{\stackrel{\ \circ}{\cal A}}{}^{\imath}(\theta)\phantom{xxx}}\right)
\equiv 0,\;
 \Theta^H_{\imath}({\cal A}(\theta)) = \displaystyle\frac{\partial_l S_L(
\theta)}{\partial{\cal A}^{\imath}(\theta)\phantom{x}} \equiv -
\frac{\partial_l S({\cal A}(\theta))}{\partial{\cal A}^{\imath}(\theta)
\phantom{xxx}} = 0\,, 
\end{eqnarray}
and GHS has the form
\renewcommand{\theequation}{\arabic{section}.\arabic{equation}\alph{lyter}}
\begin{eqnarray}
\setcounter{lyter}{1}
{} & \displaystyle\frac{d_r {\cal A}^{\imath}(\theta)}{d\theta\phantom{
xxxx}} =
\frac{\partial S_H({\cal A}(\theta), {{\cal A}^{\ast}}(\theta))}{\partial{
\cal A}^{\ast}_{\imath}(\theta)\phantom{xxxxxxxx}} =
\frac{\partial T\bigl({\stackrel{\ \circ}{\cal A}}({{\cal A}^{\ast}}(\theta))
\bigr)}{\partial{\cal A}^{\ast}_{\imath}(\theta)\phantom{xxxxxx}},\
\displaystyle\frac{d_r {\cal A}^{\ast}_{\imath}(\theta)}{d\theta
\phantom{xxxx}} = -
\frac{\partial_l S({\cal A}(\theta))}{\partial{\cal A}^{\imath}(\theta)
\phantom{xxx}}
\;, & {} \\ 
\setcounter{equation}{51}
\setcounter{lyter}{2}
{} &  \Theta^H_{\imath}({\cal A}(\theta)) = 0\;.& {} 
\end{eqnarray}
In that example the HS is the subsystem (3.51a) which has  its 1st subsystem
to be necessarily considered by solvable and then the equivalence of
GHS and HS is obvious;

\noindent
{\bf 2)} if we will ignore the dynamical Eqs.(2.3b) in obtaining
of HS (3.51a) by means of Legendre transform (3.3), (3.4) then the 2nd
subsystem in (3.51a) has the form
\begin{eqnarray}
{\stackrel{\ \circ}{\cal A}}{}^{\ast}_{\imath}(\theta) = 0\;.
\end{eqnarray}
HS being defined in question by the 1st subsystem in (3.51a) and by
Eqs.(3.52) will not be equivalent to HS (3.51a), therefore to GHS (3.51) and
to one's own GHS!

\noindent
\underline{\bf Statement 3.5} (indication for solvability of EHS (ELS))

\noindent
In order that EHS (3.31a,c), ELS (2.9) were solvable it is
sufficient that corresponding master equations (3.19b), (2.6) would be by
double zeros for solutions of HS (3.31a), LS (2.9a,b) respectively.

\noindent
\underline{Proof:} {\bf 1)} Hamiltonian formulation.\\
In force of remark after relation (3.47) it is sufficient to verify in
validity of subsystem (3.46a). Reduce the right-hand side of (3.46a)
to more convenient form for further analysis with allowance made for Jacobi
identity for antibracket (3.27c). Jacobi identity with superfunctions $X(
\theta)$ $\equiv$
$X\bigl(\Gamma(\theta), {\stackrel{\circ}{\Gamma}}(\theta),\theta\bigr)$,
$S_H(\Gamma(\theta))$, $S_{H}(\Gamma(\theta))$ leads to expression
\renewcommand{\theequation}{\arabic{section}.\arabic{equation}}
\begin{eqnarray}
((X(\theta), S_H(\Gamma(\theta)))_{\theta}, S_H(\Gamma(\theta)))_{\theta}=
\textstyle\frac{1}{2}(X(\theta), (S_H(\Gamma(\theta)), S_H(\Gamma(\theta))
)_{\theta})_{\theta}\;. 
\end{eqnarray}
Choosing as  $X(\theta)$ the coordinates $\Gamma^p(\theta)$
obtain the equivalent subsystem for (3.46a)
\begin{eqnarray}
{\textstyle\frac{1}{2}
(\Gamma^p(\theta), (S_H(\Gamma(\theta)), S_H(\Gamma(\theta))
)_{\theta})_{\theta}}_{\mid \overline{\Gamma}(\theta)} = 0\;, 
\end{eqnarray}
which, by virtue of nondegeneracy of the supermatrix $\left\|\omega^{p
q}(\theta)\right\|$ (as the ordinary matrix (!)) (3.24), is equivalent
to the system (3.39).
Having used by hypothesis of the Statement on the structure for master
equation (3.19b), therefore having the form (3.36b), we verify in
validity of the Statement for EHS.

\noindent
{\bf 2)} Lagrangian formulation.

\noindent
The solvability conditions for ELS are the system consisting of Eqs.(2.3b),
(3.43) and
\renewcommand{\theequation}{\arabic{section}.\arabic{equation}\alph{lyter}}
\begin{eqnarray}
\setcounter{lyter}{1}
\frac{d}{d\theta}\left(\chi_a\bigl(
{\cal A}(\theta), {\stackrel{\ \circ}{\cal A}}(\theta)\bigr) - \Theta_{\jmath}
\bigl({\cal A}(\theta), {\stackrel{\ \circ}{\cal A}}(\theta)\bigr)d^{
\jmath}{}_{a}\bigl({\cal A}(\theta), {\stackrel{\ \circ}{\cal A}}(\theta)
\bigr)\right){\hspace{-0.7em}\phantom{\Bigr)}}_{\mid \bar{\cal A}(\theta)}
=0\;,
\end{eqnarray}
or equivalently for  nonvanishing summands on the integral curve $
\bar{\cal A}^{\imath}(\theta)$ for LS (2.3)
\begin{eqnarray}
\setcounter{equation}{55}
\setcounter{lyter}{2}
\frac{d}{d\theta}\left(\chi_a
\bigl({\cal A}(\theta), {\stackrel{\ \circ}{\cal A}}(\theta)\bigr)\right)
{\hspace{-0.7em}\phantom{\Bigr)}}_{
\mid \bar{\cal A}(\theta)} - \frac{d}{d\theta}\left(
\frac{\partial_l S_L(\theta)}{\partial{\cal A}^{\jmath}(\theta)\phantom{x}}
\right)d^{\jmath}{}_{a}\bigl({\cal A}(\theta), {\stackrel{\ \circ}{\cal A}}(
\theta)\bigr)_{\mid \bar{\cal A}(\theta)} =
0\;.  
\end{eqnarray}
The structure of constraints $\chi_a(\theta)$ (2.7) and representation (3.55b)
permit one to state that for solvability of the system (3.43), (3.55a) it
is sufficient the solvability only for subsystem (3.43) which is equivalently
given by the expression (without $P_0(\theta)$ projector)
\begin{eqnarray}
{\stackrel{\;\circ}{{\cal A}^{\imath}}}(\theta)
\frac{\partial_l\phantom{xxx}}{\partial{\cal A}^{\imath}(\theta)}
\frac{
\partial_l S_L\bigl({\cal A}(\theta), {\stackrel{\ \circ}{\cal A}}(\theta)
\bigr)}{\partial{\cal A}^{\jmath}(\theta)\phantom{xxxxxxxx}}{
\hspace{-0.7em}\phantom{\Bigr)}}_{\mid\bar{\cal A}(\theta)} = 0\;.
\end{eqnarray}
Representation (3.56) for solvability conditions (3.43) is equivalent to
Eqs.(3.38). Having used by hypothesis of the Statement on the
Lagrangian master equation (2.6), therefore having the form (3.36a) we verify
in validity of the Statement in this case as well.$_{\textstyle\Box}$

Statement 3.5 permits to
make  more deeper inference on a solvability and on the solvability
conditions (3.46a) for EHS themselves, on ones (3.43) for ELS and etc.
Namely, the
calculation of the $k$th ($k = 1, 2,\ldots$) derivative with respect to
$\theta$ of expression
(3.46a) on the integral curve  $\overline{\Gamma}(\theta)$ value for HS leads
to the expression taking Eq.(3.48) into account
\begin{eqnarray}
\frac{d_r^k}{d{\theta}^k}
{(\Gamma^p(\theta), S_H(\Gamma(\theta)))_{\theta}}_{\mid \overline{\Gamma}(
\theta)}
 = {(\ldots((\Gamma^p(\theta), S_H(\Gamma(\theta)))_{\theta},\underbrace{S_H(
\Gamma(\theta)))_{\theta},\ldots,S_H(\Gamma(\theta))}_{k\,{\rm times}}
)_{\theta}}_{\mid \overline{\Gamma}(\theta)}
\hspace{-1em}= 0. 
\end{eqnarray}
The identical vanishing of the left-hand side of (3.57) implies the vanishing
(without $P_0(\theta)$ projector) in the superfield form the right-hand one as
well for any $k = 1,2\ldots$. That fact is really provided in force of
Statement
3.5 and Jacobi identity of the type (3.53)  with superfunctions
$\bigl(\ldots\bigl(\bigl(
\Gamma^p(\theta), S_H(\Gamma(\theta))\bigr)_{\theta},\underbrace{S_H(\Gamma(
\theta))\bigr)_{\theta},\ldots, S_H(\Gamma(\theta))}_{k - 2\,{\rm times}}
\bigr)_{\theta}$; $S_H(\Gamma(\theta))$; $S_H(\Gamma(\theta))$
\begin{eqnarray}
{} & (\ldots((\Gamma^p(\theta), S_H(\Gamma(\theta)))_{\theta},\underbrace{S_H(
\Gamma(\theta)))_{\theta},\ldots,S_H(\Gamma(\theta))}_{k\,{\rm times}}
)_{\theta} = & {} \nonumber\\
{} & \frac{1}{2}((\ldots((\Gamma^p(\theta),S_H(\Gamma(\theta)))_{\theta},
\underbrace{S_H(\Gamma(\theta)))_{\theta},\ldots,S_H(\Gamma(\theta))}_{k-2\,
{\rm times}})_{\theta},(S_H(\Gamma(\theta)), S_H(\Gamma(\theta)))_{\theta})_{
\theta}\;.
\end{eqnarray}
The analogous statement applied to ELS has the trivial form being written by
the formula
\begin{eqnarray}
\frac{d^k}{d{\theta}^k} \left(\frac{
\partial_l S_L\bigl({\cal A}(\theta), {\stackrel{\ \circ}{\cal A}}(\theta),
\theta\bigr)}{\partial{\cal A}^{\imath}(\theta)\phantom{xxxxxxxxx}}
\right){\hspace{-0.7em}\phantom{\Bigr)}}_{
\mid \bar{\cal A}(\theta)} \hspace{-0.2em}= \left(\frac{\partial}{\partial\theta} +
{\stackrel{\circ}{U}}_{+}(\theta)\right)^k \frac{
\partial_l S_L\bigl({\cal A}(\theta), {\stackrel{\ \circ}{\cal A}}(\theta),
\theta\bigr)}{\partial{\cal A}^{\imath}(\theta)\phantom{xxxxxxxxx}}{
\hspace{-0.7em}\phantom{\Bigr)}}_{
\mid \bar{\cal A}(\theta)} \hspace{-0.2em}\equiv 0,\;k \geq 2
\end{eqnarray}
by virtue of nilpotency of the operators ${\stackrel{\circ}{U}}_{+}(\theta)$,
$\frac{\partial}{\partial\theta}$ [1] and Eq.(2.5).

Finally, having
considered the 2nd derivative on $\theta $ of an arbitrary ${\cal F}(\theta)
\in D^k_{cl}$ calculated on the solution $\overline{\Gamma}(\theta)$ for HS in
scope of  Statement 3.5 validity obtain from (3.48) the formula
\begin{eqnarray}
{\frac{d_r^2 {\cal F}(\theta)}{d\theta^2\phantom{xxx}}}_{\mid \overline{
\Gamma}(\theta)}
= {(({\cal F}(\theta),S_H(\theta))_{\theta}, S_H(\theta))_{\theta}}_{\mid
\overline{\Gamma}(\theta)}\;.
\end{eqnarray}
If the left-hand side of (3.60) vanishes identically, then vanishing of the
right-hand one is provided by Jacobi identity in the form (3.53) for
$X(\theta)={\cal F}(\theta)$. Analogously to proof  of the
expression (3.57) vanishing one can show the same for any ${\cal F}(\theta)$
$\in$ $D^k_{cl}$. Thus, it is proved the technical, but very important for
formulae writing,\\
\underline{\bf Statement 3.6}\\
In fulfilling of the solvability conditions (3.46a) for EHS (3.31a,c)
the formula for translation of arbitrary superfunction ${\cal F}(\theta) \in
D^k_{cl}$ with respect to variable $\theta$ on a constant parameter $\mu \in
{}^1\Lambda_1(\theta)$ along solution $\overline{\Gamma}(\theta)$
for HS (3.31a) is written without $P_0(\theta)$ projector in
front of antibracket
\begin{eqnarray}
\delta_{\mu}{\cal F}\bigl(\Gamma(\theta), {\stackrel{\circ}{\Gamma}}(
\theta),\theta\bigr)_{\mid\overline{\Gamma}(\theta)}
\equiv \frac{d_r {\cal F}(\theta)}{d\theta\phantom{xxx}}_{\mid
\overline{\Gamma}(\theta)}\mu
= \left(\frac{\partial_r}{\partial\theta}{\cal F}(\theta) +
\bigl({\cal F}(\theta), S_H(\Gamma(\theta))\bigr)_{\theta}\right){
\hspace{-0.7em}\phantom{\Bigr)}}_{\mid
\overline{\Gamma}(\theta)}\mu\;. & {}
\end{eqnarray}
In obtaining of (3.61) the conditions (3.19) and following derivation rule
of antibracket have been made use being valid for any ${\cal F}(\theta)$,
${\cal J}(\theta) \in D^k_{cl}$
\begin{eqnarray}
\frac{\partial_r}{\partial\theta}({\cal F}(\theta),{\cal J}(\theta))_{
\theta}
 = ({\cal F}(\theta),\frac{\partial_r}{\partial\theta}{\cal J}(\theta))_{
\theta} + (-1)^{\varepsilon({\cal J}) + 1}
(\frac{\partial_r}{\partial\theta}{\cal F}(\theta),{\cal J}(\theta))_{
\theta}\;. 
\end{eqnarray}
\underline{\bf Remark:} In  Statement 3.5 (later in 3.6 too) in fact it is
considered as the integral curve for EHS (ELS) not only the
integral curve $\overline{\Gamma}(\theta)$ for HS (3.31a)
($\bar{\cal A}(\theta)$
for LS (2.3)) but such $\overline{\Gamma}(\theta)$, ($\bar{\cal A}(\theta)$)
which satisfies to SCHF (3.31c) (SCLF (2.9c,d)) and is denoted further as
\begin{eqnarray}
\hat{\Gamma}^p(\theta) = \overline{\Gamma}^p(\theta) \cap \{\chi^H_a(
\overline{\Gamma}(
\theta)) = 0\}\;\bigl(\hat{\cal A}^{\imath}(\theta) = \bar{\cal A}^{\imath}(
\theta) \cap
\{\chi_a\bigl(\bar{\cal A}(\theta), {\stackrel{\ \circ}{\bar{\cal A}}}(\theta)
\bigr) =0\}\bigr)\;.
\end{eqnarray}
\begin{sloppypar}
\section{Properties of the Differential Systems for  Lag\-ran\-gi\-an and
Hamiltonian Formulations of GSTF}
\end{sloppypar}
\setcounter{equation}{0}

\noindent
The combined analysis of  Statements 3.1, 3.3, 3.4 leads to the result
deserving of the exceptional consideration.

\noindent
\underline{\bf Definition:} We denote the model of GSTF on ${\cal M}_{cl}$ given on
$T_{odd}{\cal M}_{cl}$ ($T^{\ast}_{odd}{\cal M}_{cl}$) with
superfunction $S_L(\theta)$ ($S_H(\theta)$), with corresponding dynamic
system LS, ELS [GHS, HS, EGHS, EHS]  by the triple of
quantities $\bigl(T_{odd}{\cal M}_{cl}, S_L(\theta), A\bigr)$ $[\bigl(
T^{*}_{odd}{\cal M}_{cl}, S_H(\theta), B\bigr)]$, where
$A \in \{{\rm LS}$, ${\rm ELS}\}$ ($B \in \{{\rm GHS}$, ${\rm HS}$,
${\rm EGHS}$, ${\rm EHS}\}$).

\noindent
\underline{\bf Statement 4.1} (diagram of equivalent formulations for GSTF
model on ${\cal M}_{cl}$)

\noindent
The following commutative diagram is valid in formulating of the GSTF model
\begin{eqnarray}
\begin{array}{cccl}
\bigl(T_{odd}{\cal M}_{cl}, S_L(\theta), {\rm LS}\bigr) & \longrightarrow &
\bigl(T_{odd}{\cal M}_{cl}, S_L(\theta), {\rm ELS}\bigr) & {}\\
\downarrow & {} & \downarrow \\
\bigl(T^{\ast}_{odd}{\cal M}_{cl}, S_H(\theta), {\rm GHS}\bigr) &
\longrightarrow & \bigl(T^{\ast}_{odd}{\cal M}_{cl}, S_H(\theta), {\rm EGHS}
\bigr) & {} \\
\downarrow & {} & \downarrow & {} \\
\bigl(T^{\ast}_{odd}{\cal M}_{cl}, S_H(\theta), {\rm HS}\bigr) &
\longrightarrow & \bigl(T^{\ast}_{odd}{\cal M}_{cl}, S_H(\theta), {\rm EHS}
\bigr)& ,
\end{array} 
\end{eqnarray}
whose vertical arrows formally denote the equivalence relation (i.e.
presence,  for instance, of special isomorphic mappings between
$T_{odd}{\cal M}_{cl}$ and
$T^{\ast}_{odd}{\cal M}_{cl}$, $S_L(\theta)$ and
$S_H(\theta)$, LS and GHS (HS)), and horizontal ones are
mappings of the restriction of the left column by means of
SCLF in the 1st row or by SCHF in the 2nd and 3rd rows.

\noindent
\underline{\bf Remarks:}\\
{\bf 1)} Conducting of a detailed proof is the separate problem and requires
except for established correlation for $S_L(\theta)$ and $S_H(\theta)$ by
Legendre transform a more exact  description for $T_{odd}{\cal M}_{cl}$,
$T^{\ast}_{odd}{\cal M}_{cl}$ from geometric viewpoint;

\noindent
{\bf 2)} the right column is described by  $S_L(\theta)$, $S_H(
\theta)$ not explicitly depending on $\theta$;

\noindent
{\bf 3)} if the master equations (2.6), (3.19b) have the form
(3.36) then it
follows from Statement 3.5 the solvability of  the corresponding dynamical
systems of equations from the right column.

ELS (2.9) arises from variational problem on conditional extremum for
superfunctional (2.1) in fulfilling of Eqs.(2.9c,d)
for SCLF. On the other hand, ELS follows from variational problem on
unconditional extremum for superfunctional
\begin{eqnarray}
{} & Z^{(1)}[{\cal A}, \lambda] = \displaystyle\int d\theta
S^{(1)}_L\bigl({\cal A}(\theta), {\stackrel{\ \circ}{\cal A}}(\theta),
\lambda(\theta)\bigr),\
S^{(1)}_L\bigl({\cal A}(\theta), {\stackrel{\ \circ}{\cal A}}(\theta),
\lambda(\theta)\bigr) =
& {} \nonumber \\
{} &
S_L\bigl({\cal A}(\theta), {\stackrel{\ \circ}{\cal A}}(\theta)\bigr) +
\lambda^a(\theta)\left(\chi_a\bigl({\cal A}(\theta), {\stackrel{\ \circ}{
\cal A}}(\theta)\bigr) -
\Theta_{\imath}\bigl({\cal A}(\theta), {\stackrel{\ \circ}{\cal A}}(
\theta)\bigr)
\lambda_{a}{}^{\imath}\bigl({\cal A}(\theta), {\stackrel{\ \circ}{\cal A}}(
\theta)\bigr)\right), & {} 
\end{eqnarray}
with vanishing, after calculation of the 1st order variation for
$Z^{(1)}[{\cal A}, \lambda]$, of the superfields
$\lambda^a(\theta)$ extending the
$T_{odd}{\cal M}_{cl}\times\{\theta\}$ and being by Lagrange multipliers to
SCLF (2.9c,d).
$\lambda^a(\theta)$ are transformed on a some $J$ subrepresentation
from $T\oplus T^{\ast}$  and possess by the gradings
\begin{eqnarray}
\bigl(\varepsilon_P, \varepsilon_{\bar{J}}, \varepsilon\bigr)
\lambda^a(\theta) =
\bigl(\varepsilon_P(\chi_a(\theta)), \varepsilon_{\bar{J}}(\chi_a(\theta)),
\varepsilon(\chi_a(\theta))\bigr),\;a=(A_1, A_2)\;.
\end{eqnarray}
\underline{\bf Statement 4.2}\\
{\bf 1)} Superfunctions $S_L\bigl({\cal A}(\theta), {\stackrel{\ \circ}{\cal
A}}(\theta)\bigr)$, $S^{(1)}_L\bigl({\cal A}(\theta), {\stackrel{\ \circ}{
\cal A}}(\theta),\lambda(\theta)\bigr)$ at absence of the explicit dependence
upon $\theta$ appear by the integrals of solvable ELS (2.9);\\
{\bf 2)} On the integral curve $\hat{\cal A}^{\imath}(\theta)$ for solvable
ELS the superfunctionals $Z[{\cal A}]$ and $Z^{(1)}[{\cal A}, \lambda]$
achieve their critical values for arbitrary configurations of
$\lambda^a(\theta)$
\begin{eqnarray}
Z[\hat{\cal A}] = Z^{(1)}[\hat{\cal A},\lambda] = 0\;.
\end{eqnarray}
\underline{Proof:} {\bf 1)} Expressions
\renewcommand{\theequation}{\arabic{section}.\arabic{equation}\alph{lyter}}
\begin{eqnarray}
\setcounter{lyter}{1}
{} & {} & \hspace{-2em}{\stackrel{\,\circ}{S}}_L\bigl({\cal A}(\theta),
{\stackrel{\ \circ}{\cal A}}(\theta)\bigr)_{\mid\hat{\cal A}(\theta)} =
{\stackrel{\;\circ}{{\cal A}^{\imath}}}(\theta)\displaystyle\frac{\partial_l
S_L(\theta)}{\partial {\cal A}^{\imath}(\theta)\phantom{x}}_{\mid\hat{
\cal A}(\theta)} = 0\;,\\
\setcounter{lyter}{2}
\setcounter{equation}{5}
{} & {} & \hspace{-4em}{\stackrel{\,\circ}{S}}{}^{(1)}_L\bigl({\cal A}(
\theta),
{\stackrel{\ \circ}{\cal A}}(\theta),\lambda(\theta)\bigr)_{\mid\hat{\cal
A}(\theta)} = \hspace{-0.1em}\left(\hspace{-0.2em}{\stackrel{\,\circ}{S}}_L
\bigl({\cal A}(\theta),
{\stackrel{\ \circ}{\cal A}}(\theta)\bigr) +
{\stackrel{\circ}{\lambda}}{}^a(\theta) \chi_a(\theta) +
\lambda^a(\theta){\stackrel{\circ}{\chi}}{}^a(\theta)
(-1)^{\varepsilon(\chi_a)}\hspace{-0.2em}\right)_{\mid\hat{\cal A}(\theta)}
\hspace{-1.5em}
= 0 
\end{eqnarray}
taking the equations $\Theta_{\imath}(\theta)_{\mid\hat{\cal A}(\theta)}$ =
${\stackrel{\circ\;}{\Theta_{\imath}}}(\theta)_{\mid\hat{\cal A}(\theta)} = 0$
into account prove the 1st part of the Statement.

\noindent
{\bf 2)} Identical formulae for the critical values of $Z[{\cal A}]$,
$Z^{(1)}[{\cal A}, \lambda]$
\renewcommand{\theequation}{\arabic{section}.\arabic{equation}}
\begin{eqnarray}
Z[\hat{\cal A}] =
{\stackrel{\,\circ}{S}}_L\bigl({\cal A}(\theta),
{\stackrel{\ \circ}{\cal A}}(\theta)\bigr)_{\mid\hat{\cal A}(\theta)},\;
Z^{(1)}[\hat{\cal A}, \lambda] =
{\stackrel{\,\circ}{S}}{}^{(1)}_L\bigl({\cal A}(\theta),
{\stackrel{\ \circ}{\cal A}}(\theta),\lambda(\theta)\bigr)_{\mid\hat{\cal
A}(\theta)}\;,
\end{eqnarray}
arising from (2.1), (4.2) with allowance made for (4.5) prove the formula
(4.4) as well.${}_{\textstyle\Box}$

From Statement 4.1 it follows the possibility to formulate the proposition
for Hamiltonian formulation being analogous  to Statement 4.2. To this end
indicate that EHS can be obtained from variational problem on conditional
extremum for superfunctional $Z_H[\Gamma]$ (3.13) under condition (3.31c) or
on conditional one for $Z^{(1)}_H[\Gamma, D]$ (3.14a) in
fulfilling of SCHF (3.31c) (after variation with respect to
$\Gamma^p(\theta)$ it is necessary to put $D^{\imath}(\theta) = 0$), or on
unconditional one for the extended by superfields $\lambda^a(\theta)$
superfunctional
\begin{eqnarray}
{} & Z^{(2)}_H[\Gamma,\lambda] =
\displaystyle\int d\theta\bigl( {\stackrel{\ \circ}{\cal
A}}{}^{\imath}(\theta){\cal A}^{\ast}_{\imath}(\theta) -
S_H^{(2)}(\Gamma(\theta),\lambda(\theta))\bigr) \;, & {}\nonumber \\ {} &
\hspace{-1.5em} S_H^{(2)}(\Gamma(\theta),\lambda(\theta)) =
S_H(\Gamma(\theta)) - \lambda^a(\theta)\hspace{-0.2em}\left[\chi^H_a(\Gamma(\theta)) -
\Bigl(\displaystyle\frac{d_r\Gamma^p(\theta)}{d\theta\phantom{xxx}} -
(\Gamma^p(\theta),S_H(\theta))_{\theta}\Bigr)\lambda_{pa}(\Gamma(
\theta))\hspace{-0.1em}\right]. & {} 
\end{eqnarray}
Superfields $\lambda^a(\theta)$ in (4.7) can be chosen by the same as ones
in (4.2). In
order to obtain EHS (3.31a,c) it is necessary to put $\lambda^a(\theta) = 0$
 after variation in $\delta_1 Z^{(2)}_H$.

\noindent
\underline{\bf Statement 4.3}\\
{\bf 1)} Superfunctions
\renewcommand{\theequation}{\arabic{section}.\arabic{equation}\alph{lyter}}
\begin{eqnarray}
\setcounter{lyter}{1}
{} & {} & S_L\bigl({\cal A}(\theta),
{\stackrel{\ \circ}{\cal A}}(\theta)\bigr)_{\mid{\stackrel{\ \circ}{\cal A}}(
\theta) = {\stackrel{\ \circ}{\cal A}}(\Gamma(\theta))} =
{\stackrel{\ \circ}{\cal A}}{}^{\imath}(\theta){\cal
A}^{\ast}_{\imath}(\theta) - S_H(\Gamma(\theta))\;,\\
\setcounter{lyter}{2}
\setcounter{equation}{8}
{} & {} & \tilde{S}_L\bigl({\cal A}(\theta),
{\stackrel{\ \circ}{\cal A}}(\theta), D(\theta)\bigr)_{\mid{\stackrel{\
\circ}{\cal A}}( \theta) = {\stackrel{\ \circ}{\cal A}}(\Gamma(\theta))} =
{\stackrel{\ \circ}{\cal A}}{}^{\imath}(\theta){\cal
A}^{\ast}_{\imath}(\theta) - S_H^{(1)}(\Gamma(\theta), D(\theta))\;,\\
\setcounter{lyter}{3}
\setcounter{equation}{8}
{} & {} & {S}_L^{(1)}\bigl({\cal A}(\theta),
{\stackrel{\ \circ}{\cal A}}(\theta), \lambda(\theta)\bigr)_{\mid{\stackrel{\
\circ}{\cal A}}( \theta) = {\stackrel{\ \circ}{\cal A}}(\Gamma(\theta))} =
{\stackrel{\ \circ}{\cal A}}{}^{\imath}(\theta){\cal
A}^{\ast}_{\imath}(\theta) - S_H^{(2)}(\Gamma(\theta),
\lambda(\theta))
\end{eqnarray}
without explicit dependence upon $\theta$ are the integrals for solvable EGHS
(3.31);

\noindent
{\bf 2)} On the integral curve for solvable EGHS $\hat{\Gamma}^p(\theta)$ for
arbitrary $D^{\imath}(\theta)$, $\lambda^a(\theta)$ the superfunctionals
$Z^{(k)}_H,\;k = 0,1,2$ ($Z^{(0)}_H \equiv Z_H[\Gamma]$) achieve their
critical values
\renewcommand{\theequation}{\arabic{section}.\arabic{equation}}
\begin{eqnarray}
Z_H[\hat{\Gamma}] = Z_H^{(1)}[\hat{\Gamma},{D}] =
Z_H^{(2)}[\hat{\Gamma},{\lambda}] = 0\;.
\end{eqnarray}
\underline{Proof:} {\bf 1)} Relations
\renewcommand{\theequation}{\arabic{section}.\arabic{equation}\alph{lyter}}
\begin{eqnarray}
\setcounter{lyter}{1}
{} & {} &\hspace{-2em}\displaystyle\frac{d}{d\theta}\bigl(
{\stackrel{\;\circ}{{\cal A}^{\imath}}}(\theta){\cal
A}^{\ast}_{\imath}(\theta)\bigr)_{\mid\hat{\Gamma}(\theta)}
 = {\stackrel{\;\circ}{{\cal A}^{\imath}}}(\theta)
{\stackrel{\circ}{{\cal
A}^{\ast}_{\imath}}}(\theta)\bigr)_{\mid\hat{\Gamma}(\theta)}
(-1)^{\varepsilon_{\imath} + 1} = - \frac{1}{2}
(S_H(\hat{\Gamma}(\theta)),
S_H(\hat{\Gamma}(\theta)))_{\theta} = 0\;, \\
\setcounter{lyter}{2}
\setcounter{equation}{10}
{} & {} &\hspace{-2em}
{\stackrel{\,\circ}{S}}_H(\Gamma(\theta))_{\mid\hat{\Gamma}(\theta)} = -
(S_H(\hat{\Gamma}(\theta)),
S_H(\hat{\Gamma}(\theta)))_{\theta} = 0\;,\\
\setcounter{lyter}{3}
\setcounter{equation}{10}
{} & {} & \hspace{-2em}\displaystyle\frac{d}{d\theta}\bigl(D^{\imath}(\theta)
\Theta_{\imath}^H(\Gamma(\theta))\bigr)_{\mid\hat{\Gamma}(\theta)} =
{\stackrel{\circ}{D}}{}^{\imath}(\theta)\Theta^H_{\imath}(\hat{\Gamma}(
\theta)) +
(-1)^{\varepsilon_{\imath}}D^{\imath}(\theta){\stackrel{\,\circ}{
\Theta}}{}^H_{\imath}(\hat{\Gamma}(\theta)) = 0\;,\\
\setcounter{lyter}{4}
\setcounter{equation}{10}
{} & {} & \hspace{-2em}
\displaystyle\frac{d}{d\theta}\left(\lambda^a(\theta)\Biggl[
\chi_a^H(\Gamma(\theta)) -
\Bigl(\displaystyle\frac{d_r\Gamma^p(\theta)}{d\theta\phantom{xxx}} -
(\Gamma^p(\theta),
S_H(\theta))_{\theta}\Bigr)\lambda_{pa}(\Gamma(\theta))\Biggr]
\right){\hspace{-0.7em}\phantom{\Bigr)}}_{\mid\hat{\Gamma}(\theta)} =
 {\stackrel{\circ}{\lambda}}{}^a(\theta)\cdot 0 + \nonumber \\
{} & {} & \hspace{-2em}
(-1)^{\varepsilon(\chi^H_a)}\lambda^a(\theta)\left[{\stackrel{\circ}{
\chi}}{}^H_a(\Gamma(\theta)) + (-1)^{\varepsilon_p}\frac{1}{2}(\Gamma^p(
\theta),(S_H(\theta),
S_H(\theta))_{\theta})_{\theta}\lambda_{pa}(\Gamma(\theta))\right]_{
\mid\hat{\Gamma}(\theta)} = 0
\end{eqnarray}
being obtained with help of solvability conditions for EGHS, representation
(3.36b), formula (3.53) and Statement 3.6 prove the 1st part of the Statement.

\noindent
{\bf 2)} The validity of the 2nd one follows from formulae (4.10) and
\renewcommand{\theequation}{\arabic{section}.\arabic{equation}}
\begin{eqnarray}
{} & Z_H[\hat{\Gamma}] =
{\stackrel{\,\circ}{S}}_L\bigl({\cal A}(\theta),
{\stackrel{\ \circ}{\cal A}}(\Gamma(\theta))\bigr)_{\mid\hat{\Gamma}(
\theta)},\
 Z_H^{(1)}[\hat{\Gamma},{D}] =
{\stackrel{\,\circ}{\tilde{S}}}_L\bigl({\cal A}(\theta),
{\stackrel{\ \circ}{\cal A}}(\Gamma(\theta)),
D(\theta)\bigr)_{\mid\hat{\Gamma}(\theta)}\;, & {} \nonumber \\
{} & Z_H^{(2)}[\hat{\Gamma},{\lambda}] =
{\stackrel{\,\circ}{S}}{}_L^{(1)}\bigl({\cal A}(\theta),
{\stackrel{\ \circ}{\cal A}}(\Gamma(\theta)),
\lambda(\theta)\bigr)_{\mid\hat{\Gamma}(\theta)}.{}_{\textstyle\Box}  & {}
\end{eqnarray}
\underline{\bf Corollary:} From Statement 4.3 it follows the
$S^{(k)}_H(\theta)$, $k=0,1,2$ ($S^{(0)}_H(\theta) = S_H(\Gamma(\theta))$)
appear by the integrals for EHS as well.

Representation (3.31a,c) for EHS  implies for
$\lambda_{pa}(\Gamma(\theta)) = 0$ the following consequence for EHS under
parametrization of $T^{\ast}_{odd}{\cal M}_{cl}$ by the coordinates (3.21)
\renewcommand{\theequation}{\arabic{section}.\arabic{equation}\alph{lyter}}
\begin{eqnarray}
\setcounter{lyter}{1}
{} &  \displaystyle\frac{d_r\varphi^a(\theta)}{d\theta\phantom{xxxx}} =
(\varphi^a(\theta),S_H(\Gamma(\theta)))_{\theta},\ \
\displaystyle\frac{d_r\varphi_a^{\ast}(\theta)}{d\theta\phantom{xxxx}} = 0\;
,& {} \\
\setcounter{lyter}{2}
\setcounter{equation}{12}
{} & \chi_a^H(\Gamma(\theta)) = 0\;. & {}
\end{eqnarray}
Assume now the superfunction $S_H(\theta)$ depends explicitly upon
$\theta$. Then its translation along integral curve $\tilde{\Gamma}^p(
\theta)$
for Hamiltonian subsystem (4.12a) has the form
\renewcommand{\theequation}{\arabic{section}.\arabic{equation}}
\begin{eqnarray}
{} & \hspace{-1em}\delta_{\mu}S_H(\Gamma(\theta),\theta)_{\mid\tilde{
\Gamma}(\theta)}
\hspace{-0.2em}=
\hspace{-0.2em}\left[\displaystyle\frac{\partial_r S_H(\Gamma(\theta),\theta)
}{\partial\theta\phantom{xxxxxxxx}}\hspace{-0.1em} + \hspace{-0.1em}P_0(\theta)
\Biggl(\hspace{-0.1em}\frac{\partial_r
S_H(\theta)}{\partial\varphi^a(\theta)\phantom{x}}\frac{d_r\varphi^a(
\theta)}{d\theta\phantom{xxxx}}\hspace{-0.1em}+ \hspace{-0.1em}
\frac{\partial_r
S_H(\theta)}{\partial\varphi^{\ast}_a(\theta)\phantom{x}}
\frac{d_r\varphi^{\ast}_a(\theta)}{d\theta\phantom{xxxx}}\hspace{-0.1em}
\Biggr)\right]_{\mid\tilde{\Gamma}(\theta)}\hspace{-0.5em}\mu  & {}
\nonumber \\
{} &
=\left[\displaystyle\frac{\partial_r S_H(\tilde{\Gamma}(\theta),\theta)
}{\partial\theta\phantom{xxxxxxxx}} + \frac{1}{2}P_0(\theta)
(S_H(\tilde{\Gamma}(\theta),\theta), S_H(\tilde{\Gamma}(\theta),\theta))_{
\theta}\right]\mu\;, & {} 
\end{eqnarray}
where in view of absence of the solvability conditions  fulfilment
for (4.12a) without use of SCHF (4.12b) one can not apply the Statement 3.6.

Result (4.13) is differred from one (3.17). The requirement of  invariance for
$S_H(\Gamma(\theta),\theta)$ under such translations leads to the equation
taking place on solutions $\tilde{\Gamma}^p(\theta)$ for Eqs.(4.12a)
\begin{eqnarray}
\frac{\partial_r S_H(\tilde{\Gamma}(\theta),\theta)
}{\partial\theta\phantom{xxxxxxxx}} + \frac{1}{2}P_0(\theta)
(S_H(\tilde{\Gamma}(\theta),\theta), S_H(\tilde{\Gamma}(\theta),\theta))_{
\theta} = 0\;.
\end{eqnarray}
A solution for (4.14) will be equivalent to one for (3.18) only in
fulfilling of conditions (3.19). Moreover the transformations
$\delta_{\mu}{\cal F}(\theta)_{\mid \tilde{\Gamma}(\theta)}$,
$\forall {\cal F}(\theta) \in D^k_{cl}$ is devoided of the nilpotency
property
\begin{eqnarray}
\delta_{\mu_1}(\delta_{\mu_2}{\cal F}(\theta))_{\mid\tilde{\Gamma}(\theta)}
\neq 0\;.
\end{eqnarray}
At the same time the restriction of those transformations by SCHF (4.12b),
so that the master equation (3.36b) holds, makes them by
coinciding with  $\delta_{\mu}{\cal F}(
\theta)_{\mid \hat{\Gamma}(\theta)}$ (see (3.61), (3.63)).
\section{Operator $\Delta^{cl}(\theta)$, Master Equation, Translations on
$\theta$ along Integral Curves for HS}
\setcounter{equation}{0}
\subsection{Operator $\Delta^{cl}(\theta)$}

For an arbitrary superfunction ${\cal F}(\theta) \in D^k_{cl}$  the
expression for supercommutator is valid
\begin{eqnarray}
\left[\frac{\partial_r\phantom{xxx}}{\partial\Gamma^p(\theta)},
\frac{d_r}{d\theta}\right]_s{\cal F}(\theta) = 0\;.
\end{eqnarray}
That formula is not in general true in calculating of mentioned
supercommutator in the direction of integral curve
$\overline{\Gamma}(\theta)$ for Hamiltonian part of EHS (3.31a) which has
the form
\begin{eqnarray}
\left[\frac{\partial_r\phantom{xxx}}{\partial\Gamma^p(\theta)},
\frac{d_r}{d\theta}\right]_s{\cal F}(\theta)_{\mid \overline{\Gamma}(\theta)}
= (-1)^{\varepsilon_q}({\cal F}(\theta),(\Gamma^q(\theta),
S_H(\Gamma(\theta)))_{\theta})_{\theta}\omega_{qp}(\theta)_{\mid
\overline{\Gamma}(\theta)}\;.
\end{eqnarray}
In deriving of (5.2) it was used the formulae (3.61), (3.62) and
\renewcommand{\theequation}{\arabic{section}.\arabic{equation}\alph{lyter}}
\begin{eqnarray}
\setcounter{lyter}{1}
{} & {} &\hspace{-1em}\displaystyle\frac{\partial_r{\cal F}(\theta)}{\partial\Gamma^p(
\theta)} \equiv ({\cal
F}(\theta),\Gamma^q(\theta))_{\theta} \omega_{qp}(\theta),\;
\frac{\partial\Gamma^p(\theta)}{\partial\theta
\phantom{xxx}} = 0\;,  \\
\setcounter{lyter}{2}
\setcounter{equation}{3}
{} & {} & \hspace{-1em}(\varepsilon_P,\varepsilon_{\bar{J}},
\varepsilon)\omega_{qp}(\theta) = (\varepsilon_P(\Gamma^p)\hspace{-0.1em} +
\hspace{-0.1em}
\varepsilon_P(\Gamma^q)\hspace{-0.1em} +\hspace{-0.1em}1,
\varepsilon_{\bar{J}}(\Gamma^p) \hspace{-0.1em}+\hspace{-0.1em} \varepsilon_{
\bar{J}}(\Gamma^q), \varepsilon_p \hspace{-0.1em}+ \hspace{-0.1em}
\varepsilon_q \hspace{-0.1em}+ \hspace{-0.1em}1),\;
\omega_{qp}(\theta) = P_0(\theta)\omega_{qp}(\theta)\,,\nonumber \\
{} & {} & \hspace{-1em} \omega_{qp}(\theta) =
-(-1)^{\varepsilon_q\varepsilon_p}\omega_{pq}(\theta),\;
\omega^{dq}(\theta)\omega_{qp}(\theta) = \delta^d{}_p,\;\left\|\omega_{qp}(
\theta)\right\|=\left\|
\begin{array}{cc}
\hspace{-0.2em}0_n & \hspace{-0.3em}-1_n\\
\hspace{-0.2em}1_n & \hspace{-0.3em}0_n
\end{array} \hspace{-0.2em}\right\|,\;d,p,q = \overline{1,2n}
\end{eqnarray}
together with Jacobi identity (3.27c) for ${\cal F}(\theta),
\Gamma^q(\theta), S_H(\Gamma(\theta))$.

\noindent
\underline{\bf Remark:} The fulfilment of Eqs.(3.18) without formal writing
of $P_0(\theta)$ projector leads to the contradictory relation
\renewcommand{\theequation}{\arabic{section}.\arabic{equation}}
\begin{eqnarray}
0=\frac{d^2_r{\cal
F}(\theta)}{d\theta^2\phantom{xxx}}_{\mid\overline{\Gamma}(\theta)} =
{\left({\cal F}(\theta),
\frac{\partial_r S_H(\theta)}{\partial\theta\phantom{xxxx}} + \frac{1}{2}
(S_H(\theta),
S_H(\theta))_{\theta}\right)_{\theta}}_{\mid\overline{\Gamma}(\theta)}
\neq 0\;.
\end{eqnarray}
The contradiction is taken off in fulfilling of Eqs.(3.19) and (3.36b).

The requirement of renewal of supercommutator value in the form (5.1) on
solutions for Hamiltonian part of EGHS (3.31a) by calculating of the
supercommutator on the coordinates $\Gamma^q(\theta)$ of $T^{\ast}_{odd}{
\cal M}_{cl}$ with consequent summation under condition $p=q$ uniquely
leads to relation
\begin{eqnarray}
\left[\frac{\partial_r\phantom{xxx}}{\partial\Gamma^p(\theta)},
\frac{d_r}{d\theta}\right]_s\Gamma^p(\theta)_{\mid\overline{\Gamma}(\theta)}
& = &
(-1)^{\varepsilon_q}\omega_{qp}(\theta){(\Gamma^p(\theta),(\Gamma^q(\theta),
S_H(\theta))_{\theta})_{\theta}}_{\mid\overline{\Gamma}(\theta)} =
\nonumber\\
{} & {} &
2\Delta^{cl}(\theta)S_H(\Gamma(\theta),\theta)_{\mid\overline{\Gamma}(\theta)}
= 0\;.
\end{eqnarray}
In (5.5) it  was introduced the definition  of the 2nd order odd (with respect
to $\varepsilon_P$ and $\varepsilon$ gradings, but not on
$\varepsilon_{\bar{J}}$ one) differential operator
\begin{eqnarray}
\Delta^{cl}(\theta) =  \frac{1}{2}
(-1)^{\varepsilon_q}\omega_{qp}(\theta)(\Gamma^p(\theta),(\Gamma^q(\theta),\
\;)_{\theta})_{\theta}  \equiv  (-1)^{\varepsilon_q}\frac{1}{2}
\frac{\partial_l\phantom{xxx}}{\partial
\Gamma^q(\theta)}\left(\omega^{qp}(\theta)
\frac{\partial_l\phantom{xxx}}{\partial \Gamma^p(\theta)}\right),
\end{eqnarray}
which in Darboux coordinates $({\cal A}^{\imath}(\theta), {\cal
A}^{\ast}_{\imath}(\theta))$ and $(\varphi^a(\theta),
\varphi^{\ast}_a(\theta))$ has the form
\begin{eqnarray}
\Delta^{cl}(\theta) = (-1)^{\varepsilon_{\imath}}
\frac{\partial_l\phantom{xxx}}{\partial {\cal A}^{\imath}(\theta)}
\frac{\partial\phantom{xxxx}}{\partial {\cal A}^{\ast}_{\imath}(\theta)}
 = (-1)^{\varepsilon(\varphi^a)}
\frac{\partial_l\phantom{xxx}}{\partial \varphi^a(\theta)}
\frac{\partial_l\phantom{xxx}}{\partial \varphi_a^{\ast}(\theta)} \;. 
\end{eqnarray}
The geometric interpretation of relation (5.5) is as follows:
antisymplectic divergence of tangent vector ${\stackrel{\circ}{\Gamma}
}{}^p(\theta)$ in any point of the integral curve
$\overline{\Gamma}{}^p(\theta)$ for HS (3.5a,b) is equal to doubled value of
the result of operator $\Delta^{cl}(\theta)$ action on
$S_H(\Gamma(\theta),\theta)$ out of dependence from Eqs.(3.19), (3.31c).
Vanishing of the above divergence is equivalent to validity of the equation
\begin{eqnarray}
\Delta^{cl}(\theta)S_H(\Gamma(\theta),\theta)_{\mid\overline{\Gamma}(\theta)}
= 0\;.
\end{eqnarray}
\underline{\bf Statement 5.1}

\noindent
In calculating of the supercommutator (5.5) along integral curve
$\hat{\Gamma}^p(\theta)$ (3.63) for EHS (3.31a,c) the equation (5.8) at
absence of the explicit dependence upon $\theta$ for $S_H(\theta)$ is the
differential consequence for SCHF (3.31c) with vanishing $\lambda_{pa}(
\Gamma(\theta))$.

\noindent
\underline{Proof:} Using of the fact that action of $\Delta^{cl}(\theta)$ on
$S_H(\Gamma(\theta))$ under comparison of the formulae (5.7),
(3.31c) takes the form in question
\begin{eqnarray}
\Delta^{cl}(\theta)S_H(\Gamma(\theta))_{\mid\hat{\Gamma}(\theta)}
= - (-1)^{\varepsilon(\varphi^a)}
\frac{\partial_l\chi_a^H(\Gamma(\theta))}{\partial
\varphi_a^{\ast}(\theta)\phantom{xxx}}{\hspace{-0.7em}\phantom{\Bigr)}}_{
\mid\hat{\Gamma}(\theta)} = 0\;,
\end{eqnarray}
we arrive directly to validity of the Statement.${}_{\textstyle\Box}$

\noindent
\underline{\bf Remarks:}

\noindent
{\bf 1)} The relation to be analogous to (5.5) in the case of symplectic
geometry and usual Hamiltonian equations is identically fulfilled;

\noindent
{\bf 2)} the conservation of supercommutator value (5.5) on the solutions
$\hat{\Gamma}^p(\theta)$ for EHS in terms of Darboux coordinates
${\cal A}^{\imath}(\theta), {\cal A}^{\ast}_{\imath}(\theta)$ leads to
consequence for (5.5)
\renewcommand{\theequation}{\arabic{section}.\arabic{equation}\alph{lyter}}
\begin{eqnarray}
\setcounter{lyter}{1}
\left[\frac{\partial\phantom{xxxx}}{\partial{\cal A}^{\imath}(\theta)},
\frac{d_r}{d\theta}\right]_s{\cal
 A}^{\imath}(\theta)_{\mid\hat{\Gamma}(\theta)} =
\Delta^{cl}(\theta)S_H(\Gamma(\theta))_{\mid\hat{\Gamma}(\theta)}=0\;,
\\
\setcounter{lyter}{2}
\setcounter{equation}{10}
\left[\frac{\partial_r\phantom{xxx}}{\partial{\cal A}^{\ast}_{\imath
}(\theta)},\frac{d_r}{d\theta}\right]_s{\cal
 A}^{\ast}_{\imath}(\theta)_{\mid\hat{\Gamma}(\theta)} =
\Delta^{cl}(\theta)S_H(\Gamma(\theta))_{\mid\hat{\Gamma}(\theta)}=0
\;.
\end{eqnarray}

The requirement of Eqs.(3.19b) fulfilment
for any configuration $\Gamma^p(\theta)$ irrespective to
EHS (3.31a,c) appears by  stronger condition on the structure of
$S_H(\Gamma(\theta))$
\renewcommand{\theequation}{\arabic{section}.\arabic{equation}}
\begin{eqnarray}
(S_H({\Gamma}(\theta)), S_H({\Gamma}(\theta)))_{\theta} =
0,\;\forall\Gamma^p(\theta) \in T^{\ast}_{odd}{\cal M}_{cl}\;.
\end{eqnarray}
On the Lagrangian formalism language
the situation considered in Sec.II beginning with Eqs.(2.16a) up to (2.22)
inclusively corresponds to this equation.

\noindent
\underline{\bf Statement 5.2} (solvability criterion for Eq.(5.11))

\noindent
For existence of a solution for Eq.(5.11) it is necessary and sufficient to
realize the following representation $\forall\Gamma^p(\theta) \in
T^{\ast}_{odd}{\cal M}_{cl}$
\renewcommand{\theequation}{\arabic{section}.\arabic{equation}\alph{lyter}}
\begin{eqnarray}
\setcounter{lyter}{1}
{} & \displaystyle\frac{\partial_l S_H(\Gamma(\theta))}{\partial
\Gamma^p(\theta)\phantom{xxx}} = \left(
\frac{\partial_l S_H(\Gamma(\theta))}{\partial
\varphi^a(\theta)\phantom{xxx}},
\lambda^{ba}(\Gamma(\theta))
\frac{\partial_l S_H(\Gamma(\theta))}{\partial
\varphi^a(\theta)\phantom{xxx}}\right),\
p=\overline{1,2n}\;, & {}\\
\setcounter{lyter}{2}
\setcounter{equation}{12}
{} & \hspace{-2em}\varepsilon(\lambda^{ba}) =  \varepsilon(\varphi^b) +
\varepsilon(\varphi^a) + 1,\;\lambda^{ba} =
-(-1)^{\varepsilon(\varphi^a)\varepsilon(\varphi^b)}\lambda^{ab},\;
a,b=(A_1,A_2),(B_1,B_2) = \overline{1,n} & {} 
\end{eqnarray}
with superfunctions $\lambda^{ab}(\theta)$ $\in$ $C^{k\ast}$\footnote{
representation $\Gamma^p(\theta)=(\varphi^a(\theta), \varphi^{
\ast}_a(\theta))$ being used in (5.12) can be distinguished from adopted one
in (3.21) by value of parameter $n_1$. It is main that antibracket in terms
of $(\varphi^a(\theta),\varphi^{\ast}_a(\theta))$ has the same form as in
(3.26)}.

\noindent
\underline{Proof:}
{\bf 1)} The sufficiency by a certain way follows from representation
(3.26) for antibracket and the equality respectively
\renewcommand{\theequation}{\arabic{section}.\arabic{equation}}
\begin{eqnarray}
(S_H(\theta), S_H(\theta))^{(\varphi,\varphi^{\ast})}_{
\theta} = 2\frac{\partial_r S_H(\theta)}{\partial
\varphi^a(\theta)\phantom{x}}\frac{\partial_l S_H(\theta)}{\partial
\varphi^{\ast}_a(\theta)\phantom{x}},\
\frac{\partial_l S_H(\theta)}{\partial
\varphi^{\ast}_a(\theta)\phantom{x}} = \lambda^{ab}(\theta)
\frac{\partial_l S_H(\theta)}{\partial
\varphi^b(\theta)\phantom{x}}\;.
\end{eqnarray}
{\bf 2)} Let us represent the antibracket  in the invariant coordinate-free
form
\begin{eqnarray}
({\cal F}(\theta),{\cal J}(\theta))_{\theta} = \omega_{\theta}({\rm
osgrad}{\cal F}(\theta), {\rm osgrad}{\cal J}(\theta)),\
\forall {\cal F}(\theta), {\cal J}(\theta) \in C^{k\ast}\times\{\theta\}
\;,
\end{eqnarray}
where $\omega_{\theta}$ is the nondegenerate
closed odd differential 2-form on $T^{\ast}_{odd}{\cal M}_{cl}$. That form
directly inherits all above properties from ones for antibracket (3.25),
(3.27).  The explicit form for antibracket (3.25) in local coordinates
$\Gamma^p(\theta)$, $({\cal A}^{ \imath}(\theta),{\cal
A}^{\ast}_{\imath}(\theta))$ on $T^{\ast}_{odd}{\cal M}_{cl}$ or in
coordinates $(\varphi^a(\theta),\varphi^{\ast}_a(\theta))$ of the type
(3.26a) permits one to present $\omega_{\theta}$ by the formulae
\begin{eqnarray}
\omega_{\theta} =
\frac{1}{2}\omega_{qp}(\theta)d\Gamma^p(\theta)\land d\Gamma^q(\theta)=
d{\cal A}^{\imath}(\theta)\land d{\cal A}^{\ast}_{\imath}(\theta) =
d\varphi^a(\theta)\land d\varphi^{\ast}_a(\theta)\;,
\end{eqnarray}
where $"\land"$ is the sign of
external multiplication. Grassmann parities $\varepsilon_P$,
$\varepsilon_{\bar{J}}$, $\varepsilon$ for elements $d\Gamma^p(\theta)$,
$d{\cal A}^{\imath}(\theta)$, $d{\cal A}^{\ast}_{\imath}(\theta)$,
$d\varphi^a(\theta)$, $d\varphi^{\ast}_a(\theta)$ are defined to be
coinciding with $\varepsilon_P$, $\varepsilon_{\bar{J}}$, $\varepsilon$
gradings of the coordinates $\Gamma^p(\theta)$, ${\cal A}^{\imath}(\theta)$,
${\cal A}^{\ast}_{\imath}(\theta)$,
$\varphi^a(\theta)$, $\varphi^{\ast}_a(\theta)$ respectively.

Symbol ${\rm osgrad}{\cal F}(\theta)$ in (5.14) denotes by definition the odd
(with respect to $\varepsilon_P$, $\varepsilon$ but not
$\varepsilon_{\bar{J}}$) skew-symmetric (antisymplectic) gradient of
superfunction ${\cal F}(\theta)$. Obtain in coordinates
$\Gamma^p(\theta)$ on $T^{\ast}_{odd}{\cal M}_{cl}$ the expression
for coordinates of ${
\rm osgrad}{\cal F}(\theta)$ as the vector field on $T^{\ast}_{odd}{\cal
M}_{cl}$ in the basis
$\{\frac{\partial_l\phantom{xxx}}{\partial\Gamma^p(\theta)}\}$
with regard for (3.24)
\begin{eqnarray}
\left({\rm osgrad}{\cal F}(\theta)\right)^p =
\frac{\partial_r {\cal F}(\theta)}{\partial
\Gamma^q(\theta)\phantom{x}}\omega^{qp}(\theta)\;.
\end{eqnarray}
In every point of $T^{\ast}_{odd}{\cal M}_{cl}$ a vector fields subset
turning the form $\omega_{\theta}$ into zero  forms the isotropic
subsuperspace
$L_{\theta}$ of the maximal dimension $(k,n-k)$, where $k$ and $(n-k)$ are
dimensions of even and odd (with respect to $\varepsilon$ grading) isotropic
subsuperspaces respectively. For the case of maximal dimension $L_{\theta}$
is called by the Lagrangian superspace.

From Eq.(5.11) and relation (5.14) it follows that ${\rm osgrad}S_H(\theta)$
$\in$ $L_{\theta}$ in any point from $T^{\ast}_{odd}{\cal M}_{cl}$. In its
turn, it means that
in basis $\{\frac{\partial_l\phantom{xxx}}{\partial\Gamma^p(\theta)}\}$ in an
arbitrary point from $T^{\ast}_{odd}{\cal M}_{cl}$, with coordinates
$\Gamma^p(\theta)$, among coordinates of the vector ${\rm osgrad}S_H(\theta)$
at most $n$ are linearly independent. By virtue of supermatrix
$\left\|\omega^{qp}(\theta)\right\|$ nondegeneracy in (5.16) the last
conclusion is valid for column-vector $\frac{\partial_l S_H(\theta)}{\partial
\Gamma^p(\theta)\phantom{x}}$ as well, i.e. the representation (5.12a) with
$\lambda^{ab}(\Gamma(\theta))$ possessing by properties (5.12b)
follows from (5.11) and (5.13).${}_{\textstyle\Box}$

\noindent
\underline{\bf Remark:} Coordinates $({\cal A}^{\imath}(\theta), {\cal
A}^{\ast}_{\imath}(\theta))$ and $(\varphi^a(\theta), \varphi^{\ast}_a(
\theta))$ of the same point from $T^{\ast}_{odd}{\cal M}_{cl}$ defined in
(3.21) are equivalent in the sense that they are connected by anticanonical
transformation.

\noindent
\underline{\bf Statement 5.3} (indication for solvability of Eq.(5.11))

\noindent
For fulfilment of Eq.(5.11) it is necessary to be valid the following
condition on solutions
for equations
$\frac{\partial_l S_H(\theta)}{\partial\Gamma^q(\theta)\phantom{x}}= 0$ for
a some parametrization of $\Gamma^p(\theta)$ = $(\varphi^a(\theta),
\varphi^{\ast}_a(\theta))$
\begin{eqnarray}
{} & {\rm
rank}\left\|\displaystyle\frac{\partial_r\phantom{xxx}}{\partial
\Gamma^p(\theta)}
\frac{\partial_l S_H(\Gamma(\theta))}{\partial\Gamma^q(\theta)\phantom{xxx}}
\right\|=
{\rm rank}\left\|\displaystyle\frac{\partial_r\phantom{xxx}}{\partial
\varphi^a(\theta)}\frac{\partial_l
S_H(\Gamma(\theta))}{\partial\varphi^b(\theta)\phantom{xxx}}\right\|
= l \leq n,\ l = (l_{+},l_{-})\;.  & {} 
\end{eqnarray}
The proof follows from the results of Statement 5.2, namely from
representation (5.12).${}_{\textstyle\Box}$

Note that Lagrangian subsuperspace $L_{\theta}$ is defined ambiguously.
Relation (5.17) itself is the consequence for formula (2.22) taking account of
(2.4a), (3.3), (3.4). The comparison of the condition (5.17) with
one (2.21) expressed in terms of $S_H(\Gamma(\theta))$
and coordinates $({\cal A}^{\imath}(\theta), {\cal
A}^{\ast}_{\imath}(\theta))$ on $T^{\ast}_{odd}{\cal M}_{cl}$,
with regard for (3.35d), in the form
\begin{eqnarray}
{\rm rank}K^{\ast}(\theta) = {\rm
rank}\left\|(S_H'')^{\imath\jmath}(\Gamma(\theta))\right\| = n\;
,
\end{eqnarray}
in fact leads to independence of
$S_H(\Gamma(\theta))$ upon ${\cal A}^{ \imath}(\theta)$ that is very strong
restriction on a structure of the model. Ignoring of the condition (5.18)
after Legendre transform of $S_L(\theta)$ realization results in
completely different dynamics.

\noindent
\underline{\bf Statement 5.4}

\noindent
$S_H(\Gamma(\theta))$ satisfying to (5.11) is the integral for HS (3.5a,b)
constructed with respect to given $S_H(\Gamma(\theta))$.

Really the translation of $S_H(\Gamma(\theta))$ with respect to $\theta$
along integral curve
$\check{\Gamma}^p(\theta)$ of the pointed HS on $\mu \in {}^1\Lambda_1(
\theta)$  is given by the formula
\begin{eqnarray}
\delta_{\mu}S_H(\Gamma(\theta))_{\mid\check{\Gamma}(\theta)} =
(S_H(\check{\Gamma}(\theta)), S_H(\check{\Gamma}(\theta)))_{\theta}\mu = 0
\;.
\end{eqnarray}
This fact provides the solvability of corresponding HS (3.5a,b)
in the sense of the type (2.12) relations as well and validity of
Statement 3.6 in question.

\noindent
\underline{\bf Remark:} Without fulfilment of equation (5.11) HS itself does
not satisfy to solvability conditions (i.e. is unsolvable), and therefore
the Statement 3.6 is not valid for such HS in the sense of the Eqs.(3.8)
nonfulfilment. The removal of that contradiction is provided by restriction of
the right-hand sides of HS (3.5a,b) until their $P_0(\theta)$ components
[1]. But the latter is the explicit violation of the superfield form of
equations and  appears by the obstacle for further interpretation of HS role
in the quantization method itself.

\noindent
\underline{\bf Statement 5.5}

\noindent
There exists a parametrization of $T^{\ast}_{odd}{\cal M}_{cl}$
by Darboux coordinates $(\varphi^a(\theta)$, $\varphi^{\ast}_a(\theta))$
that in fulfilling  of Eq.(5.11) the identities are valid
\begin{eqnarray}
- \frac{\partial_l
S_H(\Gamma(\theta))}{\partial\varphi^a(\theta)\phantom{xxx}} =
(\varphi^{\ast}_a(\theta),
S_H(\Gamma(\theta)))_{\theta}^{(\varphi,\varphi^{\ast})} = 0,\;
\forall\Gamma^p(\theta) \in T^{\ast}_{odd}{\cal M}_{cl}\;.
\end{eqnarray}
HS in this coordinates has the representation
\begin{eqnarray}
\frac{d_r\varphi^a(\theta)}{d\theta\phantom{xxxx}} =
(\varphi^a(\theta),
S_H(\theta))_{\theta}^{(\varphi,\varphi^{\ast})},\ \
\frac{d_r\varphi^{\ast}_a(\theta)}{d\theta\phantom{xxxx}} = 0\;.
\end{eqnarray}

From Statements 5.2, 5.3 in fact it follows the validity of the last
Statement for $\lambda^{ab}(\Gamma(\theta))=0$, but more detailed proof is
not carried out here.

Relations (5.20) represented through antibracket are conserved under
anticanonical transformations together with form of the operator
$\Delta^{cl}(\theta)$ action on an arbitrary superfunction from
$C^{k\ast}$. Then the following equation holds in
any local coordinates on $T^{\ast}_{odd}{\cal M}_{cl}$ (connected with each
other via anticanonical transformations)
\renewcommand{\theequation}{\arabic{section}.\arabic{equation}}
\begin{eqnarray}
\Delta^{cl}(\theta)S_H(\Gamma(\theta)) = 0,\;
\forall\Gamma^p(\theta) \in T^{\ast}_{odd}{\cal M}_{cl} 
\end{eqnarray}
defining the further limitation for $S_H(\Gamma(\theta))$. Really, having
acted on Eqs.(5.20) by
$\frac{\partial_l\phantom{xxx}}{\partial\varphi^{\ast}_a(\theta)}$  obtain
the realization of the result (5.22).
Thus, Eq.(5.22) appears by the consequence of Eq.(5.11)(!) Equation
(5.22) permits among them the same interpretation as it was made for relation
(5.5).
\subsection{Translations on $\theta$ along Integral Curves for \protect \\
the Hamiltonian Systems}

Let us classify some properties of translations with respect to
$\theta$ on $\mu \in {}^1\Lambda_1(\theta)$ of an arbitrary ${\cal
F}(\Gamma(\theta),\theta)$ $\in$ $C^{k\ast} \times \{\theta\}$ along
integral curves $\overline{\Gamma}(\theta)$,
$\hat{\Gamma}(\theta)$, $\tilde{\Gamma}(\theta)$, $\check{\Gamma}(\theta)$ of
the corresponding differential systems of equations: HS (3.5a,b) (or
equivalently (3.31a)), EHS (3.31a,c) with master equation (3.36b), HS given
by (4.12a) and HS described in Statement 5.4 respectively
\renewcommand{\theequation}{\arabic{section}.\arabic{equation}\alph{lyter}}
\begin{eqnarray}
\setcounter{lyter}{1}
\hspace{-2em}
\delta_{\mu}{\cal F}(\theta)_{\mid\overline{\Gamma}(
\theta)} & = &
\overline{s}(\overline{\Gamma}(\theta)){\cal F}(\overline{\Gamma}(
\theta),\theta)\mu =
\left[\frac{\partial_r{\cal F}(\overline{\Gamma}(\theta),\theta)}{\partial
\theta\phantom{xxxxxxx}} +
({\cal F}(\overline{\Gamma}(\theta),\theta),
S_H(\overline{\Gamma}(\theta)))_{\theta}\right]\mu\,,\\
\setcounter{lyter}{2}
\setcounter{equation}{23}
\hspace{-2em}
\delta_{\mu}{\cal F}(\theta)_{\mid\hat{\Gamma}(
\theta)} & = & \hat{s}(\hat{\Gamma}(\theta)){\cal F}(\hat{\Gamma}(
\theta),\theta)\mu =
\left[\frac{\partial_r{\cal F}(\hat{\Gamma}(\theta),\theta)}{\partial
\theta\phantom{xxxxxxxx}} +
({\cal F}(\hat{\Gamma}(\theta),\theta),
S_H(\hat{\Gamma}(\theta)))_{\theta}\right]\mu\nonumber \\
{} & = &
\left[\frac{\partial_r{\cal F}(\hat{\Gamma}(\theta),\theta)}{\partial
\theta\phantom{xxxxxxxx}} + \frac{\partial_r{\cal
F}(\hat{\Gamma}(\theta),\theta)}{\partial\varphi^a(\theta)\phantom{xxxx}}
\frac{\partial_l
S_H(\hat{\Gamma}(\theta))}{\partial\varphi^{\ast}_a(\theta)\phantom{xxx}}
\right]\mu\,,\\
\setcounter{lyter}{3}
\setcounter{equation}{23}
\hspace{-2em}
\delta_{\mu}{\cal F}(\theta)_{\mid\tilde{\Gamma}(\theta)} & = &
\tilde{s}(\tilde{\Gamma}(\theta)){\cal F}(\tilde{\Gamma}(\theta),
\theta)\mu =
\left[\frac{\partial_r{\cal F}(\tilde{\Gamma}(\theta),\theta)}{\partial
\theta\phantom{xxxxxxxx}} + \frac{\partial_r{\cal
F}(\tilde{\Gamma}(\theta),\theta)}{\partial\varphi^a(\theta)\phantom{xxxx}}
\frac{\partial_l
S_H(\tilde{\Gamma}(\theta))}{\partial\varphi^{\ast}_a(\theta)\phantom{xxx}}
\right]\mu\,,\\
\setcounter{lyter}{4}
\setcounter{equation}{23}
\hspace{-2em}
\delta_{\mu}{\cal F}(\theta)_{\mid\check{\Gamma}(\theta)} & = &
\check{s}(\check{\Gamma}(\theta)){\cal F}(\check{\Gamma}(\theta),
\theta)\mu =
\left[\frac{\partial_r{\cal F}(\check{\Gamma}(\theta),\theta)}{\partial
\theta\phantom{xxxxxxxx}} + ({\cal F}(\check{\Gamma}(\theta),\theta),
S_H(\check{\Gamma}(\theta)))_{\theta}\right]\mu\,.
\end{eqnarray}
Call the differential operators of the 1st order $\bar{s}(\theta)$,
$\hat{s}(\theta)$, $\tilde{s}(\theta)$, $\check{s}(\theta)$ by the generators
of translations on $\theta$ along corresponding integral curve.

Formally formulae (5.23c,d) point to coincidence with respect to the form with
corresponding transformations (5.23b,a) but it is not in general the case. The
action of the generators on $S_H(\Gamma(\theta))$ is equal respectively
\renewcommand{\theequation}{\arabic{section}.\arabic{equation}\alph{lyter}}
\begin{eqnarray}
\setcounter{lyter}{1}
{} & {} &\overline{s}(\theta)S_H(\overline{\Gamma}(\theta)) = (
S_H(\overline{\Gamma}(\theta)), S_H(\overline{\Gamma}(\theta)))_{\theta} \neq
0\;,\\
\setcounter{lyter}{2}
\setcounter{equation}{24}
{} & {} &\hat{s}(\theta)S_H(\hat{\Gamma}(\theta)) =
(S_H(\hat{\Gamma}(\theta)), S_H(\hat{\Gamma}(\theta)))_{\theta} =
0\;, \\
\setcounter{lyter}{3}
\setcounter{equation}{24}
{} & {} & \tilde{s}(\theta)S_H(\tilde{\Gamma}(\theta)) = \frac{1}{2}
(S_H(\tilde{\Gamma}(\theta)), S_H(\tilde{\Gamma}(\theta)))_{\theta} \neq 0\;
 ,\\
\setcounter{lyter}{4}
\setcounter{equation}{24}
{} & {} & \check{s}(\theta)S_H(\check{\Gamma}(\theta)) =
(S_H(\check{\Gamma}(\theta)), S_H(\check{\Gamma}(\theta)))_{\theta}= 0\;
 .
\end{eqnarray}
Additional relation (4.12b) applied to $\tilde{s}(\theta)$ in its definition
permits to make conclusion that
\renewcommand{\theequation}{\arabic{section}.\arabic{equation}}
\begin{eqnarray}
\tilde{s}(\tilde{\Gamma}(\theta))_{\mid\chi_a^H(\tilde{
\Gamma}(\theta)) = 0} = \hat{s}(\tilde{\Gamma}(\theta))\;.
\end{eqnarray}
It is not difficult to establish the properties of
possession by nilpotency for the transformations (5.23), that is equivalently
reformulated for their generators
\renewcommand{\theequation}{\arabic{section}.\arabic{equation}\alph{lyter}}
\begin{eqnarray}
\setcounter{lyter}{1}
{} & {} &\overline{s}^2(\theta) =
- \frac{1}{2}\left((
S_H(\overline{\Gamma}(\theta)), S_H(\overline{\Gamma}(\theta)))_{\theta},
\;\;\;\right)_{\theta} \neq 0\;,
\\
\setcounter{lyter}{2}
\setcounter{equation}{26}
{} & {} &\hat{s}^2(\theta) =
- \frac{1}{2}\left((
S_H(\hat{\Gamma}(\theta)), S_H(\hat{\Gamma}(\theta)))_{\theta},
\;\;\;\right)_{\theta} = 0\;,
\\
\setcounter{lyter}{3}
\setcounter{equation}{26}
{} & {} & \tilde{s}^2(\theta) = \left(
\frac{\partial_r\phantom{xxx}}{\partial\varphi^b(\theta)}
\frac{\partial_l
S_H(\tilde{\Gamma}(\theta))}{\partial\varphi^{\ast}_a(\theta)\phantom{xxxx}}
\right)
\frac{\partial_l
S_H(\tilde{\Gamma}(\theta))}{\partial\varphi^{\ast}_b(\theta)\phantom{xxxx}}
\frac{\partial_l\phantom{xxx}}{\partial\varphi^a(\theta)} \neq 0
\;,\\
\setcounter{lyter}{4}
\setcounter{equation}{26}
{} & {} & \check{s}^2(\theta) =
- \frac{1}{2}\left((
S_H(\check{\Gamma}(\theta)), S_H(\check{\Gamma}(\theta)))_{\theta},
\;\;\;\right)_{\theta}  = 0\;.
\end{eqnarray}
Let us note that in this case in spite of equality (5.25) validity the
relations hold
\renewcommand{\theequation}{\arabic{section}.\arabic{equation}}
\begin{eqnarray}
\hat{s}^2(\tilde{\Gamma}(\theta)) \neq
\tilde{s}^2(\tilde{\Gamma}(\theta))_{\mid\chi_a^H(\tilde{
\Gamma}(\theta))=0},\;
\tilde{s}^2(\tilde{\Gamma}(\theta))_{\mid\chi_a^H(\tilde{
\Gamma}(\theta))=0}
\neq 0\;.
\end{eqnarray}
In obtaining of the formulae (5.26) the antibracket's properties
(3.25), (3.27), master equation (5.11) for (5.26d), Eq.(3.19b)
in the form (3.36b)  and its
differential consequence (3.39) for (5.26b) (see remark after (3.62)) have
been made use.

In terms of operators $\hat{s}(\theta)$, $\check{s}(\theta)$ we have proved
the

\noindent
\underline{\bf Statement 5.6}

\noindent
The nilpotency of operators $\hat{s}(\theta)$, $\check{s}(\theta)$ appears by
sufficient condition for the type (2.12) solvability  respectively for EHS
(3.31a,c) and HS of the form (3.31a), built with respect to $S_H(\Gamma(
\theta))$ being by a solution for master equation (5.11).

The generators in (5.23a,b,d) $\overline{s}(\theta)$, $\hat{s}(\theta)$,
$\check{s}(\theta)$ given through only the operator $\frac{\partial}{
\partial\theta}$ and antibracket do not depend on a concrete choice of the
coordinates $\Gamma^p(\theta)$ on $T^{\ast}_{odd}{\cal M}_{cl}$ connected
with each other via anticanonical transformations. Therefore, the translation
transformation with respect to $\theta$ along integral curve
$\check{\Gamma}(\theta)$ of HS (3.31a) given, equivalently, by the Statement
5.5. (Eqs.(5.21)) together with Eq.(5.20) is defined by a generator
$s_{\varphi}(\theta)$ coinciding with $\check{s}(\theta)$. Properties
(5.24d), (5.26d) are valid for  $s_{\varphi}(\theta)$ as well.

\noindent
\underline{\bf Remark:} Under attainment of the maximal value for rank in the
expression (5.17) the corresponding Lagrangian subsupermanifold ${\cal
M}^{L}_{cl}$ $\subset$ $T^{\ast}_{odd}{\cal M}_{cl}$ in terms of coordinates
$\varphi^{a}(\theta), \varphi^{\ast}_a(\theta)$ is defined one-valued in the
framework of the Statement 5.5 and is parametrized only by $\varphi^{\ast}_a(
\theta)$.
On the other
hand, the projection of solution $\check{\Gamma}(\theta)$ for HS of the type
(3.31a), in
fulfilling of (5.11), on subsystem of the form (5.21) without realization of
the identity (5.20) leads to unsolvable HS. That fact means the
supermanifold ${\cal M}^{L}_{cl}$ is not already parametrized
by superfields $\varphi^{
\ast}_a(\theta)$ (value of parameter $n_1$ in (3.21)  characterizing
index $a$ is now differred from formulated in Statement 5.5). Corresponding
$\tilde{s}_{\varphi}(\theta)$ being by the translation generator on $\theta$
along projection for given
$\check{\Gamma}(\theta)$ on the above-described type (5.21) subsystem
is defined by the formula
\begin{eqnarray}
\tilde{s}_{\varphi}(\theta){\cal F}(\theta) = \frac{\partial_r{\cal F}(
\theta)}{\partial\theta\phantom{xxx}} +
\frac{\partial_r{\cal F}(\theta)}{ \partial\varphi^a(\theta)}
\frac{\partial_l S_H(\check{\varphi}(\theta), \check{\varphi}^{\ast}(
\theta))}{ \partial\varphi^{\ast}_a(\theta)\phantom{xxxxxxxx}},\
{\cal F}(\theta)\in D^k_{cl}\;,
\end{eqnarray}
has
$S_H(\check{\Gamma}(\theta))$ as an eigenfunction with zero eigenvalue,
but is not nilpotent
\begin{eqnarray}
\tilde{s}_{\varphi}(\theta)S_H(\check{\varphi}(\theta), \check{\varphi}^{
\ast}(\theta)) = \frac{1}{2}{(S_H(\theta),
S_H(\theta))_{\theta}}_{\mid (\check{\varphi}(\theta),
\check{\varphi}^{\ast}(\theta))} = 0,\ \
\tilde{s}^2(\theta) = \tilde{s}_{\varphi}^2(\theta) \neq 0\;.
\end{eqnarray}
It is the same operator $\tilde{s}_{\varphi}(\theta)$ for $n_1 = 0$
($\varphi^a(\theta)$ $\equiv$ ${\cal A}^{\imath}(\theta)$) is considered in
the BV quantization method (without auxiliary, for instance, ghost-antighost
type superfields  and for $\theta=0$).

HS (3.5a,b) with $S_H({\Gamma}(\theta))$ satisfying to (5.11) corresponds in
the Lagrangian formulation to solvable LS (2.3) with
$S_L\bigl({\cal A}(\theta),{\stackrel{\ \circ}{\cal A}}(\theta)\bigr)$
satisfying to Eqs.(2.5), (2.6) for any ${\cal A}^{\imath}(\theta)$ $\in$
${\cal M}_{cl}$
\begin{eqnarray}
{\stackrel{\;\circ}{{\cal A}^{\imath}}}(\theta)
\frac{\partial_l S_L(\theta)}{\partial{\cal A}^{\imath}(\theta)} = 0,\;
\forall{\cal A}^{\imath}(\theta) \in {\cal M}_{cl}
\;.
\end{eqnarray}
Condition (5.18) for $S_H({\Gamma}(\theta))$, meaning for $S_L(\theta)$ being
one-valued constructed via Legendre transform of the given
$S_H({\Gamma}(\theta))$ with respect to ${\cal A}^{\ast}_{\imath}(\theta)$ in
view  of the supermatrix $K(\theta)$ (2.21) nondegeneracy, leads to identity
\begin{eqnarray}
\frac{\partial_l S_L\bigl({\cal A}(\theta),{\stackrel{\ \circ}{\cal A}}(
\theta)\bigr)}{\partial{\cal A}^{\imath}(\theta)\phantom{xxxxxxx}} = 0,\;
\forall{\cal A}^{\imath}(\theta) \in {\cal M}_{cl}\;
.
\end{eqnarray}
Formula (5.31) means the independence of
S$_{L}(\theta)$ upon ${\cal A}^{\imath}(\theta)$ that leads to the trivial
dynamics on $\theta$.  The latter  implies that LS (2.3) is restricted to
proper trivial subsystem (2.3a). Note that any superfields ${\cal
A}^{\imath}(\theta)$ satisfy to that subsystem and in fulfilling of
(2.5) the corresponding superfunctional $Z[{\cal A}]$ (2.1) vanishes
identically.

Eq.(5.22) for $S_{H}(\theta)$ taking into account of (5.30), (5.31)
and (3.35) means the realization of the equivalent equation for
$S_{L}\bigl({\stackrel{\ \circ}{\cal A}}(\theta)\bigr)$
\begin{eqnarray}
-(-1)^{\varepsilon_{\imath}}({S_L''}^{-1})^{\imath\jmath}\bigl({\stackrel{\
\circ}{\cal A}}(\theta)\bigr)
\frac{\partial_l\phantom{xxx}}{\partial{\stackrel{\ \circ}{\cal
A}}{}^{\jmath}(\theta)} \frac{\partial_l S_L\bigl({\stackrel{\ \circ}{\cal
A}}( \theta)\bigr)}{\partial{\cal A}^{\imath}(\theta)\phantom{xxx}} = 0\;
.
\end{eqnarray}
\section{Detailed Investigation of the Hamiltonian \protect \\
Formulation of GSTF}
\setcounter{equation}{0}

In connection with addition to LS (2.3) the SCLF $\chi_{a}\bigl({\cal A}(
\theta),{\stackrel{\ \circ}{\cal A}}(\theta)\bigr)$  (2.7) it is necessary
for Lagrangian formulation of GSTF to modify the system of initial postulates
formulated in [1]. SCLF $\chi_{a}(\theta)$ can be
linearly (functionally) dependent themselves. Taking into account of the SCLF
structure  the assumptions have the form being consistent with ones
in Ref.[1].

\noindent
$1_{M}$. There exists a configuration
$\Bigl(\tilde{\cal A}^{\imath}_0(\theta), {\stackrel{\ \circ}{\tilde{\cal
A}}}{}^{\imath}_0(\theta)\Bigr)$ $\in$ $T_{odd}{\cal M}_{cl}$ that
\begin{eqnarray}
\Theta_{\imath}
\Bigl({\cal A}(\theta),
{\stackrel{\ \circ}{\cal A}}(\theta)
\Bigr)_{\mid \Bigl(\tilde{\cal A}^{\imath}_0(\theta),
{\stackrel{\ \circ}{\tilde{\cal A}}}{}^{\imath}_0(\theta)\Bigr)} = \chi_a
\bigl({\cal A}(\theta),{\stackrel{\ \circ}{\cal A}}(\theta)\bigr)_{\mid
\Bigl(\tilde{\cal A}^{\imath}_0(\theta),
{\stackrel{\ \circ}{\tilde{\cal A}}}{}^{\imath}_0(\theta)\Bigr)} =0\;
;
\end{eqnarray}
$2_{M}$. There exists, at least locally, a smooth supersurface
${}^{1}\Sigma$ $\subset$ ${\cal
M}_{cl}$ $\Bigl[\Bigl(\tilde{\cal A}^{\imath}_0(\theta),
{\stackrel{\ \circ}{\tilde{\cal
A}}}{}^{\imath}_0(\theta)\Bigr)$ $\in$ $T_{odd}{}^1\Sigma\Bigr]$
\renewcommand{\theequation}{\arabic{section}.\arabic{equation}\alph{lyter}}
\begin{eqnarray}
\setcounter{lyter}{1}
{} &  \dim{{}^1\Sigma} = {}^1m = ({}^1m_{+}, {}^1m_{-}),\;
\dim{T_{odd}{}^1\Sigma} =  ({}^1m_{+} + {}^1m_{-}, {}^1m_{+} + {}^1m_{-})
\;,& {}\\
\setcounter{lyter}{2}
\setcounter{equation}{2}
{} & \chi_a\bigl({\cal A}(\theta),{\stackrel{\ \circ}{\cal A}}(\theta)
\bigr)_{\mid
T_{odd}{}^1\Sigma} = 0\;, & {} 
\end{eqnarray}
the such that the following condition almost everywhere on $T_{odd}{}^1\Sigma$
is fulfilled
\renewcommand{\theequation}{\arabic{section}.\arabic{equation}}
\begin{eqnarray}
{\rm rank}\left\|\frac{\delta_l
\chi_a\bigl({\cal A}(\theta),{\stackrel{\ \circ}{\cal
A}}(\theta)\bigr)}{\delta {\cal
A}^{\imath}(\theta_1)\phantom{xxxxxxx}}\right\|_{\mid T_{odd}{}^1\Sigma} =
n-{}^1m\;. 
\end{eqnarray}
\underline{\bf Remarks:}

\noindent
{\bf 1)} Assumption 2 from Ref.[1] on the rank of supermatrix
of the 2nd superfield variational derivatives with respect to ${\cal
A}^{\imath}(\theta)$ and ${\cal A}^{\jmath}(\theta_{1})$ of superfunctional
$Z[{\cal A}]$ $\Bigl[{\rm or}\;
\left\|\frac{\delta_l \Theta_{\imath}
\bigl({\cal A}(\theta),{\stackrel{\ \circ}{\cal
A}}(\theta)\bigr)}{\delta{\cal
A}^{\jmath}(\theta_1)\phantom{xxxxxxx}}\right\|\Bigr]$ remains valid without
changes;

\noindent
{\bf 2)} the rule for calculation  of the type
(6.3) supermatrices ranks had  been pointed out in [1];

\noindent
{\bf 3)} because of SCLF $\chi_{a}(\theta)$ and DCLF
$\Theta_{\imath}(\theta)$ are not in general case (functionally) independent
from each other, then in addition to assumption 2 of the work [1] and $2_{M}$
(6.2), (6.3) it is necessary to introduce the value for rank of $2{n}\times
2n$ supermatrix \begin{eqnarray} {\rm rank}\left\|\frac{\partial_l
\bigl(\Theta_{\imath}(\theta), \chi_a(\theta)\bigr)}{\partial\bigl({\cal
 A}^{\jmath}(\theta),{\stackrel{\circ}{{\cal A}^k}}(\theta)\bigr)}
\right\|_{\mid T_{odd}(\Sigma \cap {}^1\Sigma)}\;,
\end{eqnarray}
where $\Sigma$ is the smooth local supersurface described
in [1];

\noindent
{\bf 4)} the smooth supersurfaces ${}^1\Sigma$ and $\Sigma$ are defined at
least locally in a some neighbourhood of the corresponding configuration
$\tilde{\cal A}{}^{\imath}_0(\theta)$ and ${\cal A}{}^{\imath}_0(\theta)$
below.

In consequence  of SCLF $\chi_{a}(\theta)$ dependence for
${}^{1}m \ne 0$ it follows by analogy with Theorem 2 from the Ref.[1] a
possibility to realize an equivalent set of SCLF with one's own
differential operators $\hat{\cal R}^a_{\beta_1}\bigl(
{\cal A}(\theta),{\stackrel{\ \circ}{\cal
A}}(\theta),\theta;\theta'\bigr)$ $\equiv$  $\hat{\cal R}^a_{\beta_1}
(\theta ;\theta')$, $a = (A_{1},A_{2}) = \overline{1,n}$, $\beta_1
=1,\ldots ,{}^{1}m$ leading to identities among $\chi_{a}(\theta)$ [1].
The joint investigation of $\chi_{a}(\theta)$ and $\Theta_{\imath}(\theta)$,
based on the relation (6.4), together
with corresponding operators $\hat{\cal R}^{\imath}_{\alpha}(\theta;\theta')$,
$\alpha = 1,\ldots ,m$ [1] and $\hat{\cal R}^a_{\beta_1}(\theta ;\theta')$
is not developed here.

Analysis  of GCHF $\Theta_{\imath}^{H}(\Gamma(\theta))$,
SCHF $\chi_{a}^{H}(\Gamma(\theta))$ independence under condition (3.19a) one
can separately and jointly develop in ${T}^{\ast}_{odd}{\cal M}_{cl}$
directly. To this end it is necessary to reformulate the assumptions 1--3
suggested in [1] in terms of $Z_{H}[\Gamma]$, $Z_{H}^{(1)}[\Gamma,D]$, $Z_{
H}^{(2)}[\Gamma,\lambda]$, $S_{H}^{(k)}(\theta)$, $k=0,1,2$.

\noindent
$1_{H}$. There exists a configuration  $\Gamma_{0}^{p}(\theta)$ =
$({\cal A}_{0}^{\imath}(\theta), {\cal A}_{0{}\imath}^{\ast}(\theta))$
$\in$ $T^{\ast}_{odd}{\cal M}_{cl}$ that
\begin{eqnarray}
\Theta_{\imath}^H(\Gamma_{0}(\theta),\theta) = 0\;;
\end{eqnarray}
$2_{H}$. A smooth local supersurface $\Sigma \subset {\cal M}_{cl}:$ $
\Gamma_{0}^{p}(\theta)$ $\in$ $T^{\ast}_{odd}\Sigma$ exists the such that
\begin{eqnarray}
\Theta_{\imath}^H(\Gamma(\theta),\theta)_{\mid T^{\ast}_{odd}\Sigma} = 0
\;.
\end{eqnarray}
A separation of index $\imath$ [1] exists
\renewcommand{\theequation}{\arabic{section}.\arabic{equation}\alph{lyter}}
\begin{eqnarray}
\setcounter{lyter}{1}
\imath = (A,\alpha),\ A=1,\ldots,n-m,\ \alpha = n-m+1,\ldots,n,\ m=(m_{+},
m_{-})\;,
\end{eqnarray}
that the following condition with regard of (3.35f) almost everywhere on
$\Sigma$ holds
\begin{eqnarray}
\setcounter{lyter}{2}
\setcounter{equation}{7}
{} &{\rm
rank}\left\|\displaystyle\frac{\delta_l\phantom{xxxx}}{\delta\hat{ \cal
A}^{\jmath}( \theta_1)}\displaystyle\frac{\delta_l
Z_H^{(1)}[\Gamma,D]}{\delta D^{\imath}(\theta)\phantom{xxxx}}\right\|_{\mid
\Sigma} = {\rm rank}\left\|\displaystyle\frac{\delta_l\phantom{xxxx}}{
\delta\hat{\cal A}^{\jmath}(
\theta_1)}\frac{\delta_l Z_H^{(1)}[\Gamma,D]}{\delta D^{A}(\theta)
\phantom{xxx}}\right\|_{\mid \Sigma} = n-m\;; &{} 
\end{eqnarray}
\vspace{-2ex}
\renewcommand{\theequation}{\arabic{section}.\arabic{equation}}
\begin{eqnarray}
{}\hspace{-3em}3_{H}. & \hspace{5em}{\rm rank}K^{\ast}(\theta)_{\mid
T^{\ast}_{odd}V} \equiv
{\rm rank}\left\|(S_H'')^{\jmath\imath}(\Gamma(\theta),\theta)\right\|_{\mid
T^{\ast}_{odd}V}=n,\;\Sigma \subset V\subset{\cal M}_{cl}\;, &{}
\end{eqnarray}
where $V$ is a some neighbourhood of $\Sigma$ described, for instance, in [1].

\noindent
$1_{HM}$. A configuration  $\tilde{\Gamma}_{0}^{p}(\theta)$ $\in$ $T^{\ast
}_{odd}{\cal M}_{cl}$ exists that
\begin{eqnarray}
\Theta_{\imath}^H(\tilde{\Gamma}_{0}(\theta)) =
\chi_a^H(\tilde{\Gamma}_{0}(\theta)) = 0 \;
 ;
\end{eqnarray}
$2_{HM}$. A smooth local supersurface  ${}^{1}\Sigma \subset {\cal M}_{cl}:$
$\tilde{\Gamma}_{0}^{p}(\theta)$ $\in$ $T^{\ast}_{odd}{}^{1}\Sigma$ exists
the such that together with (6.2a) being valid for $T^{\ast}_{odd}{}^{
1}\Sigma$ the expression is true
\begin{eqnarray}
\chi_a^H({\Gamma}(\theta))_{\mid T^{\ast}_{odd}{}^1\Sigma} = 0 \;,
\end{eqnarray}
and almost
everywhere on ${}^{1}\Sigma$ the relation holds
\begin{eqnarray}
{\rm rank}\left\|\frac{\delta_l\chi_a^H({\Gamma}(\theta))}{ \delta\hat{\cal
A}^{\jmath}( \theta_1)\phantom{xxx}}\right\|_{\mid T^{\ast}_{odd}{}^1\Sigma} =
{\rm rank}\left\|\frac{\delta_l \phantom{xxxx}}{\delta\hat{\cal A}^{\jmath}(
\theta_1)}\frac{\delta_l Z_H^{(2)}[\Gamma,\lambda]}{\delta
\lambda^{a}(\theta) \phantom{xxxx}}\right\|_{\mid  T^{\ast}_{odd}{}^1\Sigma}
= n-{}^{1}m\;.
\end{eqnarray}
Adapting remark 3) after (6.3) to the case of the Hamiltonian formulation of
GSTF note that by analog of the formula (6.4) under a simultaneous analysis of
$\Theta_{\imath}^{H}(\theta)$ and $\chi_{a}^{H}(\theta)$ in
addition to assumptions $2_{H}$ and $2_{HM}$ it appears  the rank value
of  $2{n}\times 2{n}$ supermatrix
\begin{eqnarray}
{\rm rank}\left\|\frac{\partial_l
\bigl(\Theta^H_{\imath}(\theta), \chi_a^H(\theta)\bigr)}{\partial\bigl({\cal
 A}^{\jmath}(\theta),{\cal A}_k^{\ast}(\theta)\bigr)\phantom{x}}
\right\|_{\mid T^{\ast}_{odd}(\Sigma \cap {}^1\Sigma)}.
\end{eqnarray}
Assumptions $1_{H}$, $2_{H}$ are sufficient in separate investigating of
GCHF $\Theta_{\imath}^{H}(\theta)$ structure without SCHF and $2_{HM}$ in
investigating of $\chi_{a}^{H}(\theta)$.

In the framework of conditions $1_{H}$--$3_{H}$ the theorem being almost
analogous to Theorem 2 in Ref.[1] is formally valid.

\noindent
\underline{\bf Theorem 1} (on reduction of GCHF to equivalent
system in generalized normal form (GNF))

\noindent
A nondegenerate parametrization
for superfields ${\Gamma}^{p}(\theta)$ exists of the form
\begin{eqnarray}
{} & {} &{\Gamma}^{p}(\theta) = ({\cal A}^{\imath}(\theta), {\cal
A}^{\ast}_{\imath}(\theta))=
(\delta^{\overline{\imath}}(\theta),
\delta^{\ast}_{\overline{
\imath}}(\theta); \beta^{\underline{\imath}}(\theta),
\beta^{\ast}_{\underline{\imath}}(\theta);
\xi^{\alpha}(\theta),\xi^{\ast}_{\alpha}(\theta))\equiv \nonumber \\
{} & {} & (\varphi^A(\theta), \varphi^{\ast}_A(\theta);\xi^{\alpha}(\theta),
\xi^{\ast}_{\alpha}(\theta)),\ \imath =
(\overline{\imath},\underline{\imath},\alpha)=(A,\alpha),\ [\xi^{\alpha}]=m
\;,
\end{eqnarray}
that the system of $n$ GCHF $\Theta_{\imath}^{H}(\Gamma(\theta),\theta)$
(3.5c) being by algebraic with respect to derivatives on $\theta$ is
equivalent to one of independent algebraic equations in GHF
\begin{eqnarray}
\delta^{\ast}_{\overline{\imath}}(\theta) = \phi_{\overline{
\imath}}(\delta(\theta), \xi(\theta), \xi^{\ast}(\theta),\theta),\ \;
\beta^{\underline{\imath}}(\theta) = \kappa^{\underline{\imath}}(
\delta(\theta), \xi(\theta), \theta)\;.
\end{eqnarray}
Superfields $\xi^{\alpha}(\theta)$ and their "momenta" $\xi^{\ast}_{
\alpha}(\theta)$ are arbitrary and the number of them
($\xi^{\alpha}(\theta)$) coincides with number of algebraic (on
$\theta$) identities among GCHF
\begin{eqnarray}
{} &\hspace{-1em}\displaystyle\int d\theta\frac{\delta_r Z_H^{(1)}[\Gamma,
D]}{\delta D^{\imath}(\theta)
\phantom{xxxx}}\hat{\cal R}^{\imath}_{H{}\alpha}(\Gamma(\theta),\theta;
\theta')
= 0,\;(\varepsilon_P, \varepsilon_{\bar{J}}, \varepsilon)\hat{\cal R}^{
\imath}_{H{}\alpha}(\theta;\theta') = (1,
\varepsilon_{\imath} + \varepsilon_{\alpha},\varepsilon_{\imath} +
\varepsilon_{\alpha} + 1),& {} 
\end{eqnarray}
with a) local and b) linear independent operators $\hat{\cal R}^{\imath}_{H{}
\alpha}(\Gamma(\theta),\theta;\theta')$ $\equiv$
$\hat{\cal R}^{\imath}_{H{}\alpha}(\theta;\theta')$:
\begin{eqnarray}
{} & \hspace{-8.5em}{\rm a}) \hspace{8.5em} \hat{\cal R}^{\imath}_{H{}\alpha}(
\theta;\theta')
= \displaystyle\sum\limits_{k=0}^1
\displaystyle\left(\left(\frac{d}{d\theta}\right)^k \delta(\theta - \theta')
\right){\cal R}^{\imath}_{H{}k\alpha}(\Gamma(\theta),\theta)\;, &{} \nonumber
\\ {} & (\varepsilon_P, \varepsilon_{\bar{J}}, \varepsilon){\cal R}^{\imath}_{
H{}k\alpha}(\theta) = (\delta_{1k},
\varepsilon_{\imath} + \varepsilon_{\alpha},\varepsilon_{\imath} +
\varepsilon_{\alpha} + \delta_{1k})\;, & {} 
\end{eqnarray}
b) functional equation
\begin{eqnarray}
\displaystyle\int d\theta'\hat{\cal R}^{\imath}_{H{}\alpha}(\theta;\theta')
u^{\alpha}(\Gamma(\theta'),\theta') = 0,\ u^{\alpha}(\theta)\in C^{k\ast}
\times\{\theta\}
\end{eqnarray}
has the unique vanishing solution.

\noindent
\underline{\bf Remarks:}

\noindent
{\bf 1)} Operators $\hat{\cal R}^{\imath}_{H{}\alpha}(\theta;\theta')$ can
be chosen in a such way that from (3.5c), (2.21) and Ref.[1] it follows
the equalities
\begin{eqnarray}
{} & \hspace{-2em} \hat{\cal R}^{\imath}_{\alpha}({\cal A}(\theta),
{\stackrel{\ \circ}{\cal A}}(\theta),\theta;\theta')_{\mid
{\stackrel{\ \circ}{\cal
A}} \hspace{-0.2em}= {\stackrel{\ \circ}{\cal A}}(\Gamma(\theta),\theta)} = \hat{\cal
R}^{\imath}_{H{}\alpha}(\theta;\theta'),\;
\hat{\cal R}^{\imath}_{k\alpha}({\cal A}(\theta),{\stackrel{\ \circ}{\cal
A}}(\theta),\theta)_{\mid{\stackrel{\ \circ}{\cal
A}} = {\stackrel{\ \circ}{\cal A}}(\Gamma(\theta),\theta)}
 \hspace{-0.2em}= {\cal R}^{\imath}_{H{}k\alpha}(\theta);
\end{eqnarray}
{\bf 2)} relation (3.40) permits to
conclude on convention of Theorem 1 because the subsystem (3.5a)  of the 1st
order with respect to derivatives on $\theta$  differential equations is in
NF now. However there are the forcible arguments to consider Theorem 1 (see
corollary 1.2);

\noindent
{\bf 3)} the writing of GCHF (3.5c) in equivalent form (6.14) leads to
necessity of the type (2.12)  solvability conditions fulfilment for
Eqs.(6.14).
Realization of those conditions with necessity results in
absence of explicit dependence on $\theta$ for $\Theta_{\imath}^{H}(
\theta)$ and therefore the Eq.(3.19a) holds;

\noindent
{\bf 4)} representation of equivalent GCHF in the form (6.14) is possible,
in particular, under assumption that in realizing of Theorem 2 in Ref.[1]
additionally to postulate $3_{H}$ the following condition is valid
\begin{eqnarray}
{\rm
rank}\left\|\frac{\partial^2S_L(\theta)\phantom{xxx}}{\partial
{\stackrel{\;\circ}{\delta}}{}^{\bar{\imath}}(\theta)
\partial{\stackrel{\;\circ}{\delta}}{}^{\bar{\jmath}}(\theta)}
\right\|_{\mid T_{odd}V} =
{\rm
rank}\left\|\frac{\partial^2S_H(\theta)\phantom{xxx}}{\partial
\delta^{\ast}_{\bar{\imath}}(\theta)
\partial\delta^{\ast}_{\bar{\jmath}}(\theta)}
\right\|_{\mid T^{\ast}_{odd}V}, 
\end{eqnarray}
permitting to present ${\stackrel{
\;\circ}{\delta}}{}^{\bar{\imath}}(\theta)$ as the superfunctions being
essentially dependent upon $\delta^{\ast}_{\bar{ \imath}}(\theta)$.

The proof of Theorem 1 is out of the paper's scope. Let us indicate a some
main corollaries (being by analogs of the corollaries for Theorem 2 in
Ref.[1]).

\noindent
\underline{\bf Corollary 1.1}

\noindent
HCHF $\overline{\Theta}{}^{H}_{\imath}(\theta) \equiv \Theta_{\imath}({\cal
A}(\theta),\theta)$ in fulfilling of the conditions
\begin{eqnarray}
{\rm
rank}\left\|\frac{\partial_l \overline{\Theta}{}^H_{\jmath}(\theta)
}{\partial{\cal A}^{\imath}(\theta)\phantom{x}}\right\|_{\mid \Sigma} = n-m
\end{eqnarray}
and under nondegenerate parametrization ${\cal A}^{\imath}(\theta)$ = $(
\varphi^{A}(\theta), \xi^{\alpha}(\theta))$ are equivalent to system
of algebraically independent in a sense of differentiation on $\theta$ HCHF
\begin{eqnarray}
\overline{\Theta}{}^{H}_A(\varphi(\theta), \xi(\theta),\theta) = 0\;
.
\end{eqnarray}
The number of [$\xi^{\alpha}(\theta)$] coincides with one of the
algebraic on $\theta$ identities among $\overline{\Theta}{}^{H}_{\imath}(
\theta)$ with linearly independent operators
${\cal R}_{H{}\alpha}^{\imath}({\cal A}(\theta),\theta)$ $\in$ $C^{k}({\cal
M}_{cl} \times \{\theta\})$
\begin{eqnarray}
\overline{\Theta}{}^{H}_{\imath}(\theta)
{\cal R}_{H{}\alpha}^{\imath}({\cal
A}(\theta),\theta) = 0,\
{\cal R}_{H{}\alpha}^{\imath}({\cal A}(\theta),\theta) =
{\cal R}_{0\alpha}^{\imath}({\cal A}(\theta),\theta)\;,
\end{eqnarray}
where ${\cal R}_{0{}\alpha}^{\imath}({\cal A}(\theta),\theta)$ are the
operators satisfying to conditions of the corollary 2.1 from Ref.[1].

\noindent
\underline{\bf Corollary 1.2}

\noindent
For a model of GSTF in the Lagrangian formulation representing the
almost natural system (see Corollary 2.2 of Ref.[1]) with $S_{L}(\theta)$
$\in$ $C^{k}(T_{odd}{\cal M}_{cl}\times \{\theta\})$ and
HCLF $\Theta_{\imath}({\cal A}(\theta),\theta)$
\renewcommand{\theequation}{\arabic{section}.\arabic{equation}\alph{lyter}}
\begin{eqnarray}
\setcounter{lyter}{1}
{} & S_L \bigl( {\cal A}(\theta), {\stackrel{\
\circ}{\cal A}}(\theta), \theta\bigr) = T \bigl( {\cal A}(\theta),
{\stackrel{\ \circ}{\cal A}}(\theta)\bigr) - S\bigl( {\cal A}(\theta),
 \theta\bigr),\ \min{\rm deg}_{{\cal A}(\theta)}S(\theta) = 2 \;, & {}\\
{} & T\bigl( {\cal A}(\theta), {\stackrel{\ \circ}{\cal
A}}(\theta)\bigr) = T_1\bigl({\stackrel{\ \circ}{\cal A}}(\theta)\bigr) +
{\stackrel{\ \circ}{\cal A}}{}^{\jmath}(\theta) T_{\jmath}\bigl( {\cal A}
(\theta)\bigr),\ T_{\jmath}\bigl( {\cal A}(\theta)\bigr) = g_{\jmath
\imath}(\theta){\cal A}^{\imath} (\theta)\ , & {} \nonumber \\
\setcounter{lyter}{2}
\setcounter{equation}{23}
{} &  g_{\jmath
\imath}(\theta) = (-1)^{\varepsilon_{\jmath}\varepsilon_{\imath}}g_{\imath
\jmath}(\theta),\ g_{\jmath \imath}(\theta) = P_0(\theta)g_{\jmath
\imath}(\theta),\ \min{\rm deg}_{{\stackrel{\ \circ}{\cal
A}}(\theta)}T_1(\theta) = 2\;,
 & {}  \\
\setcounter{lyter}{3}
\setcounter{equation}{23}
{} & \hspace{-2em}T_1\bigl({\stackrel{\ \circ}{\cal A}}(\theta)\bigr)
= \frac{1}{2}
{\stackrel{\ \circ}{\cal A}}{}^{\jmath}(\theta)
\kappa_{\jmath\imath}\bigl({\stackrel{\ \circ}{\cal A}}(\theta)\bigr)
{\stackrel{\ \circ}{\cal A}}{}^{\imath}(\theta)(-1)^{\varepsilon_{\imath}},\;
\kappa_{\jmath\imath}\bigl({\stackrel{\ \circ}{\cal A}}(\theta)\bigr) =
(-1)^{(\varepsilon_{\imath} + 1)(\varepsilon_{\jmath} + 1)}
\kappa_{\imath\jmath}\bigl({\stackrel{\ \circ}{\cal A}}(\theta)\bigr)\;,
& {}\\
\setcounter{lyter}{4}
\setcounter{equation}{23}
{} &{\Theta}_{\imath}( {{\cal A}}(\theta), \theta ) =  - S,_{\imath} ({{\cal
A}}(\theta), \theta)(-1)^{ \varepsilon_{\imath}} = 0\;,& {}
\end{eqnarray}
the superfunction $S_{H}(\Gamma(\theta),\theta)$ and HCHF
$\overline{\Theta}{}^{H}_{\imath}(\theta)$ are given by sequence of
relationships
\begin{eqnarray}
\setcounter{lyter}{1}
{} & S_{H}(\Gamma(\theta),\theta) = S_{H_{0}}(\Gamma(\theta),\theta) +
S_{H{}{\rm int}}(\Gamma(\theta),\theta)\;,&{} \\
\setcounter{lyter}{2}
\setcounter{equation}{24}
{} & S_{H_{0}}(\Gamma(\theta),\theta) =
T_0(\tilde{\cal A}^{\ast}(\theta)) + S_{0}({\cal A}(\theta),
\theta),\;
{\rm deg}_{{\tilde{\cal A}^{\ast}}(\theta)}T_0(\theta) = {\rm
deg}_{{\cal A}(\theta)}S_0(\theta)=2\;, & {}
\\
\setcounter{lyter}{3}
\setcounter{equation}{24}
{} & T_0(\tilde{\cal A}^{\ast}(\theta)) =
\frac{1}{2} \left(\kappa_{0}^{-1}\right)^{\imath\jmath}(\theta)
\tilde{\cal A}^{\ast}_{\jmath}(\theta)
\tilde{\cal A}^{\ast}_{\imath}(\theta)(-1)^{\varepsilon_{\imath}},\;
\tilde{\cal A}^{\ast}_{\imath}(\theta) = {\cal
A}^{\ast}_{\imath}(\theta)- T_{\imath}({\cal A}(\theta))\;,
 & {}\\
\setcounter{lyter}{4}
\setcounter{equation}{24}
{} & \hspace{-2.8em}
\tilde{\cal A}^{\ast}_{\imath}(\theta) =
\displaystyle\frac{\partial_l T_1\bigl({\stackrel{\ \circ}{\cal
A}}(\theta)\bigr)}{\partial{\stackrel{\ \circ}{\cal
A}}{}^{\imath}(\theta)\phantom{xxx}} = \left(\kappa_{\imath\jmath}(\theta)
{\stackrel{\ \circ}{\cal A}}{}^{\jmath}(\theta) - \displaystyle\frac{1}{2}
\frac{\partial_l \kappa_{k\jmath}(\theta)}{\partial{\stackrel{\ \circ}{\cal
A}}{}^{\imath}(\theta)}
{\stackrel{\ \circ}{\cal A}}{}^{\jmath}(\theta)
{\stackrel{\ \circ}{\cal A}}{}^{k}(\theta)(-1)^{\varepsilon_{k}}\right)
(-1)^{\varepsilon_{\jmath}},
& {} \\
\setcounter{lyter}{5}
\setcounter{equation}{24}
{} & \kappa_{\imath\jmath}\bigl({\stackrel{\ \circ}{\cal A}}(\theta)\bigr) =
\kappa_{0{}\imath\jmath}(\theta) +
\tilde{\kappa}_{\imath\jmath}\bigl({\stackrel{\ \circ}{\cal
A}}(\theta)\bigr),\
\min{\rm deg}_{{\stackrel{\ \circ}{\cal A}}(\theta)}
\tilde{\kappa}_{\imath\jmath}(\theta) \geq 1,\;
{\rm deg}_{{\stackrel{\ \circ}{\cal A}}(\theta)}\kappa_{0{}\imath\jmath}(
\theta) = 0\;, & {}\nonumber\\
{} & {\rm sdet}\left\|\kappa_{\imath\jmath}(\theta)\right\| \neq 0,\
{\rm sdet}\left\|\kappa_{0{}\imath\jmath}(\theta)\right\| \neq 0\;, & {}
\\
\setcounter{lyter}{6}
\setcounter{equation}{24}
{} &\overline{\Theta}{}^H_{\imath}(\theta) =  - S,_{\imath}
({{\cal A}}(\theta), \theta)(-1)^{ \varepsilon_{\imath}} = 0\;.& {} 
\end{eqnarray}
Equation (6.24d) is resolvable with respect to
${\stackrel{\ \circ}{\cal A}}{}^{\imath}(\theta)$
\begin{eqnarray}
\setcounter{lyter}{7}
\setcounter{equation}{24}
{} & {\stackrel{\ \circ}{\cal A}}{}^{\imath}(\theta) =
h^{\imath\jmath}(\tilde{\cal A}^{\ast}(\theta))
\tilde{\cal A}^{\ast}_{\jmath}(\theta)(-1)^{\varepsilon_{\imath}},\
h^{\imath\jmath}(\tilde{\cal A}^{\ast}(\theta)) =
(\kappa_0^{-1})^{\imath\jmath}(\theta) +
\tilde{h}^{\imath\jmath}(\tilde{\cal A}^{\ast}(\theta))\;,& {}\nonumber\\
{} & \hspace{-3em}
\min{\rm deg}_{{\tilde{\cal A}^{\ast}}(\theta)}\tilde{h}^{\imath
\jmath}
\geq 1,\;\tilde{h}^{\imath\jmath}(\tilde{\cal A}^{\ast}(\theta)) =
P_0(\theta)\tilde{h}^{\imath\jmath}(\tilde{\cal A}^{\ast}(\theta)),\;
(\kappa_0^{-1})^{\jmath\imath}(\theta) = -(-1)^{\varepsilon_{\imath}
\varepsilon_{\jmath}}(\kappa_0^{-1})^{\imath\jmath}(\theta)\,.
\end{eqnarray}
In (6.24) $S_{H_{0}}(\theta)$, $T_{0}(\theta)$ are  quadratic
parts with respect to extended superantifields $\tilde{\cal A}^{\ast
}_{\imath}(\theta)$ and ${\cal A}^{\imath}(\theta)$, whereas
$S_{H{}{\rm int}}(\theta)$ is at least cubic on $\tilde{\cal A}^{\ast
}_{\imath}(\theta)$, ${\cal A}^{\imath}(\theta)$.
Condition (6.7b) and identities with linearly independent operators ${\cal
R}_{H{}\alpha}^{\imath}({\cal A}(\theta),\theta)$ are given by the relations
respectively
\renewcommand{\theequation}{\arabic{section}.\arabic{equation}}
\begin{eqnarray}
{\rm rank}\left\|S,_{\imath\jmath}({\cal A}(\theta),\theta)\right\|_{\mid
\Sigma} = n-m,\ \
S,_{\imath}({\cal A}(\theta),\theta){\cal R}_{H{}\alpha}^{\imath}( {\cal
A}(\theta),\theta)= 0\;.
\end{eqnarray}
\underline{\bf Remark:} If HCHF are resolvable then the explicit dependence
upon $\theta $ does not take place for all superfunctions in Corollaries
1.1, 1.2.

\noindent
\underline{\bf Corollary 1.3}

\noindent
If for GCHF (3.5c) (HCHF $\overline{\Theta}{}^{H}_{\imath}(\theta)$) the
conditions almost everywhere in any neighbourhood
$U \subset {\cal M}_{cl}$ are realized:
\begin{eqnarray}
{\rm rank}\left\|\displaystyle\frac{\delta_l\phantom{
xxxx}}{\delta\hat{\cal A}^{\jmath}(\theta_1)}\frac{\delta_l Z_H^{(1)}[
\Gamma,D]}{\delta D^{\imath}(\theta)
\phantom{xxxx}}\right\|_{\mid T^{\ast}_{odd}U} = n,\ \;
\left({\rm
rank}\left\|\overline{\Theta}{}^H_{\imath,\jmath}(\theta)\right\|_{\mid U}
= n \right), 
\end{eqnarray}
then  all $\Theta^{H}_{\imath}(\theta)$
($\overline{\Theta}{}^{H}_{\imath}(\theta)$) appear by linearly independent
and already are in GHF.

\noindent
\underline{\bf Corollary 1.4}

\noindent
Under a particular choice of $S_{H}(\Gamma(\theta),\theta)$ in
Corollary 1.2  being not explicitly dependent upon $\theta$ and for
$T_{\imath}({\cal A}(\theta))=0$ (natural  system) in the form
\begin{eqnarray}
S_{H}(\Gamma(\theta)) = T({\cal A}^{\ast}(\theta)) + S({\cal A}(\theta))\;,
\end{eqnarray}
the Hamiltonian subsystem in GHS (3.51)
obtained via only  Legendre transform (3.2) without regard for
dynamical equations of HCLF (2.3b) is represented by means of the system
consisting of the 1st subsystem in (3.51a) and identities (3.52)
\begin{eqnarray}
\frac{d_r{\cal A}^{\imath}(\theta)}{d\theta
\phantom{xxxx}} = \frac{\partial S_{H}(\Gamma(\theta))}{\partial{\cal
A}^{\ast}_{\imath}(\theta)\phantom{xxx}} =
 \frac{\partial T({\cal A}^{\ast}(\theta))}{\partial{\cal
A}^{\ast}_{\imath}(\theta)\phantom{xxx}},\ \;
\frac{d_r{\cal A}^{\ast}_{\imath}(\theta)}{d\theta
\phantom{xxxx}} = 0\;.
\end{eqnarray}
a) The above system (6.28) satisfies to solvability condition,
but $S_{H}(\Gamma(\theta))$ is not its integral;

\noindent
b) By adding to that system the HCHF $\overline{\Theta}{}^{H}_{\imath}(
\theta)$ the solvability conditions for given HS is conserved and
$S_{H}(\Gamma(\theta))$
appears now by integral for obtained GHS to be equivalent to LS (2.3).
However the HCHF themselves do not appear by solvable in view of nonfulfilment
of Eqs.(3.39). Corresponding integral curve according to Existence and
Uniqueness Theorems for solution of the 1st order on $\theta$ ODE [1] has
the form
\renewcommand{\theequation}{\arabic{section}.\arabic{equation}\alph{lyter}}
\begin{eqnarray}
\setcounter{lyter}{1}
{} & \bar{\cal A}^{\imath}(\theta) = {\cal A}^{\imath}_0(0) +
\displaystyle\frac{\partial T({\cal A}^{\ast}(\theta))}{\partial{\cal
A}^{\ast}_{\imath}(\theta)\phantom{xxx}} \theta,\ \;
\bar{\cal A}^{\ast}_{\imath}(\theta) = {\cal A}^{\ast}_{0{}\imath}(0)\;
,\\
\setcounter{lyter}{2}
\setcounter{equation}{29}
{} & \Theta^H_{\imath}(\bar{\cal A}(\theta)) = (-1)^{\varepsilon_{\imath}+1}
S,_{\imath}(\bar{\cal A}(\theta)) = 0\;. & {} 
\end{eqnarray}
The fulfilment of (6.29b) reflects the requirement of compatibility of
$\bar{\cal A}^{\imath}(\theta)$ with HCHF.
Eqs.(6.29b) themselves for all $\theta $
already are the Euler-Lagrange equations for $S({\cal A}(\theta))$ in an usual
sense (as in the classical field theory).

If to start from a priority of superfunctionals $Z[{\cal A}]$ (2.1), $Z_{H}[
\Gamma]$ (3.13), then they are defined with accuracy up to
$P_{0}(\theta)$ component of one's densities. This ambiguity leads
to possibility to satisfy to the nondegeneracy conditions
for $K(\theta)$ (2.21) and therefore for $K^{\ast}(\theta)$ (6.8)
by means of addition of the corresponding $P_{0}(\theta){\cal F}(\theta)$.
However above described operation can destroy a
explicit superfield form for integrands ($S_{L}(\theta)$, $S_{H}(\theta)$) in
the expressions for $Z[{\cal A}]$, $Z_{H}[\Gamma]$. If it is
necessary to hold  the structure of densities $S_{L}(\theta)$,
$S_{H}(\theta)$ by fixed (that is  more natural  from viewpoint
of physics), then $Z[{\cal A}]$ and $Z_{H}[\Gamma]$ must not
possess of the such above-mentioned ambiguity in one's definitions and
hypothesis (2.21) is not fulfilled obstructing to possibility for
the Hamiltonization (3.2) realization. It means that assumption $3_{H}$
(6.8) does not take place. In this case a corresponding problem of
the theory investigation  in Hamiltonian formulation for GSTF is  more
complicated one and requires the combined study of DCLF (2.3b) and constraints
arising on the 1st stage of Dirac-Bergmann algorithm [8]. The last
problem appears by separate subject of investigation just as a study
of  the SCHF $\chi_{a}^{H}(\theta)$ independence problem both
separately and simultaneously with GCHF $\Theta_{\imath}^{H}(\theta)$ in a
presence of the assumptions $1_{HM}$, $2_{HM}$, $3_{H}$ and condition (6.12).

Led in Secs.II--VI investigation permits to introduce  the terminology
being analogous to one for the Lagrangian formulation of GSTF [1].

\noindent
\underline{\bf Definitions:}

\noindent
{\bf 1)} The model of superfield theory of fields (mechanics) being given by
superfunction $S_{H}(\theta)\in C^{k\ast}\times\{\theta \}$, $k  \le \infty$
(or by superfunctional $Z_{H}[\Gamma]$ $\in$
$C_{FH,cl}$) under condition  (3.5c) (or by $Z_{H}^{(1)}[\Gamma, D]$)
satisfying to assumptions $1_{H}$--$3_{H}$ (6.5)--(6.8) is called the gauge
theory of general type (GThGT) of superfields $\Gamma^p(\theta)$
with nondegenerate $K^{\ast}(\theta)$ and in realizing of the 1st condition
in (6.26) for GCHF the nondegenerate theory of general type (ThGT).

\noindent
{\bf 2)} Under additional fulfilment of the  Corollary 1.1 conditions (6.20)
on HCHF $\overline{\Theta}{}^{H}_{\imath}(\theta)$ and $m>0$ let us call the
model
by the gauge theory of special type (GThST) of superfields $\Gamma^p(\theta)$
with nondegenerate $K^{\ast}(\theta)$. In realizing for HCHF of the 2nd
relation in (6.26) we will call the model by the nondegenerate theory of
special type (ThST).

\noindent
{\bf 3)} Call the formulation of GThGT, GThST with $S_{H}(\theta)$ (or
$Z_{H}[\Gamma]$)  by the Hamiltonian
formalism for description of GThGT, GThST, or equivalently the Hamiltonian
formalism (formulation) for GSTF.

\noindent
\underline{\bf Remark}: At  presence of SCHF $\chi_{a}^{H}(\Gamma(
\theta))$ in the Hamiltonian formalism for the model description  we will say
that the GThGT (GThST) is given with SCHF.

\noindent
{\bf 4)} GThGT (GThST) with or without SCHF, for which the
$S_{H}(\Gamma(\theta),\theta)$ possibly satisfies to Eqs.(3.19), (3.39)
on an integral curve of the
corresponding system (GHS, HS, EGHS, EHS), we will regard by belonging to the
I class  of gauge field theory models. The same definition we relate to GThGT
(GThST) with SCLF $\chi_{a}\bigl({\cal A}(\theta),$ $
{\stackrel{\ \circ}{\cal A}}(\theta)\bigr)$ or without them, for which the
Eqs.(2.5), (2.6), (3.38) on an integral curve of the corresponding system
(LS, ELS) in the Lagrangian formalism  for GSTF are possibly valid.

\noindent
{\bf 5)} GThGT (GThST) for which the $S_{H}(\Gamma(\theta))$ satisfies to
master equation (5.11) let us relate to gauge field theories of the II class.
The same we will say on GThGT (GThST) in the Lagrangian formalism with
$S_{L}(\theta)$ satisfying to (5.30).
\vspace{1ex}

A set of GThGT (GThST) is not divided onto
models  belonging only to the I class or only to the II one.
As the example, consider in Hamiltonian formalism the GThST belonging to the
I class with SCHF $\chi_{a}^{H}(\Gamma(\theta))$ for $n_{1}=n$
\renewcommand{\theequation}{\arabic{section}.\arabic{equation}}
\begin{eqnarray}
\frac{\partial S_H(\theta)}{\partial{\cal A}^{\ast}_{\imath}(\theta)} = 0\;.
\end{eqnarray}
Given SCHF corresponds to SCLF $\chi_{a}\bigl({\stackrel{\  \circ}{\cal
A}}(\theta)\bigr)$ defined in (2.7a). The latter  means that $S_{H}(\Gamma
(\theta))$ satisfying to (3.15), (2.18) may contain only potential
term on solutions for (6.30)
\begin{eqnarray}
S_{H}(\Gamma(\theta)) = - S({\cal A}(\theta))\;,
\end{eqnarray}
and therefore the model belongs
on solutions for SCHF (6.30) to the II class theories.

As for the generators of gauge
transformations of general type (special type) GGTGT (GGTST) in Lagrangian
formalism [1] the identities (6.15) for GThGT and (6.22) for
GThST with operators $\hat{\cal R}^{\imath}_{H{}\alpha}(
\theta;\theta')$, ${\cal R}^{\imath}_{H{}\alpha}({\cal A}(\theta),\theta)$
respectively, whose sets are complete and linearly independent, i.e.
are the bases in  corresponding linear spaces $Q(Z^{(1)}_H)$ = ${\rm
Ker}\left\{\frac{\delta_l Z^{(1)}_H[\Gamma,D]}{\delta
D^{\imath}(\theta)\phantom{xxxx}}\right\}$, $Q(S_{H})$ = ${\rm
Ker}\{\overline{\Theta}{}^{H}_{\imath}(\theta)\}$, make possible the
following interpretation for quantities
$\hat{\cal R}^{\imath}_{H{}\alpha}(\theta;\theta')$, ${\cal
R}^{\imath}_{H{}\alpha}(\theta)$.

\noindent
\underline{\bf Definitions:}

\noindent
1) Any $\hat{\cal R}_{H}^{\imath}(\Gamma(\theta), \theta)\in
C^{k\ast}\times\{\theta\}$,
${\cal R}_{H}^{\imath}({\cal A}(\theta),\theta)\in C^{k}({\cal M}_{cl}\times
\{\theta\})$ satisfying to identities
\renewcommand{\theequation}{\arabic{section}.\arabic{equation}\alph{lyter}}
\begin{eqnarray}
\setcounter{lyter}{1}
{} & {} &  \displaystyle\int d\theta
\frac{\delta_r Z_H^{(1)}[\Gamma,D]}{\delta D^{\imath}(\theta)\phantom{xxxx}}
\hat{\cal R}^{\imath}_{H}(\Gamma(\theta),\theta) = 0,\
(\varepsilon_P, \varepsilon_{\bar{J}}, \varepsilon)
\hat{\cal R}^{\imath}_{H}(\theta) = (0,\varepsilon_{\imath}, \varepsilon_{
\imath})\;,\\
\setcounter{lyter}{2}
\setcounter{equation}{32}
{} & {} & \Theta^H_{\imath}({\cal A}(\theta), \theta)
{\cal R}^{\imath}_{H}({\cal A}(\theta),\theta) = 0,\
(\varepsilon_P, \varepsilon_{\bar{J}}, \varepsilon)
{\cal R}^{\imath}_{H}(\theta) = (0,\varepsilon_{\imath}, \varepsilon_{
\imath}) \;, 
\end{eqnarray}
let us call by the GGTGT and GGTST respectively in the Hamiltonian formalism;

\noindent
2) Call superfunctions $\hat{\tau}^{\imath}_{H}(\Gamma(\theta),\theta)$,
$\tau^{\imath}_{H}({\cal A}(\theta),\theta)$ given by the formulae
\begin{eqnarray}
\setcounter{lyter}{1}
{} & {} & \hspace{-2em}\hat{\tau}^{\imath}_{H}(\Gamma(\theta),\theta) =
\displaystyle\int d\theta'
\frac{\delta_r Z_H^{(1)}[\Gamma,D]}{\delta D^{\jmath}(\theta')\phantom{xxx}}
\hat{E}^{\imath\jmath}_{H}(\Gamma(\theta),\theta;\theta'),\;
(\varepsilon_P, \varepsilon_{\bar{J}}, \varepsilon)\hat{\tau}^{\imath}_{
H}(\theta) =  (0,\varepsilon_{\imath}, \varepsilon_{\imath})\;,\\
\setcounter{lyter}{2}
\setcounter{equation}{33}
{} & {} & \tau^{\imath}_{H}({\cal A}(\theta),\theta) =
\Theta_{\jmath}^H({\cal A}(\theta),\theta){E}^{\imath\jmath}_{H}({\cal A}(
\theta),\theta),\;
(\varepsilon_P, \varepsilon_{\bar{J}}, \varepsilon){\tau}^{\imath}_{
H}(\theta) =  (0,\varepsilon_{\imath}, \varepsilon_{\imath})\;
,
\end{eqnarray}
the trivial GGTGT, GGTST respectively. Properties for superfunctions
$\hat{E}^{\imath\jmath}_{H}(\theta;\theta')$,
${E}^{\imath\jmath}_{H}(\theta)$  are the same as for the analogous
superfunctions in Lagrangian formalism for GSTF [1].

Any GGTGT, GGTST can be represented respectively in the form
\renewcommand{\theequation}{\arabic{section}.\arabic{equation}}
\begin{eqnarray}
{} & \hat{\cal R}^{\imath}_{H}(\Gamma(\theta),\theta) =
\displaystyle\int d\theta'
\hat{\cal
R}^{\imath}_{H{}\alpha}(\Gamma(\theta),\theta;\theta')\hat{\xi}^{\alpha}
(\Gamma(\theta'),\theta') + \hat{\tau}^{\imath}_{H}(\Gamma(\theta),\theta)
,\;\hat{\xi}^{\alpha}(\theta) \in C^{k\ast}\times\{\theta\}, & {} \\ 
{} & {\cal R}^{\imath}_{H}(\Gamma(\theta),\theta) =
{\cal
R}^{\imath}_{H{}\alpha}({\cal A}(\theta),\theta){\xi}^{\alpha}_0
({\cal A}(\theta),\theta) + {\tau}^{\imath}_{H}({\cal A}(\theta),\theta),\;
{\xi}^{\alpha}_0(\theta) \in C^k({\cal M}_{cl}\times \{\theta \}),\;
& {} \\
{} &
(\varepsilon_P, \varepsilon_{\bar{J}}, \varepsilon)\hat{\xi}^{\alpha}(\theta)=
(\varepsilon_P, \varepsilon_{\bar{J}}, \varepsilon){\xi}^{\alpha}_0(\theta) =
(0, \varepsilon_{\alpha}, \varepsilon_{\alpha})\;. &{}\nonumber
\end{eqnarray}
These formulae convert
$Q(Z^{(1)}_H)$ in an affine $C^{k\ast}\times \{\theta\}$-module and
$Q(S_H)$ in an
affine $C^{k}({\cal M}_{cl}\times \{\theta\})$-module.

The so-called equivalence transformations (affine transformations of modules
$Q$) are valid for the basis elements
$\hat{\cal R}^{\imath}_{H{}\alpha}(\theta;\theta')$,
${\cal R}^{\imath}_{H{}\alpha}(\theta)$ respectively
\renewcommand{\theequation}{\arabic{section}.\arabic{equation}\alph{lyter}}
\begin{eqnarray}
\setcounter{lyter}{1}
{} & \hspace{-1em}\hat{{\cal R}}'^{\imath}_{H{}\alpha}(\theta;\theta') =
\displaystyle\int d\theta_1 \left[
\hat{\cal R}^{\imath}_{H{}\beta}(\theta;\theta_1)
\hat{\xi}^{\beta}_{\alpha}(
\Gamma(\theta_1),\theta_1;\theta') +
\frac{\delta_r Z_H^{(1)}[\Gamma,D]}{\delta D^{\jmath}(\theta_1)\phantom{xxx}}
\hat{E}^{\imath\jmath}_{H{}\alpha}(\Gamma(\theta),\theta,\theta_1;\theta')
\right], & {}
\\
\setcounter{lyter}{2}
\setcounter{equation}{36}
{} & {{\cal R}}'^{\imath}_{H{}\alpha}({\cal A}(\theta),\theta) =
{{\cal R}}^{\imath}_{H{}\beta}( {\cal A}(\theta),\theta)
\xi^{\beta}_{0{}\alpha}(
{\cal A}(\theta),\theta)  +  \Theta^H_{\jmath}({\cal A}(\theta),\theta)
{E}^{\imath\jmath}_{H{}\alpha}({\cal A}(\theta),\theta)\;.{} & 
\end{eqnarray}
Properties of the superfunctions
$\hat{E}^{\imath\jmath}_{H{}\alpha}(\theta,\theta_1;\theta')$,
${E}^{\imath\jmath}_{H{}\alpha}(\theta)$,
$\hat{\xi}^{\beta}_{\alpha}(\theta;\theta')$,
$\xi^{\beta}_{0{}\alpha}(\theta)$ completely coincide with analogous
ones for corresponding superfunctions in Lagrangian formalism [1].
Relationships (6.15) for GThGT are interpreted as a consequence of invariance
for superfunctional $Z_{H}^{(1)}[\Gamma,D]$ with respect to following
transformations of $\Gamma^{p}(\theta)$, $D^{\imath}(\theta)$  in the
infinitesimal form
\renewcommand{\theequation}{\arabic{section}.\arabic{equation}}
\begin{eqnarray}
{} & \hspace{-1.9em}(\Gamma^{p}(\theta), D^{\imath}(\theta))\hspace{-0.2em}
\mapsto \hspace{-0.2em}({
\Gamma}'^{p}(\theta),
{D}'^{\imath}(\theta)) \hspace{-0.2em}= \hspace{-0.2em}(\Gamma^{p}(\theta),
D^{\imath}(\theta) + \delta D^{\imath}(\theta))\hspace{-0.2em}:\hspace{-0.2em}
\delta D^{\imath}(\theta)\hspace{-0.3em} = \hspace{-0.3em}\displaystyle\int
\hspace{-0.2em}d\theta'
\hat{{\cal R}}^{\imath}_{H{}\alpha}(\theta;\theta') \xi^{\alpha}(\theta')
{} & 
\end{eqnarray}
with arbitrary superfields $\xi^{\alpha}(\theta)$
$(\varepsilon_P, \varepsilon_{\bar{J}},
\varepsilon){\xi}^{\alpha}(\theta) =  (0, \varepsilon_{\alpha},
\varepsilon_{\alpha})$
defined on $\Lambda_{D\vert Nc +1}(z^{a},\theta; {\bf K})$.

Really the sequence of the formulae holds
\begin{eqnarray}
Z_{H}^{(1)}[\Gamma',D'] = Z_{H}^{(1)}[\Gamma,D] + \int d\theta
d\theta'\Theta_{\imath}^H(\Gamma(\theta),\theta)\hat{{\cal R}}^{\imath}_{H{
}\alpha}(\theta;\theta')\xi^{\alpha}(\theta') =
Z_{H}^{(1)}[\Gamma,D] \;.
\end{eqnarray}
Identities (6.22) for
GThST, in general, can not be  interpreted as a consequence of invariance for
$S_H(\theta)$ with respect to the infinitesimal transformations
\begin{eqnarray}
 {\cal A}^{\imath}(\theta)  \mapsto
{{\cal A}}'^{\imath}(\theta) = {\cal A}^{\imath}(\theta)
 + \delta {\cal A}^{\imath}(\theta):
\delta {\cal A}^{\imath}(\theta) = {\cal
R}^{\imath}_{H{}\alpha}({\cal A}(\theta),\theta) \xi^{\alpha}_0(\theta)
\end{eqnarray}
with arbitrary superfields $\xi^{\alpha}_0(\theta)$ having the same Grassmann
gradings as for $\xi^{\alpha}(\theta)$ in (6.37). However for superfunction
$S({\cal A}(\theta),\theta)$ in Corollaries 1.2, 1.4 (in the latter
$S(\theta)$, $\hat{\cal R}_{H{}\alpha}^{\imath}(\theta)$ do not depend
on $\theta$ explicitly) a property of its invariance holds just as in
Lagrangian formalism [1].

Formally one can call the considered infinitesimal transformations written
in the form (6.37) the GTGT for $Z_{H}^{(1)}$  and the GTST for
$S({\cal A}(\theta),\theta)$, if latters are of the form (6.39).

The concluding remarks from Sec.VI of the paper [1] on irreducible and
reducible
GThGT, GThST and on investigations of corresponding gauge algebras of GThGT,
GThST are valid for the Hamiltonian formalism of GSTF as well.
\section{Extension of Superalgebra ${\cal A}_{cl}$  up to ${\cal B}_{cl}$}
\setcounter{equation}{0}

Let us continue the action of special involution $*$ [1] from $\tilde{
\Lambda}_{D\vert Nc + 1}(z^{a},\theta;{\bf K})$ and
$C^{k}(T_{odd}{\cal M}_{cl} \times \{\theta\})$ onto $D^{k}_{cl}$ by
means of the relations
\begin{eqnarray}
\left(\Gamma^p(\theta)\right)^{\ast} = \Gamma^p(\bar{\theta}) \equiv
\overline{\Gamma^p(\theta)} \equiv \Gamma^p(- \theta),\;
\left({\stackrel{\circ}{\Gamma}}{}^p(\theta)\right)^{\ast} =
{\stackrel{\circ}{\Gamma}}{}^p(\theta)\;,
\end{eqnarray}
where $\overline{\Gamma^{p}(\theta)}$ are the superfields being $*$-conjugate
to superfields $\Gamma^{p}(\theta )$ with components
\begin{eqnarray}
\overline{\Gamma^p(\theta)} = P_0(\theta){\Gamma}^p(\theta) -
P_1(\theta){\Gamma}^p(\theta),\;P_1(\theta)\overline{\Gamma^p(\theta)} =
- P_1(\theta){\Gamma}^p(\theta)\;.
\end{eqnarray}
The subspace of superfields being invariant with respect to $*$ is formed
by the superfields
\begin{eqnarray}
P_0(\theta){\Gamma}^p(\theta) = \frac{1}{2}\left({\Gamma}^p(\theta) +
\overline{\Gamma^p(\theta)}\right),\;
{\stackrel{\circ}{\Gamma}}{}^p(\theta)\;.
\end{eqnarray}
Superfields $\overline{\Gamma^{p}(\theta)}$ are transformed with respect
to the supergroup $J$ superfield representation $\overline{T}$ being by
${\ast}$-conjugate to representation $T$: ($\overline{T}_{\vert J}
 = T_{\vert J}$, $\overline{T}_{\vert P}$ = $(T_{\vert P})^{ - 1})$.
The restrictions of involution onto subsuperalgebras
${}^{0,0}C^{k}(T^{\ast}_{odd}{\cal M}_{cl} \times \{\theta\})$,
${}^{0,0}D^{k}_{cl}$ appear by the identity mappings.

The continuation of the superalgebra {\boldmath${\cal A}_{cl}$}  [1] of the
1st order differential operators acting on $C^{k}(T_{odd}{\cal
M}_{cl}\times \{\theta \})$ being by linear span of
$\{U_{a}(\theta), U_{\pm}(\theta), {\stackrel{\circ}{U}}_{a}(\theta),
{\stackrel{\circ}{U}}_{\pm}(\theta)$, $a=0,1\}$  to superalgebra
{\boldmath${\cal B}_{cl}$}
of the 1st and 2nd orders ones acting on $D^{k}_{cl}$
through elements of the basis
\begin{eqnarray}
\mbox{\boldmath${\cal B}^b_{cl}$} = \left\{ U_a(\theta),
{\stackrel{\circ}{U}}_a(\theta), {V}_a(\theta),
{\stackrel{\circ}{V}}_a(\theta) =
\left[\frac{d}{d\theta},{V}_a(\theta)\right]_s,\;\Delta_{ab}(\theta),\;
a,b = 0,1\right\},
\end{eqnarray}
is realized with help of tensor product $\otimes_{\theta}$ introduced in [1]
and written, for instance, for
\renewcommand{\theequation}{\arabic{section}.\arabic{equation}\alph{lyter}}
\begin{eqnarray}
\setcounter{lyter}{1}
{} &  U_a(\theta)  =  \bigl(P_1(\theta){\cal A}^{\imath}(
\theta)\bigr)
{\otimes}_{\theta}\displaystyle\frac{\partial_{l}
\phantom{xxxxxxx}}{\partial P_a(\theta){\cal A}^{\imath}(\theta)} = &{}
\nonumber \\
{} & \bigl(0, P_1(\theta){\cal A}^{\imath}(\theta)\bigr)
{\otimes}_{\theta} \left(\displaystyle\frac{\partial_{l}
\phantom{xxxxxxx}}{\partial P_0(\theta) {\cal
A}^{\imath}(\theta)}\delta_{a 0}, \frac{\partial_{l}
\phantom{xxxxxxx}}{\partial P_1(\theta){\cal A}^{\imath}
(\theta)}\delta_{a 1}\right)^T  \equiv P_1(\theta){\cal A}^{\imath}(\theta)
\displaystyle\frac{\partial_{l} \phantom{xxxxxxx}}{ \partial
P_a(\theta){\cal A}^{\imath}(\theta)}\;, & {}
\\
\setcounter{lyter}{2}
\setcounter{equation}{5}
{} & {\stackrel{\circ}{U}}_{a}(\theta)  =
{\stackrel{\;\circ}{{\cal A}^{\imath}}}(\theta)
{\otimes}_{\theta}\displaystyle\frac{\partial_{l} \phantom{x
xxxxxx}}{\partial P_a(\theta){{\cal A}^{\imath}(\theta)}} = & {}
\nonumber \\
{} & \left({\stackrel{\;\circ}{{\cal
A}^{\imath}}}(\theta), 0\right)\otimes_{\theta}
\left(\displaystyle\frac{\partial_{l}
\phantom{xxxxxxx}}{\partial P_0(\theta){{\cal A}^{\imath}(\theta)}}\delta_{a0},
\displaystyle\frac{\partial_{l}
\phantom{xxxxxxx}}{\partial P_a(\theta){{\cal A}^{\imath}(\theta)}}\delta_{a1}
\right)^T
 \equiv {\stackrel{\;\circ}{{\cal
A}^{\imath}}}(\theta) \displaystyle\frac{\partial_{l} \phantom{xxxxxxx}}{
\partial P_a(\theta){{\cal A}^{\imath}(\theta)}}\;. & {} 
\end{eqnarray}
For other operators (7.4)  write compactly with allowance made for
structure of  formulae (7.5)
\begin{eqnarray}
\setcounter{lyter}{1}
{} & V_a(\theta)  =  P_1(\theta){\cal A}^{\ast}_{\imath}(\theta)
 \displaystyle\frac{\partial
\phantom{xxxxxxx}}{\partial P_a(\theta){\cal A}^{\ast}_{\imath}(\theta)},\;
{\stackrel{\circ}{V}_{a}}(\theta)  =
{\stackrel{\circ}{{\cal
A}^{\ast}_{\imath}}}(\theta) \frac{\partial \phantom{xxxxxxx}}{\partial
P_a(\theta){{\cal A}^{\ast}_{\imath}(\theta)}}\;, & {} \\
\setcounter{lyter}{2}
\setcounter{equation}{6}
{} & \hspace{-2em}\Delta_{ab}(\theta) =
 \displaystyle\frac{\partial_{l}
\phantom{xxxxxxx}}{\partial P_a(\theta){\cal A}^{\imath}(\theta)}
 \displaystyle\frac{\partial
\phantom{xxxxxxx}}{\partial P_b(\theta){\cal A}^{\ast}_{\imath}(\theta)}
(-1)^{\varepsilon_{\imath}} \equiv
 \frac{1}{2}\displaystyle\frac{\partial_{l}
\phantom{xxxxxxx}}{\partial P_a(\theta)\Gamma^{p}(\theta)}\omega^{pq}(\theta)
 \displaystyle\frac{\partial_l
\phantom{xxxxxxx}}{\partial P_b(\theta)\Gamma^q(\theta)}
(-1)^{\varepsilon_{p}}.
\end{eqnarray}
Note that $\Delta_{11}(\theta) \equiv 0$ on $D^{k}_{cl}$. Properties
of gradings for operators (7.4) are given by the table
\renewcommand{\theequation}{\arabic{section}.\arabic{equation}}
\begin{eqnarray}
\begin{array}{lcccccl}
{} & U_a(\theta)  & {\stackrel{\circ}{U}_{a}}(\theta) & V_a(
\theta) & {\stackrel{\circ}{V}_{a}}(\theta) & \Delta_{ab}(\theta) & {}\\
\varepsilon_P & 0 & 1 & 0 & 1 & 1 & {}\\
\varepsilon_{\bar{J}} & 0 & 0 & 0 & 0 & 0  & {}\\
\varepsilon & 0 & 1 & 0 & 1 & 1 & .
\end{array}
\end{eqnarray}
In terms of the coordinates $\Gamma^{p}(\theta)$ the operators
$U_{a}(\theta), V_{a}(\theta)$ and
${\stackrel{\circ}{U}}_{a}(\theta),
{\stackrel{\circ}{V}}_{a}(\theta)$ are combined  into the expressions
respectively
\renewcommand{\theequation}{\arabic{section}.\arabic{equation}\alph{lyter}}
\begin{eqnarray}
\setcounter{lyter}{1}
{} & {} & {\cal W}_a(\theta) = U_{a}(\theta) + V_{a}(\theta) = P_1(\theta)
\Gamma^p(\theta)
\displaystyle\frac{\partial_{l}
\phantom{xxxxxxx}}{\partial P_a(\theta)\Gamma^{p}(\theta)}\;, \\
\setcounter{lyter}{2}
\setcounter{equation}{8}
{} & {} & {\stackrel{\circ}{\cal W}}_a(\theta) =
{\stackrel{\circ}{U}}_{a}(\theta) + {\stackrel{\circ}{V}}_{a}(\theta) =
{\stackrel{\circ\ \;}{\Gamma^p}}(\theta)
\displaystyle\frac{\partial_{l} \phantom{xxxxxxx}}{\partial
P_a(\theta)\Gamma^{p}(\theta)}\;.
\end{eqnarray}
By means of the
basis elements (7.5), (7.6), (7.8) one can write the superfield (at least
with respect to operators of differentiation) elements of
{\boldmath${\cal B}_{cl}$} not being contained in {\boldmath${\cal A}_{cl}$}
\begin{eqnarray}
\setcounter{lyter}{1}
{} & {} & \hspace{-4em}
V_{+}(\theta) = V_{0}(\theta) + V_{1}(\theta) = P_1(\theta)
{\cal A}^{\ast}_{\imath}(\theta)
\displaystyle\frac{\partial
\phantom{xxx}}{\partial {\cal A}^{\ast}_{\imath}(\theta)}\;,
\\
\setcounter{lyter}{2}
\setcounter{equation}{9}
{} & {} & \hspace{-4em}V_{-}(\theta) = V_{0}(\theta) - V_{1}(\theta) =
P_1(\theta){\cal A}^{\ast}_{\imath}(\theta)
\displaystyle\frac{\partial
\phantom{xxx}}{\partial \overline{{\cal A}^{\ast}_{\imath}(\theta)}} =
-\left(V_{+}(\theta)\right)^{\ast},\\
\setcounter{lyter}{3}
\setcounter{equation}{9}
{} & {} & \hspace{-4em}{\stackrel{\circ}{V}}_{+}(\theta) =
 {\stackrel{\circ}{{\cal A}^{\ast}_{\imath}}}(\theta)
\displaystyle\frac{\partial \phantom{xxxx}}{\partial{\cal A}^{\ast}_{
\imath}(\theta)},\;
 {\stackrel{\circ}{V}}_{-}(\theta) =
 {\stackrel{\circ}{{\cal A}^{\ast}_{\imath}}}(\theta)
\displaystyle\frac{\partial \phantom{xxxx}}{\partial\overline{{\cal
A}^{\ast}_{\imath}(\theta)}} = \left({\stackrel{\circ}{V}}_{+}(\theta)
\right)^{\ast},\\
\setcounter{lyter}{1}
\setcounter{equation}{10}
{} & {} & \hspace{-4em}\Delta^{cl}(\theta) \equiv \Delta_{++}(\theta) =
\sum\limits_{a,b} \Delta_{ab} = (-1)^{\varepsilon_{\imath}}
\displaystyle\frac{\partial_{l} \phantom{xxx}}{\partial{\cal A}^{
\imath}(\theta)}
\frac{\partial \phantom{xxxx}}{\partial{\cal A}^{\ast}_{
\imath}(\theta)}\;
,\\
\setcounter{lyter}{2}
\setcounter{equation}{10}
{} & {} & \hspace{-4em}\Delta_{+-}(\theta) = \sum\limits_{a,b}(-1)^{
\varepsilon_{\imath}}
\displaystyle\frac{\partial_{l} \phantom{xxxxxxx}}{\partial P_a(\theta){\cal
A}^{ \imath}(\theta)} \frac{\partial \phantom{xxxxxxxx}}{\partial
\overline{P_b(\theta){\cal A}^{\ast}_{ \imath}(\theta)}} =
(-1)^{\varepsilon_{\imath}} \displaystyle\frac{\partial_{l}
\phantom{xxx}}{\partial{\cal A}^{ \imath}(\theta)} \frac{\partial
\phantom{xxxx}}{\partial\overline{{\cal A}^{\ast}_{ \imath}(\theta)}}\;,
\\
\setcounter{lyter}{3}
\setcounter{equation}{10}
{} & {} & \hspace{-4em}\Delta_{-+}(\theta) = \sum\limits_{a,b}(-1)^{
\varepsilon_{\imath}}
\displaystyle\frac{\partial_{l} \phantom{xxxxxxx}}{\partial
\overline{P_a(\theta){\cal A}^{ \imath}(\theta)}} \frac{\partial
\phantom{xxxxxxxx}}{\partial {P_b(\theta){\cal A}^{\ast}_{
\imath}(\theta)}} = (-1)^{\varepsilon_{\imath}}
\displaystyle\frac{\partial_{l} \phantom{xxx}}{\partial\overline{{\cal A}^{
\imath}(\theta)}} \frac{\partial\phantom{xxxx}}{\partial{{\cal
A}^{\ast}_{ \imath}(\theta)}} = \left(\Delta_{+-}(\theta)\right)^{\ast}
, \\
\setcounter{lyter}{4}
\setcounter{equation}{10}
{} & {} & \hspace{-4em}\Delta_{--}(\theta) = \sum\limits_{a,b}(-1)^{
\varepsilon_{\imath}}
\displaystyle\frac{\partial_{l} \phantom{xxxxxxx}}{\partial
\overline{P_a(\theta){\cal A}^{ \imath}(\theta)}} \frac{\partial
\phantom{xxxxxxxx}}{\partial \overline{P_b(\theta){\cal A}^{\ast}_{
\imath}(\theta)}} = (-1)^{\varepsilon_{\imath}}
\displaystyle\frac{\partial_{l} \phantom{xxx}}{\partial\overline{{\cal A}^{
\imath}(\theta)}} \frac{\partial\phantom{xxxx}}{\partial\overline{{\cal
A}^{\ast}_{ \imath}(\theta)}} = \left(\Delta_{++}(\theta)\right)^{\ast}
.
\end{eqnarray}
By writing of (7.9), (7.10) it is taken into consideration according to
Ref.[1] that by relationships
\renewcommand{\theequation}{\arabic{section}.\arabic{equation}}
\begin{eqnarray}
\frac{\partial_{l} \phantom{xxx}}{\partial\overline{{\cal A}^{\imath}(
\theta)}} =
\frac{\partial_{l} \phantom{xxxxxxx}}{\partial
{P_0(\theta){\cal A}^{\imath}(\theta)}}  -
\frac{\partial_{l} \phantom{xxxxxxx}}{\partial
{P_1(\theta){\cal A}^{\imath}(\theta)}}  \equiv
\left(\frac{\partial_{l} \phantom{xxx}}{\partial{\cal A}^{\imath}(\theta)}
\right)^{\ast}, \nonumber \\
\frac{\partial\phantom{xxxx}}{\partial\overline{{\cal A}^{\ast}_{\imath}(
\theta)}} =
\frac{\partial\phantom{xxxxxxxx}}{\partial
{P_0(\theta){\cal A}^{\ast}_{\imath}(\theta)}}  -
\frac{\partial\phantom{xxxxxxxx}}{\partial
{P_1(\theta){\cal A}^{\ast}_{\imath}(\theta)}}  \equiv
\left(\frac{\partial\phantom{xxxx}}{\partial{\cal A}^{\ast}_{\imath}(
\theta)}\right)^{\ast}\
\end{eqnarray}
involution ${*}$ is extended onto {\boldmath${\cal B}_{cl}$}  as well.

The superfield elements built from operators (7.8)
have the form in coordinates $\Gamma^{p}(\theta)$
\renewcommand{\theequation}{\arabic{section}.\arabic{equation}\alph{lyter}}
\begin{eqnarray}
\setcounter{lyter}{1}
{} & {} & {\cal W}_{+}(\theta) = {\cal W}_{0}(\theta) + {\cal W}_{1}(\theta)
= P_1(\theta)\Gamma^p(\theta)
\displaystyle\frac{\partial_{l} \phantom{xxx}}{\partial\Gamma^{p}(\theta)}
\;,\\
\setcounter{lyter}{2}
\setcounter{equation}{12}
{} & {} & {\cal W}_{-}(\theta) = {\cal W}_{0}(\theta) - {\cal W}_{1}(\theta)
= P_1(\theta)\Gamma^p(\theta)
\displaystyle\frac{\partial_{l}
\phantom{xxx}}{\partial\overline{\Gamma^{p}(\theta)}} = - \left( {\cal
W}_{+}(\theta)\right)^{\ast},\\
\setcounter{lyter}{3}
\setcounter{equation}{12}
{} & {} & {\stackrel{\circ}{\cal W}}_{+}(\theta) =
{\stackrel{\circ\ \;}{\Gamma^p}}(\theta)
\displaystyle\frac{\partial_{l} \phantom{xxx}}{\partial\Gamma^{p}(\theta)},\;
{\stackrel{\circ}{\cal W}}_{-}(\theta) =
{\stackrel{\circ\ \;}{\Gamma^p}}(\theta)
\displaystyle\frac{\partial_{l}
\phantom{xxx}}{\partial\overline{\Gamma^{p}(\theta)}} =
\left({\stackrel{\circ}{ \cal W}}_{+}(\theta)\right)^{\ast},
\end{eqnarray}
\vspace{-3ex}
\renewcommand{\theequation}{\arabic{section}.\arabic{equation}}
\begin{eqnarray}
{} & {} &
\displaystyle\frac{\partial_{l}
\phantom{xxx}}{\partial\overline{\Gamma^{p}(\theta)}} =
\displaystyle\frac{\partial_{l}
\phantom{xxxxxxx}}{\partial P_0(\theta)\Gamma^{p}(\theta)} -
\displaystyle\frac{\partial_{l}
\phantom{xxxxxxx}}{\partial P_1(\theta)\Gamma^{p}(\theta)} \equiv
\left(\displaystyle\frac{\partial_{l}
\phantom{xxx}}{\partial\Gamma^{p}(\theta)}\right)^{\ast}.
\end{eqnarray}

The operators $U_1(\theta)$, ${\stackrel{\circ}{U}}_{0}(\theta)$,
$V_{1}(\theta)$,
${\stackrel{\circ}{V}}_{0}(\theta)$, ${\cal W}_{1}(\theta)$, $
{\stackrel{\circ}{\cal W}}_{0}(\theta)$, $\Delta_{00}(\theta)$, $\Delta_{11}(
\theta)$ are invariant with respect to ${*}$, but the only
${\stackrel{\circ}{U}}_{\pm}(\theta)$,
${\stackrel{\circ}{V}}_{\pm}(\theta)$,
${\stackrel{\circ}{\cal W}}_{\pm}(\theta)$, $\Delta_{\pm \pm}(\theta)$,
$\Delta_{\pm\mp}(\theta)$ are defined in the superfield form with respect to
$T$, $\overline{T}$ representations.
Operators $V_{+}(\theta)$, ${\cal W}_{+}(\theta)$,
$U_{+}(\theta)$ appear by the realizations of projectors $V(\theta)$,
${\cal W}(\theta)$, $U(\theta)$ on $D^k_{cl}$ in the relation (2.37).

All algebraic properties for {\boldmath${\cal A}_{cl}$} [1] under the
composition of operators are literally transferred onto subsuperalgebra
{\boldmath${\cal B}^1_{cl}$} of the 1st
order operators in {\boldmath${\cal B}_{cl}$}. The additional properties for
analogous ones from [1] have the form
\renewcommand{\theequation}{\arabic{section}.\arabic{equation}\alph{lyter}}
\begin{eqnarray}
\setcounter{lyter}{1}
1) \hspace{2em} & {V_a}^2 = \delta_{a1}V_a\ ,\ V_{\pm}^2 = \pm
V_{\pm},\
V_{+}V_{-} = V_{-}\;,\;V_{-}V_{+} = -V_{+}\;, & {}
\\
\setcounter{lyter}{2}
\setcounter{equation}{14}
2) \hspace{2em} & [V_a , V_b]_{-} = \varepsilon_{a b}V_0\ ,\
[V_{+},V_{-}]_{-} = 2V_0\ ,\ \varepsilon_{a b} = -\varepsilon_{b
 a},\;\varepsilon_{10} = 1,\ a,b=0,1\;,  & {} \nonumber \\
 {} & [V_{+},V_{-}]_{+} =
-2V_1\ ,\ [V_{+},V_{a}]_{-} = (-1)^a V_0\ ,\ [V_{-},V_{a}]_{-} = -V_0\ , &
{}   \\
\setcounter{lyter}{3}
\setcounter{equation}{14}
3) \hspace{2em} & {\stackrel{\circ}{V}}{}_i^2 = 0\ ,\
\left[{\stackrel{\circ}{V}}_{i},{\stackrel{ \circ}{V}}_{j}\right]_{+} =0\
,\ i,j \in \{0,1,+,-\phantom{1}\hspace{-0.5em}\}\;, & {}\\
\setcounter{lyter}{4}
\setcounter{equation}{14}
4) \hspace{2em} &
\left[{{\stackrel{\circ}{V}}_{0}},
{\stackrel{\phantom{\circ}}{V}_{i}}\right]_{-} = 0\ ,\
\left[{\stackrel{\circ}{V}}_{1},{\stackrel{\phantom{
\circ}}{V}_{i}}\right]_{-} =
{\stackrel{\circ}{V}}_{i}\ ,\ \left[{\stackrel{\circ}{V}}_{\pm},
{\stackrel{\phantom{\circ}}{V}_{i}}\right]_{-} =
\pm{{\stackrel{\circ}{V}}_{i}}\;, & {} \\
\setcounter{lyter}{5}
\setcounter{equation}{14}
5) \hspace{2em} &
\left[U_{i},V_j\right]_{-} = \left[U_{i},{\stackrel{\circ}{V}}_j\right]_{-}
= \left[{\stackrel{\circ}{U}}_{i},V_j\right]_{-} = \left[{\stackrel{\circ}{
U}}_{i},{\stackrel{\circ}{V}}_{j}\right]_{+} = 0\ . & {} 
\end{eqnarray}
For subset {\boldmath$\vec{\cal B}^1_{cl}$} with basis $\{U_{a}(\theta),
V_{a}(\theta)\}$ the remarks  being analogous
for {\boldmath$\vec{\cal A}_{cl}$} [1] are valid.

For the 2nd order operators let us point out only the following formulae
\begin{eqnarray}
\setcounter{lyter}{1}
{} & {} &\hspace{-4em}\left(\Delta^{cl}(\theta)\right)^2 =
\left[\Delta^{cl}(\theta),{\stackrel{\circ}{\cal W}}_{+}(\theta)\right]_{+}
= \left[\Delta^{cl}(\theta),{\stackrel{\circ}{U}}_{+}(\theta)\right]_{+}
= \left[\Delta^{cl}(\theta),{\stackrel{\circ}{V}}_{+}(\theta)\right]_{+} =
 \Delta^2_{ab}(\theta) = 0\;, \\
\setcounter{lyter}{2}
\setcounter{equation}{15}
{} & {} & \left[\Delta^{cl}(\theta),{\cal W}_{+}(\theta)\right]_{-} =
\Delta_{1+}(\theta) + \Delta_{+1}(\theta)\;.
\end{eqnarray}
A deviation of $\Delta_{ab}(\theta)$ in order to be derivation in acting on
the product of arbitrary ${\cal F}(\theta)$, ${\cal J}(\theta)$ $\in$
$D^{k}_{cl}$ yields the definition for antibrackets
\renewcommand{\theequation}{\arabic{section}.\arabic{equation}}
\begin{eqnarray}
{} & \Delta_{ab}(\theta)({\cal F}(\theta)\cdot {\cal J}(\theta)) =
(\Delta_{ab}(\theta){\cal F}(\theta)){\cal J}(\theta) +
(-1)^{\varepsilon({\cal F})}{\cal F}(\theta)\Delta_{ab}(\theta){\cal
J}(\theta) + & {} \nonumber \\
{} & (-1)^{\varepsilon({\cal F})}\left({\cal F}(\theta),{\cal
J}(\theta)\right)_{ab},\;a,b=0,1\;,& {} \\
{} & \left({\cal F}(\theta),{\cal J}(\theta)\right)_{ab} =
\displaystyle\frac{\partial {\cal F}(\theta)\phantom{xxxx}}{\partial
{P_a(\theta){\cal A}^{\imath}}(\theta)}
\displaystyle\frac{\partial {\cal J}(\theta)\phantom{xxxx}}{\partial
{P_b(\theta){\cal A}^{\ast}_{\imath}}(\theta)}  -
(-1)^{(\varepsilon({\cal
F}) +1)(\varepsilon({\cal J}) + 1)}\left({\cal F} \longleftrightarrow {\cal
J}\right) & {} 
\end{eqnarray}
which have the gradings as in (3.25).
Antibracket $(\;,\;)_{11}$ is not obtained by means
of (7.16) from operator $\Delta_{11}(\theta) \equiv 0$. It follows from
(7.17) the definitions for antibrackets in the superfield form
\renewcommand{\theequation}{\arabic{section}.\arabic{equation}\alph{lyter}}
\begin{eqnarray}
\setcounter{lyter}{1}
{} & {} & \hspace{-4em}\left({\cal F}(\theta),{\cal J}(\theta)\right )_{++}=
\displaystyle\sum\limits_{a,b} \left({\cal F}(\theta),{\cal J}
(\theta)\right)_{ab} \equiv \left({\cal F}(\theta),{\cal J}(\theta)\right)_{
\theta}\;,
\\
\setcounter{lyter}{2}
\setcounter{equation}{18}
{} & {} & \hspace{-4em}\left({\cal F}(\theta),{\cal J}(\theta)\right)_{+-}=
\displaystyle\sum\limits_{a ,b}\left( \frac{\partial{\cal
F}(\theta)\phantom{xxxx}}{\partial {P_a(\theta){\cal A}^{ \imath}(\theta)}}
\frac{\partial{\cal J}(\theta)\phantom{xxxx}}{\partial
\overline{P_b(\theta){\cal A}^{\ast}_{\imath}(\theta)}} -
(-1)^{(\varepsilon({\cal
F}) +1)(\varepsilon({\cal J}) + 1)}\left({\cal F} \longleftrightarrow {\cal
J}\right)\right),
\\
\setcounter{lyter}{3}
\setcounter{equation}{18}
{} & {} & \hspace{-4em}\left({\cal F}(\theta),{\cal J}(\theta)\right)_{-+} =
\displaystyle\sum\limits_{a ,b}\left( \frac{\partial{\cal
F}(\theta)\phantom{xxxx}}{\partial \overline{P_a(\theta){\cal A}^{
\imath}(\theta)}} \frac{\partial{\cal J}(\theta)\phantom{xxxx}}{\partial
{P_b(\theta){\cal A}^{\ast}_{\imath}(\theta)}} -
(-1)^{(\varepsilon({\cal
F}) +1)(\varepsilon({\cal J}) + 1)}\left({\cal F} \longleftrightarrow {\cal
J}\right)\right)
 ,\\
\setcounter{lyter}{4}
\setcounter{equation}{18}
{} & {} & \hspace{-4em}\left({\cal F}(\theta),{\cal J}(\theta)\right)_{--} =
\displaystyle\sum\limits_{a,b}\left( \frac{\partial{\cal
F}(\theta)\phantom{xxxx}}{\partial \overline{P_a(\theta){\cal A}^{
\imath}(\theta)}} \frac{\partial{\cal J}(\theta)\phantom{xxxx}}{\partial
\overline{P_b(\theta){\cal A}^{\ast}_{\imath}(\theta)}} -
(-1)^{(\varepsilon({\cal
F}) +1)(\varepsilon({\cal J}) + 1)}\left({\cal F} \longleftrightarrow {\cal
J}\right)\right)
 .
\end{eqnarray}
Antibrackets (7.17), (7.18) satisfy to standard properties for odd Poisson
bracket (3.25), (3.27). In particular, Jacobi identity for corresponding
antibrackets can be obtained from action of squared (formally) operators
$\Delta_{ab}(\theta)$, $\Delta_{\pm \pm}(\theta)$, $\Delta_{\pm\mp}(\theta)$
on the product of 3 arbitrary superfunctions from $D^{k}_{cl}$.

Operators ${\stackrel{\circ}{U}}_{i}(\theta)$,
${\stackrel{\circ}{V}}_{i}(\theta)$, $\Delta_{i j}(\theta)$,
$i,j$ $\in$ $\{0,1,+,-\}$ differentiate the antibrackets $(\;,\;)_{i j}$
with corresponding (coinciding for
$\Delta_{i j}(\theta)$ and $(\;,\;)_{i j}$) indices by
Leibnitz rule. For example, for any from operators $B(\theta)$ $\in$
$\{{\stackrel{\circ}{U}}_{+}(\theta)$, ${\stackrel{\circ}{V}}_{+}(\theta)$,
$\Delta^{cl}(\theta)\}$ and for $(\;,\;)_{\theta}$ the following relation
holds
\renewcommand{\theequation}{\arabic{section}.\arabic{equation}}
\begin{eqnarray}
B(\theta)\left({\cal F}(\theta),{\cal J}(\theta)\right)_{\theta} = \left(B(
\theta){\cal F}(\theta),{\cal J}(\theta)\right)_{\theta} +
(-1)^{\varepsilon({\cal F}) +1} \left({\cal F}(\theta), B(\theta){\cal
J}(\theta)\right)_{\theta}\;.
\end{eqnarray}
In fact the relationships (7.16)--(7.18), (7.6), (7.10), (5.6), (5.7) make
having the same rights a construction of antisymplectic differential
geometry on $T^{\ast}_{odd}{\cal M}_{cl}$ both starting from the operators $
\Delta_{i j}(\theta)$, $i,j \in  \{0,1,+,-\}$, next
obtaining the antibrackets $(\;,\;)_{i j}$ and vice versa
starting from the antibrackets(!) (with exception of $(\;,\;)_{11}$
and $\Delta_{11}(\theta) \equiv 0$).

With help of antibrackets one can define a so-called transformation of the
operators from {\boldmath${\cal B}_{cl}$}.
Let us confine ourselves by the case of operators $B(\theta)$ $\in$
$\{{\stackrel{\circ}{U}}_{+}(\theta)$, ${\stackrel{\circ}{V}}_{+}(\theta)$,
$\Delta^{cl}(\theta)\}$
having required the fulfilment of the transformation rule
\begin{eqnarray}
B(\theta) \mapsto B'(\theta) = B(\theta) + \left({\cal
F}(\theta),\;\;\right)_{\theta} \equiv B(\theta) + {\rm ad}_{{\cal F}(\theta)},\;
(\varepsilon_P,\varepsilon_{\bar{J}},\varepsilon){\cal F}(\theta) = (0,0,0)\;
.
\end{eqnarray}
The condition of nilpotency conservation for $B'(\theta)$ with
allowance made for antibracket's properties
(3.27) leads to the equation on a superfunction ${\cal F}(\theta)\in
D^{k}_{cl}$
\begin{eqnarray}
B(\theta){\cal F}(\theta) + \frac{1}{2}\left({\cal F}(\theta),{\cal F}(\theta)
\right)_{\theta} =
f\bigl({\stackrel{\circ}{\Gamma}}(\theta),\theta\bigr),\;
(\varepsilon_P,\varepsilon_{\bar{J}},\varepsilon)f(\theta) = (1,0,1),\;
f(\theta)_{\mid T^{\ast}_{odd}{\cal M}_{cl}} = 0\; .
\end{eqnarray}
Transformation (7.20) permits to achieve of the vanishing for one(!)
from $B(\theta)$ with help of a special choice for ${\cal F}(\theta)$
satisfying to (7.21). So putting successively
\begin{eqnarray}
{\cal F}_{1}(\theta) = {\stackrel{\ \circ}{\cal A}}{}^{\imath}(\theta)
{\cal A}^{\ast}_{\imath}(\theta),\;
{\cal F}_{2}(\theta) = - {\stackrel{\ \circ}{\cal
A}}{}^{\ast}_{\imath}(\theta) {\cal A}^{\imath}(\theta) \;,
\end{eqnarray}
we obtain respectively
\begin{eqnarray}
{\stackrel{\circ}{U}}{}_{+}'(\theta) = 0 \Longleftrightarrow
{\stackrel{\circ}{U}}_{+}(\theta) = - {\rm ad}_{{\cal F}_1(\theta)},\ \;
{\stackrel{\circ}{V}}{}_{+}'(\theta) = 0 \Longleftrightarrow
{\stackrel{\circ}{V}}_{+}(\theta) = - {\rm ad}_{{\cal F}_2(\theta)}\;
.
\end{eqnarray}
Choosing as ${\cal F}(\theta)$ in (7.20) the sum of ${\cal F}_{1}(\theta
)$ and ${\cal F}_{2}(\theta)$ from (7.22) we derive according to (7.20),
(7.22), (7.23) the formulae
\begin{eqnarray}
{\stackrel{\circ}{U}}{}_{+}'(\theta) = -
{\stackrel{\circ}{V}}_{+}(\theta),\ \;
{\stackrel{\circ}{V}}{}_{+}'(\theta) = -
{\stackrel{\circ}{U}}_{+}(\theta)\; .
\end{eqnarray}
For arbitrary ${\cal F}_{1}(\theta),{\cal F}_{2}(\theta) \in D^{k}_{cl}$ a
requirement   of the properties (7.14e) conservation taking
account of antibracket's ones (3.27) and (7.19) leads to necessity
of the compatibility condition fulfilment
\begin{eqnarray}
{} & {\stackrel{\circ}{{V}}_{+}}(\theta){\cal F}_1(\theta) +
{\stackrel{\circ}{{U}}_{+}}(\theta){\cal F}_2(\theta) +
\left({\cal F}_1(\theta), {\cal F}_2(\theta)\right)_{\theta} =
f_1\bigl({\stackrel{\circ}{\Gamma}}(\theta),\theta\bigr)\;, & {}
\nonumber\\
{} & (\varepsilon_P,\varepsilon_{\bar{J}},\varepsilon)f_1(\theta) =
(1,0,1),\; f_1(\theta)_{\mid T^{\ast}_{odd}{\cal M}_{cl}} = 0 & {}
\end{eqnarray}
which is the additional relation to Eqs.(7.21) for
$B(\theta) \in \{{\stackrel{\circ}{U}}_{+}(\theta),
{\stackrel{\circ}{V}}_{+}(\theta)\}$.
\section{Component Formulation}
\setcounter{equation}{0}

Continue in the framework of the Hamiltonian formulation for GSTF a
programme, started in [1] for Lagrangian formalism, of the establishment of
connection between superfield and component field quantities and relations
of GSTF.

In fact all  formulae of the analogous section in Ref.[1] are valid
here as well under formal change of the form ${\cal A}^{\imath}(\theta)$
$\to$ $\Gamma^{p}(\theta)= ({\cal A}^{\imath}(\theta), {\cal A}^{\ast}_{
\imath}(\theta))$ in the corresponding ones from [1]. So from (2.47) let us
find
the expression for densities of superfunctionals on $D^{k}_{cl}$ through
latters themselves
\begin{eqnarray}
{} & {\cal F}\bigl(\Gamma(\theta),{\stackrel{\circ}{\Gamma}}(\theta),
\theta\bigr) = P_0(\theta){\cal F}(\theta) + \theta F_{H,cl}[\Gamma] \equiv
 {\cal F}\bigl(P_0\Gamma(\theta),{\stackrel{\circ}{\Gamma}}(\theta),0
\bigr) + \theta\overline{F}_{H,cl}[P_0\Gamma,{\stackrel{\circ}{
\Gamma}}] \equiv & {} \nonumber \\
{} & {\cal F}\bigl(\Gamma_0,\Gamma_1,0\bigr)
 + \theta\overline{F}_{H,cl}[\Gamma_0,\Gamma_1]\;, & {}
\\
{} & \Gamma^p(\theta) = \Gamma^p_0 + \Gamma^p_1 \theta \equiv (A^{\imath} +
\lambda^{\imath}\theta, A^{\ast}_{\imath} - \theta J_{\imath})\;. & {}
\end{eqnarray}
The formula
\begin{eqnarray}
\frac{\partial_r
{\cal F}\bigl(\Gamma(\theta),{\stackrel{\circ}{\Gamma}}(\theta), \theta
\bigr)}{\partial\Gamma^p(\theta)\phantom{xxxxxxxx}} =
\frac{\partial_r {\cal F}(\theta)\phantom{xxxx}}{\partial
P_0(\theta)\Gamma^p(\theta)} + \frac{\partial_r {\cal
F}(\theta)\phantom{xxxx}}{\partial P_1(\theta)\Gamma^p(\theta)} =
\frac{\delta_r {\cal F}(\theta)}{\delta \Gamma^p_0 \phantom{xxx}} +
\frac{\delta_r {\cal F}(\theta)}{\delta
(\Gamma^p_1\theta)}
\end{eqnarray}
establishes the
connection of the partial superfield derivative with respect to superfield
$\Gamma^{p}(\theta)$ with component ones.

The component representation for supermatrix of the 2nd partial superfield
derivatives of ${\cal F}(\theta)$ with respect to $\Gamma^{p}(\theta
)$, $\Gamma^{q}(\theta)$ is literally the same as in [1] under change of
${\cal A}^{\imath}(\theta)$, ${\cal A}^{\jmath}(\theta)$ on $\Gamma^{p}(\theta
)$, $\Gamma^{q}(\theta)$ respectively.

From formulae
\begin{eqnarray}
\frac{\partial_r \phantom{xxxxxxx}}{\partial \left(\displaystyle\frac{ d_r
{\Gamma}^{p}(\theta)}{d\theta\phantom{xxx}}\right)} =
\frac{\delta_r \phantom{x}}{\delta \Gamma^p_1} =
\left(\frac{\delta_r }{\delta\lambda^{\imath}},
\frac{\delta_r }{\delta J_{\imath}}
(-1)^{\varepsilon_{\imath} + 1}\right),
\end{eqnarray}
it seems natural to introduce the following component variational
derivatives (remind that ${\cal F}(\theta)$ for fixed $\theta$
represents the usual functional (or function) on ${}^{0,0}D^{k}_{cl})$
\begin{eqnarray}
\frac{\delta_r \phantom{xxx}}{\delta \bigl(\Gamma^{p}_1\theta\bigr)} =
\left(\frac{d_r}{d\theta}
\frac{\tilde{\delta}_r \phantom{x}}{\delta
\Gamma^p_1}\right)(-1)^{\varepsilon_p + 1} \equiv
\left(\left(
\frac{d_r}{d\theta} \frac{{\tilde{\delta}}_r \phantom{x}}{\delta
\lambda^{\imath}}\right)(-1)^{\varepsilon_{\imath} + 1},
- \left(\frac{d_r}{d\theta} \frac{{\tilde{\delta}}_r \phantom{x}}{\delta
J_{\imath}}\right)\right)
 ,
\end{eqnarray}
where ($\frac{d_r}{d\theta}\frac{\tilde{\delta}_r\phantom{x}}{\delta
\Gamma^p_1}$) are considered as a single differential object acting on
$D^{k}_{cl}$.

Connection for the left and right derivatives of the form (8.5) is yielded by
the formulae
\begin{eqnarray}
\frac{\delta_r {\cal F}(\theta)}{\delta (\Gamma^p_1\theta)}  =
(-1)^{\varepsilon_{p} (\varepsilon({\cal F}) + 1)} \frac{\delta_l {\cal
F}(\theta)}{\delta (\Gamma^{p}_1\theta)},\;
\frac{{\tilde{\delta}}_r {\cal F}(\theta)}{\delta
\Gamma^p_1\phantom{xx}} = \frac{{\tilde{\delta}}_l {\cal
F}(\theta)}{\delta \Gamma^p_1\phantom{xx}}
(-1)^{(\varepsilon_p+1) (\varepsilon({\cal F}) + 1)}.
\end{eqnarray}
The relationship of the right variational superfield derivative with respect
to $\Gamma^{p}(\theta)$ and the component variational (in an usual sense)
ones on $\Gamma_{0}^{p}$ and $\Gamma_{1}^{p}$ follows
from (2.2), (2.48b), (8.1)
\begin{eqnarray}
{} &  \displaystyle\frac{\delta_r
F_{H,cl}[\Gamma]}{\delta\Gamma^p(\theta)\phantom{xx}}  =
\frac{\delta_r P_0 {\cal
F}(\theta)}{\delta \Gamma^p_0\phantom{xxxx}} + \theta\frac{\delta_r
\overline{F}_{H,cl}[\Gamma_0,\Gamma_1]}{\delta \Gamma^p_0 \phantom{xxxxxxxx}}
+ \frac{\delta_r \overline{F}_{H,cl}[\Gamma_0,\Gamma_1]}{\delta \Gamma^p_1
\phantom{xxxxxxxx}} (-1) ^{\varepsilon (F_{H,cl})} = & {}
\nonumber \\
{} & \displaystyle\frac{\partial_r P_0 {\cal F}(\theta)}{\partial P_0
\Gamma^p(\theta)\phantom{x}} +
\theta\frac{\partial_r \overline{F}_{H,cl}\bigl[P_0 \Gamma,
{\stackrel{\circ}{\Gamma}}\bigr]}{\partial P_0 {\Gamma}^{p}(
\theta)\phantom{xxxxx}} + \frac{\partial_r \overline{F}_{H,cl}\bigl[P_0{
\Gamma}, {\stackrel{\circ}{\Gamma}}\bigr]}{\partial
\left(\displaystyle\frac{d_r {\Gamma}^{p}(\theta)}{d
\theta\phantom{xxxx}}\right)\phantom{xx}}(-1)^{\varepsilon (F_{H,cl})} = &{}
\nonumber \\
{} &\Biggl(\displaystyle\frac{\delta P_0 {\cal F}(\theta)}{\delta
A^{\imath}\phantom{xxxx}} + \theta\frac{\delta \overline{F}_{H,cl}\bigl[
\Gamma_0,\Gamma_1\bigr]}{\delta A^{\imath}
\phantom{xxxxxxxx}} +
\displaystyle\frac{\delta \overline{F}_{H,cl}\bigl[\Gamma_0,\Gamma_1\bigr]}{
\delta\lambda^{\imath}
\phantom{xxxxxxxx}}(-1)^{\varepsilon (F_{H,cl})}, & {} \nonumber \\
{} & \displaystyle\frac{\delta_r P_0 {\cal F}(\theta)}{\delta
A^{\ast}_{\imath}\phantom{xxxx}} + \theta\frac{\delta_r \overline{F}_{H,cl}
\bigl[\Gamma_0,\Gamma_1\bigr]}{\delta  A^{\ast}_{\imath}(
\theta)\phantom{xxxxxx}} + \frac{\delta_r \overline{F}_{H,cl}\bigl[\Gamma_0,
{\Gamma}_1\bigr]}{\delta J_{\imath}\phantom{xxxxxx
xx}} (-1)^{\varepsilon (F_{H,cl}) +\varepsilon_{\imath} + 1}\Biggr). & {}
\end{eqnarray}
From (8.7) it follows as in [1] the formulae in coordinates
$\Gamma ^{p}(\theta )$ and $({\cal A}^{\imath}(\theta ), {\cal A}^{*}_{
\imath}(\theta))$ respectively
\begin{eqnarray}
{} & \displaystyle\frac{d_r}{d\theta}\frac{\delta_r F_{H,cl}[\Gamma]}{\delta
{\Gamma}^{p}(\theta)\phantom{xxx}} = (-1)^{\varepsilon_{p} +
\varepsilon(F_{H,cl})}\frac{\partial_r \overline{F}_{H,cl}
\bigr[P_0\Gamma,
{\stackrel{\circ}{\Gamma}}\bigr]}{\partial
{\Gamma}^{p}(\theta)\phantom{xxxxxxx}} = & {} \nonumber \\
{} & (-1)^{\varepsilon_{p} +
\varepsilon (F_{H,cl})}\displaystyle\frac{\partial_r
\overline{F}_{H,cl}\bigl[P_0{\Gamma},
{\stackrel{\circ}{\Gamma}}\bigr]}{
\partial P_0{\Gamma}^{p}(\theta) \phantom{xxxxx}}
= (-1)^{\varepsilon_{p} + \varepsilon (F_{H,cl})}\frac{\delta_r
\overline{F}_{H,cl}[\Gamma_0,\Gamma_1]}{ \delta \Gamma^{p}_0\phantom{xxxx
xxxx}}\;,
& {} \\
{} & \displaystyle\left(\frac{d_r}{d\theta}\frac{\delta \phantom{xxxx}}{
\delta{\cal A}^{\imath}(\theta)},
\frac{d_r}{d\theta}\frac{\delta_r \phantom{xxxx}}{\delta
{\cal A}^{\ast}_{\imath}(\theta)}\right)F_{H,cl}[{\cal A},{\cal A}^{\ast}] =
\left(
(-1)^{\varepsilon_{\imath}}\frac{\partial\phantom{xxxxx}}{\partial P_0{\cal
A}^{\imath}(\theta)}, (-1)^{\varepsilon_{\imath} + 1}
\frac{\partial_r\phantom{xxxxx}}{\partial P_0{\cal A}^{\ast}_{\imath}(\theta)}
\right)\times & {} \nonumber \\
{} & (-1)^{\varepsilon (F_{H,cl})}\overline{F}_{H,cl}\bigl[P_0\Gamma,
{\stackrel{\circ}{\Gamma}}\bigr] = \displaystyle\left(
(-1)^{\varepsilon_{\imath}}\frac{\delta\phantom{x}}{\delta
{A}^{\imath}}, (-1)^{\varepsilon_{\imath} + 1}
\frac{\delta_r\phantom{x}}{\delta {A}^{\ast}_{\imath}}\right)
\overline{F}_{H,cl}[ A, A^{\ast}, \lambda, J](-1)^{\varepsilon (F_{H,cl})}.
& {}
\end{eqnarray}
Relationships being  analogous to (8.8) for a
superfunction ${\cal F}(\theta)$  have the form
\begin{eqnarray}
{} &
\displaystyle\frac{d_r}{d\theta_1}\frac{\delta_r
\phantom{xxxx}}{\delta{\Gamma}^{p}(\theta_1)} {\cal F}\bigl(
{\Gamma}(\theta), {\stackrel{\circ}{\Gamma}}(\theta),\theta\bigr)   =
\frac{d_r}{d\theta_1}\left(\delta(\theta_1 - \theta) \frac{\partial_r
\phantom{xxxx}}{\partial {\Gamma}^{p}(\theta_1)}{\cal
F}\bigl({\Gamma}(\theta_1), {\stackrel{\circ}{\Gamma}}(\theta_1),
\theta_1\bigr)\right) = & {} \nonumber \\
{} &
\displaystyle\frac{\partial_r \phantom{xxxx}}{ \partial {\Gamma}^{p}
(\theta_1)}\left({\cal F}(\theta_1) (-1)^{\varepsilon({\cal F})} +
\delta(\theta_1 - \theta)\frac{d_r}{d\theta_1}{\cal
F}(\theta_1)\right)(-1)^{\varepsilon_{p}} = & {} \nonumber\\
{} & (-1)^{\varepsilon({\cal F}) + \varepsilon_{p}}\displaystyle\frac{
\partial_r \phantom{xxxxxx}}{\partial P_0{\Gamma}^{p}(\theta_1)}{\cal
F}\bigl({\Gamma}(\theta), {\stackrel{\circ}{\Gamma}}(\theta),\theta\bigr) =
(-1)^{\varepsilon({\cal F}) + \varepsilon_{p}} \frac{\delta_r {\cal F}
(\theta)}{\delta \Gamma^{p}_0\phantom{xx}} \;.
\end{eqnarray}
In obtaining of (8.10) the identities have been made use
\begin{eqnarray}
P_0(\theta_1){\cal F}(\theta_1) \equiv  P_0(\theta){\cal
F}(\theta),\;\frac{d\phantom{x}}{d\theta_1}{\cal F}(\theta_1) \equiv
\frac{d}{d\theta}{\cal F}(\theta),\;\frac{d^2}{d\theta_1^2} \equiv 0\;.
\end{eqnarray}
All above relations are sufficient in order to get the component
expressions for operators from {\boldmath${\cal B}_{cl}$} being additional
to ones obtained in Ref.[1]. Let us point out the formulae only for basis
operators $V_{a}(\theta)$, ${\stackrel{\circ}{V}}_{a}(\theta)$,
$\Delta_{ab}(\theta)$, $a,b=0,1$
\begin{eqnarray}
V_0(\theta) & = &
-\theta J_{\imath}  \frac{\delta \phantom{x}}{\delta A^{\ast}_{\imath}} ,\
V_1(\theta) = - \theta J_{\imath} \frac{\delta \phantom {xxxx}}{\delta
(-\theta J_{\imath})} = \theta J_{\imath} \left(
\frac{d}{d\theta}\frac{\tilde{\delta} \phantom{x}}{\delta J_{\imath}}
\right)(-1)^{\varepsilon_{\imath}}\,, \\
{\stackrel{\circ}{V}}_0(\theta) & = &
- J_{\imath} \frac{\delta
\phantom{x}}{\delta A^{\ast}_{\imath}},\ {\stackrel{\circ}{V}}_1(\theta) =
- J_{\imath} \frac{\delta \phantom
{xxxx}}{\delta (- \theta J_{\imath})} = (-1)^{\varepsilon_{\imath}}
J_{\imath} \left(\frac{d}{d\theta}\frac{\tilde{\delta} \phantom{x}}{
\delta J_{\imath}} \right), \\
\Delta_{00}(\theta) & = & \frac{\delta_l \phantom{x}}{\delta A^{\imath}}
\frac{\delta \phantom{x}}{\delta A^{\ast}_{\imath}}(-1)^{\varepsilon_{
\imath}},\;
\Delta_{01}(\theta) = \frac{\delta_l \phantom{x}}{\delta A^{\imath}}
\frac{\delta \phantom{xxxx}}{\delta (-
\theta J_{\imath})}(-1)^{\varepsilon_{ \imath}} = -
\frac{\delta_l \phantom{x}}{\delta A^{\imath}}
\left(\frac{d}{d\theta}\frac{\tilde{\delta} \phantom
{x}}{\delta J_{\imath}} \right), \nonumber \\
\Delta_{10}(\theta) & = & \frac{\delta_l \phantom{xxx}}{\delta (
\lambda^{\imath}\theta)}\frac{\delta\phantom{x}}{\delta A^{\ast}_{\imath}}
(-1)^{\varepsilon_{ \imath}}   =
 \left(\frac{d}{d\theta}\frac{\tilde{\delta}_l \phantom
{x}}{\delta \lambda^{\imath}} \right)
\frac{\delta \phantom{x}}{\delta A^{\ast}_{\imath}}
(-1)^{\varepsilon_{ \imath}}\,.
\end{eqnarray}
At last for the antibrackets (7.17) calculated on arbitrary
 ${\cal F}(\theta)$, ${\cal J}(\theta)$ from $D^{k}_{cl}$ we
have
\begin{eqnarray}
\left({\cal F}(\theta), {\cal J}(\theta)\right)_{00} =
\frac{\delta {\cal F}(\theta)}{\delta A^{\imath}\phantom{xx}}
\frac{\delta {\cal J}(\theta)}{\delta A^{\ast}_{\imath}\phantom{xx}}
- (-1)^{(\varepsilon({\cal F}) + 1)(\varepsilon({\cal J}) + 1)}({\cal F}
\longleftrightarrow {\cal J})\;.
\end{eqnarray}
The antibrackets $({\cal F}(\theta), {\cal J}(\theta))_{01}$,
$({\cal F}(\theta),
{\cal J}(\theta))_{10}$, $({\cal F}(\theta), {\cal J}(\theta))_{11}$ are
yielded from (8.15) under corresponding changes for operators
$\frac{\delta\phantom{xx}}{\delta A^{\ast}_{\imath}}$ on $\frac{\delta
\phantom{xxxxx}}{\delta
(-\theta J_{\imath})}$, $\frac{\delta\phantom{xx}}{\delta A^{\imath}}$ on
$\frac{\delta\phantom{xxxx}}{\delta (\lambda^{\imath}\theta)}$ and under
their simultaneous change for antibracket $(\;,\;)_{11}$.

Formulae of this Section are sufficient in order to write  all relations
of the Hamiltonian formulation for GSTF in the component form.
\section{Models in the Hamiltonian Formulation for GSTF}
\setcounter{equation}{0}

Continue investigation of the models from work [1] made in
the Lagrangian formulation for the case of the Hamiltonian one.
\subsection{Models of Massive Complex Scalar Superfield}

Begin from introduction of the superantifields\footnote{in contrast to the
designation of the complex conjugation $\ast$ for superfields
$\varphi^{\ast}(x,\theta)$ adopted in Ref.[1], in this section the sign
$"\ast"$ is reserved for notation of the only superantifields, whereas the
complex conjugate quantity is denoted with bar: $\bar{g}(x,\theta)$ unlike of
$\theta$-conjugate quantity $\overline{g(x,\theta)}$ for $g(x,\theta)$ as in
Sec.VII}
$(\varphi_{j}(x,\theta))^{\ast }$ $\in$ $\tilde{\Lambda}_{4\mid
0+1}(x^{\mu},\theta;{\bf R})$, $j=1,2$ for real component superfields
$\varphi_{j}(x,\theta)$ forming the complex scalar superfield
$\varphi(x,\theta)$
$\in$ $\tilde{\Lambda}_{4\mid 0+1}(x^{\mu},\theta;{\bf C})$
defined on ${\cal M} = {\bf R}^{1,3}\times\tilde{P}$ [1]
\begin{eqnarray}
\varphi(x,\theta)=\varphi_1(x,\theta)+\imath\varphi_2(x,\theta) =
\varphi(x) + \lambda(x)\theta\;. 
\end{eqnarray}
The noncontradictory join of the real superantifields
$(\varphi_{j}(x,\theta))^{\ast}$ into
complex ones $(\varphi(x,\theta))^{\ast}$,
$(\overline{\varphi}(x,\theta))^{\ast}$ is realized by the formulae
\renewcommand{\theequation}{\arabic{section}.\arabic{equation}\alph{lyter}}
\begin{eqnarray}
\setcounter{lyter}{1}
{} & \hspace{-1.8em}(\varphi(x,\theta))^{\ast}\hspace{-0.2em} =\hspace{-0.2em}
(\varphi_1(x,\theta))^{\ast}\hspace{-0.25em} - \imath(\varphi_2(x,\theta))^{
\ast}\hspace{-0.2em}=\hspace{-0.2em}
(\varphi(x))^{\ast} \hspace{-0.25em}-\theta J_{\varphi}(x),\;
(\varphi_{j}(x,\theta))^{\ast}\hspace{-0.2em} =\hspace{-0.2em}
 (\varphi_{j}(x))^{\ast}\hspace{-0.25em} -\theta
J_{\varphi{}j}(x), & {} \\
\setcounter{lyter}{2}
\setcounter{equation}{2}
{} & (\overline{\varphi}(x,\theta))^{\ast} = ({\varphi}_1(x,\theta))^{\ast}-
(\imath{\varphi}_2(x,\theta))^{\ast} =
(\varphi_1(x,\theta))^{\ast} + \imath(\varphi_2(x,\theta))^{\ast}\,,&{}
\end{eqnarray}
providing the fulfilment of the relationship with bar from the left-hand
side meaning in question the complex conjugation
\renewcommand{\theequation}{\arabic{section}.\arabic{equation}}
\begin{eqnarray}
\overline{({\varphi}(x,\theta))^{\ast}}= (\overline{\varphi}(x,\theta))^{\ast}
\,, 
\end{eqnarray}
to be valid in this case in view of below following in (9.7a,b) relations.

The complex superantifields are the elements of complex irreducible massive
Poincare group representation and possess by the following Grassmann gradings
written, for instance, for $({\varphi}(x,\theta))^{\ast}$ and its components
with respect to $\theta$
\begin{eqnarray}
\begin{array}{lccccc}
{} & ({\varphi}(x,\theta))^{\ast} & J_{\varphi}(x) &
({\varphi}(x,\theta))^{\ast} & \theta J_{\varphi}(x) & {}\\
\varepsilon_P & 1 & 0 & 1 & 1  & {} \\
\varepsilon_{\Pi} & 0 & 0 & 0 & 0 & {}\\
\varepsilon & 1 & 0 & 1 &  1 & \hspace{-2em},
\end{array}
\end{eqnarray}
in accordance with connection of spin with only statistic $\varepsilon_{\Pi}$.

Superantifields $({\varphi}(x,\theta))^{\ast}$, $(\overline{\varphi}(x,
\theta))^{\ast}$ are transformed with respect to
$T^{\ast}_{\vert P}$ representation in the form
\renewcommand{\theequation}{\arabic{section}.\arabic{equation}\alph{lyter}}
\begin{eqnarray}
\setcounter{lyter}{1}
{} & {} & \delta({\varphi}(x,\theta))^{\ast} = {({\varphi
}(x,\theta))^{\ast}}'
- ({\varphi}(x,\theta))^{\ast} = -\mu
\bigl({\stackrel{\circ}{\varphi}}(x,\theta)\bigr)^{\ast} =
\mu J_{\varphi}(x)\;,\\
\setcounter{lyter}{2}
\setcounter{equation}{5}
{} & {} & \delta(\overline{\varphi}(x,\theta))^{\ast} =
{(\overline{\varphi}(x,\theta))^{\ast}}'
- (\overline{\varphi}(x,\theta))^{\ast} = -\mu \bigl({\stackrel{
\circ}{\overline{\varphi}}}(x,\theta)\bigr)^{\ast} =
\mu\overline{J}_{\varphi}(x)\;.
\end{eqnarray}
Superantifields $\bigl({\stackrel{\circ}{\varphi}}(x,\theta)\bigr)^{\ast}$,
$\bigl({\stackrel{\circ}{\overline{\varphi}}}(x,\theta)\bigr)^{\ast}$ are
the elements of mentioned Poincare group superfield representation and the
scalars with respect to action of $T^{\ast}_{\vert P}$ operators (2.26b).

In view of the fact that supermatrix $K(\theta,x,y)$ of the 2nd derivatives
of $S_L(\theta)$  with respect to ${\stackrel{\circ}{
\varphi}}(x,\theta)$,
${\stackrel{\circ}{\overline{\varphi}}}(y,\theta)$ can be chosen by
nondegenerate [1], then it is possible to pass to the Hamiltonian formulation
in  $T^{\ast}_{odd}{\cal M}_{cl}$ parametrized by
coordinates $\Gamma(x,\theta)$ = ($\varphi(x,\theta)$,
$\overline{\varphi}(x,\theta)$, $(\varphi(x,\theta))^{\ast}$,
$(\overline{\varphi}(x,\theta))^{\ast}$).  Dimensions of the supermanifolds
${\cal M}_{cl}$ and $T^{\ast}_{odd}{\cal M}_{ cl}$ with respect to
$\varepsilon$ parity read as follows, regarding as the independent
coordinates, for instance,
the only $\varphi(x,\theta)$, $(\varphi(x,\theta))^{\ast}$
\renewcommand{\theequation}{\arabic{section}.\arabic{equation}}
\begin{eqnarray}
{\rm dim}_{\bf C}{\cal M}_{cl} = (1,0),\;
{\rm dim}_{\bf C}T^{\ast}_{odd}{\cal M}_{cl} =
{\rm dim}_{\bf C}T_{odd}{\cal M}_{cl} = (1,1)\;.
\end{eqnarray}
Legendre transform (3.2), (3.3) is defined by the relations
\renewcommand{\theequation}{\arabic{section}.\arabic{equation}\alph{lyter}}
\begin{eqnarray}
\setcounter{lyter}{1}
{} & ({\varphi}(x,\theta))^{\ast}=
\displaystyle\frac{\partial_l
S_L\bigl({\varphi}(\theta),\overline{\varphi}(\theta),
{\stackrel{\circ}{\varphi}}(\theta),
{\stackrel{\circ}{\overline{\varphi}}}(\theta)\bigr)}{\partial
{\stackrel{\circ}{{\varphi}}}(x,\theta)\phantom{xxxxxxxxxxxxxx}} =
\frac{\partial_l
T\bigl({\stackrel{\circ}{{\varphi}}}(\theta),
{\stackrel{\circ}{\overline{\varphi}}}(\theta)\bigr)}{\partial
{\stackrel{\circ}{{\varphi}}}(x,\theta)\phantom{xxxxx}} =
\imath{\stackrel{\circ}{\overline{\varphi}}}(x,\theta)\;, & {}
\\
\setcounter{lyter}{2}
\setcounter{equation}{7}
{} & (\overline{\varphi}(x,\theta))^{\ast} =
\displaystyle\frac{\partial_l
S_L\bigl({\varphi}(\theta),\overline{\varphi}(\theta),
{\stackrel{\circ}{{\varphi}}}(\theta),
{\stackrel{\circ}{\overline{\varphi}}}(\theta)\bigr)}{\partial
{\stackrel{\circ}{\overline{\varphi}}}(x,\theta)\phantom{xxxxxxxxxxxxxx}} =
\frac{\partial_l
T\bigl({\stackrel{\circ}{{\varphi}}}(\theta),
{\stackrel{\circ}{\overline{\varphi}}}(\theta)\bigr)}{\partial
{\stackrel{\circ}{\overline{\varphi}}}(x,\theta)\phantom{xxxxx}} = -
\imath{\stackrel{\circ}{\varphi}}(x,\theta)\;, & {}
 \\
\setcounter{lyter}{3}
\setcounter{equation}{7}
{} & S_H(\Gamma(\theta)) =
T\bigl((\varphi(\theta))^{\ast},(\overline{\varphi}(\theta))^{\ast}\bigr) +
S_0(\varphi(\theta), \overline{\varphi}(\theta)),\;
T\bigl({\stackrel{\circ}{{\varphi}}}\bigl((\overline{\varphi}(\theta))^{\ast}
\bigr),
{\stackrel{\circ}{\overline{\varphi}}}\bigl((\varphi(\theta))^{\ast}\bigr)
\bigr) =
& {}\nonumber \\
{} & \hspace{-1em}
T\bigl((\varphi(
\theta))^{\ast},(\overline{\varphi}(\theta))^{\ast}\bigr) =
\hspace{-0.1em}\displaystyle\int \hspace{-0.2em}d^4x \displaystyle\frac{1}{
\imath}(\varphi(x,\theta))^{\ast}(\overline{\varphi}(x,\theta))^{\ast}
\equiv \hspace{-0.1em}\displaystyle\int \hspace{-0.2em}d^4x {\cal L}_{\rm
kin}^{\ast}(x,\theta) \equiv
\hspace{-0.1em} \displaystyle\int \hspace{-0.2em}d^4x {\cal L}_{\rm
kin}(x,\theta).{} &
\end{eqnarray}
Remind that this free GSTF model is the
nondegenerate (and even nonsingular [5]) ThST. In the framework of Sec.VI
terminology  the model
belongs to the I class in Lagrangian and Hamiltonian formulations  and
former one is given by  superfunction $S_{L}(\theta)$ [1] defined on
 $T_{odd}{\cal M}_{cl}$
\begin{eqnarray}
\setcounter{lyter}{1}
S_L(\theta) & \equiv & S_L\Bigl(\varphi(\theta),\overline{\varphi}(\theta),
{\stackrel{\circ}{\varphi}}(\theta), {\stackrel{\circ}{\overline{\varphi}}
}(\theta)\Bigr)
= T\Bigl({\stackrel{\circ}{\varphi}}(\theta), {\stackrel{\circ}{
\overline{\varphi}}}(\theta)\Bigr) -
S_{0}\bigl(\varphi(\theta), \overline{\varphi}(\theta)\bigr)\;, \\ 
\setcounter{equation}{8}
\setcounter{lyter}{2}
T(\theta) & \equiv &
T\Bigl({\stackrel{\circ}{\varphi}}(\theta), {\stackrel{
\circ}{\overline{\varphi}}}(\theta)\Bigr) =
\int d^4 x \displaystyle\frac{1}{\imath}{\stackrel{\circ}{\overline{\varphi}}
}(x,\theta) {\stackrel{\circ}{\varphi}}(x,\theta) \equiv
 \int d^4 x {\cal L}_{\rm kin}(x,\theta)\; , \\ 
\setcounter{equation}{8}
\setcounter{lyter}{3}
S_0(\theta) & \equiv &
S_{0}\bigl(\varphi(\theta),\overline{\varphi}(\theta)\bigr)
= \int d^4x (\partial_{\mu}
\overline{\varphi}\partial^{\mu}\varphi - m^2\overline{\varphi}\varphi)(
x,\theta)
\equiv \int d^4x {\cal L}_0 (x,\theta)\;, 
\end{eqnarray}
and therefore the Hamiltonian formulation of model (9.7c) falls literally
under conditions of Corollary 1.4 because of the Lagrangian one in this case
had satisfied to conditions of Corollary 2.2 from the work [1].

Correspending GHS has the form (6.28), (3.51b)
\begin{eqnarray}
\setcounter{lyter}{1}
{} & {} &\hspace{-2em}\displaystyle\frac{d_r \varphi(x,\theta)}{d\theta
\phantom{xxxxx}}  = \frac{\partial T(\theta)\phantom{xxx}}{\partial
(\varphi(x,\theta))^{\ast}} = \displaystyle\frac{1}{\imath}(\overline{
\varphi}(x,\theta))^{\ast},\;
\displaystyle\frac{d_r \overline{\varphi}(x,\theta)}{d\theta\phantom{xxxxx}} =
\frac{\partial
T(\theta)\phantom{xxx}}{\partial(\overline{\varphi}(x,\theta))^{\ast}} = -
\displaystyle\frac{1}{\imath}(\varphi(x,\theta))^{\ast}\;,
\\
\setcounter{lyter}{2}
\setcounter{equation}{9}
{} & {} & \hspace{-2em}\displaystyle\frac{d_r (\varphi(x,\theta))^{
\ast}}{d\theta\phantom{xxxxxxx}} = 0,\ \;
\displaystyle\frac{d_r (\overline{\varphi}(x,\theta))^{\ast}}{d\theta
\phantom{xxxxxxx}} = 0\;,
\\
\setcounter{lyter}{3}
\setcounter{equation}{9}
{} & {} &\hspace{-2em}\Theta_{\varphi}^H\bigl(\overline{\varphi}(x,
\theta),
\Box\overline{\varphi}(x,\theta)\bigr) = -
\displaystyle\frac{\partial_l
S_0(\theta)}{\partial{\varphi}(x,\theta)} =
(\Box + m^2)\overline{\varphi}(x,\theta) = 0\;,\\
\setcounter{lyter}{4}
\setcounter{equation}{9}
{} & {} & \hspace{-2em}
\Theta_{\overline{\varphi}}^H\bigl({\varphi}(x,\theta),
\Box{\varphi}(x,\theta)\bigr) = -
\displaystyle\frac{\partial_l
S_0(\theta)}{\partial \overline{\varphi}(x,\theta)} =
(\Box + m^2){\varphi}(x,\theta) = 0\;.
\end{eqnarray}
The only half of them is independent, for instance, the equations for
${\varphi}(x,\theta)$, $\overline{\varphi}(x,\theta))^{\ast}$. HS (9.9a,b)
is solvable. HCHF (9.9c,d) representing the superfield (on $\theta$)
generalization of Klein-Gordon equation and coinciding with corresponding HCLF
[1] do not appear by solvable. GHS (9.9) being equivalent to corresponding
LS [1] has the superfunction $S_H(\theta)$ (9.7c) by one's integral.

There are not of the type (6.22) identities among Eqs.(9.9c,d) although the
only Eq.(9.9d) is independent.

Equation (2.6), in force of the fact that $S_H(\theta)$ = $S_E(\theta)$ =
$T\Bigl({\stackrel{\circ}{\varphi}}(\theta), {\stackrel{
\circ}{\overline{\varphi}}}(\theta)\Bigr)$ $+$ $S_0(\theta)$ in terms of
coordinates on $T_{odd}{\cal M}_{cl}$, holds
\renewcommand{\theequation}{\arabic{section}.\arabic{equation}}
\begin{eqnarray}
\int d^4x \left[{\stackrel{\circ}{\varphi}}(x,\theta)(\Box + m^2)\overline{
\varphi}(x,\theta) + {\stackrel{\circ}{\overline{\varphi}}}(x,\theta)(
\Box + m^2){\varphi}(x,\theta)\right]_{\mid\Theta_{\varphi}(x,\theta)=
\Theta_{\overline{\varphi}}(x,\theta)= 0}=0\;.
\end{eqnarray}
Choosing SCLF and equivalent to them SCHF respectively in the form
\begin{eqnarray}
{} & {\stackrel{\circ}{\varphi}}(x,\theta)_{\mid\Theta_{\varphi}(x,\theta)=
\Theta_{\overline{\varphi}}(x,\theta)= 0}=0,\
{\stackrel{\circ}{\overline{\varphi}}}(x,\theta)_{\mid\Theta_{\varphi}(x,
\theta)=\Theta_{\overline{\varphi}}(x,\theta)= 0}=0\;, & {}\\ 
{} & (\overline{\varphi}(x,\theta))^{\ast}_{\mid\Theta_{\varphi}^H(x,\theta)=
\Theta_{\overline{\varphi}}^H(x,\theta)= 0}=0,\
({\varphi}(x,\theta))^{\ast}_{\mid\Theta_{\varphi
}^H(x,\theta)=\Theta_{\overline{\varphi}}^H(x,\theta)= 0}=0\;, & {} 
\end{eqnarray}
we get the free massive complex scalar superfield model is described on their
solutions by only $S_0\bigl({\varphi}(\theta),\overline{\varphi}(\theta)
\bigr)$ and therefore belongs to the II class ThST with respect to
terminology introduced in Sec.VI. Under choice of the SCLF, SCHF in the form
(9.11), (9.12) the corresponding master equations (3.36a,b), in fact written
in (9.10), appear by double zeros of the solutions for ELS and EGHS (9.9),
(9.12). Therefore EGHS and corresponding ELS are solvable.

The antibracket $(\ ,\ )_{\theta}$ (7.18a) given on $T_{odd}^{\ast}{\cal M}_{
cl}$ has the form
\begin{eqnarray}
{} & \left({\cal F}(\Gamma(\theta)), {\cal J}(\Gamma(\theta))\right)_{\theta}
= \displaystyle\int d^4x \Biggl[
\left(\displaystyle\frac{\partial
{\cal F}(\Gamma(\theta))}{\partial {\varphi}(x,\theta)\phantom{x}}
\displaystyle\frac{\partial
{\cal J}(\Gamma(\theta))}{\partial ({\varphi}(x,\theta))^{\ast}} +
\displaystyle\frac{\partial {\cal F}(\Gamma(\theta))}{\partial
\overline{\varphi}(x,\theta)\phantom{x}}
\displaystyle\frac{\partial
{\cal J}(\Gamma(\theta))}{\partial (\overline{\varphi}(x,\theta))^{\ast}}
\right) - & {} \nonumber \\
{} &
(-1)^{(\varepsilon({\cal F}) + 1)(\varepsilon({\cal J}) + 1)}({\cal F}
\longleftrightarrow {\cal J})\Biggr],\ {\cal F}(\theta), {\cal J}(\theta) \in
C^{k\ast}\,.{} & 
\end{eqnarray}
Partial superfield derivatives with respect to superantifields $(\varphi(x,
\theta))^{*}$,
$(\overline{\varphi}(x,\theta))^{*}$ in terms of densities on
$x^{\mu}$ have the same form that ones with respect to
superfields $\varphi(x,\theta)$, $\overline{\varphi}(x,\theta)$ in [1].
Operator $\Delta^{cl}(\theta)$ reads as follows on $T_{odd}^{\ast}{\cal M}_{
cl}$
\begin{eqnarray}
\Delta^{cl}(\theta) =  \int d^4x \left[
\frac{\partial_l \phantom{xxxx}}{\partial {\varphi}(x,\theta)}
\frac{\partial \phantom{xxxxxx}}{\partial (\varphi(x,\theta))^{\ast}}+
\frac{\partial_l \phantom{xxxx}}{\partial \overline{\varphi}(x,\theta)}
\frac{\partial \phantom{xxxxxx}}{\partial (\overline{\varphi}(x,
\theta))^{\ast}}\right]
.
\end{eqnarray}

Eqs.(5.22) and therefore (5.8) hold trivially in question. The superfunctional
$Z_{H}^{(1)}[\varphi,\overline{\varphi}$, $({\varphi})^{\ast},(\overline{
\varphi})^{\ast}$, $D,\overline{D}]$ (3.14a) with Lagrange complex scalar
multipliers  $D(x,\theta)$, $\overline{D}(x,\theta)$
(having the same Grassmann parities as for  $\varphi(x,\theta)$, $\overline{
\varphi}(x,\theta)$ [1]) leading on a basis of variational principle to the
GHS of the type (3.51), being differred from (9.9) in the
Eqs.(9.9b), has the form
\renewcommand{\theequation}{\arabic{section}.\arabic{equation}\alph{lyter}}
\begin{eqnarray}
\setcounter{lyter}{1}
{} & Z^{(1)}_H[\Gamma, D,
\overline{D}]  = \displaystyle\int d\theta\left[\displaystyle\int d^4x\left(
{\stackrel{\circ}{\varphi}}({\varphi})^{\ast}
 + {\stackrel{\circ}{\overline{\varphi}}}(\overline{\varphi})^{\ast}\right)(x,
\theta) - S_H^{(1)}(\Gamma(\theta), D(\theta), \overline{D}(\theta))\right],
& {} \\
\setcounter{lyter}{2}
\setcounter{equation}{15}
{} & S_H^{(1)}(\Gamma(\theta), D(\theta), \overline{D}(\theta)) =
S_H(\Gamma(\theta)) - \displaystyle\int d^4x\Bigl[\overline{D}
\bigl(\Box + m^2\bigr){\varphi}
 + {D}\bigl(\Box + m^2\bigr)\overline{\varphi}\Bigr](x,\theta)\;. &{} 
\end{eqnarray}
Because the GHS (9.9) and corresponding LS [1] (in fact
being given by Eqs.(9.9c,d) and ${\stackrel{\circ\circ}{\varphi}}(x,\theta)
=0$, ${\stackrel{\circ\circ}{\overline{\varphi}}}(x,\theta)=0$)
are the systems of differential superfield equations in partial derivatives
of the 2nd order
with respect to $x^{\mu}$ for GHS, LS and of the 1st (2nd) order on $\theta$
for GHS (LS), then as the independent initial conditions for LS one can choose
\renewcommand{\theequation}{\arabic{section}.\arabic{equation}}
\begin{eqnarray}
\Bigl(\varphi(x,\theta), \partial_0\varphi(x,\theta),
{\stackrel{\circ}{\varphi}}(x,\theta),\partial_0{\stackrel{\circ}{\varphi}
}(x,\theta)\Bigr)_{\mid x^0=\theta=0}=
\Bigl(\varphi_0, \varphi_1, \mbox{\boldmath${\stackrel{\circ}{
\varphi}}$}, \lambda_1\Bigr)(x^i),\
x^{\mu}=(x^0,x^i)\;.
\end{eqnarray}
Then in correspondence with Statement 3.1 the Cauchy problem (9.16) both for
GHS (9.9) and for GHS of the form (3.51) are set in
$T^{\ast}_{odd}{\cal M}_{cl} \times \{\theta\}$ equivalently,
with allowance made for (9.7a,b), by means of the independent relations
\begin{eqnarray}
{} & \Bigl(\varphi(x,\theta), \partial_0\varphi(x,\theta)\Bigr)_{
\mid x^0=\theta=0} = \Bigl(\varphi_0,\varphi_1\Bigr)(x^i),\ {} & \nonumber \\
{} & \Bigl((\varphi(x,\theta))^{\ast},\partial_0(\varphi(x,\theta))^{\ast}
\Bigr)_{\mid\ x^0=\theta=0} =\imath
\Bigl(\mbox{\boldmath${\stackrel{\circ}{\overline{\varphi}}}$},\overline{
\lambda}_1\Bigr)(x^i)\equiv \Bigl((\varphi_0(x^i))^{\ast},(\varphi_1(x^i))^{
\ast}\Bigr).{} & 
\end{eqnarray}

The results of this subsection can be directly rewritten for the model with
self-interaction described in Ref.[1]. To this end it is necessary only in
formulae (9.7c), (9.8a,c), (9.9c,d), (9.10), (9.15) to make the changes of the
form
\renewcommand{\theequation}{\arabic{section}.\arabic{equation}\alph{lyter}}
\begin{eqnarray}
\setcounter{lyter}{1}
S_H(\Gamma(\theta)) & \mapsto & S_{H{}M}(\Gamma(\theta)) =
T\bigl((\varphi(\theta))^{\ast}, (\overline{\varphi}(\theta))^{\ast}\bigr) +
S_{0{}M}\bigl(\varphi(\theta), \overline{\varphi}(\theta)\bigr)\;,
\nonumber \\
S_L(\theta) & \mapsto & S_{L{}M}(\theta) = T(\theta) - S_{0{}M}(\theta)\;,
\\ 
\setcounter{lyter}{2}
\setcounter{equation}{18}
S_0(\theta) & \mapsto & S_{0{}M}(\theta) = S_0(\theta) -
V\bigl(\varphi(\theta), \overline{\varphi}(\theta)\bigr)\;, \nonumber \\
V(\theta) & \equiv & V\bigl(\varphi(\theta), \overline{\varphi}(\theta)\bigr)
= \displaystyle\int d^4x\Bigl(\frac{\mu}{3}
\overline{\varphi}\varphi(\overline{\varphi} + \varphi) + \frac{\lambda}{2}(
\overline{\varphi}\varphi)^2 + \ldots\Bigr)(x,\theta)\;,
\\
\setcounter{lyter}{3}
\setcounter{equation}{18}
\Theta^H_{\varphi}(x,\theta) & \mapsto & \Theta^H_{\varphi{}M}(x,\theta) = -
\displaystyle\frac{\partial_l S_{0{}M}(\theta)}{\partial\varphi(x,\theta)} =
\Bigl(\Box + m^2 + \frac{\mu}{3}(\overline{\varphi} + 2\varphi)(x,\theta) +
\nonumber \\
{} & {} &
{\lambda}(\overline{\varphi}\varphi)(x,\theta) + \ldots\Bigr)\overline{
\varphi}(x,\theta) = 0\;, \nonumber \\
\Theta^H_{\overline{\varphi}}(x,\theta) & \mapsto &
\Theta^H_{\overline{\varphi}{}M}(x,\theta) = \overline{\Theta}{}^H_{\varphi{
}M}(x,\theta) = 0\;,
\\
\setcounter{lyter}{4}
\setcounter{equation}{18}
{} & {} & \hspace{-2em}
\displaystyle\int d^4x \left[{\stackrel{\circ}{\varphi}}(x,\theta)
\Theta^H_{\varphi{}M}(x,\theta) + {\stackrel{\circ}{\overline{\varphi}}
}(x,\theta)\Theta^H_{\overline{\varphi}{}M}(x,\theta)\right]_{\mid\Theta_{
\varphi{}M}(x,\theta)=
\Theta_{\overline{\varphi}{}M}(x,\theta)= 0}=0\;;
\end{eqnarray}
\vspace{-4ex}
\begin{eqnarray}
\setcounter{lyter}{1}
{} & Z^{(1)}_{H{}M}[\Gamma, D,
\overline{D}]  = \displaystyle\int d\theta\left[\displaystyle\int d^4x\left(
{\stackrel{\circ}{\varphi}}({\varphi})^{\ast}
 + {\stackrel{\circ}{\overline{\varphi}}}(\overline{\varphi})^{\ast}\right)(x,
\theta) - S_{H{}M}^{(1)}(\Gamma(\theta), D(\theta), \overline{D}(\theta))
\right], {} &
\\
\setcounter{lyter}{2}
\setcounter{equation}{19}
{} &
S_{H{}M}^{(1)}(\Gamma(\theta), D(\theta), \overline{D}(\theta)) =
S_{H{}M}(\Gamma(\theta)) - \displaystyle\int d^4x\Bigl[\overline{D}
\Theta^H_{\overline{\varphi}{}M}+ {D}\Theta^H_{\varphi{}M}\Bigr](x,\theta)\;.
{} & 
\end{eqnarray}
The other characteristics, among them the classified ones, for free model are
transferred onto self-interacting model taking the remarks made in Ref.[1]
into account.
\subsection{Models of Massive Spinor Superfield of Spin
$\frac{1}{2}$}

Let us introduce the complex superantifields $\Psi^{\ast}(x,\theta)$,
$\overline{\Psi}^{\ast}(x,\theta)$
$\in$ $\tilde{\Lambda}_{4\mid 0+1}(x^{\mu},\theta;{\bf C})$  defined on
${\cal M} = {\bf R}^{1,3}\times\tilde{P}$ [1]
\renewcommand{\theequation}{\arabic{section}.\arabic{equation}\alph{lyter}}
\begin{eqnarray}
\setcounter{lyter}{1}
{} & {} &  \hspace{-2em} \Psi^{\ast}(x,\theta)  =
\bigl(\psi^{\ast{}\alpha}(x,\theta),
\chi^{\ast}{}_{\dot{\alpha}}(x, \theta) \bigr),\; \Psi^{\ast}(x,\theta) =
\psi^{\ast}(x) + \psi^{\ast}_1(x)\theta=
\psi^{\ast}(x) -  \theta J(x)\;,  \\ 
\setcounter{lyter}{2}
\setcounter{equation}{20}
{} & {} & \hspace{-2em} \overline{\Psi}^{\ast}(x,\theta) =
\bigl(\overline{\chi}^{\ast}_{\beta}(x,\theta),
\overline{\psi}^{\ast}{}^{\dot{\beta}}(x,\theta)\bigr)^T,\;
\overline{\Psi}^{\ast}(x,\theta) = \overline{\psi}^{\ast}(x) +
\overline{\psi}^{\ast}_1(x)\theta = \overline{\psi}^{\ast}(x) -
\theta\overline{J}(x)\;,
\end{eqnarray}
which are the Dirac bispinors ($\overline{\Psi}^{\ast}(x,\theta) =
\Gamma^0\left(\Psi^{\ast}(x,\theta)\right)^{+} \equiv \overline{{
\Psi}^{\ast}}(x,\theta)$ is Dirac conjugate to $\Psi^{\ast}(x,\theta)$ as it
follows from (9.24) below)
and, therefore, the elements of $(\frac{1}{2},0) \oplus (0,\frac{1}{2})$
reducible superfield (on $\theta$) Lorentz group representation.
The superantifields are given in (9.20) among them in two-component spinor
formalism.

The Grassmann gradings for superantifields and their components it is
sufficient to define, for instance, for $\Psi^{*}(x,\theta)$ by the
table
\renewcommand{\theequation}{\arabic{section}.\arabic{equation}}
\begin{eqnarray}
\begin{array}{lcccccc}
{} & \psi^{\ast}(x) & \psi^{\ast}_1(x) & J(x) & \Psi^{\ast}(x,\theta) &
\psi^{\ast}_1(x)\theta & {}\\
\varepsilon_P & 1 & 0 & 0 & 1 & 1 & {} \\
\varepsilon_{\Pi} & 1 & 1 & 1 & 1 & 1 & {}\\
\varepsilon & 0 & 1 & 1 & 0 & 0 & \hspace{-2em},
\end{array}
\end{eqnarray}
that is corresponding to theorem on connection of spin with statistic
$\varepsilon_{\Pi}$ but not with $\varepsilon$.

Superantifields $\Psi^{*}(x,\theta)$, $\overline{\Psi}^{*}(x,\theta)$ are
transformed in a standard way (as bispinors) with respect to restriction
of representation $T$
of the supergroup $J$ onto ${\Pi}(1,3)^{\uparrow}$.
As to $T^{\ast}_{\vert P}$,
then the superantifields are transformed according to (2.26a) in the form
\renewcommand{\theequation}{\arabic{section}.\arabic{equation}\alph{lyter}}
\begin{eqnarray}
\setcounter{lyter}{1}
{} & {} & \delta\Psi^{\ast}(x,\theta) = {\Psi^{\ast}}'(x,\theta)
- \Psi^{\ast}(x,\theta) = -\mu {\stackrel{\circ}{\Psi}}{}^{\ast}(x,\theta) =
\mu \psi^{\ast}_1(x) = \mu J(x)\;,\\
\setcounter{lyter}{2}
\setcounter{equation}{22}
{} & {} & \delta\overline{\Psi}^{\ast}(x,\theta) =
{\overline{\Psi}^{\ast}}'(x,\theta) - \overline{\Psi}^{\ast}(x,\theta) = -\mu
{\stackrel{\circ}{\overline{\Psi}}}{}^{\ast}(x,\theta) =
\mu\overline{\psi}^{\ast}_1(x)=\mu\overline{J}(x)\;.
\end{eqnarray}
Superantifields ${\stackrel{\circ}{\Psi}}{}^{\ast}(x,\theta)$,
${\stackrel{\circ}{\overline{\Psi}}}{}^{\ast}(x,\theta)$ are elements of
mentioned Poincare group superfield representation as well and in addition
are scalars with respect to action of $T^{\ast}_{\vert P}$ operators (2.26b).

Because of supermatrix $K^{(1)}(\theta,x,y)$ of the 2nd derivatives of
$S^{(1)}_{L}(\theta)$ with respect to
${\stackrel{\circ}{\Psi}}(x,\theta)$,
${\stackrel{\circ}{\overline{\Psi}}}(y,\theta)$ is the
nondegenerate one [1], then it is possible according to Sec.III to pass to the
Hamiltonian formulation in $T^{\ast}_{odd}{\cal M}_{cl}$ parametrized by
coordinates $\Gamma(x,\theta)$ = ($\Psi(x,\theta)$,
$\overline{\Psi}(x,\theta)$, $\Psi^{\ast}(x,\theta)$,
$\overline{\Psi}^{\ast}(x,\theta)$).  Dimensions of ${\cal
M}_{cl}$ and $T^{\ast}_{odd}{\cal M}_{ cl}$ with respect to $\varepsilon$
parity are the same
\renewcommand{\theequation}{\arabic{section}.\arabic{equation}}
\begin{eqnarray}
{\rm dim}_{\bf R}{\cal M}_{cl} = (0,8),\;
{\rm dim}_{\bf R}T^{\ast}_{odd}{\cal M}_{cl} =
{\rm dim}_{\bf R}T_{odd}{\cal M}_{cl} = (8,8)\;
.
\end{eqnarray}
Legendre transform (3.2), (3.3) is given by the formulae
\renewcommand{\theequation}{\arabic{section}.\arabic{equation}\alph{lyter}}
\begin{eqnarray}
\setcounter{lyter}{1}
{} & \Psi^{ \ast}(x,\theta) =
\displaystyle\frac{\partial_l
S_L\bigl({\Psi}(\theta),\overline{\Psi}(\theta),
{\stackrel{\circ}{{\Psi}}}(\theta),
{\stackrel{\circ}{\overline{\Psi}}}(\theta)\bigr)}{\partial
{\stackrel{\circ}{{\Psi}}}(x,\theta)\phantom{xxxxxxxxxxxxxxx}} =
\frac{\partial_l
T\bigl({\stackrel{\circ}{{\Psi}}}(\theta),
{\stackrel{\circ}{\overline{\Psi}}}(\theta)\bigr)}{\partial
{\stackrel{\circ}{{\Psi}}}(x,\theta)\phantom{xxxxxx}} =
{\stackrel{\circ}{\overline{\Psi}}}(x,\theta)\;, & {}
\\
\setcounter{lyter}{2}
\setcounter{equation}{24}
{} & \overline{\Psi}^{\ast}(x,\theta) =
\displaystyle\frac{\partial_l
S_L\bigl({\Psi}(\theta),\overline{\Psi}(\theta),
{\stackrel{\circ}{{\Psi}}}(\theta),
{\stackrel{\circ}{\overline{\Psi}}}(\theta)\bigr)}{\partial
{\stackrel{\circ}{\overline{\Psi}}}(x,\theta)\phantom{xxxxxxxxxxxxxxx}} =
\frac{\partial_l
T\bigl({\stackrel{\circ}{{\Psi}}}(\theta),
{\stackrel{\circ}{\overline{\Psi}}}(\theta)\bigr)}{\partial
{\stackrel{\circ}{\overline{\Psi}}}(x,\theta)\phantom{xxxxxx}} =
{\stackrel{\circ}{\Psi}}(x,\theta)\;, & {}
 \\
\setcounter{lyter}{3}
\setcounter{equation}{24}
{} & S_H^{(1)}(\Gamma(\theta)) =
T\bigl(\Psi^{\ast}(\theta),\overline{\Psi}^{\ast}(\theta)\bigr) +
S_0(\Psi(\theta), \overline{\Psi}(\theta)),\;T\bigl(\Psi^{\ast}(\theta),
\overline{\Psi}^{\ast}(\theta)\bigr) =  & {}\nonumber \\
{} & \hspace{-2.0em}
T\bigl({\stackrel{\circ}{{\Psi}}}\bigl(\overline{\Psi}^{\ast}(\theta)),
{\stackrel{\circ}{\overline{\Psi}}}\bigl(\Psi^{\ast}(\theta)\bigr)\bigr) =
\hspace{-0.1em}\displaystyle\int\hspace{-0.2em} d^4x \Psi^{\ast}(x,\theta)
\overline{\Psi}^{\ast}(x,\theta)
\equiv \hspace{-0.1em}\displaystyle\int \hspace{-0.2em}d^4x {\cal L}_{\rm
kin}^{\ast{}(1)}(x,\theta) \equiv
\hspace{-0.1em}\displaystyle\int \hspace{-0.2em}d^4x
{\cal L}^{(1)}_{\rm kin}(x,\theta). 
\end{eqnarray}
That GSTF model  is the
nondegenerate ThST, belongs to the I class as in Lagrangian formalism as
in Hamiltonian one and is given by means of superfunction $S^{(1)}_{L
}(\theta)$ defined on $T_{odd}{\cal M}_{cl}$ [1]
\begin{eqnarray}
\setcounter{lyter}{1}
S^{(1)}_L(\theta) & \hspace{-0.5em}\equiv & \hspace{-0.5em}
S_L\Bigl(\Psi(\theta), \overline{\Psi}(\theta),
{\stackrel{
\circ}{\Psi}}(\theta), {\stackrel{\circ}{\overline{\Psi}}}
(\theta)\Bigr) = T\Bigl({\stackrel{\circ}{\Psi}}(\theta), {\stackrel{\circ}{
\overline{\Psi}}}(\theta)\Bigr) - S_{0}\bigl(\Psi(\theta),
\overline{\Psi}(\theta)\bigr)\;,
\\
\setcounter{lyter}{2}
\setcounter{equation}{25}
T^{(1)}(\theta) & \hspace{-0.5em}\equiv & \hspace{-0.5em}
T\Bigl({\stackrel{\circ}{\Psi}}(\theta),
{\stackrel{\circ}{\overline{\Psi}}}(\theta)\Bigr) =
\int d^4 x  {\stackrel{\circ}{\overline{\Psi}}}
(x,\theta) {\stackrel{\circ}{\Psi}}(x,\theta) \equiv
 \int d^4 x {\cal L}^{(1)}_{\rm kin}(x,\theta)\; ,
\\
\setcounter{equation}{25}
\setcounter{lyter}{3}
S^{(1)}_0(\theta) & \hspace{-0.5em}\equiv & \hspace{-0.5em}
S_{0}\bigl(\Psi(\theta),\overline{\Psi}(\theta)
\bigr)
= \int d^4x \overline{\Psi}(x,\theta) \left(\imath \Gamma^{\mu}\partial_{\mu}
- m\right){\Psi}(x,\theta) \equiv \int d^4x {\cal L}^{(1)}_0(x,\theta).
\end{eqnarray}
Therefore the Hamiltonian formulation of the model (9.24c)
satisfies to the conditions of Corollary 1.4.
Relations
(9.24a,b) establish the following correspondences in two-component spinor
formalism
\renewcommand{\theequation}{\arabic{section}.\arabic{equation}}
\begin{eqnarray}
\psi^{\ast{}\alpha}(x,\theta) =
{\stackrel{\;\circ}{\overline{\chi}}}{}^{\alpha}(x,\theta),\;
\chi^{\ast}_{\dot{\alpha}}(x,\theta) =
{\stackrel{\;\circ}{\overline{\psi}}}_{\dot{\alpha}}(x,\theta),\;
\overline{\chi}^{\ast}_{\alpha}(x,\theta)  =
{\stackrel{\;\circ}{\psi}}_{\alpha}(x,\theta),\;
\overline{\psi}^{\ast}{}^{\dot{\alpha}}(x,\theta) =
{\stackrel{\;\circ}{\chi}}{}^{\dot{\alpha}}(x,\theta)\;,
\end{eqnarray}
so that the only half from them is independent.

Corresponding GHS has the form (6.28), (3.51b) and
contains  $3\times 8$ real
component equations or equivalently  6 superfield spinor equations in
four-component spinor formalism
\renewcommand{\theequation}{\arabic{section}.\arabic{equation}\alph{lyter}}
\begin{eqnarray}
\setcounter{lyter}{1}
{} & {} &\displaystyle\frac{d_r \Psi(x,\theta)}{d\theta
\phantom{xxxxx}}  = \frac{\partial T^{(1)}(\theta)\phantom{x}}{\partial
\Psi^{\ast}(x,\theta)} = \overline{\Psi}^{\ast}(x,\theta),\;
\displaystyle\frac{d_r \overline{\Psi}(x,\theta)}{d\theta\phantom{xxxxx}} =
\frac{\partial
T^{(1)}(\theta)\phantom{x}}{\partial\overline{\Psi}^{\ast}(x,\theta)} =
{\Psi^{\ast}}(x,\theta)\;,
\\
\setcounter{lyter}{2}
\setcounter{equation}{27}
{} & {} & \displaystyle\frac{d_r \Psi^{\ast}(x,\theta)}{d\theta
\phantom{xxxxxx}} = 0,\ \
\displaystyle\frac{d_r \overline{\Psi}^{\ast}(x,\theta)}{d\theta
\phantom{xxxxxx}} = 0\;,
\\
\setcounter{lyter}{3}
\setcounter{equation}{27}
{} & {} &\Theta_{\Psi}^H\bigl(\overline{\Psi}(x,\theta),
\partial_{\mu}\overline{\Psi}(x,\theta)\bigr) = -
\displaystyle\frac{\partial_l
S^{(1)}_0(\theta)}{\partial{\Psi}(x,\theta)} = -
\left(\imath\partial_{\mu}
\overline{\Psi}(x,\theta)\Gamma^{\mu} + m\overline{\Psi}(x,\theta)\right) = 0
\;,\\
\setcounter{lyter}{4}
\setcounter{equation}{27}
{} & {} & \Theta_{\overline{\Psi}}^H\bigl({\Psi}(x,\theta),
\partial_{\mu}{\Psi}(x,\theta)\bigr) = -
\displaystyle\frac{\partial_l
S^{(1)}_0(\theta)}{\partial \overline{\Psi}(x,\theta)} = -
\left(\imath\Gamma^{\mu}\partial_{\mu} - m\right){\Psi}(x,\theta) = 0\;.
\end{eqnarray}
The only half from them is independent, for example, the equations for
${\Psi}(x,\theta)$, ${\Psi}^{\ast}(x,\theta)$. HS (9.27a,b) is solvable.
Superfunction $S^{(1)}_{H}(\theta)$ (9.24c) is the
integral of GHS (9.27) being equivalent to corresponding LS [1]. But HCHF
(9.27c,d) representing themselves the superfield (on $\theta$)
generalization of Dirac equation and coinciding with corresponding HCLF [1]
do not satisfy to the solvability condition.

As one had been already mentioned there were not of the type (6.22) identities
because of this model is not gauge although HCHF code 4 independent
degrees of freedom among 8 component equations [1].

At last, the equation (2.6), in view of the fact that
$S^{(1)}_{H}(\theta)$ = $S^{(1)}_{E}(\theta)$ = $T\bigl(
{\stackrel{\circ}{\Psi}}(\theta),
{\stackrel{\circ}{\overline{\Psi}}}(\theta)\bigr)$ + $S^{(1)}_{0}(\theta)$
in terms of coordinates on $T_{odd}{\cal M}_{cl}$,
is fulfilled and has the form
\renewcommand{\theequation}{\arabic{section}.\arabic{equation}}
\begin{eqnarray}
 - \displaystyle\int d^4x \Biggl[
\overline{\Psi}(x,\theta) \bigl(\imath
\Gamma^{\mu}\partial_{\mu} - m\bigr){\stackrel{\circ}{\Psi}}(x,\theta) +
{\stackrel{\circ}{\overline{\Psi}}}(x,\theta)
\bigl(\imath
\Gamma^{\mu}\partial_{\mu} - m\bigr){\Psi}(x,\theta)\Biggr]_{
\mid\Theta_{\Psi}(x,\theta)=0,\,\Theta_{\overline{\Psi}}(x,\theta)=0}
\hspace{-2em} = 0\;.
\end{eqnarray}
Specifying SCLF and being equivalent to them SCHF respectively in the form
\begin{eqnarray}
{} & {} & {\stackrel{\circ}{\overline{\Psi}}}(x,\theta)_{
\mid\Theta_{\Psi}(x,\theta)=0,\,\Theta_{\overline{\Psi}}(x,\theta)=0} = 0,\;
{\stackrel{\circ}{\Psi}}(x,\theta)_{
\mid\Theta_{\Psi}(x,\theta)=0,\,\Theta_{\overline{\Psi}}(x,\theta)=0} = 0\;
,\\
{} & {} &
{\Psi^{\ast}}(x,\theta)_{
\mid\Theta_{\Psi}^H(x,\theta)=0,\,\Theta_{\overline{\Psi}}^H(x,\theta)=0}
= 0,\;
\overline{\Psi}^{\ast}(x,\theta)_{
\mid\Theta_{\Psi}^H(x,\theta)=0,\,\Theta_{\overline{\Psi}}^H(x,\theta)=0}
= 0\;
,
\end{eqnarray}
obtain that on their solutions according to (9.27a) the  free
massive spinor superfield model is described by only superfunction
$S_{0}(\Psi(\theta), \overline{\Psi}(\theta))$ and therefore belongs to the
II class  ThST.
Master equations (3.36a,b), in fact written in (9.28), appear by double zeros
of the solutions for ELS and EGHS (9.27), (9.30) with SCLF, SCHF (9.29),
(9.30).
Therefore EGHS and corresponding ELS are solvable.

The antibracket $(\ ,\ )_{\theta}$ (7.18a),
operator $\Delta^{cl}(\theta)$ are defined in the
following way in  coordinates $\Gamma(x,\theta)$ on $T_{odd}^{\ast}{\cal
M}_{cl}$
\begin{eqnarray}
{} & \left({\cal F}(\Gamma(\theta)), {\cal J}(\Gamma(\theta))\right)_{\theta}
= \displaystyle\int d^4x \Biggl[
\left(\displaystyle\frac{\partial
{\cal F}(\Gamma(\theta))}{\partial {\Psi}(x,\theta)\phantom{x}}
\displaystyle\frac{\partial
{\cal J}(\Gamma(\theta))}{\partial {\Psi}^{\ast}(x,\theta)\phantom{x}} -
\displaystyle\frac{\partial_r {\cal F}(\Gamma(\theta))}{\partial
\overline{\Psi}^{\ast}(x,\theta)\phantom{x}}
\displaystyle\frac{\partial_l
{\cal J}(\Gamma(\theta))}{\partial \overline{\Psi}(x,\theta)\phantom{xx}}
\right) - & {} \nonumber \\
{} &
(-1)^{(\varepsilon({\cal F}) + 1)(\varepsilon({\cal J}) + 1)}({\cal F}
\longleftrightarrow {\cal J})\Biggr],\ {\cal F}(\theta), {\cal J}(\theta) \in
C^{k\ast}\;, & {} \\
{} & \Delta^{cl}(\theta) = - \displaystyle\int d^4x \left[
\displaystyle\frac{\partial_l \phantom{xxxxx}}{\partial {\Psi}(x,\theta)}
\displaystyle\frac{\partial \phantom{xxxxxx}}{\partial {\Psi}^{\ast}(x,\theta
)}+
\displaystyle\frac{\partial_l \phantom{xxxxx}}{\partial \overline{\Psi}(x,
\theta)}
\displaystyle\frac{\partial \phantom{xxxxxx}}{\partial \overline{\Psi}^{
\ast}(x,\theta)}\right]. {} & 
\end{eqnarray}
Applied in this case the Eq.(5.22) and therefore (5.8) are trivially valid.
Superfunctional
$Z_{H}^{(1)}[\Psi,\overline{\Psi},{\Psi}^{\ast},\overline{\Psi}^{\ast},
D^{(1)},\overline{D}{}^{(1)}]$ (3.14a) with auxiliary bispinors $D^{(1)}(x,
\theta)$, $\overline{D}{}^{(1)}(x,\theta)$
(having the same $\varepsilon_{P}$,
$\varepsilon_{\Pi}$, $\varepsilon$ gradings as for  $\Psi
(x,\theta)$, $\overline{\Psi}(x,\theta)$ [1]) from which it
follows the GHS of the type (3.51), being differred from (9.27) in the
Eqs.(9.27b), has the form
\renewcommand{\theequation}{\arabic{section}.\arabic{equation}\alph{lyter}}
\begin{eqnarray}
\setcounter{lyter}{1}
{} & \hspace{-2em}Z^{(1)}_H[\Gamma, D^{(1)},
\overline{D}{}^{(1)}]  = \displaystyle\int d\theta\left[\hspace{-0.2em}
\displaystyle\int\hspace{-0.2em}
d^4x\left(
{\stackrel{\circ}{\overline{\Psi}}}{\,}\overline{\Psi}^{\ast}
+ {\Psi}^{\ast}{\stackrel{\circ}{\Psi}}\right)(x,\theta) -
S_H^{(1)}(\Gamma(\theta), D^{(1)}(\theta), \overline{D}{}^{(1)}(\theta))
\hspace{-0.2em}\right], & {}
\\
\setcounter{lyter}{2}
\setcounter{equation}{33}
{} & S_H^{(1)}(\Gamma(\theta), D^{(1)}(\theta), \overline{D}{}^{(1)}(\theta))
= S^{(1)}_H(\Gamma(\theta)) + \displaystyle\int d^4x\Bigl[
\overline{D}{}^{(1)}(x,\theta)\bigl(\imath
\Gamma^{\mu}\partial_{\mu} - m\bigr){\Psi}(x,\theta) + & {}\nonumber \\
{} &  \overline{\Psi}(x,\theta) \bigl(\imath
\Gamma^{\mu}\partial_{\mu} - m\bigr)D^{(1)}(x,\theta)\Bigr]. & {} 
\end{eqnarray}
As far as the GHS (9.27) and corresponding LS [1] (in fact
being given by formulae (9.27c,d) and ${\stackrel{\circ\circ}{\Psi}}(x,\theta)
=0$, ${\stackrel{\circ\circ}{\overline{\Psi}}}(x,\theta)=0$) are
the systems
of differential superfield equations in partial derivatives of the 1st order
with respect to $x^{\mu}$ and of the 1st (2nd) order on $\theta$ for GHS
(LS) one can choose  as the independent the following initial conditions
for LS
\renewcommand{\theequation}{\arabic{section}.\arabic{equation}}
\begin{eqnarray}
\Psi(x,\theta)_{\mid x^0=\theta=0} = {\bf \Psi}(x^i),\
{\stackrel{\circ}{\Psi}}(x,\theta)_{\mid x^0=\theta = 0} =
{\bf {\stackrel{\circ}{\Psi}}}(x^i)\;. 
\end{eqnarray}
Then according to
Statement 3.1 the Cauchy problem both for GHS (9.27) and for GHS of the form
(3.51) are set in $T^{\ast}_{odd}{\cal M}_{cl} \times \{\theta\}$
equivalently  taking (9.24a,b) into account by means of the following
independent relations
\begin{eqnarray}
\Psi(x,\theta)_{\mid x^0=\theta=0} = {\bf \Psi}(x^i),\
\Psi^{\ast}(x,\theta)_{\mid\ x^0=\theta=0} =
{\bf {\stackrel{\circ}{\overline{\Psi}}}}(x^i) \equiv {\bf \Psi^{\ast}}(x^i)
\;. 
\end{eqnarray}

The generalization of the relationships from given subsection to the case of
massive spinor superfield model with interaction [1] is produced sufficiently
simple. To do this it is necessary to make the only following changes in
formulae (9.24c), (9.25a,c), (9.27c,d), (9.28), (9.33) respectively
\renewcommand{\theequation}{\arabic{section}.\arabic{equation}\alph{lyter}}
\begin{eqnarray}
\setcounter{lyter}{1}
S^{(1)}_H(\Gamma(\theta)) & \mapsto & S^{(1)}_{H{}M}(\Gamma(\theta)) =
T\bigl(\Psi^{\ast}(\theta), \overline{\Psi}^{\ast}(\theta)\bigr) +
S_{0{}M}\bigl(\Psi(\theta), \overline{\Psi}(\theta)\bigr)\;,
\nonumber \\
S^{(1)}_L(\theta) & \mapsto & S^{(1)}_{L{}M}(\theta) = T^{(1)}(\theta) -
S^{(1)}_{0{}M}(\theta)\;,
\\ 
\setcounter{lyter}{2}
\setcounter{equation}{36}
S^{(1)}_0(\theta) & \mapsto & S^{(1)}_{0{}M}(\theta) = S^{(1)}_0(\theta) -
V\bigl(\Psi(\theta), \overline{\Psi}(\theta)\bigr)\;, \nonumber \\
V^{(1)}(\theta) & \equiv & V^{(1)}\bigl(\Psi(\theta),
\overline{\Psi}(\theta)\bigr) = \displaystyle\int d^4x\Bigl[\frac{
\lambda_1}{2}(\overline{\Psi}{}\Psi)^2 + \frac{\lambda_2}{2}(
\overline{\Psi}\Gamma^{\mu}\Psi)(\overline{\Psi}\Gamma_{\mu}\Psi)
\Bigr](x,\theta)\;, \\
\setcounter{lyter}{3}
\setcounter{equation}{36}
\Theta^H_{\overline{\Psi}}(x,\theta) & \mapsto & \Theta^H_{\overline{\Psi}{}
M}(x,\theta) = -
\displaystyle\frac{\partial_l S^{(1)}_{0{}M}(\theta)}{\partial
\overline{\Psi}(x,\theta)} = - \Bigl[\imath
\partial_{\mu}\Gamma^{\mu} - m - \lambda_1(\overline{\Psi}{}\Psi)(x,\theta) -
\nonumber \\
{} & {} & \hspace{-2em}
\lambda_2(\overline{\Psi}\Gamma^{\mu}\Psi)(x,\theta)\Gamma_{\mu}\Bigr]{
\Psi} (x,\theta) = D_{\overline{\Psi}}(\Psi(x,\theta), \overline{\Psi}(x,
\theta),\partial_{\mu})\Psi(x,\theta) = 0\;, \nonumber \\
\Theta^H_{\Psi}(x,\theta) & \mapsto &
\Theta^H_{\Psi{}M}(x,\theta) = \Bigl(\Theta^H_{\overline{\Psi}{}M}(x,\theta)
\Bigr)^{+}\Gamma^0 = 0\;,
\\
\setcounter{lyter}{4}
\setcounter{equation}{36}
{} & {} & \hspace{-2em} \displaystyle\int d^4x \Bigl[
\overline{\Psi}
D_{\overline{\Psi}}(\Psi, \overline{\Psi},\partial_{\mu})
{\stackrel{\circ}{\Psi}} +
{\stackrel{\circ}{\overline{\Psi}}}
\Theta^H_{\overline{\Psi}{}M}\Bigr](x,\theta){\hspace{-0.5em}\phantom{\Bigr)}
}_{\mid\Theta_{\Psi{}M}(x,\theta)=0,\,\Theta_{\overline{\Psi}{}M}(x,\theta)=0}
= 0\;;
\end{eqnarray}
\vspace{-4ex}
\begin{eqnarray}
\setcounter{lyter}{1}
{} & \hspace{-2em}
Z^{(1)}_{H{}M}[\Gamma, D^{(1)},
\overline{D}{}^{(1)}]  = \hspace{-0.3em}\displaystyle\int\hspace{-0.3em}
d\theta\Bigl[\displaystyle\int d^4x\Bigl(
{\stackrel{\circ}{\overline{\Psi}}}{\,}\overline{\Psi}^{\ast}
+ {\Psi}^{\ast}{\stackrel{\circ}{\Psi}}\Bigr)(x,\theta) -
S_{H{}M}^{(1)}(\Gamma(\theta), D^{(1)}(\theta), \overline{D}{}^{(1)}(\theta))
\Bigr]\,,
{} &
\\
\setcounter{lyter}{2}
\setcounter{equation}{37}
{} & \hspace{-2.0em}
S_{H{}M}^{(1)}(\Gamma(\theta), D^{(1)}(\theta), \overline{D}{}^{(1)}(
\theta))\hspace{-0.2em} = \hspace{-0.2em}S^{(1)}_{H{}M}(\Gamma(\theta))
\hspace{-0.2em} - \hspace{-0.3em}\displaystyle\int
\hspace{-0.3em}d^4x\Bigl[\overline{D}{}^{(1)}\Theta^H_{\overline{\Psi}{}M}
\hspace{-0.2em}+ \hspace{-0.2em}
\overline{\Psi}D_{\overline{\Psi}}(\Psi, \overline{\Psi},\partial_{\mu})
D^{(1)}\Bigr](x,\theta). & {}
\end{eqnarray}
Given model remains nondegenerate ThST taking corresponding remarks from
Ref.[1] into account.
\subsection{Models of Free Vector Superfield}

The real superantifields ${\cal A}^{\ast}_{\mu}(x,\theta)$
$\in$ $\tilde{\Lambda}_{D\mid 0+1}(x^{\mu},\theta;{\bf R})$ on ${\cal M}$ = $
{\bf R}^{1,D-1} \times \tilde{P}$ have the following expansion
in powers of $\theta$ and $\varepsilon_P,{}\varepsilon_{\Pi},{}
\varepsilon$ gradings respectively
\renewcommand{\theequation}{\arabic{section}.\arabic{equation}}
\begin{eqnarray}
{\cal A}^{\ast}_{\mu}(x,\theta) = {A}^{\ast}_{\mu}(x) + {A}_1^{\ast}{}_{
\mu}(x)\theta = {A}^{\ast}_{\mu}(x) - \theta J_{\mu}(x),\
\begin{array}{lccccl}
{} & {A}^{\ast}_{\mu}(x) & {A}_1^{\ast}{}_{\mu}(x) & J_{\mu}(x) & {\cal
A}^{\ast}_{\mu}(x,\theta) &  {} \\
\varepsilon_P & 1 & 0 & 0 & 1 & {} \\
\varepsilon_{\Pi} & 0 & 0 & 0 & 0 &  {}\\
\varepsilon & 1 & 0 & 0 & 1 & \hspace{-2em}.
\end{array}
\end{eqnarray}
${\cal A}^{\ast}_{\mu}(x,\theta)$ are transformed with respect to massless
(or massive for Proca model for $D=4$) irreducible representation of
group $\bar{J}=\Pi(1,D-1)^{\uparrow}$ and have
the usual
transformation laws relative to that group. But with respect to $T^{
\ast}_{\mid P}$ the only ${\cal A}^{\ast}_{\mu}(x,\theta)$ are nontrivially
transformed according to (2.26a)
\begin{eqnarray}
\delta{\cal A}^{\ast}_{\mu}(x,\theta) = {{\cal
A}^{\ast}_{\mu}}'(x,\theta) - {\cal A}^{\ast}_{\mu}(x,\theta) = - \mu
{\stackrel{\ \circ}{\cal A}}{}^{\ast}_{\mu}(x,\theta)= - \mu{A}_{1}^{\ast}{}_{
\mu}(x)= \mu J_{\mu}(x)\;.
\end{eqnarray}
Let us remind that Lagrangian formulation of the massless model [1] had been
given by
the superfunction $S_L\Bigl({\cal A}^{\mu}(\theta), {\stackrel{\ \circ}{
\cal A}}{}^{\mu}
(\theta)\Bigr)$ $\equiv$ $S_{L}^{(2)}(\theta)$ on $T_{odd}{\cal M}_{cl}$ =
$\bigl\{\bigl({\cal A}^{\mu}(x,
\theta),
{\stackrel{\ \circ}{\cal A}}{}^{\mu}(x,\theta)\bigr)\bigr\}$
\renewcommand{\theequation}{\arabic{section}.\arabic{equation}\alph{lyter}}
\begin{eqnarray}
\setcounter{lyter}{1}
{} & S_{L}^{(2)}(\theta)
= T\Bigl({\stackrel{\ \circ}{\cal A}}{}^{\mu}(\theta)\Bigr) -
S_{0}\bigl({{\cal
 A}}^{\mu}(\theta)\bigr),\; T\Bigl({\stackrel{\ \circ}{\cal
A}}{}^{\mu}(\theta)\Bigr) = \displaystyle\int d^D x{ \cal L}_{\rm kin}^{(2)}
(x,\theta)\;,\\
\setcounter{equation}{40}
\setcounter{lyter}{2}
{} & {\cal L}_{\rm
kin}^{(2)}(x,\theta) \equiv {\cal L}_{\rm kin}^{(2)}
\Bigl({\stackrel{\ \circ}{\cal A}}{}^{\mu}(x,\theta)\Bigr) =
\frac{1}{2}\varepsilon_{\mu\nu} {\stackrel{\ \circ}{\cal
A}}{}^{\nu}(x,\theta){\stackrel{\ \circ}{\cal A}}{}^{\mu}(x,\theta),\;
\varepsilon_{\mu\nu} = - \varepsilon_{\nu\mu}\;, & {}
\\
\setcounter{equation}{40}
\setcounter{lyter}{3}
{} & S_{0}\bigl( {{\cal
A}}^{\mu}(\theta)\bigr) =  \displaystyle\int d^Dx {\cal
L}_0^{(2)}\bigl({{\cal A}}^{\mu}(x,\theta), \partial_{\nu}{{\cal
A}}^{\mu}(x,\theta)\bigr) \equiv \int d^Dx {\cal L}_0^{(2)}(x,\theta),\; & {}
\nonumber \\
{} & {\cal L}_0^{(2)}(x,\theta) =
-\frac{1}{4}F_{\mu\nu}(x,\theta)F^{\mu\nu} (x,\theta),\  F_{\mu\nu}(x,\theta)  =
\partial_{\mu}{{\cal A}}_{\nu}(x,\theta) - \partial_{\nu}{{\cal
A}}_{\mu}(x,\theta)\;,& {}
\end{eqnarray}
whose form  ($S_L^{(2)}(\theta)$) determined the GThST with the generators
of gauge transformations
\renewcommand{\theequation}{\arabic{section}.\arabic{equation}}
\begin{eqnarray}
 {\cal R}^{\mu}(x,y) = \partial^{\mu}\delta(x-y)\;.
\end{eqnarray}
For $D=2k+1, k\in \mbox{\boldmath$N$}$ the supermatrix $S_L^{(2)}{}''(
\theta,x,y) \equiv K^{(2)}(\theta,x,y)$ is always degenerate so that a
standard Hamiltonization by the prescription of Sec.III is impossible.
Choosing $\left\|\varepsilon_{\mu\nu}\right\|$ to be nondegenerate for
$D=2k, k \in \mbox{\boldmath$N$}$ we conclude that the necessary condition
for existence of Legendre transform for $S^{(2)}_L(\theta)$ with respect to
${\stackrel{\ \circ}{\cal A}}{}^{\mu}(x,\theta)$ is fulfilled.

Dimensions of supermanifolds ${\cal M}_{cl}$ and $T^{\ast}_{odd}{\cal M}_{
cl}$, in this example parametrized by superfields ${\cal A}^{\mu}(x,\theta)$
and $\Gamma^p(x,\theta)$ = $({\cal A}^{\mu}(x,\theta), {\cal A}^{\ast}_{
\mu}(x,\theta))$ respectively,  with respect to $\varepsilon$
grading are equal
\begin{eqnarray}
{\rm dim}_{\bf R}{\cal M}_{cl} = (4,0),\
{\rm dim}_{\bf R}T^{\ast}_{odd}{\cal M}_{cl} =
{\rm dim}_{\bf R}T_{odd}{\cal M}_{cl} = (4,4)\;.
\end{eqnarray}
Legendre transform (3.2), (3.3) according to (9.40) is given by the
formulae
\renewcommand{\theequation}{\arabic{section}.\arabic{equation}\alph{lyter}}
\begin{eqnarray}
\setcounter{lyter}{1}
{} & \hspace{-2em}{\cal A}^{\ast}_{\mu}(x,\theta) =
\displaystyle\frac{\partial_l
S_L\Bigl({\cal A}^{\nu}(\theta), {\stackrel{\ \circ}{\cal A}}{}^{\nu}(
\theta)\Bigr)}{\partial {\stackrel{\ \circ}{\cal
A}}{}^{\mu}(x,\theta)\phantom{xxxxxxxx}} = \varepsilon_{\nu\mu}
{\stackrel{\ \circ}{\cal A}}{}^{\nu}(x,\theta),\
S_H(\Gamma^p(\theta)) = T({\cal A}^{\ast}_{\mu}(\theta)) + S_0({\cal
A}^{\mu}( \theta))\;, & {}\\
\setcounter{equation}{43}
\setcounter{lyter}{2}
{} & T({\cal A}^{\ast}_{\mu}(\theta)) = \displaystyle\int d^D x
\textstyle\frac{1}{2}
(\varepsilon^{-1})^{\mu\nu}{\cal A}^{\ast}_{\mu}(x,\theta){\cal
A}^{\ast}_{\nu}(x,\theta),\;(\varepsilon^{-1})^{\mu\nu}\varepsilon_{
\nu\rho} = \delta^{\mu}{}_{\rho}\;.& {}
\end{eqnarray}
Therefore the Hamiltonian formulation of the model satisfies to conditions
of Corollary 1.4 and appears by GThST belonging to the I class with GGTST
(9.41) and HCHF having the form
\renewcommand{\theequation}{\arabic{section}.\arabic{equation}}
\begin{eqnarray}
\Theta^H_{\mu}(x,\theta)\equiv
\Theta^H_{\mu}({\cal A}^{\nu}(x,\theta), \partial_{\rho}{\cal A}^{\nu}
(x,\theta)) = - \frac{\partial_l
S_0({\cal A}^{\nu}(\theta))}{\partial{\cal A}^{\mu}(x,\theta)\phantom{xx}}
= - (\Box \eta_{\mu\nu} -
\partial_{\mu}\partial_{\nu}){\cal A}^{\nu}(x,\theta) = 0. 
\end{eqnarray}
Corresponding GHS (3.5) has the form (6.28), (3.51b) in correspondence with
Corollary 1.4
\begin{eqnarray}
\frac{d_r {\cal A}^{\mu}(x,\theta)}{d\theta\phantom{xxxxxx}} =
\frac{\partial
S_L({\cal A}^{\nu}(\theta),{\cal A}^{\ast}_{\nu}(\theta))}{\partial{\cal A}^{
\ast}_{\mu}(x,\theta)\phantom{xxxxxxx}} =
(\varepsilon^{-1})^{\mu\nu}
{\cal A}^{\ast}_{\nu}(x,\theta),\
\frac{d_r {\cal A}^{\ast}_{\mu}(x,\theta)}{d\theta\phantom{xxxxxx}}=0,\
\Theta^H_{\mu}(x,\theta)=0\;.
\end{eqnarray}
GHS (9.45) is equivalent to the corresponding LS [1].
Given GHS not being solvable, although
the Hamiltonian subsystem in (9.45) appears by solvable itself, has
$S_H(\Gamma^p(\theta))$ as
one's integral resulting in fulfilment of Eq.(2.6) in this case
in terms of coordinates on $T^{\ast}_{odd}{\cal M}_{cl}$
\begin{eqnarray}
\int d^D x \left[
[(\Box \eta_{\mu\nu} -
\partial_{\mu}\partial_{\nu}){\cal A}^{\nu}(x,\theta)]
(\varepsilon^{-1})^{\mu\rho}{\cal A}^{\ast}_{\rho}(x,\theta)
\right]_{\mid \Theta^H_{\mu}(x,
\theta) = 0} = 0\;
 ,
\end{eqnarray}
by virtue of the following expression
$$
S_H(\Gamma^p(\theta)) = S_E\left({\cal A}^{\mu}(\theta),
{\stackrel{\ \circ}{\cal A}}{}^{\mu}(\theta)\right){\hspace{-0.7em}
\phantom{\Bigr)}}_{\mid {\stackrel{\
\circ}{\cal A}}{}^{\mu}(x,\theta) = {\stackrel{\ \circ}{\cal A}}{}^{\mu}(
{\cal A}^{\ast}(x,\theta))} = T({\cal A}^{\ast}_{\mu}(\theta)) +
S_0({\cal A}^{\mu}(\theta))\;.
$$
Defining SCLF and SCHF, being equivalent to each other, in the form
respectively
\begin{eqnarray}
{\stackrel{\ \circ}{\cal A}}{}^{\mu}(x,\theta)_{\mid \Theta_{\nu}(x,
\theta) = 0} = 0,\
{\cal A}^{\ast}_{\mu}(x,\theta)_{\mid \Theta^H_{\nu}(x,
\theta) = 0} = 0
\end{eqnarray}
obtain that on their solutions the   free massless vector
superfield model is described by only $S_0({\cal A}^{\mu}(\theta))\in
{\cal M}_{cl}$ and, therefore belongs to the II class theory.
Master equations of the form (3.36a,b), in fact written in (9.46), appear by
double zeros for the solutions for EGHS (9.45), (9.47) and corresponding ELS.
These systems therefore are solvable.

The antibracket $(\ ,\ )_{\theta}$ and operator $\Delta^{cl}(
\theta)$ have the form for any
${\cal F}(\Gamma(\theta)), {\cal J}(\Gamma(\theta)) \in C^{k\ast}$
\renewcommand{\theequation}{\arabic{section}.\arabic{equation}\alph{lyter}}
\begin{eqnarray}
\setcounter{lyter}{1}
{} & \bigl({\cal F}(\Gamma(\theta)), {\cal J}(\Gamma(\theta))\bigr)_{
\theta} = \displaystyle\int d^D x \Biggl(
\frac{\partial {\cal F}(\Gamma(\theta))}{\partial {\cal A}^{\mu}(x,
\theta)}
\frac{\partial {\cal J}(\Gamma(\theta))}{\partial {\cal A}^{\ast}_{\mu}(x,
\theta)} -
\frac{\partial_r {\cal F}(\Gamma(\theta))}{\partial {\cal A}^{\ast}_{
\mu}(x,\theta)}
\frac{\partial_l {\cal J}(\Gamma(\theta))}{\partial {\cal A}^{\mu}(x,
\theta)}\Biggr)\;, &{}
\\
\setcounter{equation}{48}
\setcounter{lyter}{2}
{} & \Delta^{cl}(\theta) = \displaystyle\int d^D x
\frac{\partial_l \phantom{xxxxx}}{\partial{\cal A}^{\mu}(x,\theta)}
\frac{\partial\phantom{xxxxx}}{\partial{\cal A}^{\ast}_{\mu}(x,\theta)}\;.
& {} 
\end{eqnarray}
Eqs.(5.22) and hence Eqs.(5.8) trivially hold. GHS of the type (3.51),
being differred from (9.45) in the 2nd subsystem, follows from
variational problem for the superfunctional
\begin{eqnarray}
\setcounter{lyter}{1}
{} & Z^{(1)}_H[{\cal A}^{\mu},{\cal A}^{\ast}_{\mu},D^{\mu}]
= \displaystyle\int d\theta\left(\int d^Dx
{\stackrel{\ \circ}{\cal A}}{}^{\mu}(x,\theta){\cal A}^{\ast}_{\mu}(x,\theta)
 - S_H^{(1)}(\Gamma^p(\theta), D^{\mu}(\theta))\right)\;, & {}
\\
\setcounter{equation}{49}
\setcounter{lyter}{2}
{} & S_H^{(1)}(\Gamma^p(\theta), D^{\mu}(\theta)) = S_H(\Gamma^p(\theta))
+ \displaystyle\int d^Dx D^{\mu}(x,\theta)
[(\Box \eta_{\mu\nu} -
\partial_{\mu}\partial_{\nu}){\cal A}^{\nu}(x,\theta)]
\;,  & {} \nonumber \\
{} & (\varepsilon_P, \varepsilon_{\Pi}, \varepsilon)D^{\mu}(x,\theta) =
(0,0,0)\;. & {}
\end{eqnarray}
The setting of Cauchy problem, by virtue of functional dependence of HCHF
(9.44), for GHS (9.45) and for corresponding LS [1] for $D=2k, k \in
\mbox{\boldmath$N$}$ is nontrivial, and for $D=2k + 1$ is  more complicated.
It requires an additional study of HCLF and LS on the whole. GHS (9.45)
for $D=2k$ itself and corresponding LS [1] (given with help of (9.44) and
$\varepsilon_{\mu\nu}{\stackrel{\,\circ\circ}{\cal A}}{}^{\nu}(x,\theta) = 0$
for $D \geq 2$) are the systems of differential superfield equations in
partial derivatives of the 1st (2nd) order on $\theta$ for GHS (LS) and of the
2nd order with respect to $x^{\mu}$ for GHS and  LS.

At last, $\theta$-superfield model of free massive real vector superfield
described in the Lagrangian formulation for GSTF in [1] can be obtained in the
Hamiltonian one under formal change of the corresponding superfunctions and
differential operators in the relationships (9.40a,c), (9.43a),
(9.44)--(9.46), (9.49) onto following ones
\begin{eqnarray}
\setcounter{lyter}{1}
S^{(2)}_L(\theta) & \mapsto & S^{(2)}_{L{}m}(\theta) =
T\Bigl({\stackrel{\ \circ}{\cal A}}{}^{\mu}(\theta)\Bigr) -
S_{0{}m}\bigl({\cal A}^{\mu}(\theta)\bigr)\;,
\nonumber \\
S_H(\Gamma^p(\theta)) & \mapsto & S_{H{}m}(\Gamma^p(\theta)) =
T\bigl({\cal A}^{\ast}_{\mu}(\theta)\bigr) +
S_{0{}m}\bigl({\cal A}^{\mu}(\theta)\bigr)\;,
\\ 
\setcounter{lyter}{2}
\setcounter{equation}{50}
S_{0{}m}\bigl({\cal A}^{\mu}(\theta)\bigr) & = &  \int d^4x
{\cal L}_0^{(2)}(x,\theta),\ {\cal L}_0^{(2)}(x,\theta) = \Bigl(
-\frac{1}{4}F_{\mu\nu}F^{\mu\nu} + \frac{m^2}{2}{\cal A}_{\mu}
{\cal A}^{\mu}\Bigr)(x,\theta)\;, \\
\setcounter{lyter}{3}
\setcounter{equation}{50}
\Theta^H_{\mu}(x,\theta) & \mapsto & \Theta^H_{m{}\mu}(x,\theta) = -
 \Bigl((\Box + m^2)\eta_{\mu\nu} -
\partial_{\mu}\partial_{\nu}\Bigr){\cal A}^{\nu}(x,\theta) = 0\;, \\
\setcounter{lyter}{4}
\setcounter{equation}{50}
{} & {} & -\displaystyle\int d^4x \Theta^H_{m{}\mu}(x,\theta)
(\varepsilon^{-1})^{\mu\rho}{\cal A}^{\ast}_{\rho}(x,\theta)
{\hspace{-0.5em}\phantom{\Bigr]}}_{\mid \Theta^H_{m{}\mu}(x,
\theta) = 0} = 0\;;
\end{eqnarray}
\vspace{-4ex}
\begin{eqnarray}
\setcounter{lyter}{1}
{} & Z^{(1)}_{H{}m}[{\cal A}^{\mu},{\cal A}^{\ast}_{\mu},D^{\mu}]
= \displaystyle\int d\theta\left(\int d^Dx
{\stackrel{\ \circ}{\cal A}}{}^{\mu}(x,\theta){\cal A}^{\ast}_{\mu}(x,\theta)
 - S_{H{}m}^{(1)}(\Gamma^p(\theta), D^{\mu}(\theta))\right), & {}
\\
\setcounter{lyter}{2}
\setcounter{equation}{51}
{} & S_{H{}m}^{(1)}(\Gamma^p(\theta), D^{\mu}(\theta)) = S_{H{}m}(
\Gamma^p(\theta)) - \displaystyle\int d^Dx D^{\mu}(x,\theta)
\Theta^H_{m{}\mu}(x,\theta)\;. & {}
\end{eqnarray}

As it had been noted in Ref.[1] there are not both differential identities
among $\Theta^H_{m{}\mu}(x,\theta)$ (9.50c) and therefore any gauge
transformations. This model belongs to the class of nondegenerate ThST, being
singular in correspondence with terminology from Ref.[8] in view of 2
second-class constraints presence.
\section{Conclusion}

The second composite part of GSQM for gauge theories in the Lagrangian
formalism in a natural way continuing the ideas and elaborations of the
Lagrangian formulation for GSTF have been proposed in the work. On
space $T^{\ast}_{odd}{\cal M}_{cl} \times \{\theta\}$ the superfunction
$S_H(\theta)$ is defined, equivalently just as $S_L(\theta)$ in Lagrangian
formulation, coding  all information about given GSTF model in Hamiltonian
formalism. More complicated scheme of corresponding dynamical
systems investigation on a basis of satisfaction of the solvability
conditions has led to supplement of the additional special constraints aside
from ones following from variational principle. This investigation trend has
led to definite inverse connection, namely, to modification of Lagrangian
formulation description for GSTF. In particular, it was expressed in
introduction of concepts for theories of the I and II classes. GThGT and (or)
GThST of the II class represent themselves the models of gauge theories with
master equations being valid, in contrast to the theories of the I class, in
the whole $T^{\ast}_{odd}{\cal M}_{cl}$, $T_{odd}{\cal M}_{cl}$ in
correspondence with the kind of applied formalism.

Note that GThGT of the II class with
$S_H(\Gamma(\theta)) \in C^k\bigl(T^{\ast}_{odd}{\cal M}_{cl}\bigr)$ has as
consequence of the master equation the property of mentioned
superfunction cancellation by means of the operator $\Delta^{cl}(\theta)$
action.
Various constructions of the translation operations with respect to $\theta$
along integral curves for HS, GHS, EGHS, EHS together with their properties
were examined.

Important problems of the study of the 1st and 2nd orders differential
operators superalgebras and the component formulation for GSTF have been
considered.
Statements and conclusions for general formalism of the work have been
demonstrated on the examples of  free and
self-interacting massive complex scalar superfield of
spin 0 models, free and self-interacting massive spinor superfield of spin
$\frac{1}{2}$ models in $D=4$
and  free massless and massive (Proca model) vector superfield ones in $D=2k,
k\in \mbox{\boldmath$N$}$.
These examples appear by the starting models for construction of interacting
superfield (on $\theta$) GThGT in Hamiltonian formulation  such as
$\theta$-superfield scalar or spinor or vector (for complex vector superfield
in $D=4$) electrodynamics on a basis of the gauge principle application [6]
to models above. Note that nonlinear dynamical equations in fact written in
HCHF for self-interacting models do not coincide with respect to form with
each other for $P_0(\theta)$ and $P_1(\theta)$ components of the corresponding
superfields in contrast to free models. For the latter examples the component
on $\theta$ linear equations of motion have the same form and formally
describe the same corresponding particles.

Natural algorithm to constructing from the models of usual relativistic field
theory their superfield generalization in the form of natural system,
described in the conclusion of Ref.[1] for Lagrangian formulation for GSTF,
can be literally applied to analogous construction of their superfield
generalization in Hamiltonian formalism for GSTF.

At last, let us continue the analogy [1] between quantities and relations
from classical mechanics in its usual Hamiltonian formulation with even
with respect to supergroup $P$ ($\varepsilon_P \equiv 0$) objects and ones
from Hamiltonian formulation for GSTF.
\begin{eqnarray*}
\begin{array}{|l|l|} \hline
\phantom{xxxxxxx}
{\bf  usual \phantom{x} classical \phantom{x} mechanics} & \phantom{xxxxxxx}
{\bf Hamiltonian \phantom{x} GSTF} \\ \hline
1.\;  p_a(t) - {\rm generalized \phantom{x} momenta } &
{\cal A}^{\ast}_{\imath}(\theta) - {\rm superantifields}  \\
\phantom{1.\;}
({\varepsilon}_{\bar{J}}, \varepsilon)p_a(t) = (\varepsilon_a, \varepsilon_a)
& {}\\ \hline
2.\;  T^{\ast}{\cal M} = \{(q^a, p_a)\} - {\rm phase \phantom{x} space} &
T^{\ast}_{odd}{\cal M}_{cl} = \{({\cal A}^{\imath}(\theta),
{\cal A}^{\ast}_{\imath}(\theta))\} - {\rm odd} \\
\phantom{2.\;} {} &   {\rm phase \phantom{x} space} \\
\hline
3.\; {p_{a}}(t) = \frac{\partial_l L(t)\phantom{x}}{\partial \dot{\;q^a}(t)},
\;H(p,q,t) = \dot{\;q^a}p_a - L(t) &
{\cal A}^{\ast}_{\imath}(\theta) = \frac{\partial_l S_L(\theta)}{\partial
{\stackrel{\ \circ}{\cal A}}{}^{\imath}(\theta)\phantom{x}},\;S_H\bigl({\cal
A}(\theta),{\cal A}^{\ast}(\theta),\theta) =  \\
\phantom{3.\;} - {\rm Hamiltonian} &
{\stackrel{\ \circ}{\cal A}}{}^{\imath}(\theta){\cal A}^{\ast}_{\imath}(
\theta) - S_L(\theta) - {\rm superfunction \phantom{x} of} \\
\phantom{3.\;}{} & {\rm classical \phantom{x} Hamiltonian \phantom{x}
action}\\ \hline
4.\;  \{f(q,p,t), g(q,p,t)\} - {\rm even \phantom{x}Poisson} & ({\cal
F}(\theta),{\cal J}(\theta))_{\theta} - {\rm antibracket, }\\
\phantom{4.\;}  {\rm bracket},\; f,g\in C^k(T^{\ast}{\cal M}
\times \{t\}) & {\cal F}(\theta), {\cal J}(\theta) \in
C^k(T^{\ast}_{odd}{\cal M}_{cl} \times \{\theta\})\\ \hline
5.\; \frac{d_r{\Gamma}^c(t)}{d t\phantom{xxxx}} = \{\Gamma^c(t), H(t)\} -
{\rm Hamilton} &
\frac{d_r{\Gamma}^p(\theta)}{d\theta\phantom{xxxx}} =
(\Gamma^p(\theta),S_H(\theta))_{\theta} - {\rm odd \phantom{x} Hamilton}\\
\phantom{5.\;} {\rm equations \phantom{x}of \phantom{x}motion} &
{\rm equations \phantom{x}of \phantom{x}motion}\\ \hline
6.\; S_H[\Gamma] = \int dt \bigl(\dot{\;q^a}p_a - H(q, p, t)\bigr)  &
Z_H[\Gamma]= \int d\theta \bigl(
{\stackrel{\ \circ}{\cal A}}{}^{\imath}(\theta){\cal A}^{\ast}_{\imath}(
\theta) - S_H(\Gamma(\theta),\theta)\bigr)\\
\phantom{6.\;} - {\rm Hamiltonian \phantom{x} action \phantom{x}functional} &
 - {\rm odd \phantom{x} Hamiltonian \phantom{x}superfunctional}\\
\hline
\end{array}
\end{eqnarray*}
{\bf Acknowledgments:}
I would like to thank Bogdan Mishchuk for discussions of some results of the
present paper.
\vspace{1ex}
\begin{center}
{\large{\bf References}}
\end{center}
\begin{enumerate}
\item A.A. Reshetnyak, General Superfield Quantization Method. I. General
Superfield Theory of Fields: Lagrangian Formalism, hep-th/0210207.
\item C. Becchi, A. Rouet and R. Stora, Phys. Lett. B 25 (1974) 344; Comm.
Math. Phys. 42 (1975) 127;\\
I.V. Tyutin, P.N. Lebedev Physical Inst. of the
RF Academy of Sciences, preprint, No.39 (1975).
\item I.A. Batalin and G.A. Vilkovisky,
Phys. Lett. B 102 (1981) 27; Phys.  Rev. D 28 (1983) 2567.
\item V.N. Shander, Funct. analysis and its applications (on russian) 14
(1980) No.2 91; 17 (1983) No.1 89.
\item ${\rm \ddot{O}}$.F. Dayi, Mod.
Phys. Lett. A 4 (1989) 361; A 8 (1993) 811; ibid 2087; Int. J. Mod. Phys. A
11 (1996) 1.
\item C.N. Yang and R. Mills, Phys. Rev. 96 (1954) 191.
\item E. Noether, Nachr. K${\rm \ddot{o}}$nig. Gesellsch. Wissen, G${\rm
\ddot{o}}$ttingen, Math-Phys. Klasse. (1918) 235, in English: Transport Theory
and Stat. Phys. 1 (1971), 186.
\item D.M. Gitman and I.V. Tyutin,
Quantization of Fields with Constraints (Springer-Verlag, Berlin and
Heidelberg, 1990).
\end{enumerate}
\end{document}